\documentclass[rmp,aps,amssymb,twocolumn,nofootinbib,floatfix]{revtex4}

\usepackage{amsmath,amssymb,bm}
\usepackage{accents}
\usepackage{graphics}
\usepackage{graphicx}
\usepackage[colorlinks,linkcolor=blue,citecolor=blue]{hyperref}
\usepackage{dcolumn}
\usepackage{rotating}
\usepackage{multirow}
\usepackage{epstopdf}
\usepackage{epsfig}
\usepackage{color}
\usepackage{float}
\usepackage{setspace}
\usepackage{natbib}
\usepackage{enumitem}
\usepackage{empheq}
\usepackage{color,soul}
\usepackage{mathrsfs}
\usepackage{hyperref}
\usepackage[caption=false]{subfig} 
\usepackage{tabularx}
\usepackage{comment}
\usepackage{booktabs}
\usepackage{textgreek}
\usepackage[utf8]{inputenc}

\bibpunct{[}{]}{,}{n}{}{;}

\begin{document} 

\bibliographystyle{plainnat}

\title{Modeling the spatial growth of cities}

\author{Ulysse Marquis}
\email{umarquis@fbk.eu}
\affiliation{Fondazione Bruno Kessler, Via Sommarive 18, 38123 Povo (TN), Italy}
\affiliation{Department of Mathematics, University of Trento, Via Sommarive 14, 38123 Povo (TN), Italy}

\author{Marc Barthelemy}
\email{marc.barthelemy@ipht.fr}
\affiliation{Universit\'e Paris-Saclay, CNRS, CEA, Institut de Physique Th\'eorique, 91191, Gif-sur-Yvette, France}
\affiliation{Centre d'Analyse et de Math\'ematique Sociales
(CAMS, UMR 8557 CNRS-EHESS)\\
Ecole des Hautes Etudes en Sciences
 Sociales, Paris, France}
\affiliation{Complexity Science Hub, Vienna, Austria}

\begin{abstract}  

  The growth of cities has traditionally been studied from a population perspective, while urban expansion—its spatial growth—has often been approached qualitatively. However, characterizing and modeling this spatial expansion is crucial, particularly given its parallels with surface growth extensively studied in physics. Despite these similarities, approaches to urban expansion modeling are fragmented and scattered across various disciplines and contexts. In this review, we provide a comprehensive overview of the mathematical modeling of this complex phenomenon. We discuss the key challenges hindering progress and examine models inspired by statistical physics, economics and geography, and theoretical ecology. Finally, we highlight critical directions for future research in this interdisciplinary field.

\end{abstract}


\maketitle
\tableofcontents

\section{Urban expansion}
\label{chap:1}

\subsection{Urban expansion and spatial growth of cities}

\begin{figure}
    \centering
    \includegraphics[width=1\linewidth]{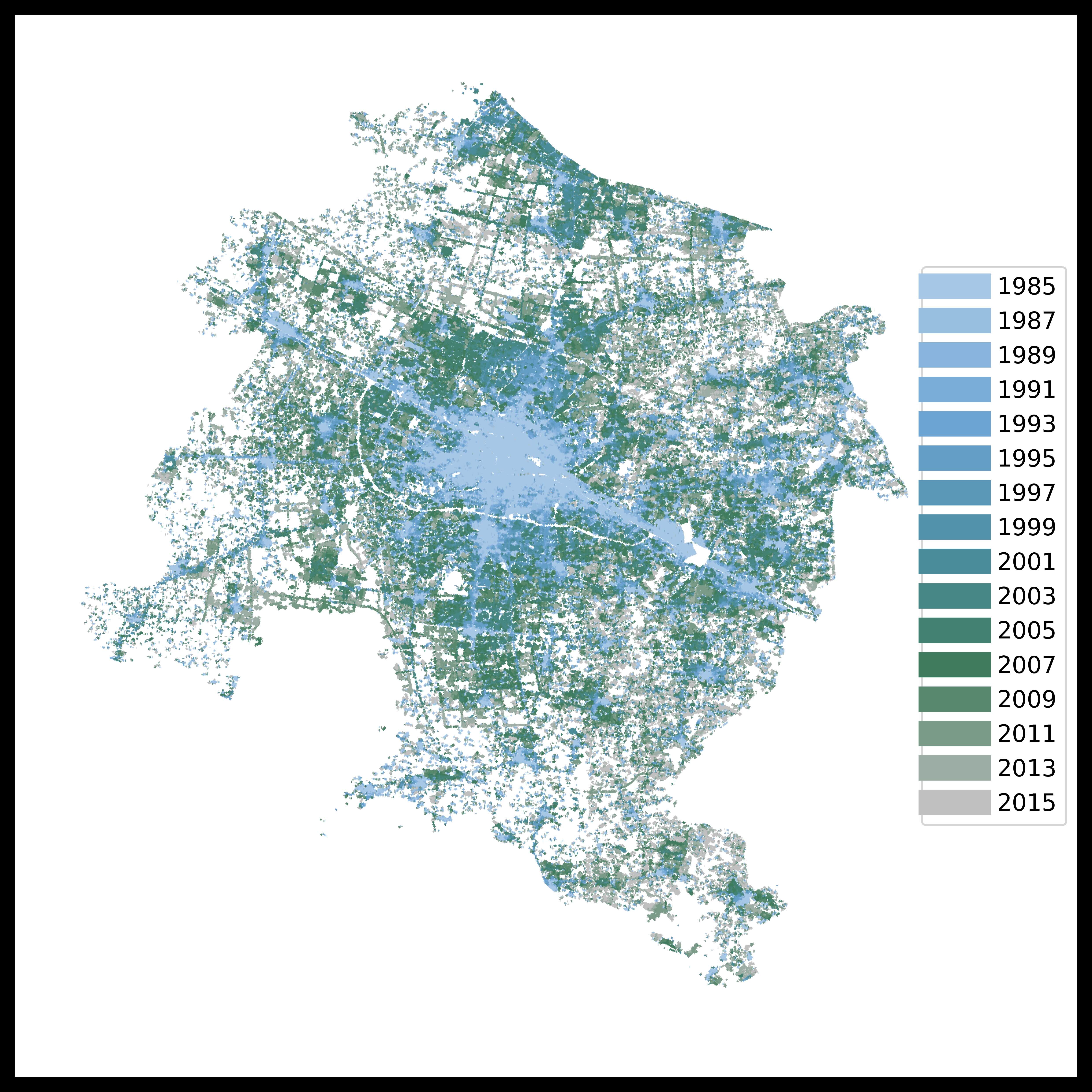}
    \caption{Example of urban expansion: growth of Changzhou (China) between 1985 and 2015 (data from \cite{wsfevo}). Several disconnected urban patches were already present before 1985 (light blue) and later became progressively connected through expansion. The main component exhibits a branch extending diagonally from northwest to west. Density decreases anisotropically with distance from the main urban component.}
    \label{fig:sprawlexample}
\end{figure}

Cities are dynamic systems whose spatial footprint evolves continuously over time. Urban expansion refers to the increase in the physical extent of built-up areas beyond their historical boundaries as population grows, economic activities develop, and 
transportation infrastructures expand. This process transforms previously rural or undeveloped land into urbanized landscapes that include residential neighborhoods, 
industrial zones, and commercial centers. An example of such spatial growth is shown in Fig.~\ref{fig:sprawlexample}, which illustrates the progressive expansion of the 
built-up area of Changzhou (China) over three decades.

\begin{figure*}
\centering
\includegraphics[width=0.90\textwidth]{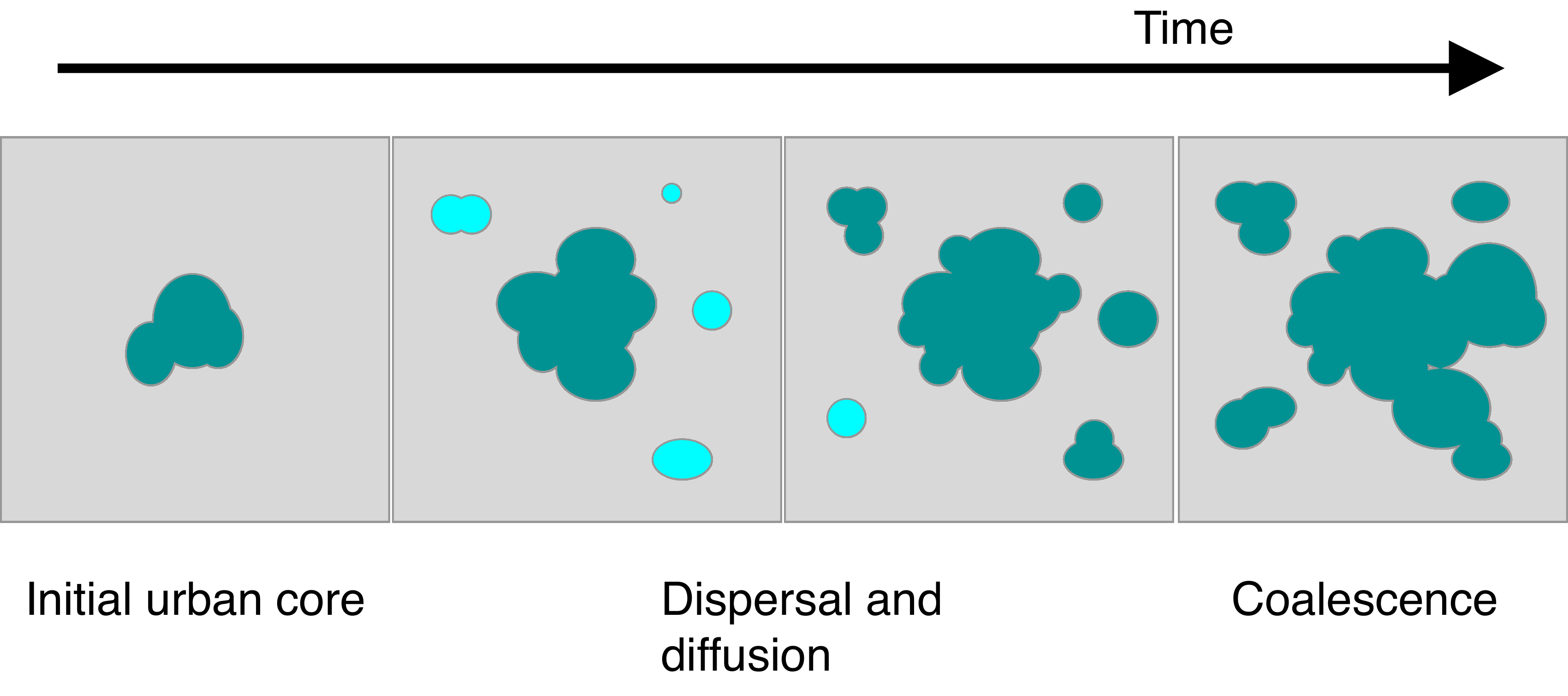}
\caption{Typical sequence of urban growth. An initial urban core expands and generates 
secondary clusters (light blue). Both the core and the peripheral clusters grow through 
a diffusion-like process and may eventually coalesce. Adapted from \cite{bhatta2010analysis}.}
\label{fig:mechanisms}
\end{figure*}

Urban expansion generally occurs through several spatial mechanisms (see Sec.~II.E). Existing urban areas may grow outward through \emph{edge expansion}, where development progressively spreads from the urban boundary into surrounding land. Growth may also occur through \emph{infilling}, whereby vacant or underutilized parcels within the 
urban fabric are developed, increasing density without significantly enlarging the urban footprint. In addition, cities may expand through the emergence of new, initially disconnected urban patches that later merge with the main urban area as development continues. This process of coalescence between multiple urban nuclei is commonly observed in rapidly growing metropolitan regions. A typical trajectory involves a compact urban core expanding radially while intermittently generating disconnected urban fragments. These fragments may subsequently grow and merge with the main core—a coalescence process \cite{marquis2025universalroughnessdynamicsurban,carra2017coalescing} that contributes to the emergence of polycentric urban forms (see Fig.~\ref{fig:mechanisms}).

Urban expansion is therefore a general and historically ubiquitous phenomenon associated with urbanization. However, the spatial form that this expansion takes can vary considerably depending on planning policies, transportation systems, land markets, and institutional contexts. In some cases, urban growth produces relatively compact and continuous urban fabrics. In other cases, it develops in a more dispersed and fragmented manner. One of the most widely discussed forms of such dispersed urban growth is known as \emph{urban sprawl}. Although the terms urban expansion and urban sprawl are sometimes used interchangeably in the literature, they refer to distinct concepts. Urban expansion denotes the increase in the spatial extent of built-up areas, whereas urban sprawl refers to a particular form of expansion characterized by low density, spatial fragmentation, and strong dependence on automobile transportation.

\subsection{Urban sprawl as a specific form of urban expansion}

Among the different forms that urban expansion can take, urban sprawl has received particular attention in the literature because of its prevalence in many metropolitan regions and the numerous environmental, economic, and social challenges it generates. 
For this reason, it is useful to examine this form of urban growth in more detail.

Urban sprawl is a multifaceted concept that refers to a particular pattern of urban expansion characterized by low-density development, spatial fragmentation, and strong dependence on automobile transportation. Rather than reinforcing the continuity of 
the urban fabric, sprawling development tends to produce scattered and discontinuous patterns of land use that spread across large areas of land. 

The European Environment Agency provides a widely cited definition of urban sprawl, describing it as a pattern of low-density, market-driven urban expansion that often occurs in the absence of strong or coordinated spatial planning mechanisms \cite{european2006urban}. According to this perspective, sprawl is typically facilitated by several interrelated socio-economic factors. These include widespread preferences for suburban lifestyles—often associated with larger housing, increased privacy, and perceived improvements in quality of life—as well as the availability of relatively inexpensive land at the urban periphery. 

Transportation infrastructure also plays a key role in enabling urban sprawl. The development of highways, road networks, and commuter rail systems increases the accessibility of peripheral areas, making them attractive locations for residential and commercial development. As a consequence, urban growth may extend far beyond the historical urban core, progressively transforming surrounding rural or agricultural landscapes.

Implicit in the concept of urban sprawl is a critique of the uncontrolled or poorly coordinated nature of this form of expansion. Sprawling urban development often produces inefficient spatial configurations that increase infrastructure costs, energy consumption, and dependence on private automobiles. Understanding the characteristics and drivers of urban sprawl is therefore essential for designing resilient urban development strategies.

\subsubsection{Main characteristics of urban sprawl}

Urban sprawl encompasses a variety of spatial, social, and economic processes that collectively shape the structure of expanding metropolitan areas. Several characteristic features are commonly identified in the literature.

\paragraph*{Low-density expansion:}
One of the defining traits of urban sprawl is its tendency to distribute population over large geographic areas with relatively few people per unit of land. This expansion frequently occurs at the expense of farmland, forests, and other natural or semi-natural areas, leading to the permanent conversion of open space into developed land. The dispersed nature of this growth pattern reduces the efficiency of public service delivery and increases per capita infrastructure costs \cite{habibi2011causes}.

\paragraph*{Decline of urban cores:}
As investment and population shift toward suburban and peri-urban areas, central parts of cities may experience economic stagnation, declining property values, and reductions in the local tax base. This phenomenon, often referred to as urban decline or inner-city shrinkage, has been widely studied in urban geography and planning. Batty highlighted the weakening of city centers in sprawling metropolitan regions \cite{batty2007cities}, while Beauregard \cite{beauregard2013voices} and Couch \& Karecha \cite{couch2006controlling} examined the socio-economic and policy dimensions of core decline in European and North American contexts. These studies emphasize how structural economic changes, suburbanization, and policy decisions can interact to undermine the vitality of urban cores.

\paragraph*{Leapfrog development:}
A particularly problematic aspect of sprawl is leapfrog development, where developers bypass parcels of undeveloped or underutilized land to build in more distant locations. This produces a fragmented urban landscape, with large gaps of vacant or partially developed land separating clusters of built-up areas. Such fragmentation complicates coherent urban planning, strains transportation networks, and often requires the extension of infrastructure and utilities across larger distances \cite{habibi2011causes,glockmann2022quantitative}.

\paragraph*{Automobile dependency:}
The spatial separation between residential areas, workplaces, shopping centers, and recreational facilities in sprawling urban forms makes the use of private automobiles almost indispensable. Limited public transportation options and the prevalence of single-use zoning further reinforce this dependency. As a consequence, sprawling cities typically experience higher levels of traffic congestion, air pollution, greenhouse gas emissions, and overall energy consumption, while also offering fewer opportunities for active modes of transportation such as walking or cycling \cite{ewing1997los}.

\paragraph*{Inadequate planning and governance:}
Urban sprawl is often associated with fragmented or weak planning frameworks in which multiple jurisdictions and private actors operate without a unified strategy for urban development. The absence of coordinated long-term planning can lead to inconsistent zoning regulations, redundant infrastructure investments, and segregated land-use patterns. These dynamics may exacerbate social and economic inequalities while contributing to inefficient land consumption \cite{downs1999realities}.

Urban sprawl thus represents a specific and widely discussed form of urban expansion characterized by dispersed development, automobile dependence, and fragmented land-use patterns. Understanding its mechanisms and consequences is crucial for developing urban policies that balance growth with environmental considerations, infrastructure efficiency, and social equity.

\subsubsection{Impacts of urban sprawl}
\label{sec1A}

Urban sprawl has significant environmental, economic, and social consequences. Because sprawling development is characterized by low density, spatial fragmentation, and strong dependence on automobile transportation, it generates a range of environmental pressures and socio-economic inefficiencies.

Environmentally, sprawl causes habitat loss, biodiversity decline, and farmland conversion. 
The reliance on private vehicles increases air pollution, greenhouse gas emissions, and  energy use \cite{brueckner2000urban}. Sprawl also exacerbates stormwater runoff, contributing 
to flooding and ecosystem fragmentation. Economically, the dispersed urban layout raises 
infrastructure costs and reduces service efficiency. Public transit, emergency services, 
and waste management become more expensive and less effective \cite{pendall1999land}. 
Socially, sprawl weakens community ties: segregated land uses and increased travel distances 
reduce social interactions and reinforce inequalities in access to services such as education 
and healthcare \cite{gordon2000critiquing}. Walkability and active transportation decline, 
commuting times increase, and overall urban livability may deteriorate \cite{galster2001wrestling}.

Johnson \cite{johnson2001environmental} outlined several key environmental consequences of urban sprawl. These include the loss of ecosystems and farmland, increased pollution and energy consumption, heightened flooding risks due to stormwater runoff, and greater fragmentation of ecosystems that ultimately reduce biodiversity. Importantly, these burdens often fall disproportionately on low-income and vulnerable communities. The paper \cite{johnson2001environmental} identifies priorities for future research:
\begin{itemize} 
    \item Collecting better data and developing region-specific studies.
    \item Enhancing economic models to capture externalities and risks.
    \item Integrating environmental justice into urban planning.
    \item Developing decision-support tools that incorporate visualization, monetization, and stakeholder negotiation.
\end{itemize}

Seto et al. \cite{seto2012} provided global projections of urban expansion and its direct impacts on biodiversity and carbon pools. Genovese \cite{Genovese_2023} emphasized the growing recognition of links between urban sprawl and public health concerns.

Effective responses to urban sprawl involve promoting compact urban forms, encouraging mixed-use development, and investing in robust public transit systems. Policy interventions may include disincentivizing private vehicle use through taxation, delineating urban growth boundaries to limit peripheral expansion, and requiring greater resident contributions to infrastructure costs. Revitalizing underutilized central areas can foster higher urban densities, while so-called “smart growth” strategies aim to integrate residential, commercial, and recreational functions within more cohesive and efficient urban layouts.

Johnson \cite{johnson2001environmental} identified several research priorities to address the complex challenges posed by sprawl. These include developing region-specific environmental assessments, improving economic models to better account for externalities and long-term risks, and incorporating environmental justice considerations into urban planning frameworks. Additionally, enhanced decision-support tools that integrate visualization, monetization, and stakeholder engagement are crucial to guide urban development strategies.

\subsection{Drivers of urban expansion and sprawl}

Urban expansion and urban sprawl emerge from the interaction of multiple socio-economic, 
political, and infrastructural forces. They are best understood not as the result of a 
single dominant cause, but as the collective response of the urban system to changes in 
its boundary conditions—such as population growth, rising income, declining transport 
costs, and regulatory environments. While some of these factors primarily drive the 
overall spatial expansion of cities, others influence the *form* that this expansion 
takes, determining whether urban growth remains relatively compact or develops into 
more dispersed patterns commonly described as urban sprawl.

Rosni et al.~\cite{rosni2016review} conducted a large-scale bibliometric and content 
analysis of more than 4,300 publications and categorized the contributing factors into 
six broad dimensions: socio-demographic, economic, political, physical, environmental, 
and transportation-related.

Without entering into too much detail, the main drivers of urban expansion and sprawl 
can be summarized as follows \cite{habibi2011causes}:

\begin{itemize}
    \item \textbf{Economic factors:} Rising incomes and economic growth increase demand 
    for larger housing units and more living space, which are often more easily available 
    in peripheral areas where land is abundant and cheaper.

    \item \textbf{Demographic dynamics:} Population growth, household fragmentation, 
    and heterogeneous housing preferences contribute to the outward expansion of cities 
    and to declining average densities.

    \item \textbf{Transportation infrastructure:} The development of road networks and 
    improvements in commuting infrastructure reduce the effective cost of distance, 
    making peripheral residential locations more accessible and enabling decentralized 
    settlement patterns.

    \item \textbf{Inner-city constraints:} High taxes, aging infrastructure, congestion, 
    and limited public amenities in central areas can push residents and firms toward 
    suburban and peri-urban zones.

    \item \textbf{Land markets:} Low land prices at the urban fringe—sometimes resulting 
    from market imperfections or weak land-use regulation—make peripheral development 
    financially attractive.
\end{itemize}

Despite broad agreement on these mechanisms, there is no consensus on their relative 
importance. Seto et al.~\cite{seto2011} conducted a meta-analysis showing that income 
growth tends to dominate in high-income countries, whereas population growth plays a larger role in developing regions. Habibi and Asadi~\cite{habibi2011causes} emphasized 
economic expansion, transportation infrastructure, and land price differentials as key structural drivers. 

In Switzerland, Weilenmann et al.~\cite{weilenmann2017} found that accessibility was a stronger determinant of sprawl than population growth, highlighting the role of infrastructure-induced changes in spatial accessibility. Travisi et al.~\cite{travisi2010} linked sprawl to the spatial decoupling of residential and employment locations, arguing that commuting patterns and employment geography play a central role in shaping urban form. Finally, Bertaud~\cite{bertaud_book} emphasized the importance of land market failures—such as the underpricing of peripheral land and weak coordination of land-use planning—as key contributors to excessive spatial expansion.

From a modeling perspective, this diversity of drivers suggests that urban sprawl should be viewed as an emergent property of a coupled system involving population dynamics, 
income levels, infrastructure development, land prices, and institutional settings. One of the central objectives of modeling is therefore to identify which mechanisms 
dominate under different urban regimes and to quantify their relative contributions to observed urban structures. This requires linking microscopic decisions—such as 
household location choices, firm location decisions, and commuting behavior—to macroscopic observables such as density gradients, spatial fragmentation, and the expansion of the urban perimeter.

\subsection{Key difficulties}

\subsubsection{Purpose and goal of modeling}

In many cases, modeling has been used to explain and reproduce specific phenomena. For example, numerous studies (discussed in the next chapters) have focused on a simple empirical stylized fact identified in the 1950s \cite{Clark:1951}, which observed that population density decreases exponentially from the city center. The objective was to develop a model and formulate an equation describing the evolution of population density that results in a solution consistent with this exponential decay.  However, we believe this approach is somewhat reversed. Instead, one should begin with fundamental `first' principles to describe the evolution of a city, ensuring that the resulting model is validated by multiple predictions that align with empirical observations. \\

Reaction-diffusion approaches, on the other hand, have primarily explored the evolution of urban structures, such as the spatial dynamics of different socioeconomic groups over time. More recent studies \cite{capel2024angiogenic} have shifted attention to the co-evolution of transportation networks and population density. \\

It is worth noting that many early models in regional science and spatial economics have relied heavily on equilibrium assumptions. These frameworks—ranging from monocentric city models (such as those developed by Alonso~\cite{alonso1964}) to more elaborate general equilibrium formulations (such as those by Fujita~\cite{Fujita} and Krugman~\cite{krugman1995self})—have provided valuable insights into the structure of cities at a given moment in time, but generally say little about how urban forms evolve over time. Moreover, these models rely on extremely strong assumptions--utility maximization, perfect market knowledge--that are weakly connected to empirical evidence, both in their parametrization and in the testing of their predictions, distancing themselves from the empirical foundations that underpin the physical sciences~\cite{Bouchaud_2008}. Dynamic processes such as infrastructure development, population migration, and socio-spatial segregation typically fall outside the scope of these static models. Nevertheless, given their historical importance and the theoretical insights they offer, these models will be briefly discussed in Chapter~\ref{chap:3}.

Moreover, the methodologies and tools employed differ significantly across disciplines. For economists, equilibrium-based analytical models or computable general equilibrium simulations are standard. For geographers, agent-based and rule-based models are more common. In contrast, physicists and applied mathematicians tend to seek a more formal, often minimal, mathematical description of the underlying mechanisms. In the best cases, this results in partial differential equations that govern the evolution of key quantities of interest, such as population density.

In this review, we will focus primarily on this latter class of models: those rooted in dynamic, often continuum-based approaches that describe urban evolution in terms of time-dependent equations. These models not only provide a natural framework for capturing spatial and temporal dynamics, but also lend themselves to theoretical analysis and empirical testing.

\subsubsection{Data sources for urban expansion}

Empirical analysis of urban expansion relies critically on the availability of spatially and temporally resolved data describing the expansion of built-up areas. In recent decades, the development of remote sensing technologies has revolutionized the study of urban growth. Satellite-derived datasets provide global, consistent, and relatively high-resolution observations of urban land cover, enabling systematic comparisons across regions and time periods.

One of the most widely used sources is the Landsat satellite program, which offers imagery at 30-meter resolution dating back to the 1970s. These data have been processed into various urban land cover products, such as the Global Human Settlement Layer (GHSL) developed by the European Commission \cite{pesaresi2013global,florczyk2019ghsl}, and the Global Urban Footprint (GUF) produced by the German Aerospace Center (DLR) \cite{esch2013urban}. These products enable the tracking of urban expansion at annual to decadal scales, providing a robust empirical foundation for analyzing spatial patterns of sprawl, including leapfrogging, edge expansion, and infilling. Another valuable source is the World Settlement Footprint Evolution dataset (WSF Evolution) \cite{wsfevo}, which provides annual global maps of built-up areas from 1985 to 2015. The dataset lists points corresponding to built-up infrastructure at a resolution of $30\,\text{m} \times 30\,\text{m}$. This fine-grained, time-resolved dataset allows detailed studies of urban growth dynamics over three decades.

\begin{figure}
\includegraphics[width=0.49\textwidth]{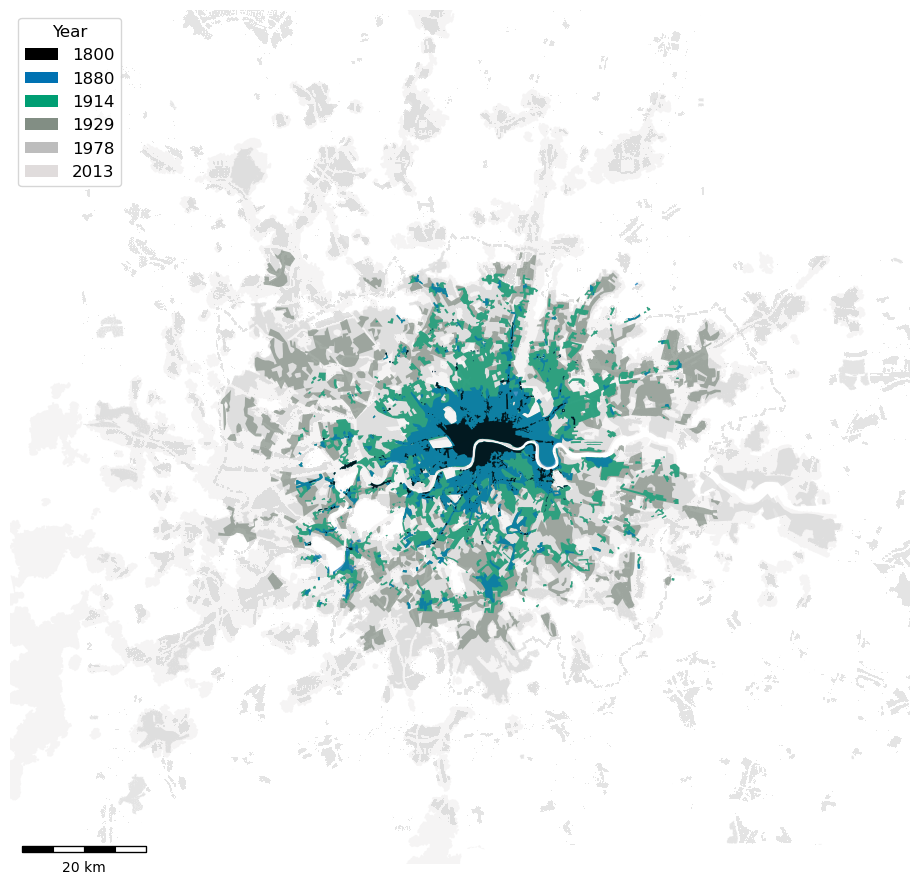}
\caption{Illustration of urban expansion in the city of London from 1800 to 2013. Data for the period 1800-1978 are from \cite{angel2012atlas} and for 2000 and 2013 from \cite{angel2016atlas} (see also \cite{AtlasUrbanExpansionHistorical} for a video documenting the historical evolution of London and many other cities worldwide).}
\label{fig:sa}
\end{figure}

In addition to remote sensing, historical data sources such as cadastral maps, municipal land-use plans, archival aerial photographs, and population censuses offer valuable insights into long-term urban dynamics (see for example \cite{geohistorical}). While these sources can be rich in detail, their coverage is often limited to specific cities or regions, and they typically require labor-intensive digitization and georeferencing efforts. To address the need for globally comparable historical data, the Atlas of Urban Expansion project, led by NYU, UN-Habitat, and the Lincoln Institute of Land Policy, has compiled a harmonized dataset for over 200 cities around the world \cite{angel2016atlas}. This atlas provides measures of built-up area, average density, and road infrastructure from as early as 1990, and in some cases includes data from the 1970s. It serves as a crucial empirical benchmark for evaluating urban models, especially those aiming to capture the mechanisms and dynamics underlying sprawl. In Fig.~\ref{fig:sa}, we show an illustrative example of historical urban growth of London from 1800 to 2013 using data from \cite{angel2012atlas,angel2016atlas}.

Overall, the increasing availability of spatial datasets—ranging from high-resolution satellite imagery to harmonized global databases—offers unprecedented opportunities for the quantitative analysis of urban form and its evolution. These empirical inputs are essential not only for describing and classifying urban sprawl, but also for testing theoretical models and identifying the mechanisms driving the spatial organization of cities.

\subsubsection{Choice of variables}


An important key issue is the choice of an appropriate time parameter. Population $P$ naturally serves as a clock for tracking urban evolution \cite{bhatta2010analysis,batty2008}. Unlike chronological time, it helps mitigate the effects of external disruptions—such as wars, epidemics, or short-term economic fluctuations—on growth dynamics. A quantitative analysis of urban expansion should therefore examine the evolution of spatial variables (e.g., built-up area, average radius) as functions of $P$.

Equally important is the selection of spatial variables used to characterize urban form. A common approach is to represent the city as a continuous population density field $\rho(\vec{r})$, providing detailed spatial information about the distribution of inhabitants or structures. For monocentric or quasi-radial cities, polar coordinates $(r,\theta)$ centered on the urban core offer a natural framework for analysis. This allows the definition of radial density profiles $\rho(r)$—which often display exponential or piecewise decaying behavior \cite{Clark:1951}—and angular heterogeneity $\rho(r,\theta)$, which reveals sectorial anisotropies.

Alternatively, several scalar descriptors are widely used to summarize urban structure. These include the total built-up area $A(P)$, the average population density $\bar{\rho}(P) = P/A(P)$, and a characteristic radius $\overline{r}(P)$, which may be computed as the root mean square distance from the city center:
\begin{equation}
\overline{r}(P) = \left( \frac{1}{P} \int r^2 \rho(\vec{r})\, d^2r \right)^{1/2}.
\end{equation}
Such quantities are particularly useful for comparing cities across different regions or timescales. In the case of an isotropic, exponentially decreasing population density of the form $\rho(r)\propto \mathrm{exp}(-r/r_0)$, we obtain $\overline{r}\propto r_0$.

\begin{figure}
\includegraphics[width=0.4\textwidth]{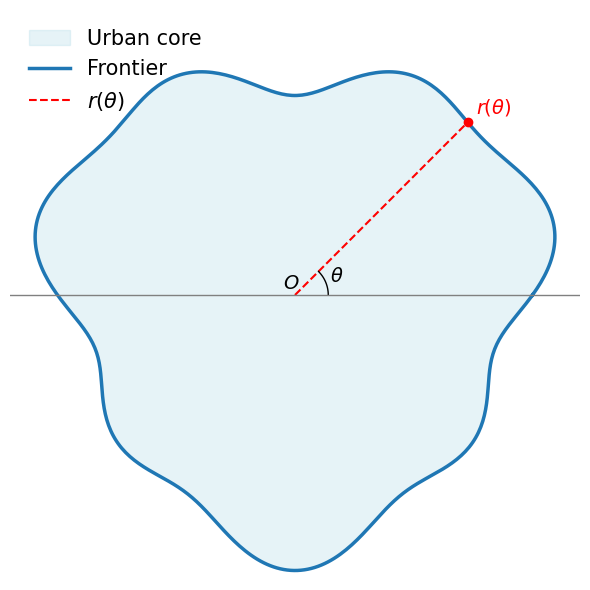}
\caption{Illustration of the quantity $r(\theta)$ for the frontier of the giant component of a city \cite{lemoy2021,marquis2025universalroughnessdynamicsurban}.}
\label{fig:polar}
\end{figure}

As we will see below, another possibility is to focus on the largest connected component as a robust and well-defined spatial entity for studying urban growth \cite{rozenfeld2008laws,lemoy2021,marquis2025universalroughnessdynamicsurban}. In this case, the radius $r(\theta)$ of this component—measured as a function of angle $\theta$ in polar coordinates—is well-defined (see Fig.~\ref{fig:polar} for an illustration), and the key question becomes how this quantity evolves over time or with population size.

\subsubsection{Components of cities} \label{subsec:component}

\begin{figure}[h!]
\centering
\includegraphics[width=0.5\textwidth]{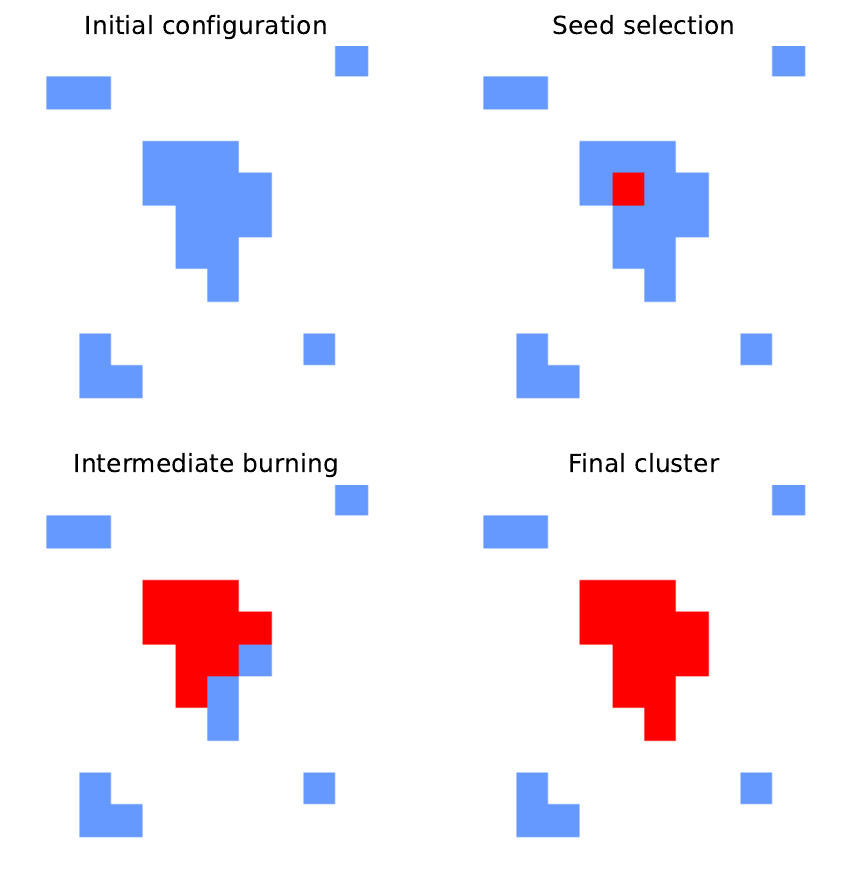}
\caption{\textbf{Illustration of the CCA algorithm.} To identify urban clusters, the City Clustering Algorithm (CCA) considers as connected all adjacent grid cells with nonzero population (in blue). The process begins by selecting an arbitrary populated cell, in the red seed in the top-right corner figure. (the final outcome is independent of this initial choice). A cluster is then grown iteratively by including all nearest neighbors of the current boundary that also have strictly positive population. This continues until no further neighboring populated cells remain. The procedure is repeated for each remaining unvisited populated cell until all such cells are assigned to a cluster. Adapted from \cite{rozenfeld2008laws}.}
\label{fig:cca}
\end{figure}

A persistent challenge in urban studies is the definition of what constitutes a `city' or `urban agglomeration.' Traditional approaches such as the Metropolitan Statistical Areas (MSAs) rely on administrative and functional criteria that vary between countries and are not always consistent. To overcome these limitations, Rozenfeld \textit{et al.}~\cite{rozenfeld2008laws} introduced the \emph{City Clustering Algorithm} (CCA), a morphological and data-driven method to identify urban clusters from high-resolution population data (see Fig.~\ref{fig:cca}).

The CCA operates on gridded population datasets, where each cell $i$ has an associated population $n_i(t)$ at time $t$. It can also be applied to point data and is conceptually related to the Hoshen--Kopelman cluster-labeling algorithm~\cite{hoshen1976percolation}. The method defines cities as spatially connected components of populated cells. The algorithm proceeds as follows:
\begin{enumerate}
  \item Overlay a regular grid on the geographic area of interest. Typical cell sizes used are 200~m for Great Britain, 2~km for the United States, and 8~km for Africa.
  \item Identify all populated cells (cells where $n_i(t) > 0$).
  \item Select a populated seed cell and assign it to a new cluster.
  \item Recursively add all neighboring cells with population $> 0$ to the cluster using a `burning' algorithm.
  \item Repeat the process with the next unvisited populated cell until all such cells are assigned to clusters.
\end{enumerate}

The population $S_i(t)$ of a cluster $i$ at time $t$ is then computed as
\begin{align}
S_i(t) = \sum_{j=1}^{N_i} n^{(i)}_j(t),
\end{align}
where $n^{(i)}_j(t)$ is the population of the $j$-th cell in cluster $i$, and $N_i$ is the number of cells in that cluster.

This algorithm is conceptually similar to percolation models and forest-fire dynamics~\cite{Stauffer:1992}, and it provides a scale-independent way of identifying cities as connected components in space, unconstrained by political boundaries.

The CCA provides a reproducible, scalable, and cross-country comparable method for defining urban clusters. When applied to a given urban area, the CCA typically reveals that the built environment consists of multiple disconnected components. Among these, there is generally a \emph{giant component}—the largest cluster—which corresponds to the urban core, surrounded by many smaller clusters of various sizes.

\begin{figure}
    \centering
    \includegraphics[width=1\linewidth]{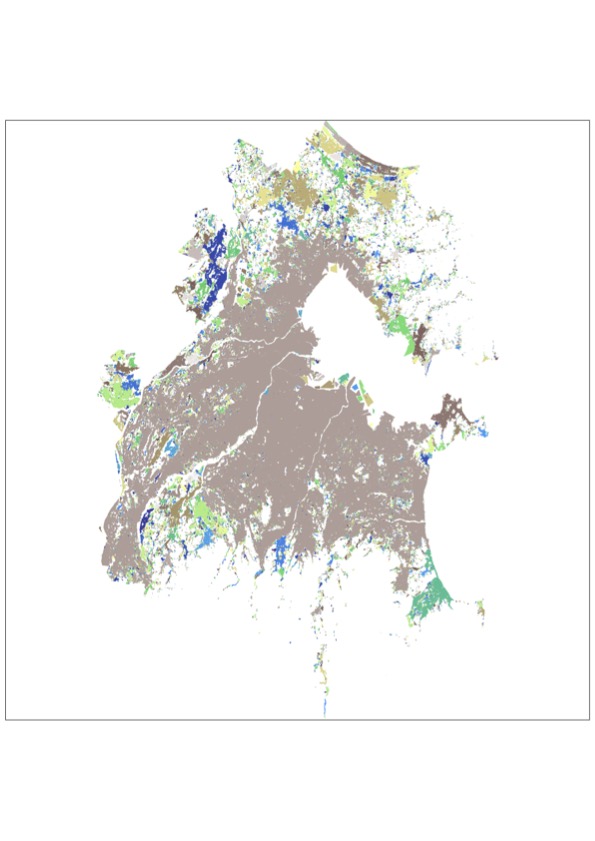}
    \caption{\textbf{Connected components (CC) of built-up area in the Tokyo urban area, 1985.} Connected components of multiple scales are visible : the giant cluster in light brown is several orders of magnitude larger than the any other clusters of macroscopic sizes, which are themselves much larger than the multitude of microscopic clusters. Each CC is assigned a random color to facilitate their visualization. Data from \cite{wsfevo}.}
    \label{fig:tokyo_mapclusters}
\end{figure}

Figure~\ref{fig:tokyo_mapclusters} shows the spatial distribution of these connected components in the Tokyo metropolitan area in 1985. Each color represents a distinct cluster identified by the CCA.

\begin{figure}[htp]
    \centering
    \includegraphics[width=1\linewidth]{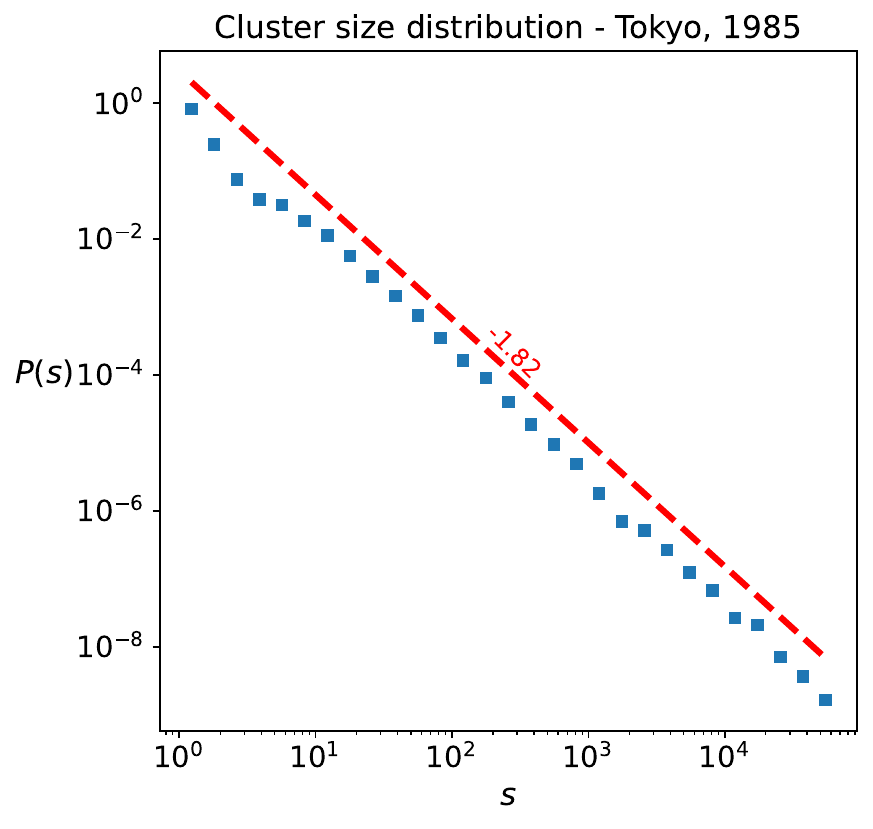}
\caption{\textbf{Distribution of cluster sizes in Tokyo (1985).}
The largest component is excluded. The cluster-size distribution shown here is consistent with a power-law with exponent $\tau = 1.82$. Data from~\cite{wsfevo}.}
    \label{fig:tokyo-clustersizedistrib}
\end{figure}

The clusters surrounding the giant component exhibit a broad size distribution, which is often well-described by a power law of the form
\begin{align}
    P(s)\sim s^{-\tau}
\end{align}
with an exponent typically close to $\tau\approx 2$~\cite{rozenfeld2016areapopulationcitiesnew} though this exponent can vary over time~\cite{Makse2}. 
In Figure~\ref{fig:tokyo-clustersizedistrib}, we present the distribution of cluster sizes for Tokyo in 1985. The data exhibit a power-law decay with exponent $\tau\approx 1.82$.

As urban areas grow, new clusters emerge, increase in size, and may eventually merge with neighboring clusters. These dynamical processes alter the structure of the urban fabric and can lead to temporal variations in the exponent, or even to deviations from power-law behavior. Only a few models (such as \cite{schweitzer1998estimation,carra2017coalescing}) discuss the formation of these clusters and the resulting size distribution (or equivalently the rank--size plot), and further studies are clearly needed to better understand the origin and value of this exponent $\tau$.

\subsection{Detailed outline}

The detailed outline of the review is the following. The first section introduces the phenomenon of urban expansion, defining it and discussing its multifaceted impacts, underlying causes, and methodological challenges. It also outlines the selection of relevant variables and city components necessary for quantitative modeling. The section~\ref{chap:2} focuses on empirical regularities and stylized facts, such as the spatial decay of density, surface roughness, and the spread of road networks, with both national and global perspectives. The section~\ref{chap:3} bridges geography, economics, and spatial modeling, detailing cellular automata, microeconomic frameworks, and urban economics models such as Alonso-Mills-Muth. Section~\ref{chap:statphys} delves into statistical physics approaches, including fractals, percolation, and growth models such as Eden and DLA. Section~\ref{chap:5} explores dispersal mechanisms through ecological analogies, while Section~\ref{chap:6} introduces diffusion-based frameworks that consider isotropy, migration, services, and the coevolution of networks and population. Section~\ref{chap:7} examines reaction-diffusion models that involve two species. In particular, we discuss the Gray-Scott model and its application to urban growth. The final section identifies open problems and avenues for further research, emphasizing the interdisciplinary nature of urban modeling. Together, these sections aim to provide a coherent and integrative understanding of the quantitative description of urban expansion.

\section{Empirical results and stylized facts}
\label{chap:2}


This section reviews key empirical findings and stylized facts that characterize urban systems across different spatial and temporal scales. We begin by examining the classic observation of density decline with distance from city centers and explore how cities exhibit homothetic scaling and diverse morphological shapes. We then quantify urban area expansion through typologies of growth and assess inequalities in built-up volumes. A detailed surface analysis follows, investigating area growth rates, anisotropy, underlying growth mechanisms, and roughness exponents that capture the complexity of urban interfaces. Next, we turn to street-network sprawl (here meant as `expansion'), tracing its evolution over more than a century in the United States, its manifestations worldwide, and constraints imposed by planning interventions such as green belts. Finally, we discuss how transportation infrastructures shape and modulate these urban growth patterns. Together, these empirical insights provide a robust foundation for understanding and modeling the dynamics of urbanization. Note that many studies have focused on econometric approaches, particularly on estimating various elasticities \cite{weilenmann2017, oueslati2015}. We refer the interested reader to this literature for further details.

\subsection{Density gradient}
\label{subsec:clark}

The earliest models \cite{ishikawa1980new,bracken1992simple} were primarily developed to explain and replicate one of the most significant empirical observations of the time: the decline in population density from the city center, typically described by an exponential function $\exp(-br)$ \cite{Clark:1951}.

Clark analyzed 20 cities worldwide, spanning Australia, the British Isles, continental Europe, and the United States, over different historical periods, primarily from 1800 to 1940. He assumed that cities were monocentric and isotropic\footnote{Something that is isotropic has the same size or physical properties when it is measured in different directions. ("Isotropic." Cambridge Dictionary, Cambridge University Press,~\url{https://dictionary.cambridge.org/dictionary/english/isotropic}.)}--assumptions that, in light of contemporary urban studies, no longer accurately represent most urban structures. 

Specifically, Clark's exponential model \cite{Clark:1951} describes the local population density $\rho(r)$ at a distance $r$ from the city center as
\begin{equation}
\rho(r) = \rho_0 \mathrm{e}^{-r/r_0},
\end{equation}
where $\rho_0$ is the central population density, and $r_0$ is the characteristic decay length, often referred to as the density gradient. For the cities studied by Clark, $r_0$ ranged from 0.7 km to 8 km, with an average of 3.2 km, while $\rho_0$ varied between 7,700 and 300,000 inhabitants per km$^2$, with an average of 70,000 inhabitants per km$^2$. It is worth noting that the exceptionally high maximum value likely results from an erroneous estimate for London in 1841.

\begin{figure}
    \centering
    \includegraphics[width=0.95\linewidth]{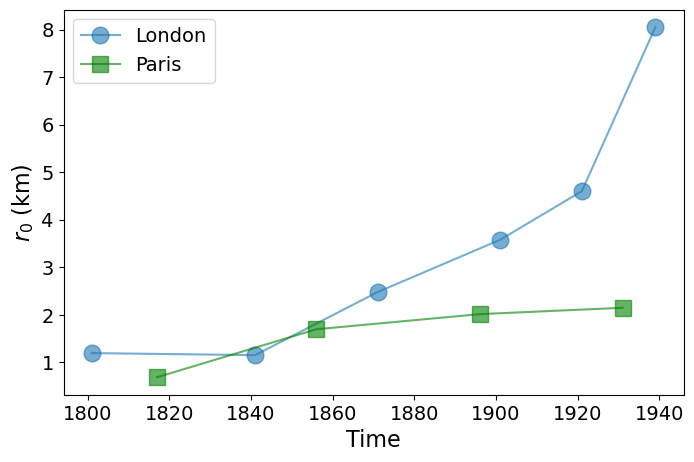}
    \caption{Time evolution of $r_0$ for London and Paris. Data from \cite{Clark:1951}.}
    \label{fig:clark2}
\end{figure}

Figure~\ref{fig:clark2} presents the temporal evolution of $r_0$ for the only two cities in Clark’s dataset with more than two recorded time points: Paris and London. The results indicate a clear trend in the variation of the characteristic decay length over time, reflecting the spatial expansion of these cities.
Notably, the data supports the well-documented demographic trajectory in which London overtook Paris in population size around 1850. This observation aligns with historical urbanization patterns, as London experienced rapid growth during the Industrial Revolution, driven by economic expansion, infrastructural development, and large-scale migration \cite{schwarz1992london}. Meanwhile, Paris followed a different trajectory, with significant urban restructuring occurring later in the 19th century, particularly during Haussmann's transformations (see for example \cite{barthelemy2013self} and references therein). The increasing values of $r_0$ over time for both cities suggest a progressive decentralization, with population density gradients becoming less steep—a characteristic feature of urban expansion observed in many growing metropolitan areas.

While small cities typically follow well the exponential decay form, larger metropolitan areas tend to have a peak in population density slightly detached from the city center, where instead space can be used for office space, administration buildings and such. This dip on the density-distance curve is called the `density crater'. Newling \cite{newling1969spatial} extended the exponential decay form to account for this with a quadratic exponential
\begin{equation}
\rho(x) = \rho_0 \mathrm{e}^{b x - c x^2},
\end{equation}
introducing a crest at $x = b/2c$. However, its predictive accuracy remains inconsistent. The total population in the city described by Clark's law is then given by (in two dimensions)
\begin{align}
\nonumber
    P&=\int \rho_0\mathrm{e}^{-br}d\theta rdr\\
    &=2\pi \rho_0/b^2
\end{align}

In one of the first mathematical approach to this problem (see Chapter~\ref{subsec:ishikawa} and \cite{ishikawa1980new}), the author discusses the problems of the simple Clark's approach. In particular:
\begin{enumerate}[noitemsep]
    \item The larger a city population becomes the smaller the density gradient is (and the larger $r_0$).
    \item The density gradient ($1/r_0$) is a decreasing function of time.
    \item In a large metropolitan area, density gradient for daytime population is larger than that for resident population
    \item A more essential problem is that Clark's description can never reproduce the `density crater' for resident population density in a large metropolitan area. The density crater means that maximum density of population occurs in a ring surrounding the city center (which can appear in Newling's description). 
\end{enumerate}

We note that Thrall \cite{thrall1988} provides an in-depth examination of the statistical and theoretical foundations underpinning the urban population density function, with particular emphasis on Clark’s seminal result, which empirical studies across many cities have confirmed as offering a good exponential fit. Thrall highlights that despite its widespread acceptance, this model is prone to significant biases. Structural specification errors arise when the monocentric assumption fails to capture real urban landscapes that are often polycentric, or when key externalities and environmental variables are omitted. Estimation biases also occur when ordinary least squares (OLS) is naively applied without accounting for potential discontinuities or `kinks' in the density gradient, which are better modeled using spline or switching regressions. While Clark’s exponential law has long stood as a central empirical regularity, Thrall stresses that rigorous model specification—including allowances for multiple centers, externalities, and appropriate estimation techniques—is essential to avoid overestimating the effect of distance and underestimating the roles of other urban factors in shaping population density.

\subsection{Homothetic scaling} \label{subsec:homo}

Lemoy and Caruso \cite{lemoy2021,lemoy2020} present an analysis of the radial structure and scaling laws of artificial land use and population density across 300 European functional urban areas, using high-resolution land use data (Urban Atlas 2006/2012) combined with Geostat population data. Assuming a monocentric, isotropic structure, they show that the share of artificial land $\eta(r)$ decreases approximately exponentially with distance $r$ from the city center, following 
\begin{align}
    \eta(r,P) \approx a(P) \mathrm{e}^{-r/\ell(P)},
\end{align}
where $a(P)$ denotes the central share and $\ell(P)$ the characteristic decay distance. A key finding is that $\ell(P)$ scales with total population $P$ as 
\begin{align}
    \ell(P) \sim \sqrt{P},
\end{align}
which implies that radial land use profiles collapse under the transformation 
\begin{align}
    r \to \frac{r}{\sqrt{P}} \,.
\end{align}
Such similarity ipon rescaling are called homotheties\footnote{In ancient Greek, `homo' means similar; `thesis' is position -- it translates similarity of shapes upon rescaling and re-orientation.}: larger cities appear as proportionally scaled versions of smaller ones (at least in terms of built-up areas). Consequently, the total artificial area satisfies 
\begin{align}
    A(P) = 2\pi a(P) \ell(P)^2 \propto P,
\end{align}
indicating that artificial area per capita remains roughly constant across city sizes. This further suggests that the average density (population per unit area) remains approximately constant during urban growth (see also section~\ref{subsec:surfacegrowth}). The authors show that non-linear fitting of the exponential model, minimizing absolute errors, outperforms traditional log-linear regressions by better capturing the dense urban core. They propose a parsimonious one-parameter formulation fixing $a(P)=1$, i.e.,
\begin{align}
    \eta(r,P) = \mathrm{e}^{-r/\ell(P)},
\end{align}
which successfully fits most cities and preserves the scaling $\ell(P)\approx \ell_1\, P^{1/2}$ with the length scale of an unit ($P=1$) city $\ell_1 \approx 7 \,$m, providing evidence of simple scale-invariant geometry in urban land use.

In \cite{lemoy2020}, the authors further analyze the scaling of population density, finding that the radial density profile satisfies
\begin{align}
    \rho(r) = P^{1/3} g\!\left(\frac{r}{P^{1/3}}\right),
\end{align}
where $\rho(r)$ scales with the cube root of population, indicating that larger cities are not only wider but also effectively `taller' in population concentration, reflecting a volumetric homothetic scaling. They also study fluctuations around these mean profiles and show that while radial fluctuations are relatively homogeneous with distance, angular fluctuations increase faster than expected from simple spatial averaging. Specifically, instead of scaling like $\sigma_n(r) \sim \sqrt{n}$ for wedges of aperture $2\pi/n$, they find $\sigma_n(r) \sim n^c$ with $c\approx 0.7$ for London and $c\approx 0.9$ for Paris. Together, these results provide an interesting quantitative baseline for interpreting urban form, emphasizing that European cities display remarkably consistent internal structure across sizes, challenging views that larger cities necessarily achieve greater land-use efficiency. This perspective is challenged by the fact that it assumes cities are monocentric and isotropic—a reasonable approximation for small to medium-sized cities developing in relatively unconstrained geographic settings. However, this assumption breaks down for larger urban systems, which often exhibit strong spatial anisotropy and a polycentric structure. 

\subsection{Fractal dimension and multifractality}

\subsubsection{Fractal cities} \label{subsec:fractal}

Numerous empirical studies have shown that urban spatial structures often exhibit fractal-like properties over a range of scales. Measurements based on built-up areas, using techniques such as box-counting or perimeter–area scaling, typically yield fractal dimensions in the range $D \approx 1.6$ to $1.8$, indicating that cities occupy space in a clustered manner that lies between a line ($D=1$) and a fully filled plane ($D=2$) \cite{BattyLongley1994, Frankhauser1998, Thomas2008}. The boundaries of urbanized areas generally display lower fractal dimensions, around $D \approx 1.2$ to $1.4$, reflecting irregular but less space-filling contours.

These observations have motivated analogies with aggregation phenomena studied in statistical physics, such as diffusion-limited aggregation or Eden growth models, which also generate clusters with non-trivial fractal dimensions (see Chapter~\ref{chap:statphys}). At the same time, the fractal perspective has proven useful in urban studies by providing a compact way to summarize the hierarchical, irregular, and scale-dependent structure of cities \cite{batty2007cities}. For instance, differences in fractal dimension can be related to variations in density gradients, the degree of expansion, or the balance between compact core development and more dispersed suburban growth \cite{frankhauser1994fractalite, white1993}.

It is important to emphasize, however, that cities are not strict mathematical fractals. The estimated fractal dimensions are approximate descriptors that depend on scale, resolution, and measurement method, and often apply only across limited spatial ranges \cite{Frankhauser1998, Salingaros1998}. Moreover, urban growth processes involve socio-economic, political, and infrastructural drivers absent from purely physical aggregation models. Thus, fractal measures should be viewed less as exact universal laws than as empirical signatures capturing some aspects of spatial complexity in urban form. In this sense, they are valuable for comparing different cities, historical periods, or planning strategies, and for linking observed morphologies to generative mechanisms of growth.

\subsubsection{Multifractality of London's street networks}

In many natural and social systems, spatial patterns are highly irregular and non-uniform. While a single fractal dimension can capture the overall scaling behavior of a system, it often fails to describe local variations in density. This is especially true for cities, where built-up areas, road networks, and population densities exhibit strong spatial heterogeneity. We emphasize that the analysis performed in that work corresponds to a classical multifractal analysis of spatial patterns, rather than to a fractal analysis of complex networks (see \cite{wang2012multifractal} and references therein).

A classic fractal set (subset of $\mathbb{R}^n$) is characterized by a scaling law for the number of boxes 
$N(\varepsilon)$ of side length $\varepsilon$ needed to cover it
\begin{equation}
    N(\varepsilon) \sim \varepsilon^{-D},
\end{equation}
where $D$ is the (box-counting) fractal dimension (a fit of this relation for small $\varepsilon$ allows then to determine empirically the value of $D$). This assumes that the points are distributed uniformly across the set. In reality, the local concentration of points may vary strongly. To capture this, one can introduce a measure $\rho(x)$, which represents some quantity distributed over space (such as built-up area or intersection density). For a ball ($B(x,r)$) of radius $r$ centered at $x$, we define the local mass
\begin{equation}
    \rho_r(x) = \int_{B(x,r)} \rho(z)\, dz.
\end{equation}
If, as $r \to 0$, the mass scales as
\begin{equation}
    \rho_r(x) \sim r^{\alpha_x},
\end{equation}
then the exponent $\alpha_x$ describes the local scaling behavior at point $x$. Different regions may have different values of $\alpha_x$, reflecting variations in density.

We can group together all the points that share the same exponent $\alpha$, and denote their set by
\begin{equation}
    E_\alpha = \{ x : \alpha_x = \alpha \}.
\end{equation}
Each of these sets may have its own fractal dimension 
$f(\alpha)$. The function
\begin{equation}
    \alpha \mapsto f(\alpha)
\end{equation}
is called the multifractal spectrum, and it tells us how common each local scaling behavior is in the system. When the spectrum reduces to a single value, the system is monofractal.

To study this more quantitatively, one considers the $q$-th moments of the local masses over all regions
\begin{equation}
    Z_r(q) = \sum_i \rho_r(x_i)^q,
\end{equation}
where the sum runs over small boxes of size $r$. When $r \to 0$, these moments typically scale as
\begin{equation}
    Z_r(q) \sim r^{-\tau(q)}.
\end{equation}
The function $\tau(q)$ is called the mass exponent and encodes the scaling of dense versus sparse regions. It is related to the multifractal spectrum via
\begin{equation}
    \tau(q) = q\,\alpha(q) - f(\alpha(q)).
\end{equation}

This relation is analogous to thermodynamic Legendre transforms, and in fact \( \tau(q) \) plays a role similar to a free energy~\cite{bunde_havlin1991}
\begin{equation}
    \tau(q) = - \frac{\log Z_r(q)}{\log r}.
\end{equation}
From this, one can define the generalized dimensions $D_q$
\begin{equation}
    D_q = \frac{1}{q-1} \lim_{r\to0} \frac{\log Z_r(q)}{\log r}.
\end{equation}
These provide a spectrum of exponents: $D_0$ is the box-counting dimension, $D_1$ the information dimension, and $D_2$ the correlation dimension. Measurements techniques are discussed extensively in~\cite{salat2017}. 
The multifractal spectrum gives a compact way to describe this complexity. It reveals, for instance, whether urban growth favors the emergence of large homogeneous areas, or whether it leads to a patchy, hierarchical structure. In this sense, multifractals provide a bridge between geometry, scaling behavior, and the social and infrastructural processes shaping cities.

Cities are not uniform objects: they have dense centers and sparse outskirts, highly connected cores and peripheral cul-de-sacs. This non-uniformity makes them good candidates for multifractal analysis. Recent studies have shown that urban street networks, especially the distribution of intersection points, can be well described by a multifractal structure. For example, Murcio et al.~\cite{murcio2015} analyzed the growth of London's street network between 1786 and 2010. Using multifractal tools, they found that the geometry of the street intersection point pattern (SIPP) changes over time and becomes more complex as the city grows. 

\begin{figure*}
    \centering
    \includegraphics[width=0.7\linewidth]{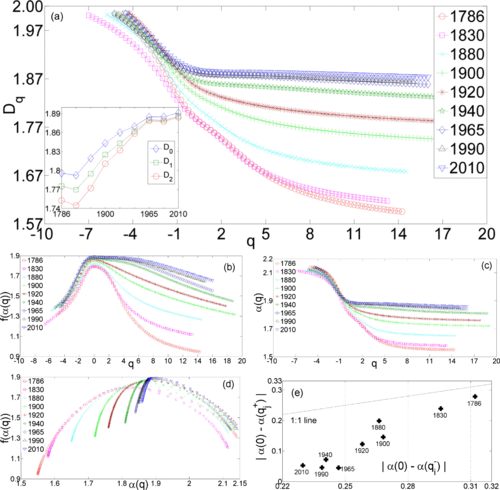}
    \caption{Multifractal analysis of Greater London's street network intersection point pattern, between 1786 and 2010. (a) Generalized fractal dimensions against~$q$ (inset : temporal evolution of~$D_q$ for selected values~$q=0,1,2$). (b) Fractal dimension of~$\alpha(q)$-sets in function of~$q$. (c)~$\alpha(q)$ against~$q$. (d) Multifractal spectrum. (e) Asymmetry of the relative spectrum. Source : From~\cite{murcio2015}.}
    \label{fig:multifractality}
\end{figure*}

In Fig.~\ref{fig:multifractality} the multifractal diagnostics of the London street network are shown for different years. Fig.\ref{fig:multifractality}(a) displays the curves $D_q$ as functions of $q$. These are characterized by an asymmetry between the negative and positive $q$ ranges, which is most pronounced in the earlier years. Since negative $q$ amplifies the contribution of low-density regions, this asymmetry signals the coexistence of both very sparse and very dense spatial zones, a hallmark of multifractality. In addition, $D_q$ values increase systematically for $q>0$, reflecting that as the city grows, the most densely built regions contribute disproportionately to the scaling. For a truly monofractal set, the $D_q$ curves would be flat, independent of $q$, so the observed decrease in slope over time constitutes a first sign of convergence toward monofractality. 

Fig.~\ref{fig:multifractality}(b) shows $f[\alpha(q)]$ as a function of $q$. For negative $q$, corresponding to low-density areas, the curves remain nearly invariant across time, indicating that the sparse periphery retains similar scaling properties throughout London’s growth. The authors argue that this can be explained by the conservation of certain areas, such as parks. By contrast, for positive $q$, which emphasizes high-density regions, the curves are widely separated in the early maps and progressively collapse in the later ones. This indicates that heterogeneities among dense areas shrink with time: once diverse and irregular, dense clusters of intersections have become increasingly uniform.

The same trend is visible in Fig.~\ref{fig:multifractality}(c), which displays $\alpha(q)$ versus $q$. In the early snapshots, $\alpha(q)$ decreases sharply for $q>0$, signaling the presence of a few highly concentrated zones of intersections with scaling properties distinct from the rest of the city. By contrast, in the later years (1965--2010) the curves almost overlap, showing that the scaling exponents have become uniform across the territory. This uniformity is precisely what is expected in a monofractal structure. Moreover, the range of $\alpha$ values narrows significantly, further confirming the loss of multifractal heterogeneity.  

Finally, Fig.~\ref{fig:multifractality}(d) presents the multifractal spectrum $f(\alpha)$. For the oldest SIPP, the spectrum is broad and symmetric, with a wide left tail corresponding to dense urban cores and a right tail corresponding to the sparse periphery. Such broad spectra embody multifractality: different regions of the city are characterized by different scaling exponents. Over time, however, the spectrum narrows substantially, resembling more and more a single point, that would represent a perfect monofractal structure, but never reaching it (see for instance the 2010 curves). The collapse of the multifractal spectrum constitutes direct evidence of a transition from multifractal to monofractal behavior. More specifically, preserved areas, aforementioned, such as parks or water bodies, remain of similar low-density, while the rest of of the city condensates through time as its density becomes more and more homogeneous. This can be observed in the 2010 curve in Fig.~\ref{fig:multifractality}(d), where the right part (corresponding to~$q<0$) is practically time-invariant, while the left part (~$q>0$) gets more and more concentrated around a point of abscissa~$\alpha(q)\approx1.9$.

To understand how spatial constraints affect urban growth patterns, Murcio et al. simulated a diffusion-limited aggregation (DLA) model within a green belt. They found that the presence of an impermeable boundary causes a transition from a multifractal to a monofractal structure. Once the growing cluster reaches the boundary, the generalized dimensions 
$D_q$ become nearly constant ($D_q \approx 2$) across all $q$, and the multifractal spectrum $f(\alpha)$ collapses to a single point—indicating a loss of heterogeneity in local scaling behavior.

Beyond synthetic models, empirical studies confirm that real-world street networks also exhibit multifractal organization. In an analysis of 12 major Chinese cities, Long and Cheng \cite{long2021multifractal} demonstrated that urban traffic networks follow consistent multifractal scaling laws.


To understand how spatial constraints affect urban growth patterns, Murcio et al. simulated a diffusion-limited aggregation (DLA) model within a green belt. They found that the presence of an impermeable boundary causes a transition from a multifractal to a monofractal structure. Once the growing cluster reaches the boundary, the generalized dimensions $D_q$ become nearly constant ($D_q \approx 2$) across all $q$, and the multifractal spectrum $f(\alpha)$ collapses to a single point—indicating a loss of heterogeneity in local scaling behavior.

\subsection{Number of buildings} \label{subsec:buildings}

\begin{figure}
    \centering
    \includegraphics[width=1\linewidth]{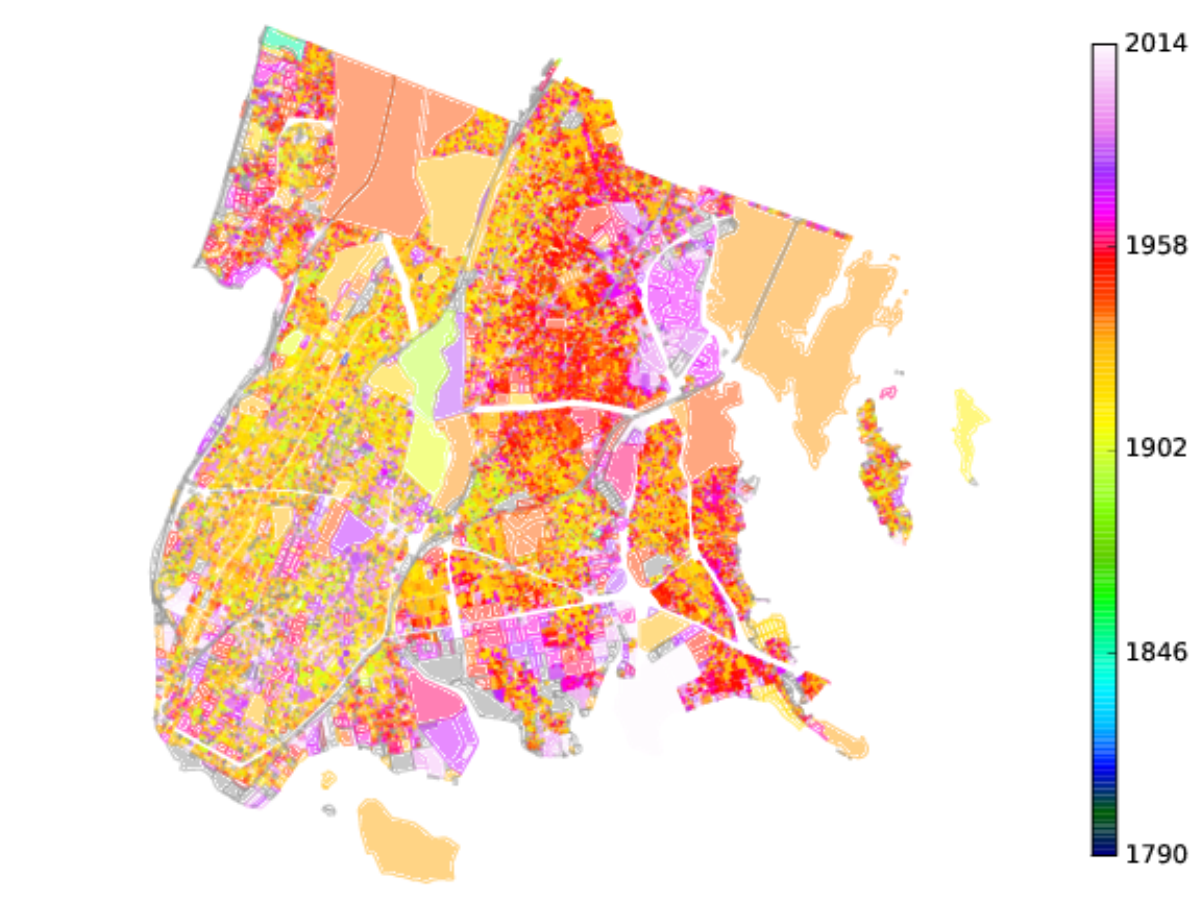}
    \caption{Map of buildings construction date for the case of the Bronx (New York City, US). Most of the buildings were constructed during the beginning of the 20th century, followed by the construction in some localized areas of buildings in the second half of the 20th century. Source: From \cite{carra2019}.}
\label{fig:carramap}
\end{figure}

\begin{figure}
    \centering
    \includegraphics[width=1\linewidth]{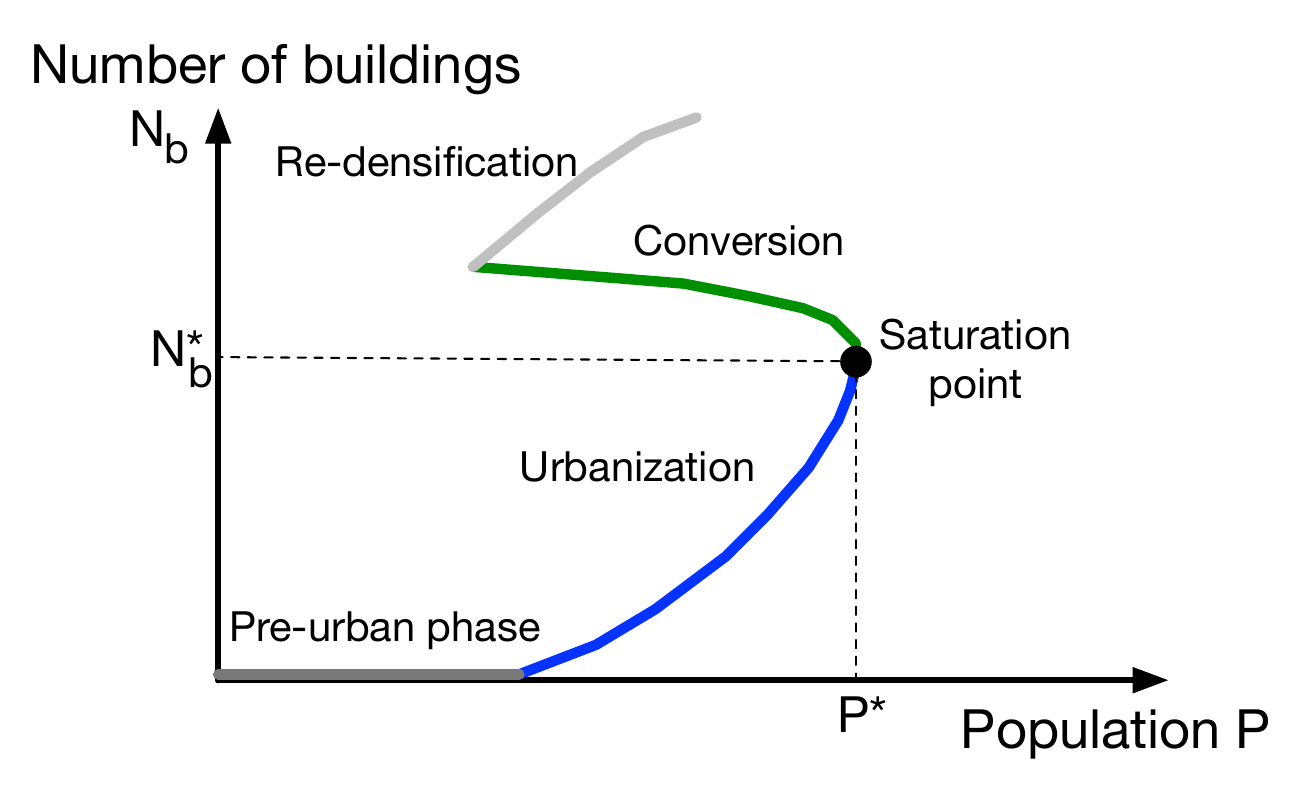}
    \caption{Schematic representation of the `fundamental diagram of urbanization'. The typical district growth curve is represented here and is characterized by three main phases: after a pre-urbanization
period, there is first an urbanization phase with a positive
growth rate $dN_b/dP$ that stops at the ‘saturation point’
$(P^*,N_b^*)$. A second ‘conversion’ phase follows, during which the population decreases. Finally, we observe a last redensification phase where both the population and the number of buildings increase. Source: From \cite{carra2019}.}
    \label{fig:fundamentaldiagram}
\end{figure}

\begin{figure*}
    \centering
    \includegraphics[width=1\linewidth]{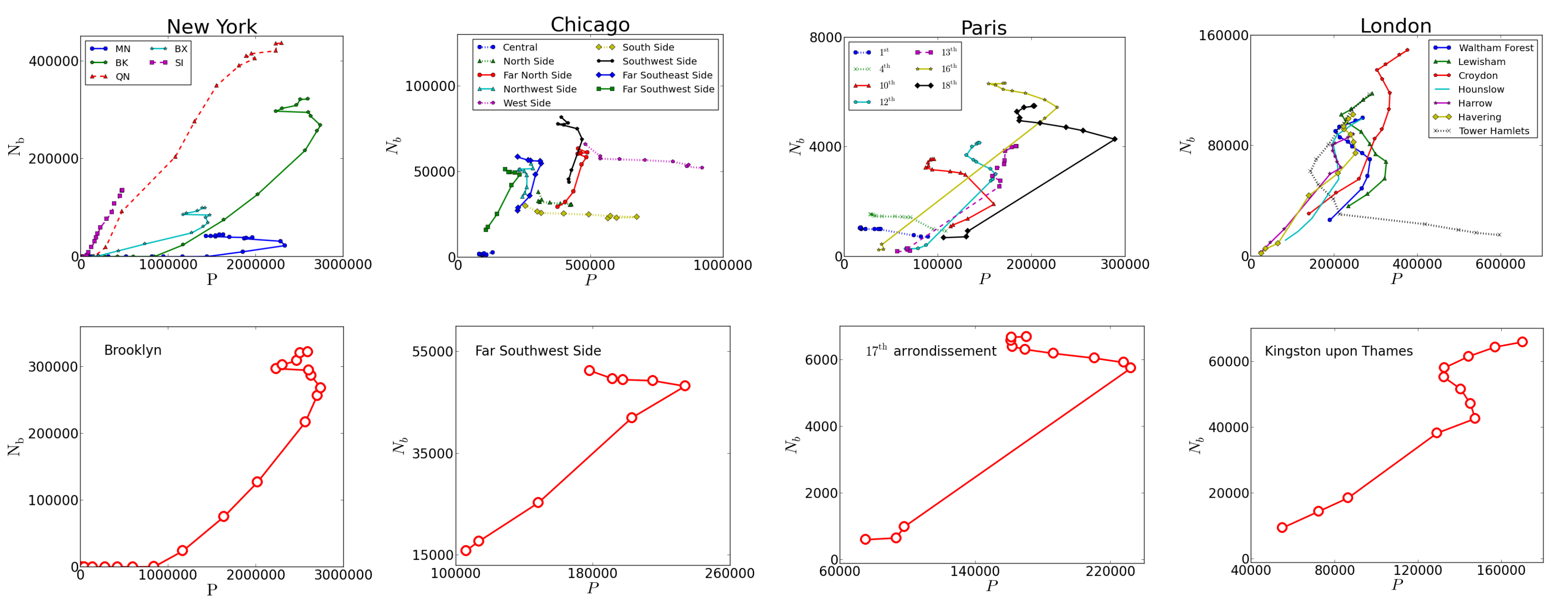}
    \caption{Number of buildings versus population. The districts that have reached their density peak are shown with continuous lines, and with dashed lines for  districts that are still in the growing phase. Dotted lines are for the districts that reached
the density peak before the first year available in the dataset. (Top panels) Results for districts in the cities studied here. (Bottom panels) Examples illustrating the  ‘universal’ diagram for districts in different cities that display all the regimes described in the text. Source: From  \cite{carra2019}.}
    \label{fig:carratrajectories}
\end{figure*}

The number of buildings is a good indicator of urbanization (see Fig.~\ref{fig:carramap} for the map of building construction dates for New York City, USA).
By looking at the relationship between number of buildings against population per neighborhood, in the cities of Chicago, London, New York City and Paris, the study \cite{carra2019} found similar patterns across cities, consisting of four phases : pre-urbanization, urbanization, conversion and re-densification. Firstly, buildings appear on vacant lots until reaching a saturation point (blue curve in Fig.~\ref{fig:fundamentaldiagram}). 
Beyond this point, the functions of buildings change, for instance from residential to commercial. Hence the population of the neighborhood decreases (green curve in Fig.~\ref{fig:fundamentaldiagram}). Finally, neighborhoods re-densify (see for instance the cases of NY, Paris and London in the bottom row of Fig.\ref{fig:carratrajectories}); this phase seems to be driven by exogenous factors. 

\subsection{Urban shape} \label{subsec:shape}

Understanding the shape of cities is central to urban growth studies, as the spatial configuration of built environments profoundly affects land consumption, infrastructure costs, commuting patterns, accessibility, and ultimately the productivity and welfare of urban residents. While classical urban economic models, such as the Alonso--Muth--Mills framework \cite{alonso1964}, predict declining densities with distance from a central business district (CBD), real-world cities often exhibit more complex forms including polycentric structures, fragmented developments, and irregular geometries shaped by both economic forces and natural constraints \cite{giuliano1991subcenters,louail2014mobile, bertaud_book, angel2016atlas}. These deviations are not merely aesthetic: they directly influence commuting distances, infrastructure provision, and spatial mismatches in labor and housing markets.

\subsubsection{Typology of urban expansion}

Urban expansion patterns display a high degree of spatial heterogeneity, shaped by planning policies, geography, and path-dependent processes. Cities may grow through continuous outward expansion, leapfrogging, or the emergence of scattered peripheral clusters \cite{bhatta2010analysis}. A typical trajectory involves a compact urban core expanding radially while intermittently generating disconnected urban fragments, which may subsequently grow and merge with the main core--a process \cite{marquis2025universalroughnessdynamicsurban,carra2017coalescing} that contributes to the development of polycentric urban forms (see Fig.~\ref{fig:mechanisms}).

\begin{figure*}
\includegraphics[width=0.99\textwidth]{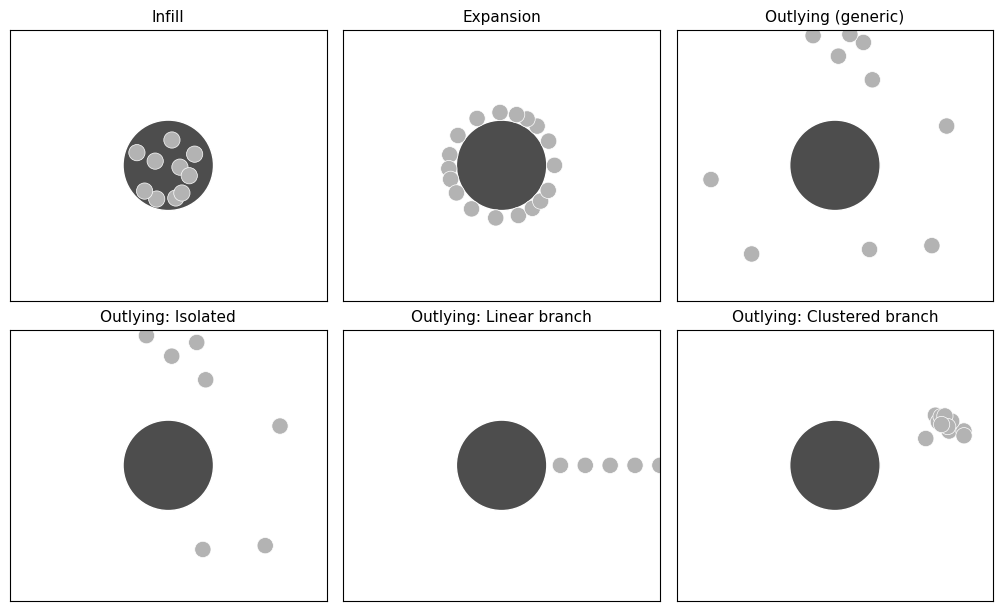}
\caption{Urban growth modes based on \cite{wilson2003development}: Infill, Expansion, and Outlying growth (further divided into Isolated, Linear branch, and Clustered branch patterns).}
\label{fig:typology}
\end{figure*}

To systematically characterize the spatial modes of urban growth, Wilson et al. \cite{wilson2003development} proposed the following classification: infill, expansion, and outlying growth. These categories serve to distinguish compact densification processes from discontinuous forms of peripheral development. Infill growth corresponds to the development of previously unbuilt parcels located within the interior of the urban fabric. These spaces are typically surrounded by existing structures and represent the most efficient use of land and infrastructure. Expansion growth, in contrast, manifests as the outward extension of the urban edge through spatially contiguous development on adjacent non-urban land. Finally, outlying growth arises in locations spatially disconnected from the main urban mass and includes a range of morphologies. It may take the form of isolated development, where individual structures emerge in low-density areas; linear branch growth, which follows infrastructural axes such as roads or railway lines, producing filamentary extensions of the urban field; or clustered branch growth, in which compact satellite agglomerations form in a leapfrog fashion beyond the urban boundary. These types are illustrated schematically in Fig.~\ref{fig:typology}.
Their diversity underscores the necessity of quantitative methods for detecting and modeling urban forms—particularly approaches grounded in spatial statistics, fractal geometry, or non-equilibrium growth processes.

\subsubsection{Quantitative characterization}

\begin{figure}
\centering
\includegraphics[width=0.9\linewidth]{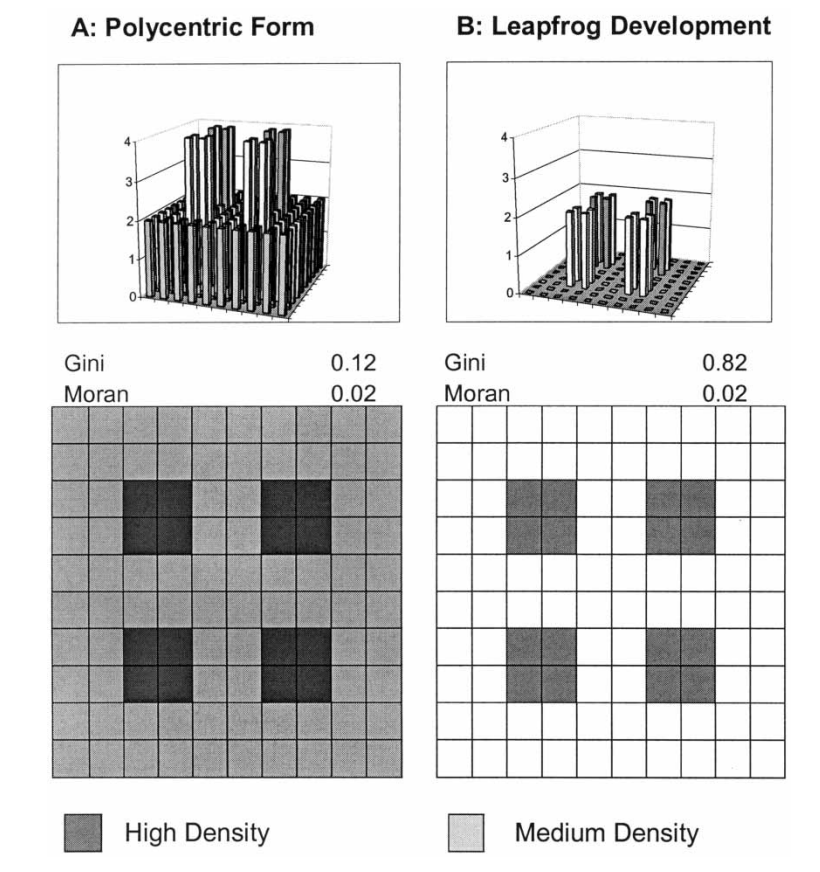}
\caption{Illustration of a polycentric structure and leapfrog development, characterized by distinct values of the Gini and Moran coefficients. Source: From \cite{tsai2005quantifying}.}
\label{fig:tsai}
\end{figure}

A quantitative characterization of urban form typically involves multiple dimensions. Tsai~\cite{tsai2005quantifying} decomposes metropolitan morphology into three components: density, distributional inequality, and spatial clustering. Inequality is commonly measured using the Gini coefficient~\cite{dixon1987bootstrapping}:
\begin{align}
G = \frac{1}{2N^2 \bar{n}} \sum_{i=1}^N \sum_{j=1}^N |n_i - n_j|,
\end{align}
where $N$ is the number of zones, $n_i$ their population (or employment), and $\bar{n}$ the mean. Spatial clustering can be captured by Moran’s $I$ statistic~\cite{moran1950notes}:
\begin{align}
I = \frac{N}{W} \frac{\sum_{i=1}^N \sum_{j=1}^N w_{ij}(n_i - \bar{n})(n_j - \bar{n})}{\sum_{i=1}^N (n_i - \bar{n})^2},
\end{align}
where $w_{ij}$ are spatial weights (often distance-based) and $W = \sum_{i,j} w_{ij}$. Fig.~\ref{fig:tsai} illustrates how these measures distinguish a polycentric structure—characterized by both low inequality and low clustering—from leapfrog development, where high inequality combines with weak spatial correlation. Tsai’s framework has inspired a large literature on urban morphology and its indicators (see \cite{zhang_2023} for a review).

A study by Harari \cite{harari2020}, focusing on the shape of 351 cities in India mostly during the time interval 1950-2010, moves beyond these by studying the internal geometry of city footprints, emphasizing that compactness has profound economic implications. Using a disconnection index $S$ defined by 
\begin{align} 
S = \frac{1}{n(n-1)} \sum_{i=1}^n \sum_{j=1}^n d_{ij}, 
\end{align} 
where $d_{ij}$ is the Euclidean distance between~$n$ randomly sampled interior points, Harari shows that more irregular, sprawling footprints (higher $S$) lead to longer average trips within the city, potentially raising commuting costs and weakening agglomeration economies.

Other empirical work has linked compactness and shape to environmental outcomes: urban compactness reduces energy consumption~\cite{kaza_2020}, urban form accounts for up to 70\% of heat inequalities~\cite{mashhoodi_2024}, and urban morphology critically influences water access~\cite{prietocuriel2025urbansprawlassociatedreduced}. Shape can also be characterized through anisometry, defined as the eccentricity of the ellipse enclosing the urban cluster. Zhou et al.~\cite{zhou2017urbanheat} show that higher fractal dimension correlates with stronger urban heat island (UHI) intensity $\Delta T$, whereas greater anisometry is associated with lower $\Delta T$.

\begin{figure*}[htp]
\centering
\includegraphics[width=0.9\linewidth]{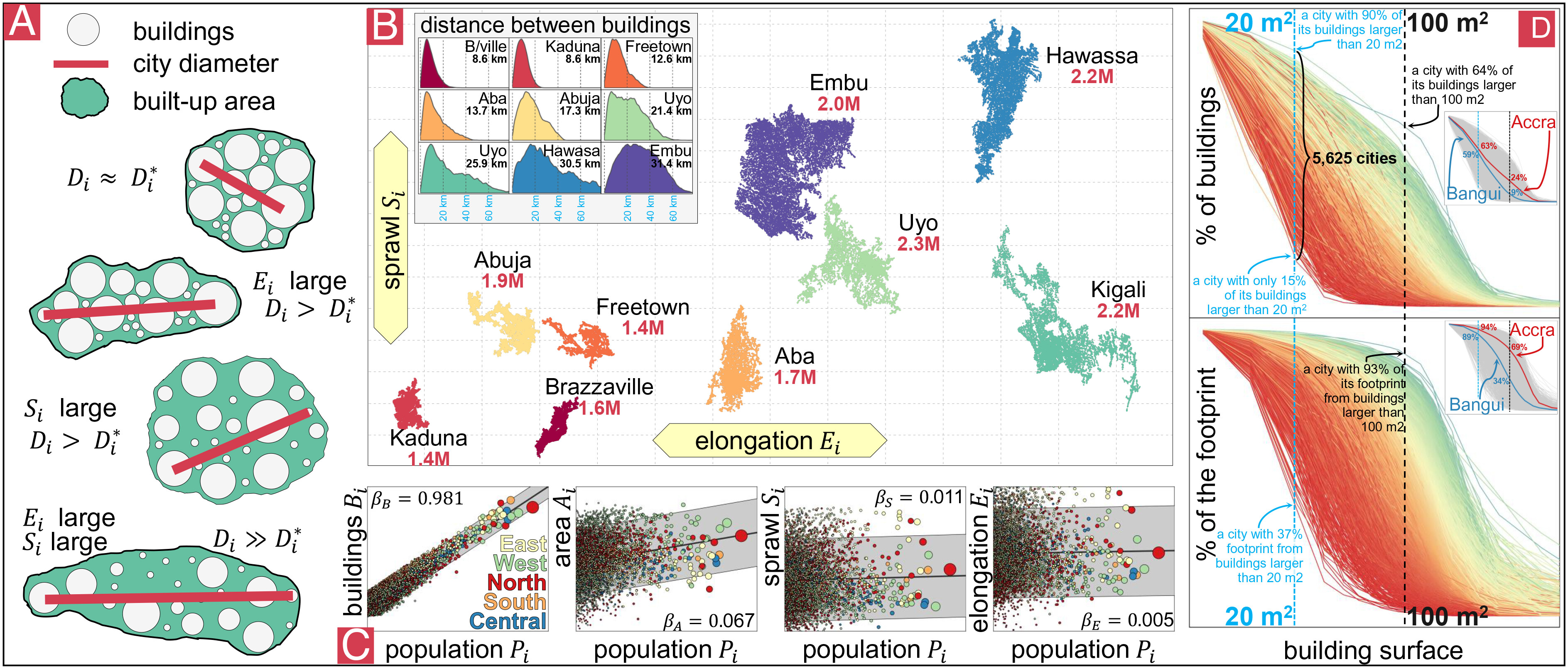}
\caption{\textbf{Framework of Prieto-Curiel et al.} 
A) Illustration of the BASE indicator. If buildings are compactly packed within the built-up area, which has a quasi-circular shape, then $\psi \approx 1$. The fragmentation index increases as either anisometry or inter-building spacing increases. B) Examples of cities in the $(E,S)$ state space. Inset: distribution of inter-building distances. C) Scaling relations for the four BASE indicators against population~$P$, yielding 
$\beta_D = (\beta_B + \beta_A + \beta_S + \beta_E)/2 \approx 0.532$. D) Cumulative proportion of buildings (top) and cumulative covered ground area (bottom) as functions of building surface. 
Source: \cite{prietocuriel2023}.}
\label{fig:scaling_africa}
\end{figure*}

A recent contribution by Prieto-Curiel et al.~\cite{prietocuriel2023} develops a unified framework (illustrated in Fig.~\ref{fig:scaling_africa}) for urban morphology and applies it to nearly 6,000 African cities using Google Open Buildings data. They propose the so-called BASE indicators for quantifying the morphology of cities: $B$ is the number of buildings, $A$ is the mean area of the footprint of buildings, $S$ is a sprawl index, capturing the average spacing between buildings, and $E$ characterizes the elongation of a city (i.e., the eccentricity of the enclosed ellipse~\cite{zhou2017urbanheat}).

Based on the fact that the mean distance $\langle d\rangle$ between two random points in a disk of radius $R$ is 
\begin{align}
\langle d \rangle = \frac{128}{45\pi} R,
\end{align}
Prieto-Curiel et al. define an indicator of the average distance between buildings as
\begin{equation} \label{eq:prieto_base}
    D_i = \frac{128}{45 \pi} \, \sqrt{B_i A_i S_i E_i},
\end{equation}
where $A_i$ denotes the area (so that $\sqrt{A_i}$ plays the role of an effective radius) and 
$B_i$, $S_i$, and $E_i$ are dimensionless factors. This observable can be compared to its benchmark value $D_i^*$ for a perfectly compact ($S=1$) circular city ($E=1$). The authors propose the fragmentation index $\psi = D_i / D_i^*$ as a metric to measure the compactness of cities. Moreover, they find that the mean distance between buildings scales with the population as $D \sim P^{\beta_D}$, with $\beta_D \approx 0.532$, slightly above but not inconsistent with the geometric benchmark of $1/2$. 

Taken together, evidence from Indian cities~\cite{harari2020} and US metropolitan areas~\cite{tsai2005quantifying} suggests that urban form systematically evolves with growth, shifting from compact, monocentric structures to more fragmented and elongated geometries. These morphological transitions increase commuting distances, reduce accessibility, and affect environmental outcomes. Ultimately, the shape of cities results from a complex interplay of economic trade-offs, agglomeration forces, transportation networks, natural constraints, and planning decisions. Quantitative frameworks that integrate density gradients, concentration indices, clustering measures, and scaling laws---as reviewed in~\cite{rybski2021modeling}---provide a rigorous basis for comparing urban forms across regions and over time, offering critical insights into how urban growth patterns influence mobility, energy demand, and development trajectories.

\subsection{The vertical dimension}

\subsubsection{Global patterns: upward versus outward}

The study by Mahtta et al.~\cite{Mahtta_2019} presents a large global dataset characterizing both two-dimensional (horizontal) and three-dimensional (vertical) urban growth for 478 cities with populations exceeding one million. Using remote sensing data from the Global Human Settlement Layer (GHSL, 2000–2014) for built-up areas and SeaWinds scatterometer data (2001–2009) for vertical structure, the authors quantified both outward expansion and upward densification. Cities were defined by aggregating 11$\times$11 km² grid cells centered on known urban cores, and a k-means clustering algorithm was applied to classify growth trajectories into five distinct typologies. The study pursued three main objectives: (i) to identify global trends in the balance between vertical and horizontal urban expansion; (ii) to construct a typology of urban growth that captures intra-urban variation in both dimensions; and (iii) to analyze how these growth patterns vary across regions and relate to population density distributions.

\begin{figure}[h!]
    \centering
    \includegraphics[width=1\linewidth]{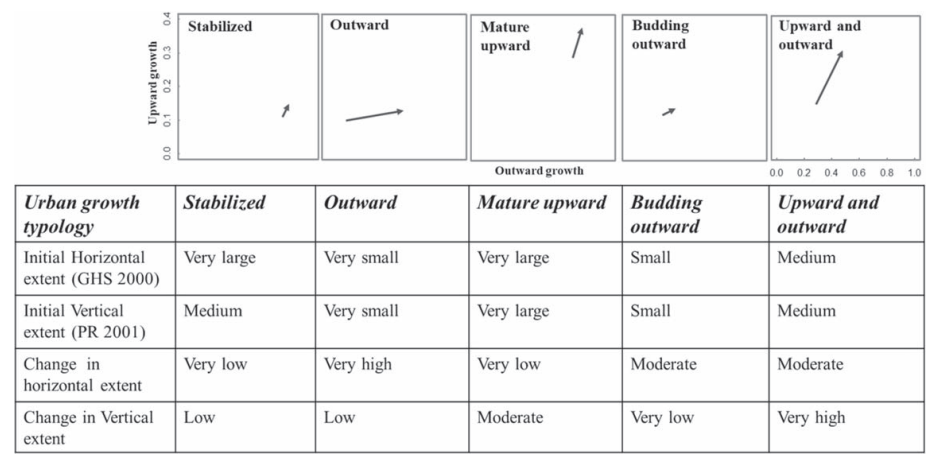}
    \caption{Urban growth typologies proposed in \cite{Mahtta_2019}, shown as mean vectors for each urban growth trajectory. Each arrow represents the change in urban extent—both outward and upward—for the corresponding typology. The x-axis indicates outward growth of urban built-up area based on the percentage urban cover in GHSL between 2000 and 2014, while the y-axis shows upward growth derived from structural backscatter power ratio (PR) between 2001 and 2009. The tail of each arrow marks the initial state (2001 for PR, 2000 for GHSL) and the head marks the final state (2009 for PR, 2014 for GHSL). Source: From \cite{Mahtta_2019}.}
    \label{fig:typoseto1}
\end{figure}

Figure~\ref{fig:typoseto1} illustrates the five urban growth typologies identified through this cluster analysis, displayed as mean vectors in the two-dimensional space defined by outward growth (change in built-up area percentage) on the x-axis and upward growth (change in backscatter power ratio) on the y-axis. Each arrow represents the average trajectory of an urban pixel belonging to a given typology, with its tail indicating the initial state (circa 2000--2001) and its head showing the state at the end of the study period (2009--2014). This graphical representation captures not only the magnitude but also the directional tendencies of urban growth, providing clear evidence that cities are composed of multiple, coexisting growth processes that differ markedly in their spatial and volumetric characteristics.

The analysis identifies five principal urban growth typologies:
\begin{enumerate}
    \item \textbf{Stabilized}: High initial built extent with negligible outward or upward growth, common in North and Central-South America and Europe (e.g. Los Angeles).
    \item \textbf{Outward}: Small initial extent, very high horizontal expansion, low vertical change, prevalent in Africa and India.
    \item \textbf{Budding outward}: Small initial extent with moderate outward growth and low upward growth, the most widespread pattern representing approximately 46\% of global urban land area.
    \item \textbf{Mature upward}: Large initial extent and verticality with minor horizontal change and moderate upward growth, found mainly in Japan, Taiwan, and parts of Europe.
    \item \textbf{Upward and outward}: Medium initial extent coupled with simultaneous high outward and upward growth, concentrated in China, South Korea, and the UAE.
\end{enumerate}

\begin{figure*}
    \centering
    \includegraphics[width=1\linewidth]{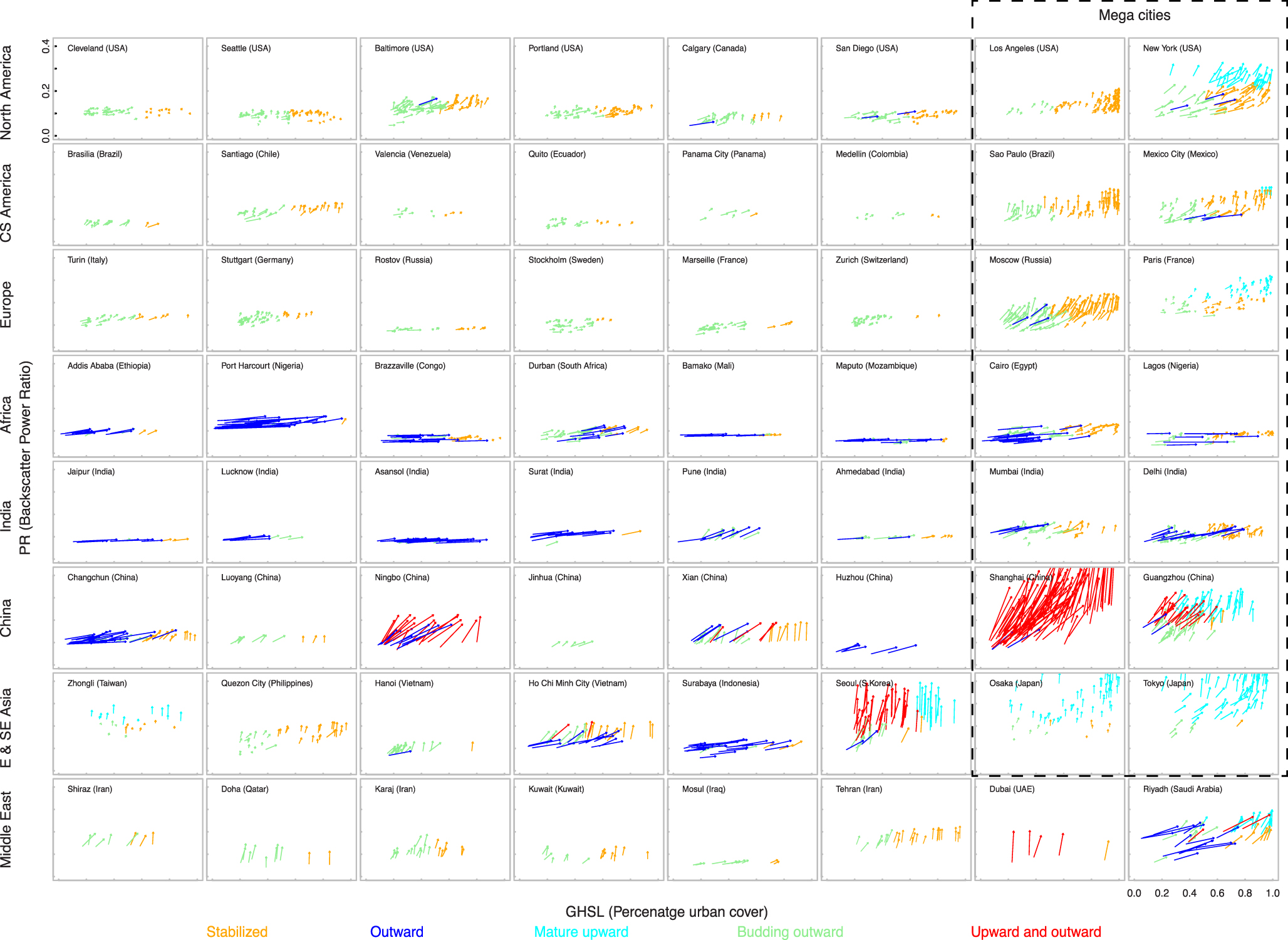}
    \caption{\textbf{Illustration of intra- and inter-urban variations in expansion patterns.}  
Urban growth typically results from a combination of processes. Arrows correspond to the pixels analyzed in an $11 \times 11$ grid around each city’s center; the tail represents the year 2001 for PR and 2000 for GHSL, while the head corresponds to 2009 for PR and 2014 for GHSL. Arrow colors indicate the five clusters identified in this study, each associated with a distinct growth process (see the legend at the bottom).  In African and Indian cities, expansion is largely dominated by outward development (blue arrows). Cities in the Middle East exhibit pronounced vertical growth. Very large cities (right-most columns) also show a marked tendency for vertical development.  Source: From \cite{Mahtta_2019}.}
    \label{fig:uporout}
\end{figure*}

Figure~\ref{fig:uporout} presents another visualization from the same study, highlighting the diversity of urban growth patterns and the heterogeneity within metropolitan regions. Middle Eastern cities exhibit predominantly vertical development, driven by multiple factors including competition for higher skylines~\cite{barr2024cities}. European, Central, and South American cities tend to expand outward, as do North American cities. In East Asia and China, regional variations are pronounced: Changzhou grows mostly outward, whereas Ningbo, Shanghai, and Xi'an show a mix of horizontal and vertical growth, with a strong tendency toward vertical development.

Several key findings emerge from this work. Every city examined contains multiple growth typologies, with approximately $82\%$ exhibiting at least two or three types. The dominant mode globally, by area, is the budding outward pattern, indicating slow, low-density expansion with significant potential to steer future urban growth towards higher-density trajectories. In contrast, mature upward typologies show the highest population densities, exceeding 7,000 persons/km$^2$ in 2015, whereas budding outward and outward typologies remain below 2,200 persons/km$^2$. 

These results have profound implications for urban trajectory forecast. Since most urban land expansion is still in early stages (budding outward), there exists a substantial opportunity to influence urban form towards more resource-efficient outcomes. However, the persistence of low-density outward growth raises concerns about `carbon lock-in' as it often establishes car-oriented infrastructure and energy-intensive spatial patterns that are difficult to reverse. Moreover, the coexistence of different growth typologies within the same city underscores the need for spatially differentiated urban planning policies. See also \cite{seto2011} for a broader discussion of global urban expansion, and related studies such as \cite{liu2020,Frolking2024} for further insights into evolving three-dimensional urban morphologies.

\subsubsection{The volume of cities}

Over the past century, cities have become increasingly vertical, yet most analytical frameworks remain rooted in a horizontal perspective. Recent scholarship emphasizes the need to account for the volumetric properties of urban form, recognizing cities as multi-layered, three-dimensional systems \cite{bruyns2020}. A vertical approach is essential for capturing the morphology of dense and interconnected urban environments, moving beyond surface-based measures toward a richer understanding of spatial configurations, functions, and interactions.

The height distribution of buildings was analyzed in \cite{batty2008}, where the authors reported approximate Zipf-like power law exponents of
$\nu=0.377, 0.288, 0.478$ for Tokyo, London, and New York, respectively. These results imply that the probability density function of building heights scales as
\begin{align}
P(h)\sim \frac{1}{h^{1+1/\nu}}\sim \frac{1}{h^\mu}
\end{align}
where $\mu\approx 3-4$ indicating a rapid decay with height. This steep decay is expected, since the maximum height of buildings is subject to strong physical, economic, and regulatory constraints, and thus one should not expect building heights to span many orders of magnitude. The authors also compute the correlation functions for the $100{,}000$ tallest buildings in the Greater London area, considering distances up to approximately three kilometers. They observe that the tails of the two-point functions decay as $d^{-\gamma}$, with~$\gamma\approx0.23$ for buildings perimeter, area, height, and volume, indicating long-range correlations of buildings geometry, and henceforth that fractal patterns extend to the vertical dimension of cities, a feature that has largely been neglected -- a notable exception is~\cite{molinero2021}.

\begin{figure}[h!]
    \includegraphics[width=0.8\linewidth]{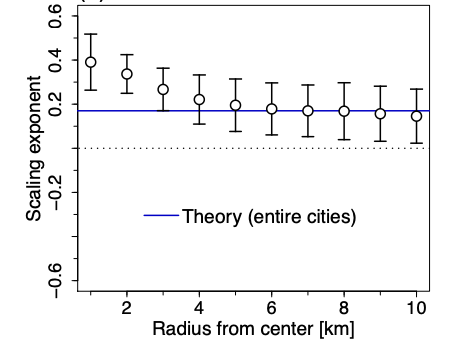}
    \caption{Scaling exponent $\beta$ for the relation between
    average building height, $\overline{h}$, and city population size $P$, plotted as a function of the radius $d$ from the city center. For the whole city ($d$ large), the exponent approaches the theoretical prediction $\beta=1/6$ \cite{schlapfer2015urbanskylinesbuildingheights}. Error bars denote the $95\%$ confidence interval. Figure adapted from \cite{schlapfer2015urbanskylinesbuildingheights}.}
    \label{fig:height}
\end{figure}

How building height varies with population was investigated in \cite{schlapfer2015urbanskylinesbuildingheights} using an extensive dataset of building shapes, covering about 5 million individual buildings across 12 major North American cities. The authors found the result shown in Fig.~\ref{fig:height}.
These results indicate that the average building height $\overline{h}$ scales with population $P$ as
\begin{align}
    \overline{h} \sim P^\beta,
\end{align}
with $\beta \approx 1/6 \simeq 0.17$ when the entire city is considered. Schläpfer et al. \cite{schlapfer2015urbanskylinesbuildingheights} provide a theoretical argument for this value; here we present a simpler alternative argument. Let $A$ denote the surface area of the city and $V$ its total built volume. By definition,
\begin{align}
    V = \overline{h} \, A,
\end{align}
and assuming that each household occupies a fixed volume $v_0$, the population is
\begin{align}
    P = \frac{\overline{h} \, A}{v_0}.
\end{align}
This relation implies that the average height scales as $\overline{h} \sim P/A$, i.e. proportional to the average population density. In other words, as a city's density increases, its built volume must grow accordingly. For the United States, the urbanized area scales with population as $A \sim P^\delta$ with $\delta \approx 0.85$ \cite{Barthelemy:2016}, leading to
\begin{align}
    \overline{h} \sim P^{1-\delta} \sim P^{0.15},
\end{align}
in excellent agreement with the empirical value (this argument was rediscussed in depth in \cite{ribeiro2025}).

In the study \cite{zhou2022}, Zhou et al. develop a high-resolution global atlas of urban built-up heights circa 2015, providing critical insights into the three-dimensional form of cities and stark inequalities in urban infrastructure. Using Sentinel-1 Ground Range Detected (GRD) data, the authors estimate mean built-up heights within 500~m grids worldwide, validated against detailed datasets from cities in the US, Europe, China, Brazil, Canada, and Germany. Their work reveals that while urban areas globally are dominated by low-density, horizontally expansive developments, there are sharp regional contrasts: cities in East Asia and parts of Europe frequently exhibit substantial verticality, whereas cities in North America, despite similar urban extents, tend to have much lower mean heights.

\begin{figure*}[htp]
    \centering
    \includegraphics[width=0.99\linewidth]{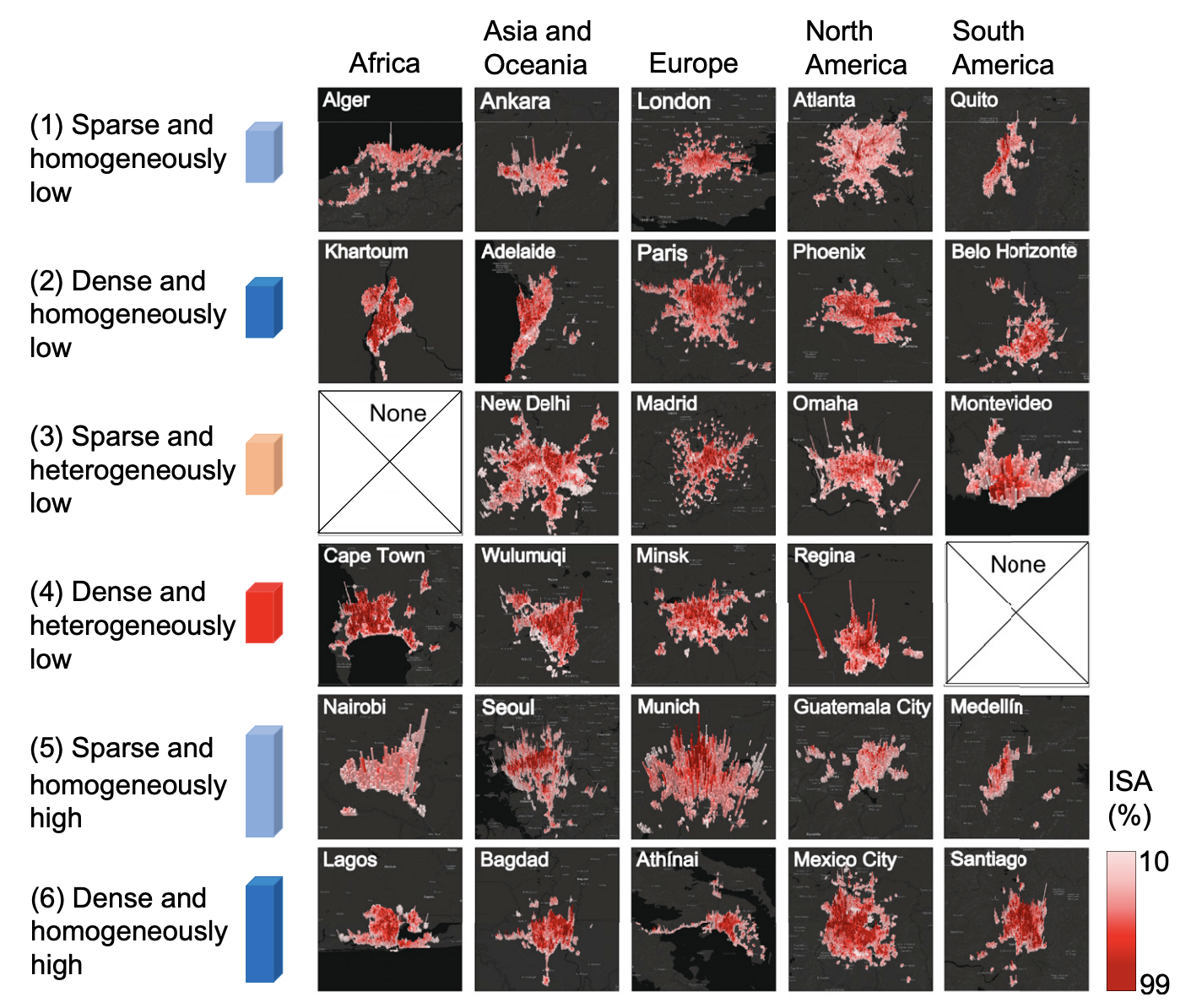}
    \caption{Three-dimensional views of representative cities categorized by urban form. The six types combine urban density and built-up height variation: (1) sparse and homogeneously low, (2) dense and homogeneously low, (3) sparse and heterogeneously low, (4) dense and heterogeneously low, (5) sparse and homogeneously high, and (6) dense and homogeneously high. Source: From \cite{zhou2022}.}
    \label{fig:zhou_fig2}
\end{figure*}

To systematically compare urban form, Zhou et al. jointly analyzed urban density (impervious surface area) and built-up height metrics (mean and quartile coefficient of dispersion) to classify global cities into six typologies. These range from sparse and homogeneously low-rise to dense and homogeneously high-rise forms. Figure~\ref{fig:zhou_fig2} illustrates representative three-dimensional views of cities in each category across different continents, underscoring not only the variation across regions but also within individual continents. The results show, for example, that many East Asian cities combine high mean heights with low internal height variation, while US cities such as Atlanta display concentrated high-rise cores amidst otherwise low-rise sprawl.

\begin{figure}
    \centering
    \includegraphics[width=1\linewidth]{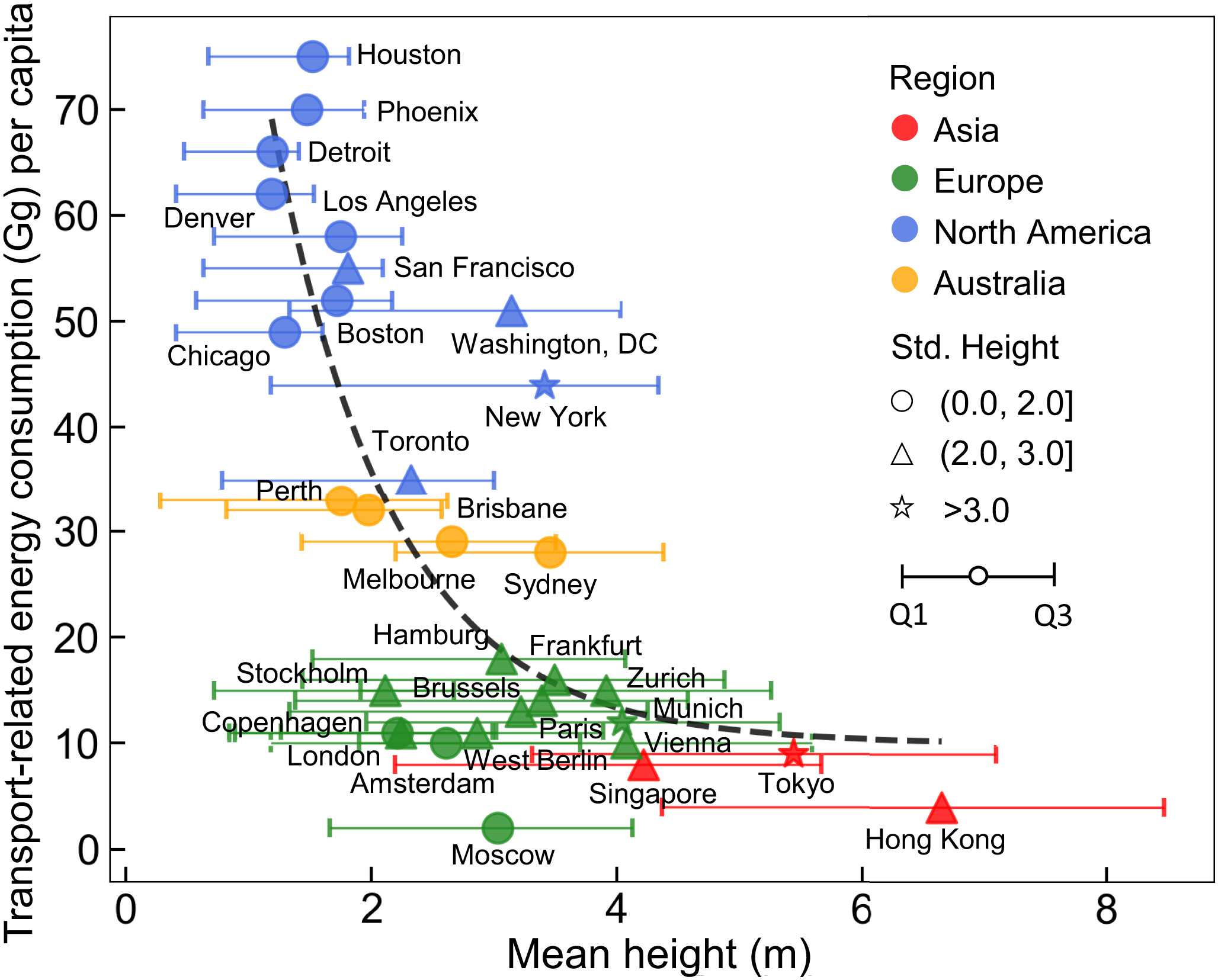}
    \caption{Relationship between cities' mean built-up heights and per capita transport-related energy consumption. Horizontal lines show interquartile ranges of built-up heights for each city, while the dashed curve is a fitted nonlinear regression. From \cite{zhou2022}.}
    \label{fig:zhou_fig6}
\end{figure}

A major contribution of the study is linking urban form to energy use. Using transport-related energy consumption data for 31 global cities, Zhou et al. demonstrate a negative relationship between mean built-up height and per capita transport energy consumption. Figure~\ref{fig:zhou_fig6} shows that cities with lower built-up heights generally consume more transport energy per person, reflecting the increased travel distances and dispersed spatial layouts typical of sprawling urban forms. This suggests that strategic vertical growth could help reduce transport energy use by concentrating populations closer to employment centers and transit, though the authors caution that such benefits must be weighed against the higher material and operational energy demands of tall buildings.

This global mapping of built-up heights not only fills a critical gap in data on urban vertical structure—particularly for the Global South—but also exposes the vast disparities in infrastructure that exist worldwide. The findings indicate that meeting future demands in the Global South could entail enormous increases in building material use, embodied energy, and greenhouse gas emissions. Moreover, the observed relationship between built-up height and transport energy highlights the need for integrated urban strategies that balance density, vertical growth, and infrastructure investment to achieve durable and equitable urbanization.

\subsection{Surface growth analysis} 
\label{subsec:surfacegrowth}

Most studies have focused on the total urbanized area, with comparatively little attention to the interface of built areas. Yet, the morphology of urban form carries important information. In particular, the largest connected component (LCC) of the built environment provides a well-defined object that can be analyzed using tools from the physics of surface growth. This framework makes it possible to quantify morphological features such as anisotropy and to compute roughness exponents.

\subsubsection{Area growth and anisotropy}

\begin{figure*}
    \centering
    \includegraphics[width=1\linewidth]{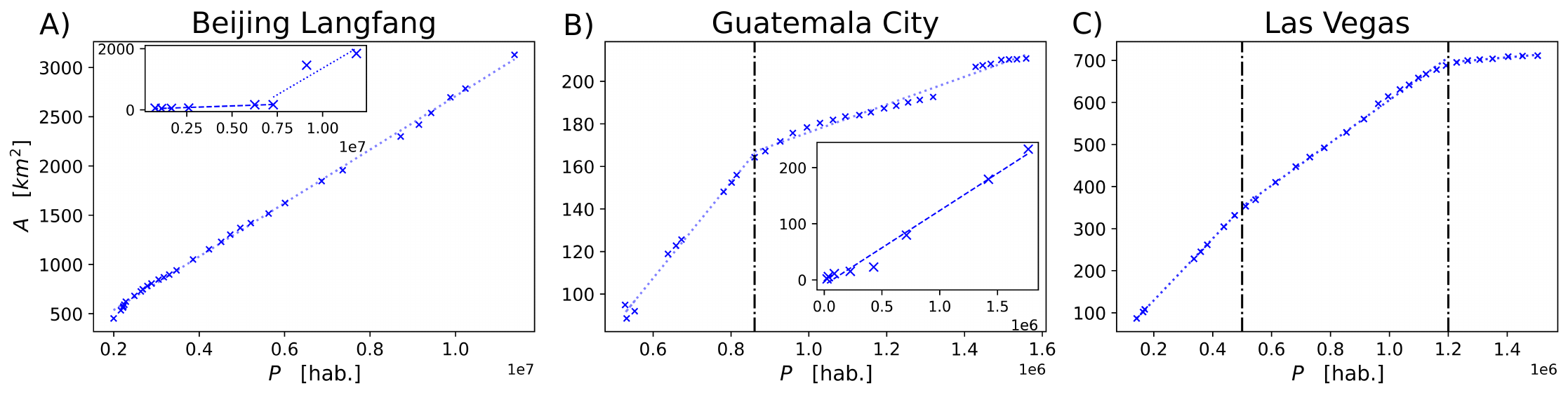}
    \caption{Illustration of the three growth area against population patterns (A) linear growth (B) piecewise linear growth with density breaking point (C) saturation. The dashed lines mark the breaking points. Insets : area against total population for historical data (from \cite{atlasurbanexpansion}).  Source: From \cite{marquis2025universalroughnessdynamicsurban}.}
    \label{fig:area_vs_pop_linear}
\end{figure*}

Marquis et al.~\cite{marquis2025universalroughnessdynamicsurban} applied the CCA algorithm (described in Chapter~\ref{chap:1}) to data from the World Settlement Footprint Evolution dataset~\cite{wsfevo}, decomposing the built-up area into connected clusters. Assuming a homogeneous population distribution, the population of the LCC is approximated as $P \approx A \, \frac{P_\text{tot}}{A_\text{tot}}$. The relationship between the area of the largest connected component and its population is then analyzed across 19 cities. This analysis identifies three distinct patterns, as shown in Fig.~\ref{fig:area_vs_pop_linear}.
The first pattern, seen in Beijing (Fig.~\ref{fig:area_vs_pop_linear}(A)), shows an affine relation between area and population, $A = aP+b$, with the giant component growing at constant density. The second, exemplified by Guatemala City (Fig.~\ref{fig:area_vs_pop_linear}(B)), follows a piecewise linear trend: growth remains linear but shifts to a higher density (lower slope) after a break point. The third, as in Las Vegas (Fig.~\ref{fig:area_vs_pop_linear}(C)), corresponds to saturation of built-up area, constrained by topographic or political boundaries.

Beyond total built-up area, Marquis et al. investigate deviations from isotropic growth, where—under uniform density—the city radius is expected to scale as $\overline{r}(P) \sim P^{1/2}$. Indeed, the average population density is given by $\rho = P/A$, and for circular cities with area $A =\pi r^2$, this yields 
$r \sim \sqrt{P}$.

For non-isotropic cities, one must distinguish between different orientation angles (measured from the center of the core area) and study the angular growth profile
\begin{equation}
    r(\theta, P) \sim P^{\mu(\theta)} \,,
\end{equation}
which represents the average radius within the angular interval $[\theta, \theta + \delta\theta]$. The resulting exponents span a wide range, from $0$ (corresponding to a pinned interface) to about $2.5$ (indicating strong super-linearity). To quantify deviations from the isotropic expectation $\mu = 1/2$, they introduce the dispersion measure  
\begin{equation}
    \Delta = \frac{1}{2}\sqrt{\frac{1}{N}\sum_{i=1}^N (\mu(\theta_i)-1/2)^2} \,,
\end{equation}
which captures the relative dispersion around $1/2$. Their analysis reveals substantial variability: some cities, such as Changzhou or Chengdu in China, exhibit nearly isotropic growth ($\Delta < 1/2$), while others, like Paris, France, display strong anisotropies with large deviations ($\Delta > 1.5$).



\subsubsection{Growth mechanisms}

Herold et al.~\cite{herold2005} suggest that cities grow through two main processes: local development and the absorption of previously built settlements. Marquis et al. evaluate the relative contribution of each mechanism, writing the change in built-up area as
\begin{align}
    \delta A(t+1) = C_o(t) + C_n(t) \,,
\end{align}
where $C_o(t)$ denotes the contribution from coalesced (absorbed) clusters, and $C_n(t)$ the contribution from newly built settlements. These contributions are analyzed as a function of the demographic pressure
\begin{align}
g = \frac{\delta P_{\text{tot}}}{P_{\text{tot}} \, \delta t}  \,.
\end{align}

\begin{figure}[htp]
     \centering
     \includegraphics[width=1\linewidth]{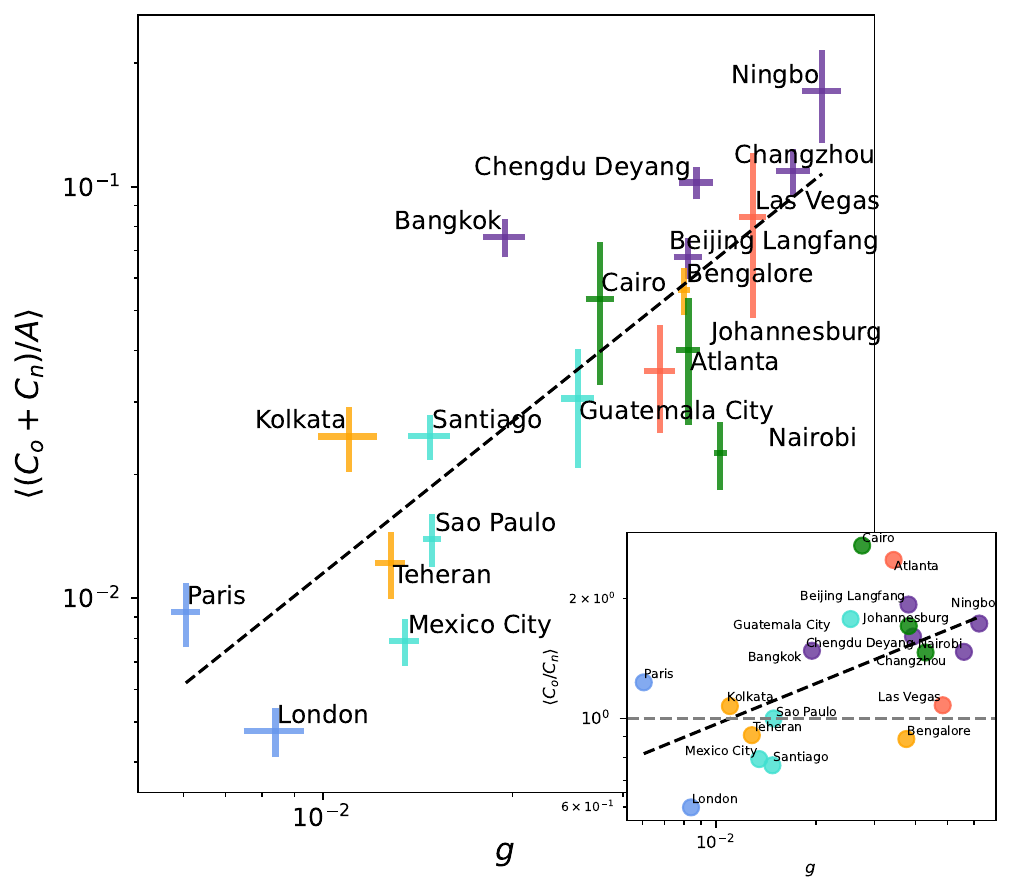}
     \caption{
     Average relative growth~$\delta A/A$ against demographic pressure~$g$, with error bars quantifying the standard error for the mean. Dashed line : power-law of exponent~$1.22$ ($R^2=0.71$). Inset : ratio of coalesced growth against local growth~$\langle C_o / C_n \rangle$ against~$g$. Dashed line : power-law of exponent~$0.33$. Source : From~\cite{marquis2025universalroughnessdynamicsurban}.}
     \label{fig:clustersize_vs_pressure}
\end{figure}

The authors show that $\langle \delta A / A \rangle \sim g^{1.22}$, with substantial fluctuations (Fig.~\ref{fig:clustersize_vs_pressure}), indicating that demographic pressure drives growth in a super-linear manner. Furthermore, coalescence becomes, on average, the dominant mechanism once $g \gtrsim 10^{-2}$, as illustrated in the inset of Fig.~\ref{fig:clustersize_vs_pressure}. These results are significant: the physics underlying local growth processes (diffusion, deposition) differ fundamentally from those governing aggregation~\cite{barabasi1995,odor2004}.


\subsubsection{Roughness exponents}

For a growing surface, it is natural to apply tools from the physics of surface growth, particularly to study the roughness of the interface, which can be characterized by scaling exponents. The standard framework to characterize interface growth is the Family-Vicsek scaling ansatz~\cite{family1984, barabasi1995}. Given an interface~$h(x,t)$ evolving on a system of linear dimension~$L$, its macroscopic roughness~$w(L,t) = \sqrt{\langle h(x,t)^2-\langle h(x,t) \rangle^2\rangle}$ ($\langle . \rangle$ here denotes the spatial average) obeys the following ansatz
\begin{equation}
    w(L,t) = t^\beta f(L t^{-1/z}) \,,
\end{equation}
with~$f(u \ll 1) \approx u^{\beta z}$ and~$f(u \gg 1) \sim 1$.~$\beta$ is the \textit{growth} exponent, characterizing the evolution at small times, while~$1/z$ rules the evolution of the correlation length exponent~$\xi \sim t^{1/z}$. When the correlation reaches the size of the system~$\xi\sim L$, at time scale~$t_{\times} \sim L^z$, the roughness stops following a power-law growth and only fluctuates around a value~$t_{\times} \sim L^{\beta z}$.~$\alpha=\beta z$ defines the \textit{roughness} exponent. This framework allows to characterize typical surfaces in \textit{universality classes}~\cite{barabasi1995, odor2004}, defined by a set of two independent exponents in this case. Typical universality classes include Edwards-Wilkinson, Kardar-Parisi-Zhang and Mullins-Herring~\cite{barabasi1995, odor2004}. Note that not all processes follow stricto sensu this framework : consider the random deposition model~\cite{barabasi1995}, which does not exhibit horizontal correlation, and hence~$\alpha$ and~$z$ are ill-defined.

The case of cities poses additional challenges. First, the substrate geometry is not a band with periodic boundary conditions, but rather radial. Second, the observed anisotropy implies that the system-size average $\overline{r(P)}$ loses its meaning—along with the associated definition of system-size roughness.
 In order to overcome these issues, Marquis et al. propose a local estimator for the width
\begin{align} \label{eq:localwidth}
    w^2(\overline{\ell},P)=\frac{1}{N}\sum_{i=1}^N\langle[r(\theta,P)-\langle r\rangle_i]^2\rangle_i \,,
\end{align}
where~$N$ denotes the number of sectors of aperture~$\Delta \theta = 2 \pi / N$, $\overline{\ell}(\theta)=\langle r\rangle_i\Delta\theta$ is the average arc length and~$\langle . \rangle_i$ denotes the average over sector~$i$. Using a scaling ansatz proposed by Ramasco et al.~\cite{ramasco2000}
\begin{equation} \label{eq:ramasco}
    w(\overline{\ell},P) = t^\beta F(\overline{\ell}P^{-1/z}) \,,
\end{equation}
where~$F(u \ll 1) \sim u^{\alpha_{\text{loc}}}$ introduces a novel local roughness exponent~$\alpha_{\text{loc}}$, independent from~$\beta$ and~$z$, governing the evolution of roughness at small scales.

\begin{figure*}[htp]
    \centering
    \includegraphics[width=0.9\linewidth]{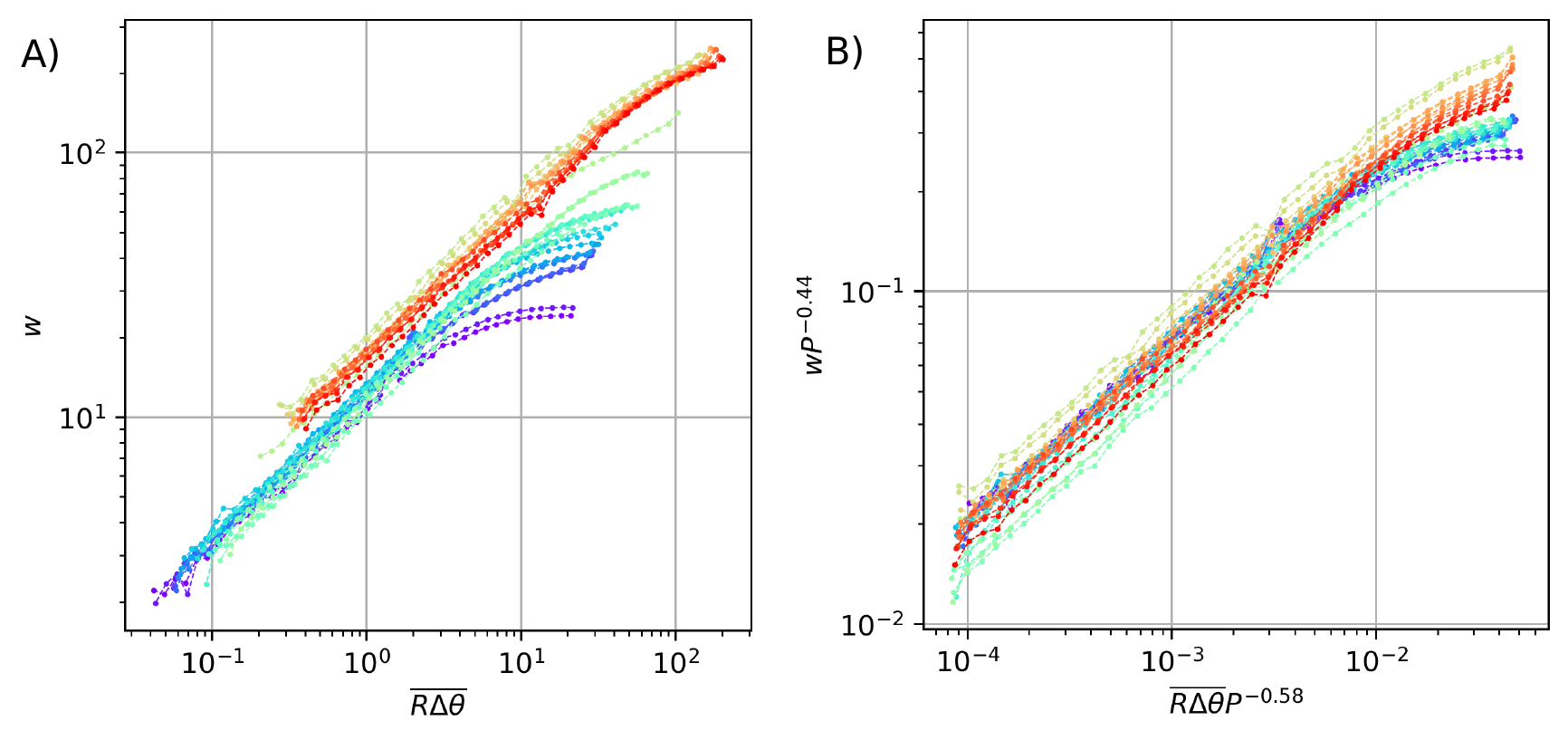}
    \caption{
    (A) Roughness of interface against~$\overline{\ell} = \overline{R \Delta \theta}$, for Ningbo, China, for each year between 1985 (purple) and 2015 (red). (B) Rescaled curves according to the Eq.~\ref{eq:ramasco} scaling ansatz. Source : From~\cite{marquis2025universalroughnessdynamicsurban}.}
    \label{fig:collapse}
\end{figure*}

\begin{figure}[htp]
    \centering
    \includegraphics[width=1\linewidth]{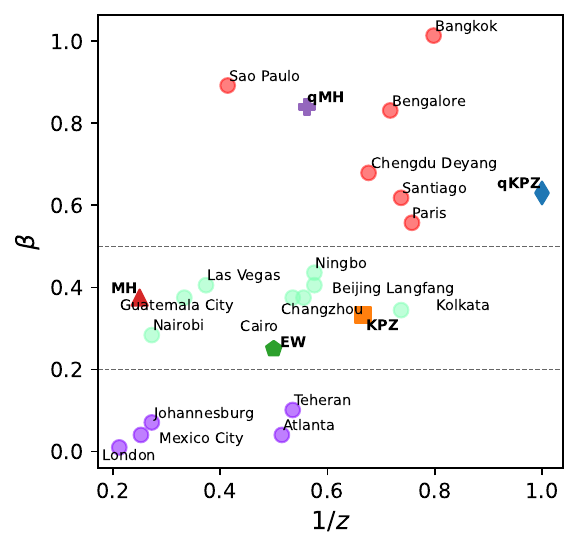}
    \caption{
    Measured exponents~$\beta$, $1/z$ obtained from the data collapse. Points are separated according to the value of~$\beta$ (in purple,~$\beta \approx 0$, $0.25<\beta<0.5$ in green and $\beta>0.5$ in red), according to the grey dashed lines. Non-circular symbols : classical universality classes (EW, KPZ, MH) and quenched universality classes (qMH, qKPZ) are represented. Source : From~\cite{marquis2025universalroughnessdynamicsurban}.}
  \label{fig:exponents}
\end{figure}

If the scaling described by equation~\ref{eq:ramasco} holds, plotting~$w(\overline{\ell},P)P^{-\beta}$ against~$\ell P^{-1/z}$, for different population should result in a collapse on the master curve~$F$, hence allowing for measurements of~$\beta$ and~$1/z$\cite{Bhattacharjee_2001}, as illustrated for Ningbo in Fig.~\ref{fig:collapse}. Once the data collapse is achieved, the exponent~$\alpha_{\text{loc}}$ can be measured in the regime~$\overline{\ell}P^{-1/z} \ll 1$. The authors find, quite surprisingly given the variety of morphologies and types of growth, an universal exponent
\begin{align}
    \alpha_{\text{loc}} \approx 0.54 \pm 0.03
\end{align}
across cities and time, while~$\beta$ and~$1/z$ vary widely. Marquis et al. propose a classification of cities in three groups as illustrated in Fig.~\ref{fig:exponents}, characterized by~$\beta\approx 0$ (purple points in Fig.~\ref{fig:exponents}), indicating interfaces with little susceptibility to population growth, $1/4 <\beta < 1/2$ (green points in.~\ref{fig:exponents}), including the universality classes associated with the usual EW, KPZ and MH (thermal) equations, and ~$\beta>1/2$ (red points in Fig.~\ref{fig:exponents}), with anomalously high roughness, reminding of quenched interface growth~\cite{barabasi1995, odor2004}.

Beyond the second moment of the height distribution, higher-order moments are essential for understanding surface growth. In the context of the KPZ universality class, a key result is that fluctuations are universal, following the Tracy–Widom distributions (for a summary of the history of the KPZ equation, see~\cite{Halpin_Healy_2015, doussal2025dynamicscalinggrowinginterfaces} and references therein). Given the intrinsically anomalous nature of urban interfaces, it becomes necessary to modify the form of the fluctuations under study relative to the standard normalization. Unlike the standard Family–Vicsek scaling, local and global fluctuations exhibit distinct scaling behaviors. Specifically, for a city with population $P$, the local width scales as
\begin{align}
    w(\ell, P) \sim \ell^{\alpha_{\text{loc}}} P^{\beta^*},
\end{align}
for observation lengths $\ell \ll \xi$, where $\xi$ denotes the correlation length. This formulation allows for a comparative analysis of fluctuations across cities of different sizes and at varying observation scales $\ell(\Delta\theta)$ by introducing the rescaled variable
\begin{align}
    x = \frac{r - \langle r \rangle_{\Delta \theta}}{\ell^{\alpha_{\text{loc}}} P^{\beta^*}},
\end{align}
where $\langle \cdot \rangle_{\Delta \theta}$ represents the average over an angular sector of size $\Delta \theta$, corresponding to a length scale $\ell(\Delta\theta)$. This approach is valid provided that the correlation length $\xi$ exceeds the segment length over which the fluctuations are measured. Then, once the statistics of~$x$ are collected for different populations and length scales, dividing~$x$ by its standard deviation allows to compare cities.

\begin{figure}
    \centering
    \includegraphics[width=1\linewidth]{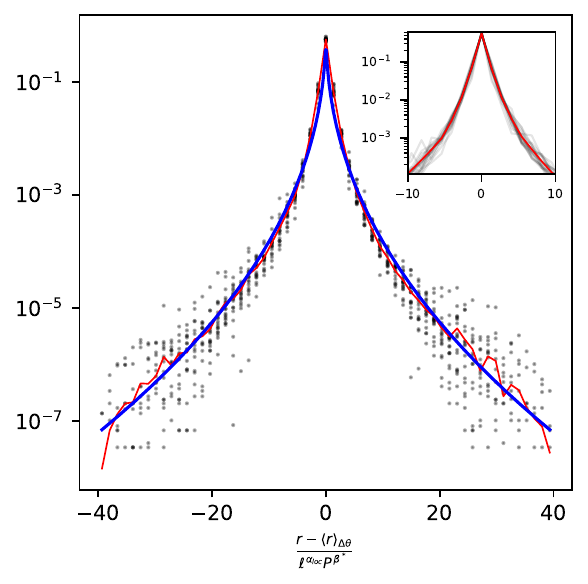}
    \caption{{Standardized distribution of fluctuations across cities (grey points).} The red line represents the average (over city) profile, and the blue line the best stretched exponential fit ~$\sim \exp(-b |x|^{0.46})$. Inset : zoom on~$-10<x<10$. Source : From \cite{marquis2025universalroughnessdynamicsurban} (Supplemental Material).} 
    \label{fig:fluctuations}
\end{figure}

Marquis et al. also report that the fluctuations collapse onto a symmetric curve, which can be approximated by a stretched exponential with exponent~$b \approx 1/2$, as shown in Fig.~\ref{fig:fluctuations}. This distribution contrasts sharply with the Tracy–Widom  typically found in empirical and simulated systems~\cite{takeuchi2010,Takeuchi_2012,prahofer2002}, being symmetric and characterized by heavier tails.


\subsection{Street-network sprawl} \label{subsec:street}

\subsubsection{A century of sprawl in the United States}

Barrington-Leigh and Millard-Ball investigated the evolution of the street network in the United States between 1920 and 2012~\cite{barringtonleigh2015}.  
Their study provides a comprehensive, quantitative history of urban sprawl -here essentially meant as spatial expansion-, arguing that the form of the street network---in particular, its degree of connectivity---is a defining and persistent feature of sprawl, with long-term implications for transportation behavior, emissions, and land-use patterns. The authors introduce the first high-resolution, century-long time series of street-network sprawl across US urbanized areas.  

\begin{figure*}
    \centering
    \includegraphics[width=0.90\linewidth]{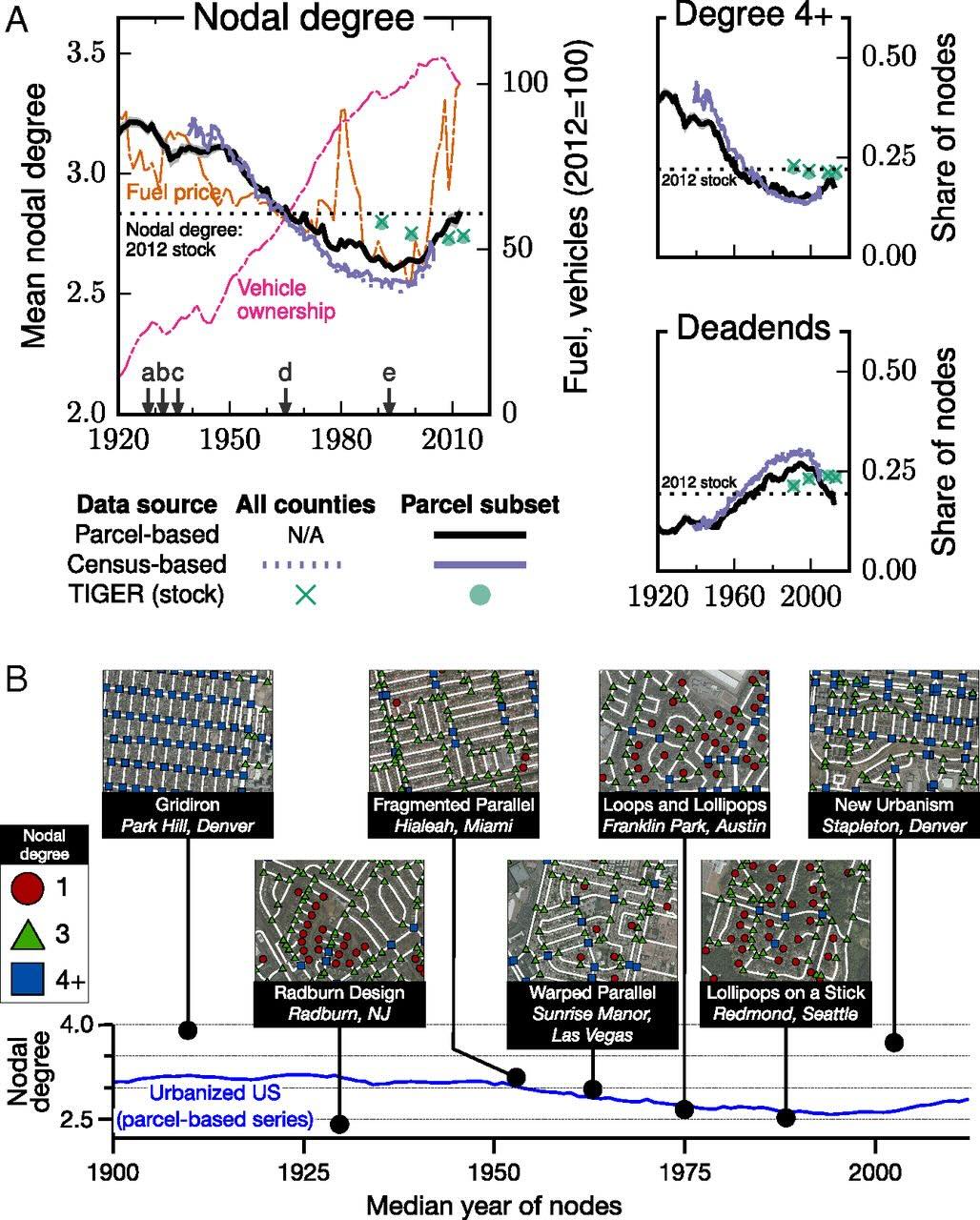}
    \caption{Trends in US urbanized areas, 1920--2012. 
    (A) The three sprawl measures show similar dynamics: street networks became increasingly sprawl-like after 1950, peaking in 1994, followed by a decline. Shaded areas show $95\%$ confidence intervals. The preferred parcel-based series (solid black) aligns with two alternative series. Key policy events (a--e) from ref.~\cite{barringtonleigh2015} are indicated. 
    (B) Empirical examples of archetypal street patterns illustrate nodal degrees consistent with these trends. Cases include the 1928 Radburn design and the recent New Urbanist development of Stapleton, highlighting extremes in street connectivity. 
    Source: \cite{barringtonleigh2015}.}
    \label{fig:centuryurbansprawl}
\end{figure*}

The expansion is quantified through three key metrics of street connectivity:  
(1) the average degree,  
(2) the proportion of dead-end streets, and  
(3) the proportion of intersections with degree greater than~$4$.  The evolution of these quantities in funciton of time are shown in Fig.~\ref{fig:centuryurbansprawl}A. Sprawling areas are characterized by low average degree, a high prevalence of culs-de-sac, and a scarcity of grid-like intersections~\cite{strano2012elementary}.  

The analysis shows that sprawl began well before the widespread adoption of the automobile, with early signs of suburban-style development emerging in the 1920s and accelerating after 1950. The peak of sprawl, in terms of newly constructed streets, occurred around 1994, when the average degree reached a minimum ($\approx 2.60$). Since then, new developments have become more connected and grid-like, reaching $\approx 2.83$ by 2012. This turnaround suggests a potential shift in planning priorities and development norms, even in the absence of coordinated national policy.  

Spatial analysis further reveals strong regional variation. Older, gridded cities such as New York, San Francisco, and Denver have maintained relatively high connectivity, while sprawling cities including Atlanta, Charlotte, and many Sunbelt metros have continued to expand disconnected networks -- see Fig.~\ref{fig:centuryurbansprawl}B) for illustrations of such street networks. In contrast, regions such as Dallas--Fort Worth and parts of the Pacific Northwest have shown notable increases in connectivity, likely influenced by local policies promoting grid layouts and New Urbanist principles -- a movement born in the 80s in the United States whose emphasis on walkable neighborhoods, mixed-use development, and human-scale streetscapes encourages denser, more interconnected urban forms (see~\cite{anderson2017,garde2020} for discussions on the New Urbanist movement).  

A key insight of the study~\cite{barringtonleigh2015} is the persistence of sprawl: areas that initially developed with low-connectivity networks tend to preserve those patterns over time. This path dependence reflects both physical constraints (e.g., the difficulty of retrofitting culs-de-sac) and institutional inertia, as well as market dynamics. Even over multiple decades, relative rankings in sprawl levels between cities remain largely unchanged.  

The study also emphasizes the policy implications of these findings. While the reversal in street-network sprawl is modest in scale, it could have significant long-term consequences due to the quasi-permanence of the built environment. The authors argue that recent policy changes---such as the adoption of connectivity standards and grid-oriented development codes---may already be influencing the trajectory of suburban growth. Because street patterns are highly persistent and shape transportation emissions for decades, they conclude that urban form should be a central component of development strategies.  

\subsubsection{Global sprawl}

In another paper \cite{barringtonleigh2020} Barrington-Leigh et al. present the first time series of global sprawl of street networks, using data from OpenStreetMap \cite{OpenStreetMap} and remote sensing. Utilizing the Street-Network Disconnectedness Index (SNDi), a metric introduced in \cite{barringtonleigh2019}, the authors analyze temporal patterns of sprawl across various scales (country size and city size). 

\begin{figure*}
    \centering
    \includegraphics[width=1\linewidth]{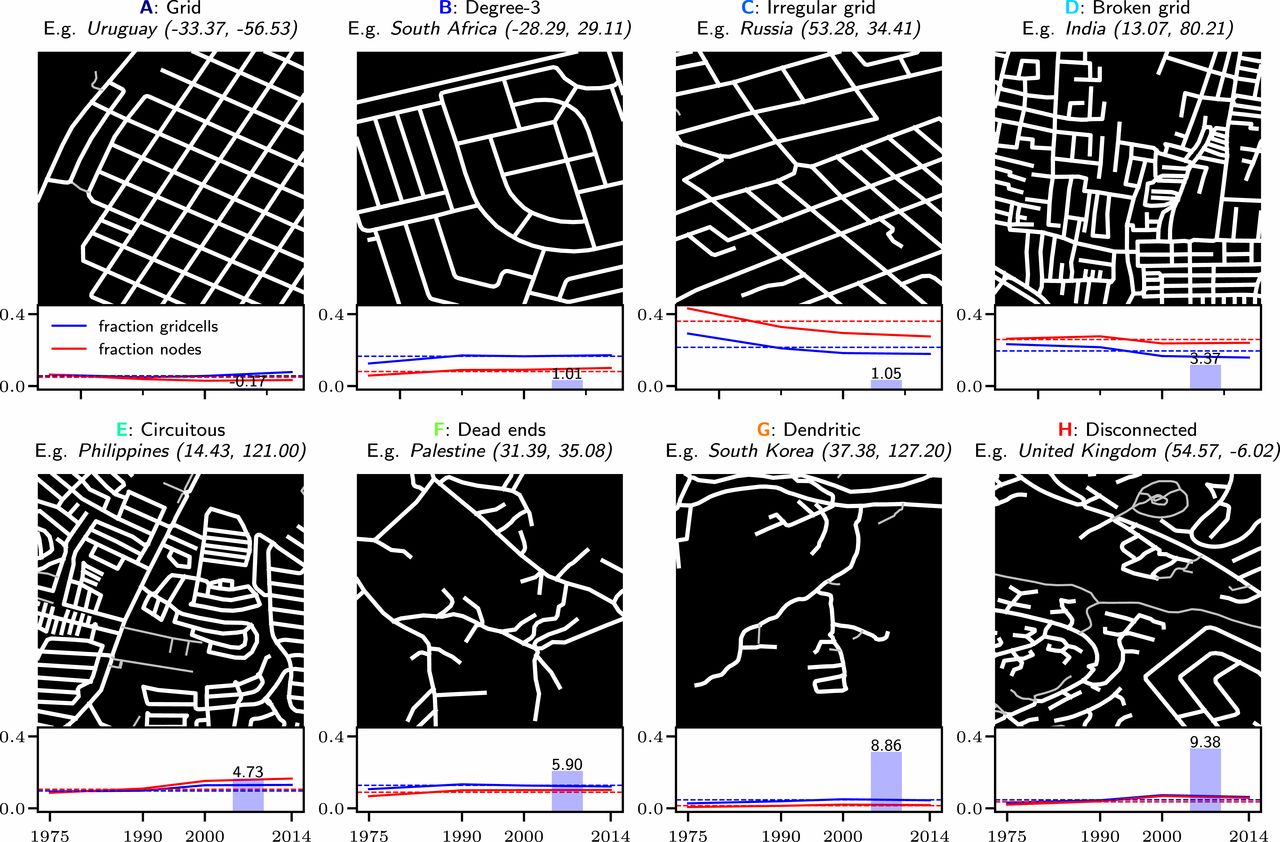}
    \caption{
    Empirical street-network types. For each network type, the figure displays an example of street patterns within a grid cell located near the centroid of the corresponding cluster. Below each map, a line plot shows the temporal evolution of the share of grid cells (in blue) and the share of street nodes (in red) belonging to that type. The types are ordered by decreasing Street Network Discontinuity Index (SNDi), indicated by the height of the blue bar and the associated index value. The caption above each plot specifies the geographic coordinates (latitude, longitude) and country of the example grid cell. The dashed horizontal line represents the overall fraction of grid cells and nodes in the complete stock of urban street networks. Source: From \cite{barringtonleigh2020}}
    \label{fig:globalsprawl}
\end{figure*}

Using street-network data, the Street-Network Disconnectedness index (SNDi) is derived from the following set of metrics:
\begin{itemize}

    \item \textbf{Intersection degree.}  
    For each node $i$, the degree $k_i$ counts the number of incident edges.  
    The mean degree is 
    \begin{align}
    \langle k \rangle = \frac{1}{N} \sum_{i=1}^N k_i ,
    \end{align}
    where $N$ is the number of nodes.  
    The fraction of dead-ends (degree~1 nodes) is
    \begin{align}
    f_{\text{dead}} = \frac{1}{N} \sum_{i=1}^N \mathbf{1}_{\{k_i = 1\}} ,
    \end{align}
    and the fraction of high-degree intersections (degree $>4$) is
    \begin{align}
    f_{k>4} = \frac{1}{N} \sum_{i=1}^N \mathbf{1}_{\{k_i > 4\}} .
    \end{align}

    \item \textbf{Dendricity.} 
    Edges in graphs can be partitioned in four categories: bridges, dead-ends, self-loops, or part of a cycle. In order to quantify the dendricity of street networks -- their similarity to topological trees -- the authors computed the fraction of edges belonging to each category.


    \item \textbf{Circuity.}  
    A local version of the detour index~\cite{barthelemy2011}, defined as
    \begin{align}
    C_{ij} = \frac{\ell_{ij}}{d_{ij}} ,
    \end{align}
    where $\ell_{ij}$ is the network distance between nodes $i$ and $j$, and $d_{ij}$ is their Euclidean distance.  
    The overall circuity is typically the average over all pairs:
    \begin{align}
    \langle C \rangle = \frac{1}{N(N-1)} \sum_{i \neq j} \frac{\ell_{ij}}{d_{ij}} .
    \end{align}

    \item \textbf{Sinuosity.}  
    For each edge $e$ with geometric length $\ell_e$ and end-to-end (Euclidean) distance $d_e$, the sinuosity is
    \begin{align}
    S_e = \frac{\ell_e}{d_e}.
    \end{align}
    The network-level sinuosity is the mean over edges:
    \begin{align}
    \langle S \rangle = \frac{1}{M} \sum_{e \in E} \frac{\ell_e}{d_e}.
    \end{align}

\end{itemize}
These metrics capture both topological (e.g degree) and geographical (e.g sinuosity) information. SNDi, a global measure of sprawl, is defined as the principal component of the PCA of these metrics. Using this metric, the authors noticed a global decline in street connectivity in $90\%$ of the analyzed countries since 1975. Only $29\%$ of countries showed increased SNDi since 2000. Additionally, the amount of `gated communities', characterized by high circuity and dead-end-heavy patterns, doubled between 1975 and 2014. 
These kind of constructions hinder walkability, public transport efficiency, and urban resilience.
The authors classify global street grids into 8 types illustrated in Fig.~\ref{fig:globalsprawl}. These types were obtained through $k$-means clustering on the set of metrics described above, following the approach of~\cite{barringtonleigh2019}. Alternative classification schemes have also been proposed--for example, based on the conditional probability distribution of block shape factors~\cite{louf2014} (see also Chapter 4 of~\cite{marshall2004streets} for a discussion on street pattern typologies).

The most sprawling type (Type E—circuitous and often gated) nearly doubled in frequency globally. Gridded street types (especially irregular grids) have declined sharply, despite their benefits. Moreover, through a path-dependence analysis, it was shown, both for cities and countries, that patterns were already disconnected in earlier decades tended to build even more disconnected streets later—indicating strong spatial and institutional inertia. Roads shape long-term urban form and are rarely altered. Poor connectivity today can lock cities into high carbon, car-dependent futures. The authors suggest to implement regulations, such as the `cul-de-tax', a tax on cul-de-sac and 3-ways intersection, to avoid gated communities and finally for administrators to promote pedestrian and transit permeability.



\subsubsection{Limited expansion : the effect of a green belt}

In \cite{Masucci_2013}, Masucci et al. investigated the dynamics of the street network of the Greater London Area (GLA) between 1786 and 2010. The box-counting analysis of the location of street intersections gives $N_B(R) \sim R^{-D_F}$ with $D_F \approx 1.78$ and is constant in time. The analysis is realized both on the core and on the whole GLA and while the exponents characterizing the distributions are identical, the distributions do not overlap for the core area while they do for the whole GLA. Moreover, persistently in time, city blocks sizes follow approximately lognormal statistics, with in particular an exponential tail. A scaling analysis reveals that the total length of the street network scales sublinearly with the number of intersections, $L(N) \propto N^{0.68}$ (which is not inconsistent with the argument showing that $L\sim \sqrt{AN}$, see \cite{barthelemy2022spatial}). In a second part, by assuming that the street-network of London is approximately planar at all times, the authors argue that the network grows in a space-filling fashion. Given that the spatial growth of London is bounded by the \textit{green belt}, intersections (and therefore streets) grow up until the capacity limit imposed by the belt. A simple logistic growth for the number $f(t)$ of intersections or street segments defining the network is given by
\begin{equation}
    \frac{df(t)}{dt} = r f(t) \left( 1 - \frac{f(t)}{C} \right),
\end{equation}
where the $C$ denotes the capacity and $r$ is the growth rate. In general, a finite $C$ yields
\begin{equation}
    f(t) = \frac{C}{1+\exp(-r(t-t_0))}
\end{equation}
where $t_0$ denotes the unique time at which the second temporal derivative is null. Fitting the number of intersections $N(t)$ and edges $E(t)$ allows to predict long-time limits
\begin{align}
    \begin{cases}
        N_\infty \approx 85123\\
        E_\infty = 115615
        \end{cases}
\end{align}
and
\begin{equation}
    \langle k  \rangle_\infty = \frac{2 E_\infty}{N_\infty} \approx 2.72.
\end{equation}
Manipulation of the logistic growth formulation allows to write the growth of the number of edges $E$ as a function of the number $N$ of intersections
\begin{equation}
    E(N) = \frac{E_\infty}{\left[ 1+a \left( \frac{N_\infty}{N}-1 \right)^{\frac{r_E}{r_N}} \right]},
\end{equation}
where $a\ = \exp \left( r_E(t_{0E}-t_{0N}) \right)$ is constant and $r_E/r_N\approx 1.07 \gtrsim 1$. This expression allows then for the computation of $\langle k(N) \rangle=2E(N)/N$. Its non-increasing nature indicates that the network changes from a more loopy topology to a more tree-like structure, reminding of space-filling phenomena which first establish large loopy structures to ensure functionality of the system and then fills space with branches.

\subsection{Effects of transportation infrastructures} \label{subsec:impact}

Transportation infrastructures act as control parameters for the spatial organization of cities. Their introduction modifies the cost structure of commuting and shipping, thereby reshaping the equilibrium distribution of population and employment. From a physics perspective, these infrastructures function like (quenched) fields that alter accessibility landscapes, lowering effective transport resistance along particular directions and introducing new boundary conditions that cities must adapt to.

In a series of studies \cite{baumsnow2007, baumsnow2020, baumsnow2005, baumsnow2016, baumsnow2019}, Baum-Snow and coauthors have systematically quantified these effects (see Chapter III. D). A first striking empirical regularity concerns U.S. metropolitan areas: the construction of a single radial highway reduced central city population by about $18\%$, compared to an $8\%$ increase in the absence of such infrastructure \cite{baumsnow2007}. Thus, each new radial connection acts as a perturbation that displaces density away from the core.

The case of rail infrastructure presents a complementary picture. Between 1970 and 2000, more than 25 billion dollars were invested in new rail lines across sixteen U.S. metropolitan areas, yet the aggregate transit share of commuting continued to decline. A careful difference-in-differences analysis \cite{baumsnow2005} showed that new rail lines modestly increased transit ridership among suburban residents living near stations but had almost no effect near the city center. The effect was strongest in compact, high-density systems such as Washington, D.C., but even there the impact decayed quickly after construction. A spatial equilibrium model showed that most new riders were former bus users, with relatively few switching from car travel unless substantial time savings were realized. The key conclusion is that decentralization of activity reduces the efficiency of new rail: the same infrastructure is more effective when the underlying urban density is centralized, but much less so in a sprawled configuration.

The influence of network geometry becomes even clearer in the Chinese context. An analysis of transport infrastructure between 1990 and 2010 \cite{baumsnow2016} shows that the configuration of highways—radial versus ring—controls the displacement of urban activity. Each additional radial highway was associated with a $4\%$ shift of central city residents toward suburbs, while the construction of ring roads reduced central populations by $20$–$25\%$. Radial railroads produced industrial decentralization, with a $24\%$ decline of central city industrial GDP, while ring roads pushed this effect above $50\%$. In dynamical systems terms, the type of link (radial or circumferential) selects the mode of decentralization: radial links decentralize residential and service activities, while rail links and ring roads expel manufacturing and associated housing to the periphery. Crucially, these redistributions occurred without significant changes in total GDP or population at the prefecture level, indicating that infrastructure reorganizes the internal spatial phase space of activity rather than driving net regional growth.

A broader analysis of the U.S. interstate highway expansion between 1960 and 2000 \cite{baumsnow2019} further quantifies the magnitude of this reorganization. Each additional radial highway reduced central city working residents by $14$–$16\%$ and jobs by $4$–$6\%$, across nearly all private-sector industries. Wholesale and retail jobs were most affected, while finance, insurance, and real estate were least so. A calibrated spatial equilibrium model showed that agglomeration externalities operate most strongly at sub-metropolitan scales: central city total factor productivity increased by $0.04$–$0.09$ for each $1\%$ relative increase in central employment compared to suburban employment. Model simulations suggest that each radial highway increased real income by up to $2.4\%$, reduced housing costs by about $1.3\%$, and lowered land rents by $4$–$9\%$. Importantly, the dominant driver of decentralization was not productivity change but greater land consumption for housing enabled by reduced commuting costs. In physical terms, highways expand the accessible configuration space, allowing households to occupy more land at lower cost, thereby lowering central density.

Finally, the expansion of the Chinese highway network \cite{baumsnow2020} reveals heterogeneous, non-linear effects at the regional scale. On average, new local roads had small or even negative effects on prefecture-level GDP and population. Yet the averages mask sharp asymmetries: regional primate cities gained population and output at the expense of their hinterlands. Improved regional highways induced primates to specialize in manufacturing and services, while peripheral areas shifted toward agriculture. Enhanced access to international ports generally increased population, GDP, and wages, particularly in hinterland prefectures. Thus, rather than uniformly stimulating growth, highways acted as a symmetry-breaking mechanism, amplifying primacy and deepening spatial inequalities. From a physics standpoint, this resembles a redistribution of mass in a coupled system: rather than adding to the total, the infrastructure perturbs the potential landscape and drives flows from weaker to stronger attractors.

Overall, these results show that transportation infrastructures do not simply accelerate urban growth but instead reorganize the internal spatial distribution of population and employment. They act as external fields reshaping density profiles, altering equilibrium configurations, and in some cases destabilizing central cores. The analogy with statistical physics is direct: infrastructure modifies the effective interaction kernel governing urban dynamics, thereby shifting the balance between agglomeration (short-range attraction) and dispersion (long-range repulsion).

\section{Spatial dynamics in geography and economics}
\label{chap:3}

In this section, we review a range of models developed at the interface of urban economics and spatial geography. These approaches differ in their assumptions, mathematical structure, and modeling goals, but all aim to capture the emergence and evolution of urban spatial patterns.

We begin with cellular automata (CA) and agent-based models (ABM), which represent cities as spatially extended systems governed by local interaction rules. These models are especially suited for simulating long-term urban growth, land-use transitions, and the influence of transport infrastructure. Despite their simplicity, they reproduce a variety of empirical morphologies and support the exploration of feedback loops between population, space, and mobility.

We then examine CA extensions that incorporate microeconomic foundations. These models introduce explicit local utility maximization or cost-minimization behaviors, bridging the gap between rule-based dynamics and standard economic theory. They illustrate how decentralized decision-making can generate complex spatial structures through local interactions alone.

Next, we turn to analytical modeling frameworks. We first revisit central place theory in a dynamical setting, where population redistribution and entrepreneurial activity give rise to evolving hierarchies of urban centers. This dynamical systems perspective connects spatial interaction theory with economic geography.

This naturally leads to a broader discussion of urban economics models. We introduce the classical Alonso-Mills-Muth (AMM) monocentric city model, which offers a baseline description of land use, commuting, and population density in equilibrium cities. We then present several important extensions: the incorporation of transportation networks (notably the Baum-Snow model for radial highways), evolving land-use dynamics in growing cities, and a myopic growth model where households sequentially choose locations based on local utility gradients. These frameworks offer theoretical insights into the spatial allocation of households, endogenous city boundaries, and the interaction between transport accessibility and suburbanization.

We conclude with the edge-city model, which departs from the monocentric paradigm and considers a polycentric structure where employment and services relocate toward decentralized locations. This framework captures the emergence of multi-nodal urban forms and provides a foundation for modeling modern metropolitan regions beyond the monocentric idealization.

\subsection{Cellular automata and agent-based models}
\label{sec2A}


Cellular automata (CA) are widely used tools across disciplines such as biology, physics, and computer science. Their application to urban science began in the 1990s \cite{white1993, batty1999}, and has since evolved to incorporate stochastic processes, socio-economic variables, and integration with geographic information systems (GIS), as well as extensions involving Markov models and artificial intelligence techniques. For a comprehensive review, see \cite{li2016}; for models addressing urban shrinkage, refer to \cite{schwarz2010}. Beyond CA, agent-based models (ABMs) offer a microscopic perspective, capturing agent heterogeneity and market influence. In the context of urban residential choice, ABMs are reviewed extensively in \cite{huang2014}.

\subsubsection{Modeling the dynamics of urban expansion}

In \cite{batty_1999_workingpaper} (see also Chapter 9 of \cite{batty2007cities}), the authors argue that the city expansion is driven by two main factors: the space availability at the urban fringe and the aging of buildings. A simple image is that one of a city growing outward from a central seed, with its fringe advancing with time, at a speed depending on factors such as the demography or the attractivity of the city, yet primarily constrained by the available space. Assuming that buildings have a finite lifetime, urban developments inevitably deteriorate over time, and a subsequent wave of demolition follows the advancing urban fringe. This notion of land changing states is reminiscent of compartmental models in epidemiology, where the `built' condition parallels the infected state. More specifically, the dynamics of urban growth can be represented as a process with three components:
\begin{itemize}
\item \textbf{Vacant Land ($A(t)$)}: Land susceptible to development.
\item \textbf{New Development ($N(t)$)}: Areas transitioning from vacant to developed states.
\item \textbf{Established Development ($M(t)$)}: Mature areas undergoing aging and potential redevelopment.
\end{itemize}
Since these quantities collectively account for all land, their sum equals the total land budget, $C$. Other components can also be included, such as a fringe $F(t)$, representing the city’s periphery where new developments occur. In this model, land transitions through the sequence of states $A \ (\rightarrow F) \rightarrow N \rightarrow M$, with $F$ and $N$ acting as intermediate filters. This formulation is analogous to compartmental models in epidemiology, such as the SI(R)(S) models, where built areas (infected) “infect” vacant land (susceptible). Using this analogy, the time evolution of each component can be written as
\begin{equation} \label{eq:compartmental_sprawl_1}
    \begin{aligned}
     \frac{dA(t)}{dt} &= -\alpha N(t) A(t), \\
     \frac{dN(t)}{dt} &= \alpha N(t) A(t) - \gamma N(t),\\ 
     \frac{dM(t)}{dt} &= \gamma N(t).
    \end{aligned}
\end{equation}

Here $\alpha$ quantifies how much new development units generate the creation of new development units, and $\gamma$ the rate at which new development units become established development units. In this first simplified system, established development does not age. In reality, cities also follow redevelopment cycles : already-built sites might be destroyed and novel constructions be built upon. A more realistic approach takes account of aging and introduces a rate $\lambda$ for transitions from established to vacant land $M \rightarrow A$, corresponding to destruction. Eq. \ref{eq:compartmental_sprawl_1} become
\begin{equation} \label{eq:aging_aggregate}
\begin{aligned} 
 \frac{dA(t)}{dt} &= \lambda M(t)-\alpha N(t) A(t), \\
     \frac{dN(t)}{dt} &= \alpha N(t) A(t) - \gamma N(t),\\  
     \frac{dM(t)}{dt} &= \gamma N(t) - \lambda M(t).
\end{aligned}
\end{equation}
These aggregate quantities give rise to simple systems of differential equations that can be analyzed analytically, for example through the self-consistent equation for the asymptotic vacant land, $A_\infty = A(t \to \infty)$,
\begin{equation}
A_\infty = A(0) \exp\Big(-\frac{\alpha}{\gamma}(C - A_\infty)\Big) \,.
\end{equation}
However, this  it gives no information about the geometry of the spatial expansion. To translate this model in space, it is necessary to incorporate the vicinity of the growth of urban areas onto vacant land. This can be achieved through local diffusion to neighboring sites
\begin{equation}
\begin{aligned}
 \frac{\partial A(x, y, t)}{\partial t} &=  - \alpha N(x, y, t) A(x, y, t) + D_A \nabla^2N(x,y,t) \,, \\
 \frac{\partial N(x, y, t)}{\partial t} &=  \alpha N(x,y,t) A(x,y,t) - \gamma N(x,y,t),\\
    \frac{\partial M(x, y, t)}{\partial t} &= \gamma N(x, y, t). \\
\end{aligned}
\end{equation}

This model takes advantage of creation of available land close to newly developed units, quantified by the diffusion coefficient $D_A$. Then, new development units arise when enough vacant land is available locally. In order to study this model, a cellular automata implements its different mechanisms. On a square lattice, this discretized version introduces transitions based on local neighborhoods:
\begin{enumerate}
    \item \textbf{Diffusion:} Vacant land becomes available based on nearby development.
    \item \textbf{Transition:} Available land becomes developed randomly ($A \to N$).
    \item \textbf{Aging:} New development becomes established if no adjacent vacant land exists ($N \to M$).
\end{enumerate}
In order to break spatial symmetry, the authors assume that transitions from vacant land to developed land ($A \rightarrow N$) happen at random with rate $1-\Phi$. At this stage, the model incorporates spatial dynamics, but building aging is once again neglected. To include the decay of buildings, Eqs. \ref{eq:aging_aggregate} could be extended to account for space
\begin{equation*} 
\begin{aligned} 
\frac{\partial M(x, y, t)}{\partial t} &= \gamma N(x, y, t) - \lambda M(x, y, t), \\ 
\frac{\partial A(x, y, t)}{\partial t} &= \lambda M(x, y, t) - \alpha N(x, y, t) A(x, y, t) \\ 
& +D_A \nabla^2N(x,y,t) \,, 
\end{aligned}
\end{equation*}
but given the CA formulation, a simple rule where established development becomes vacant $M \to A$ after an age limit~$\tau$ suffices.

The authors investigate the long-term dynamics of a cellular automaton model, initialized with a compact seed representing a newly developed urban core. For early times ($t < \tau$), growth radiates outward from the central seed in an approximately circular pattern, forming a radially expanding front. At $t = \tau$, a transition occurs: although the initial outward expansion persists, earlier-developed sites begin to decay, initiating a secondary wave of redevelopment. As cycles of development and redevelopment continue, spatial irregularities accumulate and the initial morphological order gradually deteriorates. Ultimately, the initially well-defined circular pattern is lost, yielding a more disordered urban form.
%
%
This evolution is also evidenced by the transformation of the spatial distribution of new development in time. Initially, histograms of the distances of newly built sites from the urban center show a sharp peak, reflecting a well-defined growth front. As the first wave of decay occurs, the distributions become bimodal, indicating the coexistence of a development and redevelopment front. Over time, the distribution flattens and the clear growth fronts disappear, reflecting the loss of spatial order in the growth dynamics.

At this stage, realistic urban expansion dynamics cannot be captured without incorporating a crucial element: disorder in the substrate. Cities develop within natural environments featuring rivers, lakes, mountains, and other terrain variations, which strongly influence urban morphology by creating ``holes" that themselves exhibit fractal properties \cite{frankhauser1994fractalite}. To account for this, it is appropriate to introduce a vacant state, $V$, representing unbuildable land. The CA rules are adjusted to account for the vacant state: if a site is either available for development or already vacant, it becomes or remains vacant with probability $1 - \Gamma$. Importantly, this type of disorder resembles spatio-temporally correlated noise rather than quenched noise.

\begin{figure*}
    \centering
    \includegraphics[width=0.9\linewidth]{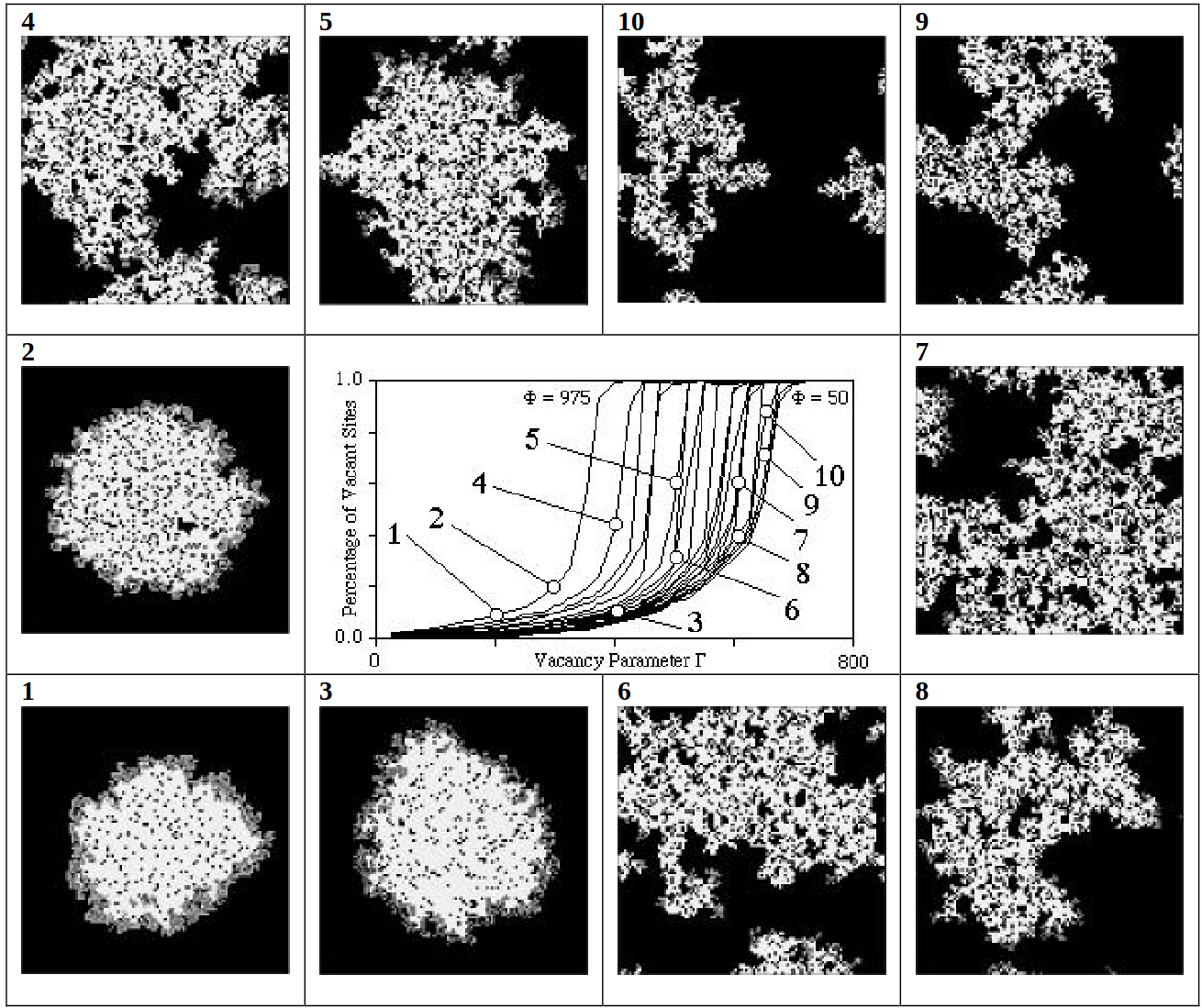}
    \caption{Morphologies obtained from Batty et al's model when sampling the parameter space $\{ \Phi, \Gamma \}$. $1-\Phi$ controls the rate of the transition $A \to N$, while $1-\Gamma$ controls the rates of transitions towards the vacant state. White : established development, gray : new development, black : vacant land. Varying the parameters brings from roughly circular, isotropic growth, to highly fragmented, disconnected patterns. Inset: variation of the percentage of vacant land versus $1000\Gamma$.  Source: From \cite{batty_1999_workingpaper}.}
    \label{fig:batty_dynamics}
\end{figure*}

\begin{figure*} 
    \centering
    \includegraphics[width=0.9\linewidth]{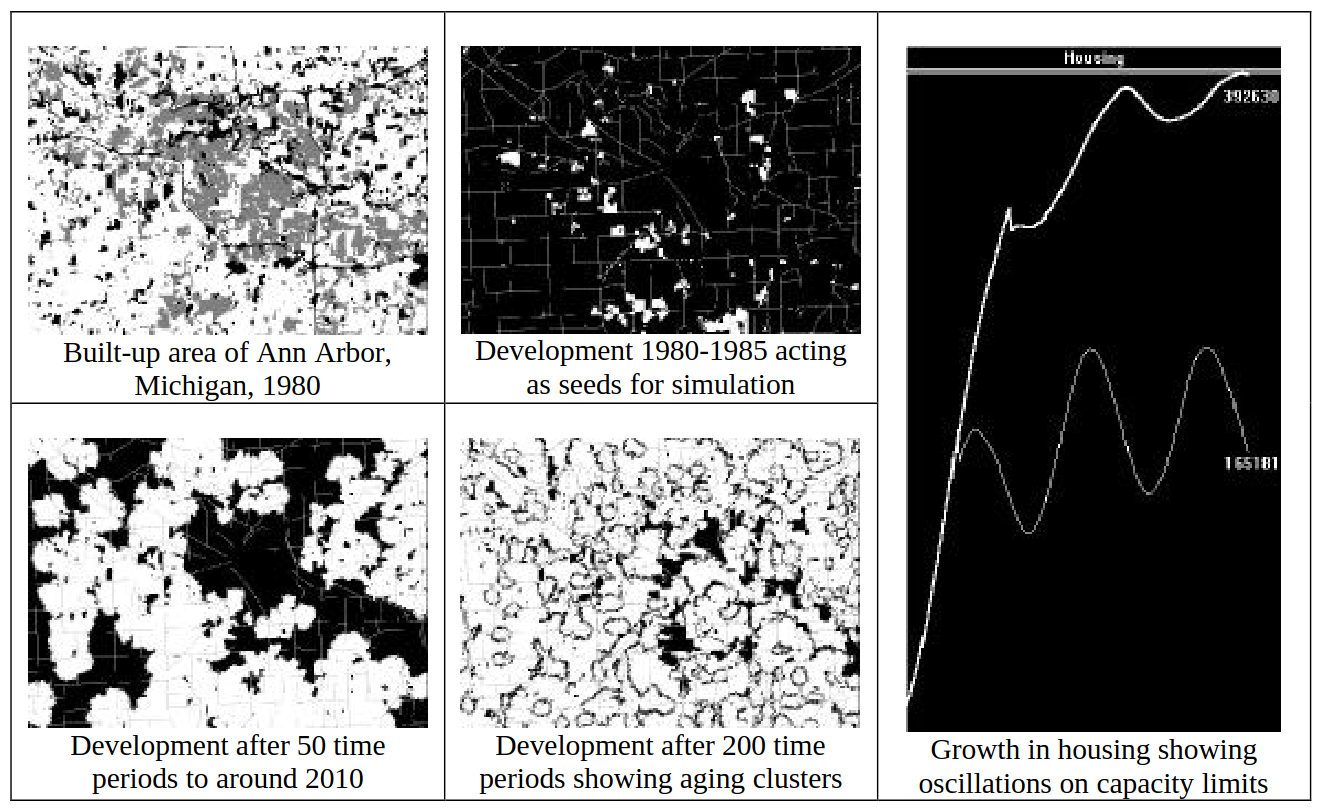}
    \caption{Results of simulations on Ann Arbor historical data. Fig.14 from \cite{batty_1999_workingpaper}.}
    \label{fig:annarbor}
\end{figure*}

Examples of the resulting morphologies are shown in Fig.~\ref{fig:batty_dynamics}. The central panel highlights an abrupt transition in the proportion of vacant sites, $V(t)$, occurring around $\Gamma = \Gamma^*(\Phi)$. Morphologies 1--2--3 in Fig.~\ref{fig:batty_dynamics} (corresponding to $\Gamma < \Gamma^*(\Phi)$ and large $\Phi$, i.e small $A \to N$ rate) are approximately circular and compact. In contrast, morphologies 7--8--9--10 (for $\Gamma \sim \Gamma^*(\Phi)$ and small $\Phi$, corresponding to large $A \to N$ rate) exhibit fragmented patterns, characterized by multiple disconnected components and intricate boundaries.
Using a framework based on these principles, described in \cite{batty1999}, but accounting for additional features such as land use or transport network, the cellular automata approach is validated on historical data of Ann Arbor, Michigan (see Fig.~\ref{fig:annarbor}). Initial seeds for the automata are built areas from 1980-1985 (see Fig.~\ref{fig:annarbor}, upper central figure) and long-time runs were analyzed. After long enough development, initial sites start aging and disappear before redeveloping (Fig.~\ref{fig:annarbor} lower left figure). At long time, only residual bands of development remain (see Fig.\ref{fig:annarbor} lower central figure). They are result of repeated waves of aging-redevelopment cycles. These simulations yield additional evidence that urban morphologies result of a combination of initial conditions, representing characteristics of a particular city, and a general development process, similar for all towns. 

This study examined the interplay between urban development and the aging of built areas, highlighting how these processes jointly shape the morphology of cities. These findings relate back to the stylized facts discussed in~\ref{subsec:fractal} and~\ref{subsec:shape}. The initial model was intentionally simple, relying only on two parameters, $\phi$ and $\gamma$, to capture the core dynamics. However, introducing additional features and realism inevitably increases the number of parameters and complicates the analysis. This type of model remains fundamental for urban policy making, as they can be used to evaluate interventions such as green belts, assessing how constraints on urban expansion might influence the long-term spatial morphology of cities.

\subsubsection{Simulations of urban land use patterns}

Beyond generic built-up area expansion, cities also display intricate land-use patterns. The cellular automaton model proposed by White and Engelen \cite{white1993} seeks to replicate this fundamental property of urban form. The model operates on a discrete lattice where each cell can exist in one of four states: vacant (V), housing (H), industrial (I), or commercial (C). Urban growth is imposed exogenously through fixed growth rates $N_i$ for each land-use type $i \in {\mathrm{H}, \mathrm{I}, \mathrm{C}}$. At every simulation step, $N_i$ vacant cells are selected and converted into the corresponding land use. Conversion decisions are guided by a potential function that combines local spatial context with stochastic perturbations. Interactions extend over a neighborhood of radius six, subdivided into distance bands to capture distance-decay effects. Importantly, the model assumes a monotone land-use hierarchy $V < H < I < C$: once a cell changes state, it cannot revert to a lower one, thereby representing the irreversibility of land conversion.
For all possible transitions, at each time, a transition potential is computed
\begin{equation}
    \Pi_{ij} = S \left( 1 + \sum_{h,k,d} m_{kd} I_{hd}  \right)
\end{equation}
where $m_{kd}$ is the weight of cells of type $k$ at distance $d$, $h$ is the index of cells at a given distance band, $I_{hd}$ is 1 if $h=k$ otherwise $0$. The quantity $S$ is a stochastic perturbation of the form $S = 1 +\left( - \ln R  \right)^\alpha$ where $R$ is uniformly distributed in $[0,1]$ and $\alpha$ controls the size of the perturbation. At each iteration, the~$N_i$ cells with the highest potential $\Pi_{ij}$ for land use~$i$ are selected (in order: C, then I, then H). If a cell appears in the top list for multiple uses, a secondary stochastic rule is applied to resolve conflicts while ensuring the correct number of transitions for each type.

Simulations of the model for different values of $\alpha$ have been examined using fractal analysis, focusing on mass–radius relations, cluster-size distributions, and perimeter scaling. A box-counting analysis of the built area, $A(r) \sim r^D$, reveals a clear bifractal structure: inner zones tend to grow compactly, while outer zones develop into more fragmented, dendritic patterns. For example, with disorder parameter $\alpha = 2.5$, the estimated exponents are $D_{\text{inner}} \approx 1.93$ and $D_{\text{outer}} \approx 1.23$. Comparable bifractal behavior is observed when analyzing individual land-use categories \cite{white1993}. In particular, as expected we have 
\begin{equation}
    D_{\text{commerce}} < D_{\text{industry}} < D_{\text{housing}} \,.
\end{equation}
However, for the sake of empirical comparison, only a few cases (industry for Cincinnati and Milwaukee, commerce for Atlanta) correspond to bi-fractal relationships, while all the other categories of land uses for the 16 cities under study show mono-fractal behavior.

Second, the distribution of cluster sizes--defined as connected sets of cells corresponding to a particular land use--exhibits power-law behavior of the form
\begin{equation}
    N_c(s) \propto s^{-\gamma},
\end{equation}
where $N_c(s)$ is the number of clusters of size $s$. The exponent $\gamma$ is found to lie in the range $1.33$ to $1.71$. Comparisons with four U.S. cities show good agreement with this scaling behavior, except for Atlanta, which displays an anomalously large cluster and very few intermediate-sized clusters. Beyond the presence of (bi)fractal densities and scale-free cluster size distributions, the simulated cellular cities also exhibit fractal perimeter properties. The fractal dimension of cluster boundaries is found to lie in the range $D_p \approx 1.41$ to $1.45$, suggesting irregular, self-similar urban edges. These values are compared to the fractal dimension of the boundary of Cardiff, around~$1.2 - 1.3$, found in \cite{batty_longley_1987_fractal}.

The spatial complexity of White and Engelen's generated morphologies--namely bifractal site distributions, power-law cluster distribution, and fractal perimeter-length relation--emerges from local transition rules, mirroring cities' self-organized evolution, perturbed by stochasticity. This relates to the stylized facts discussed in~\ref{subsec:component} and~\ref{subsec:fractal}. The model's behavior critically depends on the noise parameter $\alpha$, which interpolates between regular, over-determined geometries (low $\alpha$), and disordered, structureless configurations (high $\alpha$). The model succeeds in reproducing several city-like characteristics, most notably the fractal and spatially dispersed nature of urban form. A particularly appealing aspect is the incorporation of interactions at distance, combined with stochastic effects, which echoes real urban dynamics. Nonetheless, a stronger connection to empirical data would be desirable. Indeed, a key limitation lies in the parameterization: the model requires a large number of parameters--one for each pair of land-use type and distance band--which makes it difficult to link them directly to observable quantities. Furthermore, this intricate parameter space also complicates interpretation, making it difficult to draw robust conclusions, even when multiple simulations are performed.

\subsubsection{Stochastic cellular model and transport network}

Ward et al. \cite{ward2000stochastically} propose a cellular automata (CA) model for urban growth that integrates local decision-making processes with stochastic constraints to account for broader-scale urban development factors. The model is applied to simulate urban growth in South East Queensland, Australia.

This model is structured as a two-dimensional CA system with the following components. As usual in CA modeling each cell represents a land unit with a specific state (e.g., developed or undeveloped) and transition rules govern how the development depends on local neighborhood configurations, and on the access to roads and infrastructures. In addition, they introduce stochastic constraints that represent socio-economic factors that introduce randomness in the system. The urban growth takes place on a two-dimensional grid of size $L$ and each cell $i$ is in a state $S_t(i)\in\{0,1\}$ which corresponds to an undeveloped ($S_t=0$) or developed cell ($S_t=1$). A new development (birth) occurs at $j$ at time $t+1$ if a neighboring cell $i$ is developed:
\begin{equation}
    B_{ji}^{t} = 
    \begin{cases}
        1, & \text{if } u > \beta, \\
        0, & \text{otherwise}.
    \end{cases}
\end{equation}
where $\beta$ is an intrinsic growth rate and $0<u<1$ is a random uniform variable. The underlying assumption is that if a cell $i$ is already developed, then the necessary infrastructure (e.g., utilities and services) is available, enabling the development of neighboring cells. A key constraint of the model is that any newly developed unit must have access to the transportation network. This access condition is encoded in the variable $T_k$, defined as
\begin{align}
T_k = 
\begin{cases}
1, & \text{if cell } k \text{ belongs to the transport network}, \\
0, & \text{otherwise}.
\end{cases}
\end{align}
A candidate cell $j$ is considered accessible if it has at least one and at most $n$ neighbors that are on the transport network. This access condition is formalized by the function
\begin{equation}
A_t(j) = 
\begin{cases}
1, & \text{if } 1 \leq \sum\limits_{k \in \Omega(j)} T_k \leq n, \\
0, & \text{otherwise},
\end{cases}
\end{equation}
where $\Omega(j)$ denotes the neighborhood of cell $j$, which may be either the Moore neighborhood (8 cells) or the von Neumann neighborhood (4 cells), depending on the implementation.

Given that cell $i$ is already developed, the probability $p_{ij}(t)$ that a neighboring cell $j \in \Omega(i)$ will develop depends on its distance $d_{ij}$ to cell $i$. The influence of cell $i$ on the development of cell $j$ decays with distance, and is modeled as
\begin{equation}
p_{ij}(t) = \left(1 - \frac{d_{ij}}{d_{\max}}\right)^\alpha,
\end{equation}
where $d_{\max}$ is the maximum distance in the system and $ \alpha > 0$ is a tunable parameter controlling the strength of the decay.

To ensure no directional bias in the choice of the developing cell, the actual selection of the new development location $ L_t(j)$ is made using a cumulative probability rule:
\begin{equation}
L_t(j) = 
\begin{cases}
1, & \text{if } \sum\limits_{z=1}^{j-1} p_{iz}(t) \leq u \leq \sum\limits_{z=1}^{j} p_{iz}(t), \\
0, & \text{otherwise},
\end{cases}
\end{equation}
where $u \in [0,1]$ is a random number drawn from a uniform distribution.

Once a specific neighboring cell $j$ is selected for development, its state is updated at the next time step according to
\begin{equation}
S_j(t+1) = A_j(t) \, L_j(t) \, B_i(t) \, S_i(t),
\label{eq:caevol}
\end{equation}
where $B_i(t)$ represents an additional condition (e.g., buildability or suitability), and $S_i(t)$ is the current state of the initiating cell $i$. This equation captures the coupled dynamics of accessibility, spatial influence, and stochasticity in the urban growth process.

\begin{figure*}
    \centering
    \includegraphics[width=0.9\linewidth]{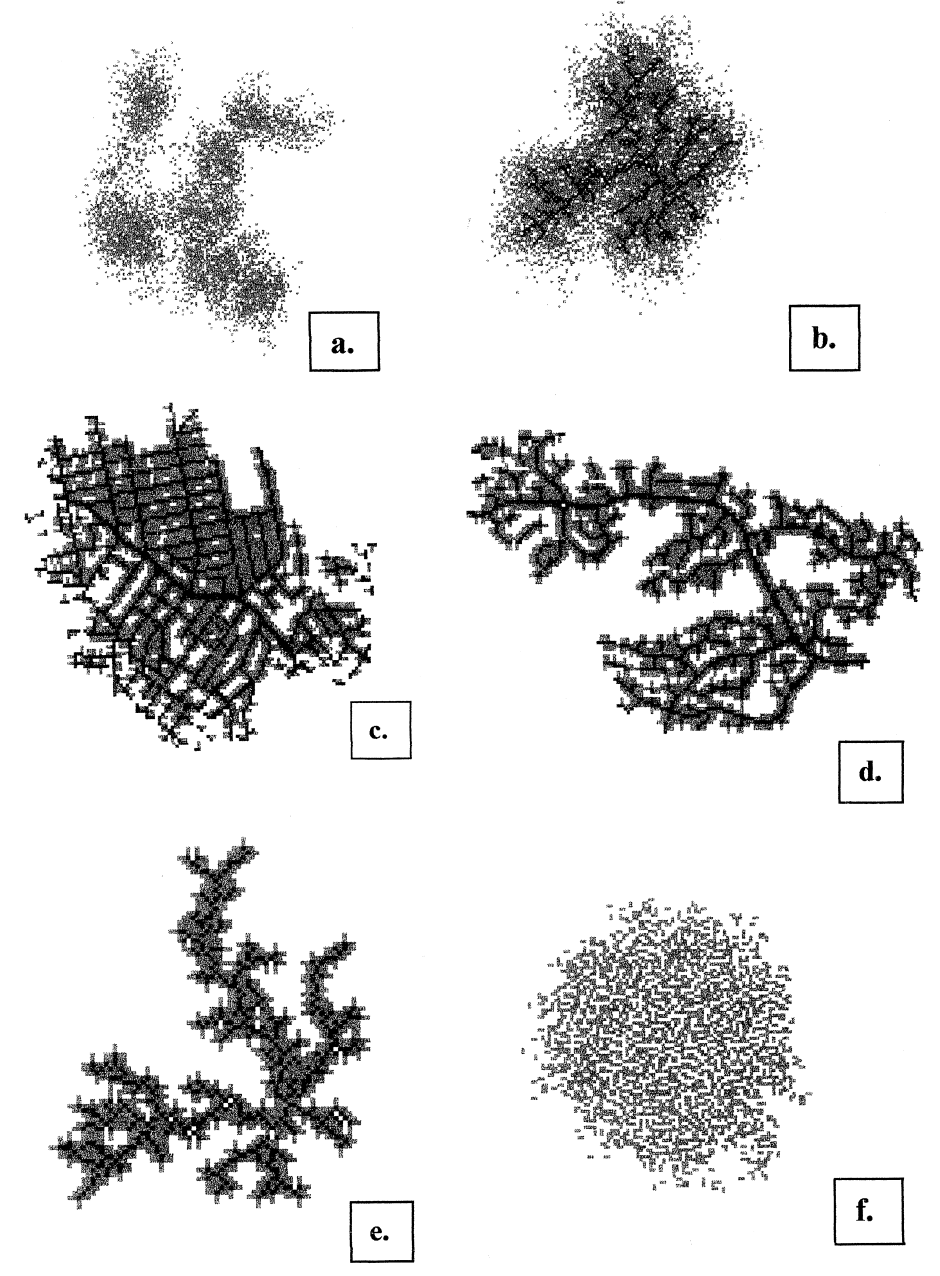}
    \caption{Simulated urban growth patterns constrained by different transport networks: (a) no network; (b) no network with growth seeded with a diffusion-limited aggregation (DLA) network; (c) actual regular grid network; (d) actual culdesac network; (e) DLA-derived network; (f) access-constrained growth with no network. Source: from \cite{ward2000stochastically}.}
    \label{fig:ward}
\end{figure*}

This evolution equation \ref{eq:caevol} can be used to simulate different types of growth shown in Fig.~\ref{fig:ward}.
In Fig.~\ref{fig:ward}(a,b) there is no transport network and the access constrain term $A_j(t)$ is always set equal to 1. The difference of these two patterns lies in the initial condition that is uniformly randomly distributed cells for 
Fig.~\ref{fig:ward}(a), while for Fig.~\ref{fig:ward}(b), the initial pattern is obtained with a DLA process (see section \ref{subsec:dla}). Figures~\ref{fig:ward}(c,d) illustrate simulated residential development on two real transport networks extracted from digital maps of Brisbane, Australia: a regular grid and a cul-de-sac layout, respectively. These simulations use the model described by Eq.~\ref{eq:caevol}, in which detailed knowledge of the underlying transport infrastructure is essential. Since the transport network is taken as a fixed input—determined once and for all—there is no feedback loop between urban expansion and infrastructure growth. This coupling, discussed in Section~VI.D, prevents the model from capturing the co-evolutionary dynamics often observed in real urban systems.

To investigate broader morphological effects, additional simulations were conducted on synthetic networks. Figure~\ref{fig:ward}(e) shows growth on a network generated using diffusion-limited aggregation, preserving the same initial conditions and neighborhood rules as in earlier simulations (e.g., Fig.~\ref{fig:ward}(a)), with a block size of two cells.

The patterns observed in Figs.~\ref{fig:ward}(a--d) demonstrate that plausible urban morphologies can emerge from CA models with rules based on local access and spatial correlation. However, these configurations do not exhibit genuine emergent properties associated with self-organization. In particular, the morphologies in Figs.~\ref{fig:ward}(a) and \ref{fig:ward}(b) follow expectations under friction-of-distance constraints, while those in Figs.~\ref{fig:ward}(c,d) reflect development patterns shaped by access to a pre-existing transport network.

Only the configuration in Fig.~\ref{fig:ward}(f) begins to show weak signs of self-organization. In this case, no predefined transport network is assumed, and the access rule is modified to require that each new development unit maintains at least one adjacent vacant cell, interpreted as a proxy for future transport access. Orientation is randomly assigned among the four cardinal directions, provided that the selected site is unoccupied. The resulting pattern resembles a random tiling of units with minimum spacing. However, since there is no explicit rule for ensuring global connectivity of the vacant cells (i.e., no emergent transport network), the resulting urban morphology remains unrealistic.

The overarching aim of such simulations is not to replicate fine-scale urban layouts, but rather to provide a coarse-grained planning tool for exploring regional urban expansion under varying demographic and land-use constraints. For such applications—e.g., evaluating plausible growth patterns under future population scenarios—capturing regional-scale morphology is often sufficient. 


Urban expansion results from the combined effects of physical, economic, and institutional constraints. Natural barriers such as slopes or water bodies limit development. Zoning laws and planning regulations shape its direction, while economic factors—like transport access or land values—affect where growth is most likely to occur. To integrate these large-scale influences, the model introduces a constraint vector $\mathbf{C}_n$ for each site $j$, defined as:
\begin{equation}
\mathbf{C}_n = 
\begin{bmatrix}
    I_j \\ 
    E_j \\ 
    G_j 
\end{bmatrix},
\end{equation}
where $I_j$ denotes institutional constraints (e.g., zoning exclusions), $E_j$ encodes prohibitive physical constraints (e.g., presence of water or steep slope), and $G_j$ captures modifying constraints (e.g., distance to roads, commercial hubs, or employment centers).

The development probability of cell $j$ is then modeled as:
\begin{equation}
g_j = W_m \prod_{m=1}^{N} I_{jm},
\end{equation}
where $I_{jm}$ represents the influence of the $m^{\text{th}}$ constraint and is computed via
\begin{equation}
P_{jm} = \exp\left(-\lambda \frac{q_{jm}}{q_{\max}}\right).
\end{equation}
Here, $q_{jm}$ is the constraint value (such as distance to road), $q_{\max}$ is the maximum over all cells, and $\lambda$ is a decay parameter that tunes the sensitivity to the constraint.

Incorporating these constraint probabilities, the transition rule for the state of cell $j$ at time $t+1$ becomes
\begin{equation}
S_j^{t+1} = A_j^t L_j^t B_i^t S_i^t \prod_{n=1}^{N} C_{jn},
\end{equation}
where $A_j^t$ and $L_j^t$ are, respectively, the accessibility and location preference terms, $B_i^t$ the buildability condition, and $S_i^t$ the state of an influencing neighbor.

This modeling framework was applied to the urban dynamics of the Gold Coast, Australia—a region that experienced rapid expansion, with urban area increasing by $32\%$ between 1988 and 1995. The model is data-driven and requires:
\begin{itemize}
    \item The initial urban footprint, obtained from Landsat TM imagery;
    \item Spatially explicit zoning and planning regulations (institutional constraints);
    \item Modifying constraints, such as slope, road proximity, and accessibility to services.
\end{itemize}

Two classes of simulation scenarios were explored. The first emphasized transport-network-based growth, leveraging detailed representations of the road infrastructure. The second considered constrained growth driven by friction-of-distance effects, highlighting the role of spatial accessibility.

Model performance was evaluated by comparing simulated and observed urban growth patterns in selected subregions. In the Helensvale region, $65\%$ of urban change was correctly reproduced, with a $34\%$ omission error. In the Labrador region, accuracy improved to $74\%$, with a 26\% omission. Across the broader Gold Coast, the overall agreement reached $63\%$. 

These results demonstrate that cellular automata incorporating constraint-based rules can effectively simulate regional-scale urbanization. The inclusion of stochastic constraint terms captures the uncertainty and heterogeneity inherent in real-world urban development. Such models provide a quantitative framework for evaluating alternative land-use policies, simulating growth under different planning regimes, and supporting decision-making in spatial planning. These are essential considerations, in the light of the empirical studies discussed in~\ref{subsec:shape},~\ref{subsec:street} and~\ref{subsec:impact}.

Overall, this approach frames urban growth as a self-organizing process influenced by environmental, institutional, and economic factors. The stochastic treatment of constraints allows for realistic modeling of complex urban trajectories and facilitates scenario-based exploration of future spatial dynamics.


\subsubsection{Modeling the co-evolution of urban form and transport networks}

In contrast to earlier approaches where infrastructure is assumed fixed or exogenous, the model proposed by Raimbault et al.~\cite{raimbault2016hybridnetworkgridmodelurban} introduces a co-evolutionary framework in which urban growth and the development of transportation networks are mutually dependent. In this model, transport infrastructure not only facilitates human mobility and access to urban functions, but also evolves dynamically in response to urban expansion. Conversely, the spatial pattern of settlements is shaped by both the structure and the accessibility provided by the network. By explicitly modeling this feedback loop, the system can exhibit emergent organization, capturing stylized facts such as radial or corridor-based urban morphologies observed in real cities.

Space is represented by a square lattice of size $N \times N$, and time progresses in discrete steps. Each site $(i,j)$ can be either developed or empty, as indicated by a binary state variable $\delta(i,j,t) \in \{0,1\}$. In parallel, the transport network is represented by a temporal graph $G(t) = (V(t), E(t))$, initialized at $t=0$ by a set of urban centers $C_0$ endowed with functional activities such as residential, commercial, or industrial land use. 

To guide urban development, the model introduces a land value function $v(i,j,t)$ that quantifies the relative attractiveness of each cell. This value is computed from four explanatory spatial variables: the local density $d_1$ of developed cells in a radius $\rho$ around site $(i,j)$, the Euclidean distance $d_2$ to the nearest road, the network distance $d_3$ to the closest urban center.
An addition measure of functional accessibility is introduced and is defined as the aggregated network distance to centers offering each activity type
\begin{equation}
    d_4(i,j,t) = \left(\frac{1}{a_{\max}} \sum_{a=1}^{a_{\max}} d_3(i,j,t;a)^{p_4} \right)^{1/p_4},
\end{equation}
where $d_3(i,j,t;a)$ denotes the network distance from $(i,j)$ to the closest center offering activity $a$, and $p_4 \geq 1$ is a tunable parameter. These four variables are linearly combined to produce the normalized land value:
\begin{equation}
    v(i,j,t) = \frac{1}{\sum \alpha_k} \sum_{k=1}^4 \alpha_k \cdot \frac{d_{k,\max}(t) - d_k(i,j,t)}{d_{k,\max}(t) - d_{k,\min}(t)},
\end{equation}
where $\alpha_k$ are parameters determining the weight of each explanatory factor. Intuitively, sites that are closer to roads, activity centers, or existing development will score higher (i.e. will have a higher value of $v$)  and are therefore more likely to be selected for growth.

The simulation proceeds iteratively. At each time step, the land value function is computed for all undeveloped sites. A fixed number of sites with the highest $v$ values are selected for development. If a newly developed site lies beyond a given threshold distance $\theta_2$ from the existing road network, it is connected by adding an orthogonal link to the nearest road node. In this way, the transport network grows endogenously in response to settlement expansion.

The presence of feedback is a defining feature of this model. As new sites are developed, they modify the local density, affect accessibility, and induce network expansion, which in turn alters the value landscape and modifies the dynamics of future growth. This endogeneity is expected to generate non-trivial spatial structures.

The model’s behavior is studied using several global metrics. The overall density $D(t)$ and Moran’s $I(t)$ quantify the degree of spatial aggregation. The transport network is characterized by the detour index $S(t)$, which measures the relative efficiency of paths in the network~\cite{barthelemy2011}, and a global accessibility indicator defined as
\begin{equation}
    A(t) = \left( \frac{1}{\sum_{i,j} \delta(i,j,t)} \sum_{\delta(i,j,t)=1} \left( \frac{d_4(i,j,t)}{d_{4,\max}(t)} \right)^{p_A} \right)^{1/p_A}.
\end{equation}

The authors explore the parameter space defined by the weights $\alpha_k \in \{0, 0.2, \ldots, 1\}$, simulating urban growth over 30 time steps. This scan reveals the emergence of three main morphological archetypes, when inspected in the~$(D,I)$ space. When the land value is dominated by local density ($\alpha_1 \gg \alpha_k$), growth results in dispersed rural-like patterns. When distance to the road network dominates ($\alpha_2$ large), growth follows linear patterns along transport corridors. When network distance to the center is the principal factor ($\alpha_3$ large), radial city structures emerge. These outcomes recall Le Corbusier's typology of urban forms. 
At very low density,~$D\approx0$, configuration of streets and buildings remind rural communities, characterized by high fragmentation. When distance to the street only is optimized, linear morphologies appear, while when mixing optimization of distance to street and distance to center, and radial forms emerge.


While flexible, this model also presents some limitations. First, it involves a relatively large number of parameters, many of which interact in non-trivial ways. This can make interpretation and calibration challenging, and may affect the robustness of the resulting morphologies. Second, fitting the model to real-world urban data is not straightforward: empirical metrics such as density or spatial autocorrelation may provide useful constraints, but are often insufficient to uniquely determine parameters or reproduce a given city’s form. Even so, the framework provides a valuable proof of concept for modeling co-evolving systems in urban contexts. It emphasizes the role of transport networks not only as constraints but also as active agents in shaping urban form -- see also section~\ref{subsec:shape} and~\ref{subsec:street}. Its ability to generate diverse morphologies from simple ingredients, and to highlight regimes where feedbacks dominate, makes it a useful case study for physicists interested in spatially extended dynamical systems, complex networks, and emergent phenomena.


\subsection{Microeconomic models }

\subsubsection{Periurban spatial configurations}

In their study on periurbanisation, the process through which urban expansion extends into rural areas and produces mixed urban–rural landscapes at the urban fringe, Caruso et al.~\cite{caruso2007} develop a hybrid model that couples microeconomic land-use theory with a dynamic cellular automata framework. The goal is to simulate the emergence of residential-agricultural configurations at the urban fringe, and to understand how preferences for environmental and social amenities shape these mixed landscapes. This approach extends classical monocentric models, incorporating both endogenous land rent dynamics and localized neighborhood interactions.

The core of the model is a bid-rent framework that combines agricultural and residential land uses. Farmers produce agricultural goods and sell them at the central business district (CBD). Their willingness to pay for land—the agricultural bid-rent function—is assumed to decay linearly with distance from the CBD:
\begin{equation}
    \Phi(d) = \Phi_0 - b d,
\end{equation}
where $b$ is the unit transport cost and $d$ the Euclidean distance to the CBD.

Households, by contrast, derive utility from three components: consumption of a composite good $Z$, environmental externalities $E$ (such as green space), and social externalities $S$ (such as access to services or transit). Their utility function is of Cobb-Douglas form:
\begin{equation} \label{eq:caruso2007_utility}
    U(Z, E, S) = Z \cdot E^\beta \cdot S^\gamma,
\end{equation}
where $\beta$ and $\gamma$ are preference parameters. Utility functions are commonly used in economics to represent agents’ preferences (see Section \ref{subsec:amm} for a brief discussion of parallels with physical energy, and e.g., \cite{Lemoy_2011} for an example of how economic concepts can be mapped onto statistical physics frameworks. The household budget constraint is given by
\begin{equation}
    Y = a d + Z + R,
\end{equation}
where $Y$ is income, $a$ is the commuting cost per unit distance, and $R$ is the rent paid for land. Substituting into the utility maximization problem under this constraint, the maximum rent a household is willing to pay to reach utility level $\bar{U}$ at location $d$ is
\begin{equation}
    \Psi(d) = Y - a d - \bar{U} \cdot E^{-\beta} \cdot S^{-\gamma}.
\end{equation}

At each time step, a new household arrives from outside the system. Migration occurs only if the utility at some location exceeds the reservation utility $\bar{U}$. Since land is owned by multiple competing farmers, the household pays the agricultural rent $\Phi(d)$. The resulting realized utility is then:
\begin{equation}
    U^t = (Y - a d - \Phi(d)) \cdot E^\beta \cdot S^\gamma.
\end{equation}
Relocation within the urban fringe is allowed at no cost. Households will switch to locations where their utility increases, until an equilibrium is reached in which all households achieve the same utility as the most recent migrant. The implied bid-rent in this short-term equilibrium is:
\begin{equation}
    \Psi^t = Y - a d - U^t \cdot E^{-\beta} \cdot S^{-\gamma}.
\end{equation}
Long-run equilibrium is obtained once the marginal utility of the last migrant equals the reservation utility $\bar{U}$, and no further migration occurs.

This microeconomic framework is embedded within a spatial cellular model. The space is discretized as a square lattice centered on the CBD. Each cell can be either occupied by a household or a farmer. Dynamics are governed by local interactions. For each cell $(i,j)$, a neighborhood $\mathcal{N}_{ij}$ is defined, typically consisting of sites within a radius $\hat{f}$. Each pair of sites $(k,l)$ is associated with a weight $w_{kl}$ that controls the interaction strength. The residential density around $(i,j)$ is given by a weighted average
\begin{equation}
    \rho_{ij} = \frac{\sum_{(k,l) \in \mathcal{N}_{ij}} w_{kl} H_{kl}}{\sum_{(k,l) \in \mathcal{N}_{ij}} w_{kl}},
\end{equation}
where $H_{kl}$ equals 1 if the site is occupied by a household, and 0 otherwise.

Externalities are then expressed as functions of this local density. Environmental quality decreases with residential density:
\begin{equation}
    E_{ij} = \exp(-\rho_{ij}^\theta),
\end{equation}
while social amenities increase with it:
\begin{equation}
    S_{ij} = \exp(\rho_{ij}^\phi).
\end{equation}
These effects combine in the utility function~\eqref{eq:caruso2007_utility} as:
\begin{equation}
    L_{ij} = \exp\left(\gamma \rho_{ij}^\phi - \beta \rho_{ij}^\theta\right),
\end{equation}
where $L_{ij}$ represents the total externality effect at site $(i,j)$. The optimal neighborhood density $\rho^*$ that maximizes $L_{ij}$ satisfies:
\begin{equation}
    \rho^* = \left( \frac{\phi \gamma}{\theta \beta} \right)^{1/(\theta - \phi)}.
\end{equation}

\begin{figure}
    \centering
    \includegraphics[width=\linewidth]{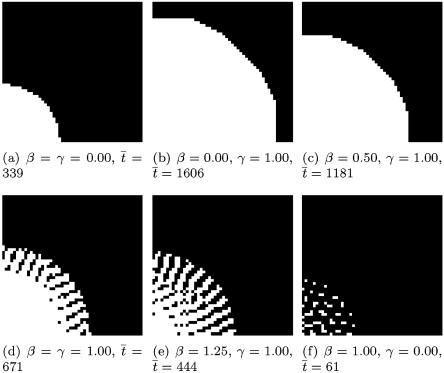}
    \caption{\textbf{Emergent urban forms in the Caruso et al. model.} Simulated residential-agricultural patterns at the urban fringe, under varying strengths of green and social amenity preferences. Fragmented or compact morphologies emerge depending on parameter values. Source: From \cite{caruso2007}.}
    \label{fig:caruso_2007_morphos}
\end{figure}

The simulation results demonstrate that under a wide range of parameter values, a mixed periurban belt spontaneously emerges between the dense urban core and surrounding agricultural land. This belt consists of residential clusters interspersed with farmland. The precise morphology of this fringe is controlled by the trade-off between social and environmental preferences, as shown in Figure~\ref{fig:caruso_2007_morphos}. When social externalities dominate ($\gamma \gg \beta$), compact, polygonal urban shapes emerge. Conversely, when green space preferences dominate ($\beta \gg \gamma$), development becomes fragmented, forming scattered clusters with high landscape heterogeneity. So far, the description of these boundary geometries remains qualitative. Quantitative measures, such as fractal dimensions, interface roughness, or spatial dispersion, would allow for a systematic mapping of the resulting morphologies and enable a more precise analysis of how the parameters and their interactions shape urban form.

A notable feature of the model is that the spatial extent of the mixed periurban belt can be analytically predicted from the underlying microeconomic framework. Let $\tilde{d}$ denote the outer edge of the traditional monocentric city (where $\beta = \gamma = 0$), and define $d_u$ and $d_c$ as the inner and outer limits of the mixed belt. These are given by:
\begin{equation}
    d_c = \tilde{d} + \frac{\bar{U} (L(\rho^*) - 1)}{(a - b) L(\rho^*)}, \quad 
    d_u = \tilde{d} + \frac{\bar{U} (L(\rho=1) - 1)}{(a - b) L(\rho=1)}.
\end{equation}

The simulations reproduce a wide variety of stylized periurban morphologies, from compact forms to dispersed, leapfrogging development. At early stages, urbanization tends to leap over agricultural zones, forming residential islands. These are eventually filled in, yielding more contiguous development. The rent dynamics are rich: rents respond not only to commuting distance, but also to changing neighborhood structure and the dynamic competition between residential and agricultural uses.

One of the key contributions of this model is its ability to reproduce fragmented urban patterns endogenously (see~\ref{subsec:shape}), without requiring external shocks or heterogeneity in land quality. The emergence of mixed periurban belts is shown to be a direct consequence of household-level utility maximization under amenity trade-offs. The model further challenges the traditional view that dispersed development is necessarily suboptimal. Indeed, some configurations of sprawl can be welfare-enhancing, reflecting voluntary household choices to trade off longer commutes for improved environmental or social amenities.

However, this framework is limited by its reliance on a relatively large set of parameters ($\beta$, $\gamma$, $\theta$, $\phi$, $\hat{f}$), which may be difficult to calibrate, and by assumptions (e.g. linear commuting costs) whose empirical grounding is often unclear and not easily related to observable data. These limitations are compounded by strong and somewhat ad hoc hypotheses built into the framework, such as perfect tenant mobility or the use of specific functional forms for externalities expressed as simple functions of local density. While such assumptions ensure analytical and computational tractability, they risk narrowing the generality of the conclusions. Moreover, to assess whether the resulting spatial morphologies represent genuine features of the dynamics or artifacts of these modeling choices, one would need to systematically quantify the generated patterns through suitable observables—such as cluster-size distributions, fringe roughness, or correlation lengths—and test whether the qualitative results persist under variations in system size and small perturbations of parameter values.

\subsubsection{Dendritic cities and dielectric breakdown}

Following the previous framework, in \cite{Caruso2011}, Caruso et al. develop a dynamic microeconomic model of urban growth that exhibits striking morphological similarities with dielectric breakdown patterns. The model combines standard household utility maximization with evolving road infrastructure and endogenous land-use dynamics on a 2D lattice.

The spatial structure consists of a square grid, where each cell can be in one of three states: agricultural, residential, or road. The CBD is located at the center, with two orthogonal roads intersecting there and serving as the initial transport infrastructure. Over time, the lattice progressively fills with new developments, but only through incremental improvements—no demolition or reuse of land is allowed. At each time step, a new household enters the system and competes for land with existing farmers. Landlords allocate land to the highest bidder, which can be either a farmer or a resident.

Farmers produce food with constant returns to scale (that is, operate under constant efficiency as their production increases) and generate green space externalities for neighboring residents. Households, arriving sequentially, choose locations by maximizing a Cobb–Douglas utility function,
\begin{equation}
    U = k Z^\delta H^\alpha E^\beta S^\gamma,
\end{equation}
where $Z$ denotes consumption of a reference good (\textit{num\'eraire} in economic jargon), $H$ represents housing, $E$ is an environmental (green space) externality, and $S$ is a social or public goods externality (e.g., services, transport access). Notice the addition of the housing in the utility function, and of parameters~$\alpha$ and~$\delta$ with respect to the framework presented in the previous section. The exponents $\alpha, \beta, \gamma$ encode household preferences, and $\delta = 1 - \alpha$ ensures constant returns in expenditure.

The household budget constraint is
\begin{equation}
    Y - \theta d = Z + S R,
\end{equation}
where $Y$ is income, $\theta$ is the unit cost of commuting, $d$ is the distance to the CBD, $R$ is the unit land rent, and $S$ (used both as a variable and a utility component) reflects the required land size for housing. Substituting the constraint into the utility function yields an indirect utility:
\begin{equation} \label{eq:carruso-iuf}
    V = (Y - \theta d) R^{-\alpha} E^\beta S^\gamma.
\end{equation}

Each arriving household selects a cell $l$ that maximizes this indirect utility, taking into account the commuting cost $\theta d_l$, the land rent $R_l$, and the levels of environmental and public good externalities $E_l$ and $S_l$. The latter are endogenous and evolve with local density. Specifically, the environmental externality decreases with density:
\begin{equation}
    E_l^t = \exp(-\rho_l^{t-1}),
\end{equation}
while the public goods externality increases with it:
\begin{equation}
    S_l^t = \exp(\sqrt{\rho_l^{t-1}}),
\end{equation}
where $\rho_l^t$ is the local residential density around cell $l$ at time $t$.

Due to competition among landlords, the rent paid by each new resident equals the agricultural rent $\Phi$, leading to a simplified form of the indirect utility:
\begin{equation}
    V_l^t = (Y - \theta d_l^t) \Phi^{-\alpha} \left( E_l^t \right)^{\beta} \left( S_l^t \right)^{\gamma}.
\end{equation}
The city grows as long as the maximum utility $V_l^t$ exceeds a threshold $\bar{V}$, which reflects the attractiveness of the city relative to the outside world.

Urban expansion requires access to the transport network. The model includes two rules governing road infrastructure. The first is a connection rule: a newly built residential site must connect to an existing road, and any new road must also connect to another road. The second is a minimum expropriation rule: if a newly chosen residential site is not adjacent to the road network, the public authority expropriates the minimal number of intermediate cells to build a road connection.

\begin{figure}
    \centering
    \includegraphics[width=1\linewidth]{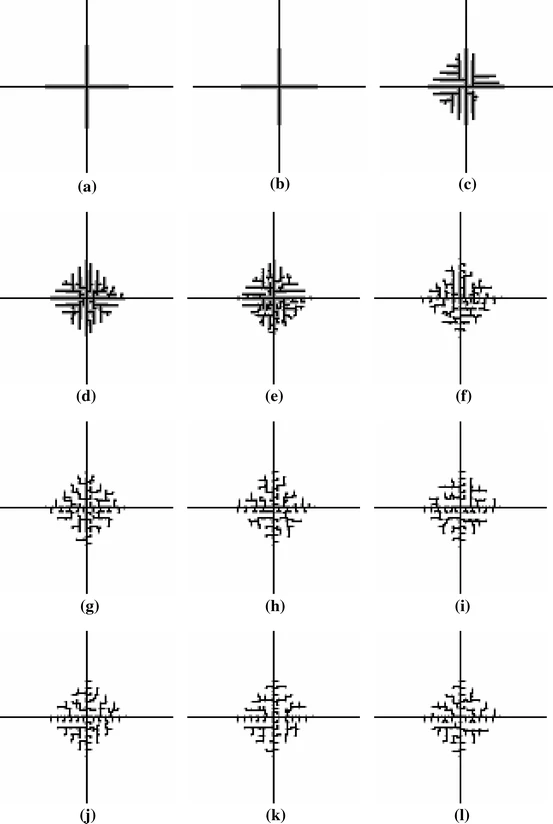}
    \caption{Resulting morphologies as $\beta$ varies with $\gamma = 0$. a) $\beta$ = 0.00, b) 0.12, c) 0.13, d) 0.22, ..., l) 2.50. Dendritic patterns emerge abruptly around $\beta \approx 0.13$ and persist with increasing $\beta$. Source : From \cite{Caruso2011}.}
    \label{fig:caruso_varyingbeta}
\end{figure}

The model enables exploration of the influence of key parameters on urban morphology at equilibrium. Figure~\ref{fig:caruso_varyingbeta} shows the effect of varying $\beta$, which controls the strength of preference for green space, while keeping $\gamma = 0$. When $\beta = 0$, the classical Alonso linear development is recovered: growth occurs along roads. As $\beta$ increases, dendritic structures emerge abruptly around $\beta \approx 0.13$, reminiscent of a non-equilibrium phase transition, giving rise tree-like growth and leapfrogging developments. Beyond this threshold, the morphology stabilizes into a dendritic regime with a constant fractal dimension $D_f \approx 1.75$, reminiscent of the dielectric breakdown model (DBM) with growth exponent $\eta = 1$ (see Section~\ref{sec:dbm}).

\begin{figure}
    \centering
    \includegraphics[width=1\linewidth]{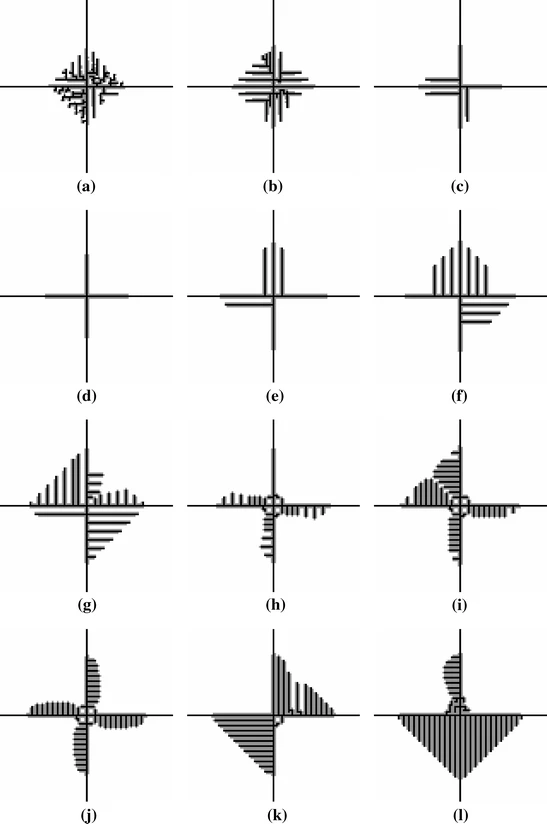}
    \caption{Long-run equilibrium morphologies as $\gamma$ varies for fixed $\beta = 0.25$. a) $\gamma$ = 0.00, b) 0.09, c) 0.11, ..., l) 0.45. Increasing $\gamma$ suppresses dendritic growth and induces linear and grid-like morphologies. Source : From \cite{Caruso2011}.}
    \label{fig:caruso_varyinggamma}
\end{figure}

To investigate the interplay between green space and public goods, the authors fix $\beta = 0.25$ and vary $\gamma$. Figure~\ref{fig:caruso_varyinggamma} illustrates the resulting urban forms. At $\gamma = 0$, dendritic structures dominate. As $\gamma$ increases, leapfrogging dissipates and linear developments re-emerge near $\gamma \approx 0.11$, again resembling the Alonso structure. Around $\gamma = 0.34$, new lateral roads begin to appear, creating grid-like structures interspersed with green corridors. For $\gamma > 0.42$, symmetry breaking emerges across city quadrants, driven by path-dependent growth along unidirectional development lines. These resulting configurations often retain large green patches near the CBD, highlighting a potential urban planning benefit of such self-organized development.

\begin{figure}
    \centering
    \includegraphics[width=1\linewidth]{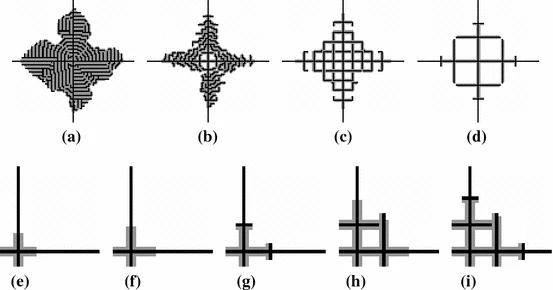}
    \caption{Urban morphologies as neighborhood radius $\hat{x}$ varies, with $\beta = 0$ and $\gamma = 0.5$. a) $\hat{x} = 4$, b) 7, c) 8, d) 20. Panels e)–i) show intermediate growth stages for $\hat{x}=8$. Source : From \cite{Caruso2011}.}
    \label{fig:caruso_varyingrad}
\end{figure}

The third parameter explored is the neighborhood radius $\hat{x}$ over which externalities are perceived. Figure~\ref{fig:caruso_varyingrad} displays the resulting morphologies. Small values of $\hat{x}$ lead to compact and dense cities, as households respond to short-range externalities. For larger $\hat{x}$, development tends to occur in patches that preserve green space. Time series snapshots reveal how migrants sequentially occupy regions around green patches, and when $\hat{x}$ becomes large enough, growth extends perpendicular to the main roads to maintain access to distant public goods.

This model, grounded in a standard microeconomic utility framework, demonstrates how urban morphologies such as leapfrogging, dendritic, and compact grid-like patterns -- see~\ref{subsec:shape} and~\ref{subsec:street}--can emerge endogenously from local preferences and accessibility constraints. The analogy with dielectric breakdown is particularly compelling: as in diffusion-limited aggregation and DBM, the structure of the growing city reflects a competition between branching and compactification forces, here driven by household preferences for green and social amenities. Remarkably, without explicit optimization or centralized control, the model reproduces both undesirable sprawl-like patterns and efficient green-integrated urban forms, offering important insights into the spatial consequences of utility-driven decision-making and infrastructure constraints. It is grounded in the same framework as the model presented in the previous section and suffers from similar pitfalls -- however, it is worth noticing the effort put by the authors to study the variation of fractal dimensions systematically to quantify changes in morphology, as well as the attention put on the phase transition points in various parameters.


\subsection{Dynamical model of central place systems}

Allen and Sanglier \cite{allensanglier} proposed  a dynamic extension of the central place theory, introduced by Christaller \cite{Christaller}, in order to demonstrate how various factors such as competition between entrepreunarial forces, demand, spatial constraints or commuting costs affect growth and decay of urban systems.

In this model, the coupling between the spatial distribution of population and the location of employment opportunities drives self-organization. The framework incorporates both deterministic mechanisms and stochastic fluctuations. As a result, urban centers can spontaneously emerge from favorable local conditions (e.g., entrepreneurial activity and population density), while existing centers may vanish through competitive interactions. In particular, the temporal evolution of the population $P_i$ at location $i$ is described by 
\begin{align}
    \begin{aligned}
    \frac{dP_i}{dt} &= bP_i(N+\sum_kS_i^{(k)}-P_i) -m P_i \\
    &+ \tau \sum_{j \ \text{neighbors}} (P_j^2 -P_i^2),
    \end{aligned}
\end{align}
where $b$ and $m$ are birth and death rates. $N$ is the natural carrying capacity (i.e in the absence of any economic attractivity) and $\sum_kS_i^{(k)}$ is the employment potential of all the present functions $k$ (a function represents a producible good or service). The additional term $\tau \sum_{j \ \text{neighbors}} (P_j^2 -P_i^2)$ accounts for spatial competition for labor. More precisely, a labor at location $i$ can be done by residents of neighboring location $j$. The addition of this simple mechanism provokes complex internal dynamics similar to urban sprawl.  

The employment opportunity offered by function $k$ at location $i$ evolves like
\begin{equation}
    \begin{aligned}
    \frac{dS_i^{(k)}}{dt} &= \alpha S_i^{(k)}(E_i^{(k)} - S_i^{(k)})
    \end{aligned}
\end{equation}
where the potential employment capacity is assumed to be
\begin{equation}
    \begin{aligned}
    E_i^{(k)} &= n(k) D_i^{(k)}
    \end{aligned}
\end{equation}
where $n(k)$ is the number of jobs involved in producing a unit of $k$, and $D_i^{(k)}$ is the demand for $k$ arriving at the point $i$. The quantity $D_i^{(k)}$ is assumed to be proportional to the density of population at and around $i$, as well as to the ‘attractivity’ of the point $i$, felt by the surrounding population, as compared to that of other centres which offer the function $k$. In the paper \cite{allensanglier}, the demand is given as a complicated function of many variables such as the quantity of $k$ demanded per individual at unit price, the cost of production of $k$ at the point $i$, the cost of transport for $k$ per unit distance, etc. In particular, they also introduce the population threshold $P_{\text{th}}$ above which appears the function $k$ at a location.
%
%
%
It should be noted that this assumption regarding the emergence of functions is quite strong; in contrast, models incorporating random knowledge diffusion or technological spillovers \cite{duranton2023} allow for more gradual or probabilistic function appearance. Moreover, this set-up introduces a set of arbitrary thresholds, increasing the complexity of the model. 

\begin{figure}
    \centering
    \includegraphics[width=1\linewidth]{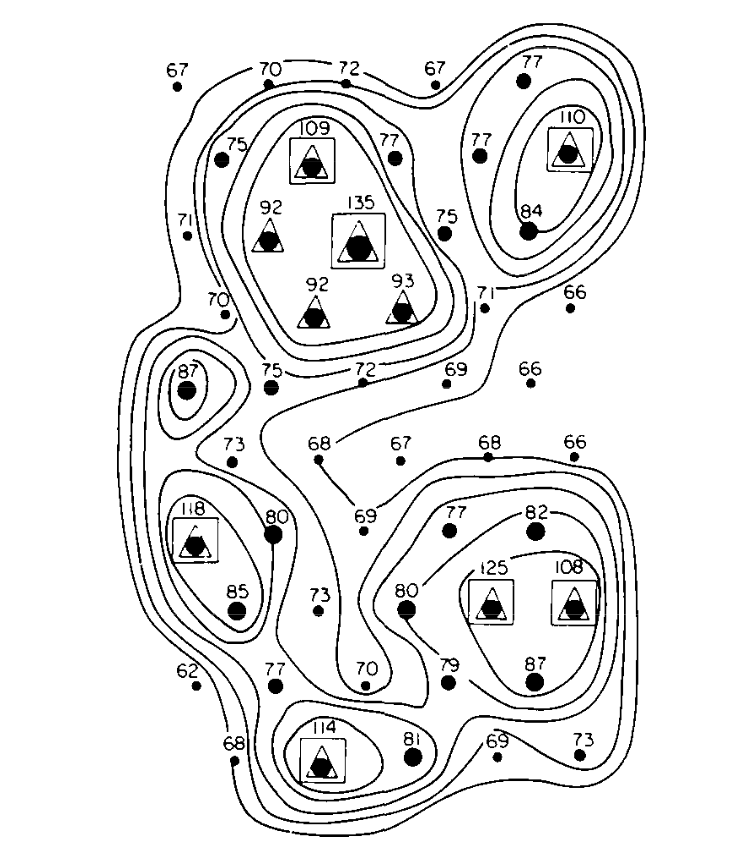}
    \caption{Snapshot for $t=46$ of the dynamic central place model. Functions appear when a site reaches the required population levels. Centers are represented by the black nodes, and functions by symbols (square, triangle, etc.), Source: from \cite{allensanglier}.}
    \label{fig:allensanglier-evolution}
\end{figure}

Allen and Sanglier performed a numerical simulation of their model. In this simulation, sites are arranged on a triangular lattice, and new urban functions emerge once local population thresholds are crossed: the second function appears at $P_\text{th}=68$, the third at $P_\text{th}=84$, and the fourth at $P_\text{th}=100$. By $t=12$, five large centers have emerged, which will become the nuclei of future metropolitan areas. Their rapid population growth allows them to attract additional functions earlier than smaller sites, thereby gaining a comparative advantage. One center in particular reaches high population levels quickly, triggering suburban development and decentralization of the labor market around it. At this point, the core reaches its maximum population and subsequently redistributes growth to surrounding sites. Overall, the dynamics can be interpreted in four stages. First, population growth is concentrated within urban cores (`central urbanization'). Second, continued core growth is accompanied by suburban development. Third, population in the cores stabilizes or declines while growth shifts outward (`counter-urbanization'). Finally, in the mature phase, competition between centers shapes the distribution of population and functions across the system. We show the graphical representation of the state of the population and functions at $t=46$ in Figure \ref{fig:allensanglier-evolution}. 



This dynamic central place framework is valuable in that it captures both deterministic mechanisms and the role of path dependence in shaping urban hierarchies. It was further developed in \cite{allen1981} to explore the model’s potential for informing decision-making strategies. However, it also presents several limitations. The model introduces numerous parameters—such as thresholds, attractivity functions, and commuting costs—that are difficult to calibrate or link directly to empirical data. Its core assumptions, including ad hoc attractivity forms, neighbor-based fluxes, and sharp functional thresholds, strongly constrain the dynamics without systematic justification. Most critically, the absence of analytical results or systematic numerical exploration limits the analysis to a qualitative level, making it unclear whether the reported urbanization phases reflect robust model behavior or specific parameter choices and initial conditions.

\subsection{Spatial growth within urban economics}

In urban economic theories \cite{Fujita_1989}, households, companies, and sometimes developers optimize an utility function under functional constraints. Households and companies compete for space. The former's utility traditionally depends on commuting costs, position in the city, rent and salary. Here, we focus solely on households and their repartition in the city. The first model, the monocentric city model, was proposed by Alonso, Muth and Mills \cite{brueckner1986structure}. In the 80s, Fujita and Ogawa \cite{fujita1982} proposed a model where the number of centers is determined endogeneously. Recently, Louf and Barthelemy~\cite{louf2013modeling} utilized an argument of Fujita and Ogawa augmented with congestion to demonstrate that the number of centers was a sub-linear function of the population. 
Urban economic theories propose a fundamental approach to describe the organization of cities : it is able to describe analytically the spatial distribution of rent prices and the position of the urban fringe for example. Moreover, it is useful to study the influence of particular features on the structure of the cities (see Duranton and Puga \cite{duranton2014} review how housing durability, amenities, transportation infrastructures,agglomeration economies, human capital and shocks affect the fate of cities). 
As these approaches focus mostly on economic and demographic aspects of the city (or of systems of cities) \cite{duranton2023}, they yield little understanding of spatial complexity. Subsequently, we give a brief formal introduction by discussing the monocentric unidimensional city, before discussing two approaches based on Alonso-Mills-Muth model tackling the spatial structure of cities.

\subsubsection{The Alonso-Mills-Muth (AMM) model} \label{subsec:amm}

The AMM model assumes a number of now standard simplifications: all households are identical, the city is at equilibrium, monocentric, isotropic, and everybody commutes to a single Central Business District (CBD) (located at $x=0$), where all jobs and occupations are concentrated. This model, developed in the 1960s by Alonso, Mills, and Muth \cite{alonso1964,muth1969cities, mills1967aggregative} forms the basis of much of classical urban economics. The city is embedded in a homogeneous Euclidean space, and the only spatial variables are land rent and commuting distance.

Households are assumed to evaluate their situation through a utility function, a concept commonly used in economics. This utility is not a physical energy, but it plays an analogous role: it is a scalar function that encodes preferences, ranking different combinations of housing, other goods, and city-wide amenities. One may think of utility as similar to a free energy that aggregates many microscopic choices into a single effective measure of `satisfaction'.

In the AMM framework, utility depends on two variables: land consumption $s$ (e.g., apartment size) and the consumption of a composite good $z$, defined as the remaining income after paying for transportation and housing. The utility function $U(z,s)$ satisfies
\begin{equation}
\frac{\partial U}{\partial s} > 0, \quad \frac{\partial U}{\partial z} > 0
\end{equation}
expressing that individuals always prefer more space and more consumption.

The budget constraint is written as
\begin{equation}
Y = z + T(x) + R(x)s
\end{equation}
where $Y$ is income, $T(x)$ is the commuting cost from location $x$ to the CBD, and $R(x)$ is the rent per unit land area at $x$. The agent maximizes utility subject to this constraint
\begin{equation}
\max_{z,s} U(z,s) \quad \text{subject to} \quad Y = z + T(x) + R(x)s
\end{equation}

Replacing $z = Y - T(x) - R(x)s$, we reduce the problem to maximizing $U(Y - T(x) - R(x)s, s)$ over $s$. The optimality condition $dU/ds=0$ becomes
\begin{equation}
\frac{\partial U}{\partial s} = R(x)\frac{\partial U}{\partial z}
\label{eq:Rfirst}
\end{equation}
which relates the marginal utilities to the rent at a given distance. The utility $U^*$ achieved at equilibrium must be spatially constant; otherwise, agents would relocate. This yields the condition $dU^*/dx=0$ which reads
\begin{equation}
\frac{dU^*}{dx} = \frac{\partial U}{\partial z} \left(-T'(x) - s \frac{dR}{dx} - R \frac{ds}{dx} \right) + \frac{\partial U}{\partial s} \frac{ds}{dx} = 0
\end{equation}
Inserting Eq.~\eqref{eq:Rfirst} into this expression gives the key result of the AMM model
\begin{equation}
\frac{dR}{dx} = -\frac{T'(x)}{s(x)}
\label{eq:AMMgrad}
\end{equation}
This equation resembles a force-balance relation: the spatial gradient of the rent is set by the local commuting cost per unit area. It implies that the rent must decrease with distance, faster in low-density zones.

To see how income heterogeneity affects spatial patterns, consider two income classes (rich and poor), with land consumptions $s_R$, $s_P$ and transport costs per unit distance $t_R$, $t_P$. Assuming linear costs $T(x) = tx$, the group with higher rent gradient $|dR/dx| = t/s$ dominates near the center. For the poor to occupy central areas, we require
\begin{equation}
\frac{t_P}{s_P} > \frac{t_R}{s_R}
\end{equation}
Otherwise, wealthier individuals will outbid them for central locations.

To solve the model analytically, economists define the bid-rent function $\Psi(x,u)$ as the maximum rent per unit land area an individual can pay while maintaining utility level $u$:
\begin{equation}
\Psi(x,u) = \max_{z,s} \left\{ \frac{Y - T(x) - z}{s} \;|\; U(z,s) = u \right\}
\end{equation}
From the implicit curve $z = z(u,s)$, one finds the tangent condition
\begin{equation}
-\frac{dz}{ds} = \Psi(x,u)
\label{eq:bidrent}
\end{equation}
which shows that the bid-rent is the slope of the indifference curve.

The market rent $R(x)$ must equal the highest bid at location $x$, i.e., the envelope of all $\Psi(x,u)$ curves. For a given income $Y$, utility-maximizing households solve
\begin{equation}
V(R,I) = \max_{z,s} \left\{ U(z,s) \;|\; z + Rs = I \right\} = \max_s U(I - Rs, s)
\end{equation}
with net income $I = Y - T(x)$. The condition
\begin{equation}
u = V(\Psi(x,u), Y - T(x))
\end{equation}
links the utility to the bid-rent function.

From the first-order condition,
\begin{equation}
R \frac{\partial U}{\partial z} = \frac{\partial U}{\partial s}
\end{equation}
and taking the derivative of $V$ with respect to $R$ gives
\begin{equation}
\frac{\partial V}{\partial R} = -s \frac{\partial U}{\partial z} < 0
\end{equation}
which implies that utility decreases as rent increases. The household maximizes utility at $x^*$ such that
\begin{equation}
u^* = V(R(x^*), Y - T(x^*)) \geq V(R(x), Y - T(x))
\end{equation}
leading to:
\begin{equation}
R(x^*) = \Psi(x^*, u^*)
\end{equation}
and more generally $R(x) \geq \Psi(x, u^*)$ elsewhere.

This equilibrium condition implies that the market rent equals the bid-rent of the household occupying the site. Locations are allocated to those who value them most, leading to:
\begin{align}
\nonumber
R(x) &= \Psi(x, u^*) \quad \text{(within the city)}\\
R(x) &= R_A \quad \text{(outside)}
\end{align}
where $R_A$ is the agricultural land rent at the fringe. In a closed-city setting (fixed population $P$), the city boundary and equilibrium utility are determined by:
\begin{equation}
\int_{-b}^{b} \frac{dx}{s(x,u^*)} = P, \quad R(b) = R_A
\end{equation}

Let us now consider a concrete example: a circular two-dimensional city, with CBD at $r = 0$, and the specific utility function
\begin{equation}
U(z,s) = \alpha \log z + \beta \log s
\end{equation}
with $\alpha + \beta = 1$. 
Assuming linear transportation cost $T(r) = ar$, the budget becomes $Y = z + R(r)s + ar$. The indifference curve is:
\begin{equation}
z(u,s) = e^{u/\alpha} s^{-\beta/\alpha}
\end{equation}
Using Eq.~\eqref{eq:bidrent}, the condition for optimal $s$ becomes:
\begin{equation}
-s \frac{dz}{ds} = Y - ar - z \Rightarrow \beta e^{u/\alpha} s^{-\beta/\alpha} = Y - ar - e^{u/\alpha} s^{-\beta/\alpha}
\end{equation}
which yields:
\begin{equation}
s(r) = \alpha^{-\alpha/\beta} e^{u^*/\beta} (Y - ar)^{-\alpha/\beta}
\end{equation}
and the corresponding rent profile
\begin{equation}
R(r) = \beta \alpha^{\alpha/\beta} e^{-u^*/\beta} (Y - ar)^{1/\beta}
\end{equation}
Imposing the closed city constraint
\begin{equation}
\int_0^{r_f} \frac{1}{s(r)} 2\pi r\, dr = P
\end{equation}
and the boundary condition $R(r_f) = R_A$, we find (assuming $R_A \approx 0$)
\begin{align}
R(r) &= \frac{P \beta a^2}{2\pi Y^{2 + \alpha/\beta} B(1, \alpha/\beta)} (Y - ar)^{1/\beta} \\
\rho(r) &= (Y - ar)^{\alpha/\beta}
\end{align}
where $B(a,b)$ is the Beta function. The model thus predicts a decreasing density profile and increasing land consumption with distance, with a maximum commuting range $r_f = Y/a$.

Finally, one may explore the effect of alternative utility functions. For example, choosing $U(z,s) = z + \alpha \log s$ leads to an exponential density decay:
\begin{equation}
\rho(r) \propto e^{-ar/\alpha}
\end{equation}
which corresponds to the classic exponential profile observed in many monocentric cities \cite{Clark:1951}, as discussed in~\ref{subsec:clark}. This illustrates how the mathematical form of the utility function acts analogously to a free energy landscape: it governs equilibrium distributions and spatial structure, just as energy functionals do in physical systems.

\subsubsection{Extending AMM to evolving cities}

Classical urban economic models, such as the AMM framework, describe cities as static equilibrium systems in which residential density declines monotonically with distance from the central business district (CBD). In these frameworks, equilibrium density gradients emerge from a trade-off between commuting costs and land consumption. However, they ignore the durability of housing capital and the fact that development decisions are inherently forward-looking. 

Wheaton \cite{wheaton1982urban} introduced a dynamical model of urban growth under perfect foresight, extending Alonso’s rent-maximization framework to multiple periods. The central assumption is that residential capital is perfectly durable, so that development is irreversible. Urban growth thus becomes a sequence of intertemporal land allocations designed to maximize the present value of rents.
Wheaton \cite{wheaton1982urban} extends the Alonso--Muth--Mills framework to a dynamic setting in which urban development occurs incrementally over time rather than through an instantaneous equilibrium adjustment. Time is divided into discrete periods, with the final one extending indefinitely. In each period $i$, households face market conditions summarized by construction cost $c_i$, income $y_i$, commuting cost per mile $k_i$, population $N_i$, agricultural rent $s_i$, and utility level $u_i$. Household utility depends on consumption $x$, land $q$, and housing capital $h$,
\begin{align}
u_i = u(x_i,q,h).
\end{align}
From the budget constraint, the bid rent function at distance $t$ is
\begin{align}
R_i(t) = y_i - k_i t - u^{-1}(u_i,q,h).
\end{align}

The central concept is the present value of development: the discounted value today of all future rents minus construction costs. If $r$ denotes the discount rate and $\delta_j = e^{-rT_j}$ are the discount factors, then the present value of developing land in period $i$ is
\begin{align}
B_i = \sum_{j=1}^{i-1} \delta_j R_j 
    + \sum_{j=i}^{n} \delta_j R(y_j,k_j,u_j,t,q,h) 
    - c_i h e^{-rT_{i-1}}.
\end{align}
The optimal choice of density $q_i$ and housing capital $h_i$ satisfies
\begin{align}
\frac{\partial B_i}{\partial q}=0, \qquad \frac{\partial B_i}{\partial h}=0,
\end{align}
so that land is developed in whichever period $i$ maximizes $B_i$. Each parcel is allocated to the period that yields the highest present value bid,
\begin{align}
L_i = \{t: B_i \geq B_k \;\;\forall k \neq i\},
\end{align}
subject to the constraint that the developed land in period $i$ must accommodate exactly the increment in population,
\begin{align}
\int_{L_i} q_i(t)\,dt = P_i - P_{i-1}.
\end{align}

This framework produced the following results. Land price, measured as the present value of rents, always declines continuously with distance, even though residential density may not. The direction of development can be either inside-out or outside-in, depending on the historical trajectory of income, transport costs, and population growth. Rising incomes, falling transport costs, or rapid demographic expansion generate the conventional inside-out pattern, whereas falling incomes or rising transport costs may induce leapfrogging or outside-in development. Within each period, density decreases with distance, but at the boundaries between periods density may jump upward or downward, creating a `sawtooth' pattern. 


Wheaton illustrates these mechanisms with numerical simulations. When only population grows, density increases with distance until late periods, when terminal effects cause a decline. Increasing the rate of population growth steepens the density gradient by raising interior densities. Declining transport costs have a similar effect, leading to steeper gradients as commuting becomes cheaper. Rising incomes likewise produce steeper declines in density with distance, while falling incomes reverse the spatial sequence of development so that peripheral areas develop before central ones. These simulations emphasize that density gradients are determined by historical trajectories rather than by static equilibrium conditions.

Despite these advances, dynamic AMM models such as Wheaton's retain fundamental flaws. One of the core issues lies in the treatment of land and development. In the original AMM formulation, every site within the city is used at its optimal intensity: there are no vacant parcels, no obsolete or underused structures, and no scope for redevelopment. The model lacks a development industry, and there are no landowners making profit-based decisions about when to demolish and rebuild. Land is continuously and instantaneously reallocated, as though directed by a benevolent social planner \cite{murray2017throw}. This assumption becomes especially problematic in comparative static applications, where any marginal change---such as the arrival of a new resident, an increase in construction efficiency, or a shift in agricultural land value---implies that the entire city is instantly rebuilt from scratch. All buildings are simultaneously demolished and replaced with a new optimal configuration, which is clearly at odds with how real cities evolve. In reality, urban change is incremental, the housing stock is persistent, and adjustments occur gradually rather than instantaneously. In this context, the AMM framework---even when extended dynamically with perfect foresight---struggles to capture the irregularities, frictions, and historical contingencies that shape real urban growth. Its equilibrium assumptions and monocentric spatial structure make it ill-suited for analyzing complex, decentralized, and path-dependent urban dynamics.


\subsubsection{Impact of radial infrastructures}

To understand how transportation infrastructure affects the shape of cities, Baum-Snow introduced a model incorporating radial highways into the AMM framework \cite{baumsnow2007model}. The idea is to analyze how high-speed access corridors modify equilibrium land rents, densities, and urban spatial structure.

The model retains the core AMM assumptions: the city is monocentric, isotropic, and in equilibrium, with all jobs located at a Central Business District (CBD) at $x=0$. Households are identical and commute daily to the CBD. However, the key novelty is the presence of radial highways—fast connections radiating outward from the CBD. These highways create spatial heterogeneity in commuting speeds and therefore break the radial isotropy of the baseline AMM model.

Households evaluate locations based on a utility function $U(z,s)$, where $z$ is the consumption of other goods, and $s$ is the land consumption (e.g., apartment size). As in the standard AMM model, the utility is maximized subject to a budget constraint. The physical analogy is useful here: utility is not an energy, but it plays an analogous role as a scalar potential, encoding individual preferences and summarizing many microscopic decisions into a macroscopic satisfaction level.

In this extended model, the street network consists of a continuum of surface streets with speed $v_s = 1/b$, and a set of radial highways with higher speed $v_h = 1/(b\gamma)$, where $0 < \gamma < 1$ is the ratio of speeds. A household at polar coordinates $(r,\phi)$ can access the CBD using one of three possible paths: (i) Perpendicular access to the highway (slow then fast); (ii) Via concentric streets (slow then slow); or (iii) A linear path minimizing total commuting time (combined path). 

The total (dimensionless) commuting cost is written as $b r \tilde{L}(\phi)$, where $\tilde{L}(\phi)$ depends on the chosen route:
\begin{equation}
    \tilde{L}(\phi) = 
    \begin{cases}
        \gamma \cos \phi + \sin \phi & \text{(i)} \\
        \gamma + \phi & \text{(ii)} \\
        \gamma \cos \phi + \sqrt{1 - \gamma^2} \sin \phi & \text{(iii)}
    \end{cases}
\end{equation}
When the highway is not used, the effective commuting cost is simply $br$. The budget constraint is modified to account for time lost in commuting:
\begin{equation}
    z + R(r,\phi)s = Y\left[1 - L(r,\phi)\right]
\end{equation}
where $Y$ is the income (wage) and $L(r,\phi) = b r \tilde{L}(\phi)$ is the effective commuting time from $(r,\phi)$ (normalized by income). Households allocate remaining resources between land and other goods.

Following the AMM logic, the bid-rent function—which expresses the maximum rent a household is willing to pay while maintaining a given utility level $u$—becomes:
\begin{equation}
    \Psi[L(r,\phi), u] = \max_s \left\{ \frac{w [1 - L(r,\phi)] - Z(s,u)}{s} \right\}
\end{equation}
where $Z(s,u)$ is the expenditure on other goods needed to achieve utility $u$ at land consumption $s$, i.e., $U(Z(s,u), s) = u$.

Let $\bar{\phi}$ denote the maximum angular range around a highway where it is beneficial to use the highway to access the CBD. That is, for $\phi < \bar{\phi}$, households use the highway; for $\phi > \bar{\phi}$, they rely on surface streets. Let $M$ be the number of radial highways. The maximum commuting distance at angle $\bar{\phi}$ defines the radius $\bar{r}_f^M$ of the urban fringe (where rent equals agricultural rent $R_a$):
\begin{equation}
    \Psi(b \bar{r}_f^M, u^M) = \frac{w [1 - b \bar{r}_f^M] - Z(\tilde{s}(R_a, u^M), u^M)}{\tilde{s}(R_a, u^M)} = R_a
\end{equation}
Here, $u^M$ is the equilibrium utility level for $M$ rays, and $\tilde{s}(R,u)$ is the optimal land consumption at rent $R$ for utility $u$.

For directions where the highway is used ($\phi < \bar{\phi}$), the urban fringe is given by:
\begin{equation}
    r_f^M(\phi) = \frac{\bar{r}_f^M}{\tilde{L}(\phi)}
\end{equation}
In contrast, for $\phi > \bar{\phi}$, we retain the circular AMM fringe $r = \bar{r}_f^M$.

Each highway influences commuting speeds within an angular wedge of width $2\bar{\phi}$. The total number of inhabitants $N$ is obtained by integrating over all directions:
\begin{align}
    P &= 2M \int_0^{\bar{\phi}} \int_0^{\frac{q(u^M)}{\tilde{L}(\phi)}} \frac{r\, dr\, d\phi}{\tilde{s}[\Psi(br \tilde{L}(\phi), u^M), u^M]} \\
    &\quad + (2\pi - 2M \bar{\phi}) \int_0^{q(u^M)} \frac{r\, dr}{\tilde{s}[\Psi(br, u^M), u^M]}
\end{align}
where $q(u^M) = \bar{r}_f^M$ ensures that the integral domain matches the maximum urban extent.

This model predicts that the construction of a highway reshapes the city. In the angular sector influenced by the highway, commuting times are reduced. As a result:
\begin{enumerate}
    \item Land becomes more accessible, lowering rents and increasing supply.
    \item The fringe $r_f(\phi)$ extends farther out, enabling suburbanization in that sector.
    \item The average utility level $u^M$ increases, reflecting improved accessibility.
\end{enumerate}
\begin{figure}[ht]
    \centering
    \includegraphics[width=1\linewidth]{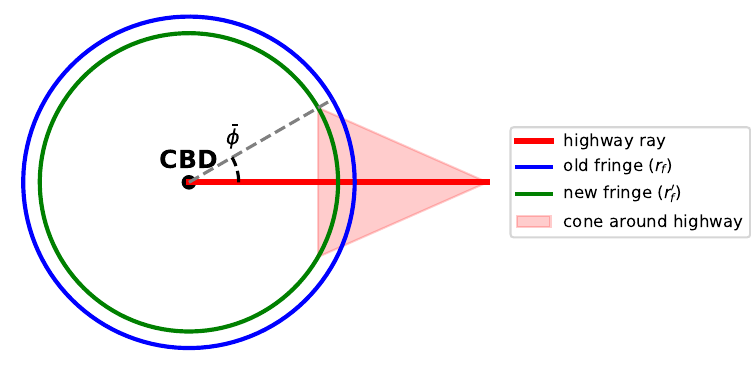}
    \caption{Outside of the range of influence of the highway, the city contracts ($r_f' < r_f$). In contrast, where households benefit from high-speed access to the CBD, the urban fringe extends, forming a wedge of lower rent and longer commuting distances (shown in red). Inspired by Fig.~1 of \cite{baumsnow2007model}.}
    \label{fig:baumsnow_ray}
\end{figure}

Baum-Snow explores this model via simulations using the utility function $U(z,s) = z + \alpha \log s$. In a representative metropolitan area, where initially half the population lives within the central city, introducing the first highway reduces the central city's population share by approximately 13\%. The second and third rays cause additional declines of 11\% and 9\%, respectively. Beyond the third, the marginal effect diminishes, with each additional highway decreasing the central share by about 1 percentage point less than the previous one. This result suggests a nonlinear saturation effect: early rays reshape the city significantly, but additional rays yield diminishing returns in terms of suburban expansion. These findings help understanding the empirical observations discussed in~\ref{subsec:impact}.

\subsubsection{Myopic growth}

To address the limitations of static monocentric city models, Anas proposed in 1976 a dynamic framework for residential urban growth \cite{anas1976}, in which the urban form evolves through a sequence of short-run spatial equilibria. Rather than assuming a steady-state equilibrium city, this model introduces time-dependent exogenous growth and endogenous spatial adjustments—resulting in what can be interpreted as a discrete-time dynamical system for urban morphology.

In the static version of the model, all households are identical, with income $W$. The commuting cost $T(x)$ and housing market price $ \eta(x)$ depend only on the radial distance $x$ from the CBD. Households derive utility from a Cobb–Douglas function (a specific form often used in economics)
\begin{align}
U = Z^\alpha H^\beta
\end{align}
where $Z$ is the consumption of goods, $H$ is the amount of housing consumed, and $\alpha + \beta = 1$. The budget constraint reads
\begin{align}
  W = Z + \eta(x)H + T(x),  
\end{align}
reflecting a trade-off between commuting, housing, and other consumption.

The supply of housing is modeled via a production function of the form
\[
H = K^a L^b N^c,
\]
where \( H \) is the quantity of housing produced (e.g., floor area), and \( K \), \( L \), and \( N \) respectively denote the capital invested (e.g., construction materials and equipment), the land used, and the labor employed in construction. The exponents \( a, b, c \) are positive constants satisfying \( a + b + c = 1 \), implying constant returns to scale: doubling all inputs doubles output. This function plays the role of a technological constraint--analogous to a production `equation of state'—that governs how physical resources are transformed into usable housing stock.

To determine how much land is worth at a given location $x$, the model assumes a perfectly competitive market among housing developers. In such a market, no firm earns economic profit: the total revenue from selling the housing just covers the costs of inputs. The revenue per unit land area is $\eta(x) H$, and the costs are the expenditures on capital and labor: $\rho K$ and $s P$, where $\rho$ and $s$ are the unit prices of capital and labor, respectively.

Enforcing the zero-profit condition leads to an expression for the land rent $R(x)$, defined as the residual revenue per unit of land left after paying for capital and labor
\begin{align}
R(x) = \frac{\eta(x) H - \rho K - s N}{L}.
\end{align}
This formula captures how much a developer is willing to pay for land at location $x$, given market prices and the local housing demand.

Finally, the city's spatial extent and population density profile are determined by how demand and supply equilibrate across space. In a closed city, the total population is fixed exogenously, and prices adjust to accommodate this constraint. In an open city, population flows in and out until a common utility level is achieved across locations, making the equilibrium population an endogenous outcome of the economic environment.

The dynamic extension of the model introduces several key assumptions. Housing is durable—once built, it persists—so the city evolves through layers of development that reflect past economic conditions. Households are mobile but short-sighted: they make decisions based on current conditions without anticipating future changes. As population grows over time, new housing is added in concentric rings, each representing a historical phase with its own prevailing income levels, transport costs, and construction prices.

Because residents respond to current conditions, rents and prices across the city adjust dynamically, even if the housing stock was built under different circumstances. The model describes how consumption and housing expenditure change as the city evolves, and how rents adjust over time based on shifts in income, commuting costs, and the durability of existing housing.

By modeling change in continuous time, the framework captures urban transformation as a truly dynamic process rather than a series of static snapshots. It explains how older housing gradually loses value and may be abandoned, and shows that rent gradients can reverse—land near the center may depreciate relative to the periphery, despite lower commuting costs. This helps reproduce observed patterns like the decline of American downtowns. Finally, the model highlights the importance of welfare dynamics. In rapidly growing but low-income cities, welfare declines and densities remain high. In slower-growing cities with rising incomes, inner neighborhoods may hollow out unless policies intervene. Overall, the framework accounts for diverse urban patterns—such as suburban expansion, core decline, and complex rent and density profiles—that static models cannot explain.

\subsection{The Edge-City Model}

Krugman \cite{krugman1995self} introduced the edge-city model as a dynamic extension of the Fujita–Ogawa framework \cite{fujita1982}, in which the spatial structure of cities emerges from the interplay between agglomeration and dispersion forces among firms. The model describes the self-organization of economic activity and explains how multi-centered urban forms can arise endogenously from local interactions.

The city is modeled as a one-dimensional space, with a time-dependent business density $\rho_b(x,t)$, normalized such that
\begin{align}
\int \rho_b(x,t)\,dx = 1
\end{align}
for all $t$. The `attractiveness' or `market potential' $ \Pi(x,t)$ of location $x$ is defined as the spatial convolution
\begin{equation}
    \Pi(x,t) = \int K(x - z)\, \rho_b(z,t)\, dz,
\end{equation}
where the kernel $K(x) = A(x) - B(x)$ captures the net spatial interaction. The function $A(x)$ represents agglomeration effects (such as customer access, shared suppliers, or knowledge spillovers) and is a positive, non-increasing function of distance. The function $B(x)$ models dispersion effects (e.g., competition, land scarcity), is also non-increasing in $|x|$. The overall effect is that firms are attracted to regions with high market potential but also experience repulsive interactions that spread them out.

The evolution of the business density is governed by
\begin{equation} \label{eq:edgecity_rhodyn}
    \frac{\partial \rho_b(x,t)}{\partial t} = \gamma \left[ \Pi(x,t) - \bar{\Pi}(t) \right],
\end{equation}
where $\gamma > 0$ sets the adjustment rate, and the average potential is given by
\begin{equation}
    \bar{\Pi}(t) = \int \Pi(x,t)\, \rho_b(x,t)\, dx.
\end{equation}
This equation ensures conservation of total density: areas with above-average attractiveness grow in activity, while those below the average decline, akin to a selection or fitness-driven diffusion.

The homogeneous state $\rho_b(x,t) = \rho_0$ is a fixed point of the system. To analyze its stability, we consider small perturbations $\delta\rho(x,t)$ and linearize Eq.~\eqref{eq:edgecity_rhodyn}, yielding
\begin{equation}
    \frac{\partial \delta\rho_b(x,t)}{\partial t} \approx \gamma \int K(x - z)\, \delta\rho_b(z,t)\, dz.
\end{equation}
We define the usual Fourier transform
\begin{equation}
    \delta\rho_b(k,t) = \int e^{ikx} \delta\rho_b(x,t)\, dx,
\end{equation}
which evolves according to
\begin{equation}
    \delta\rho_b(k,t) \sim e^{\gamma \hat{K}(k)\,t},
\end{equation}
where \( \hat{K}(k) \) is the Fourier transform of \( K(x) \). The sign of \( \hat{K}(k) \) thus determines whether perturbations of wavelength \( \lambda = 2\pi/k \) grow or decay.

For an explicit kernel of the form
\begin{equation}
    K(x) = A\, e^{-|x|/r_1} - B\, e^{-|x|/r_2},
\end{equation}
with $r_1 < r_2$. The first exponential term corresponds to attraction acting on a shorter characteristic length scale, while the second term describes repulsion acting on a larger spatial scale.
The Fourier transform leads to the dispersion relation:
\begin{equation}
    \frac{d \delta\rho_b(k,t)}{dt} = 2\gamma \Lambda(k)\, \delta\rho_b(k,t),
\end{equation}
with the growth rate:
\begin{equation}
    \Lambda(k) = \frac{A r_1}{1 + (r_1 k)^2} - \frac{B r_2}{1 + (r_2 k)^2}.
\end{equation}

The function $\Lambda(k)$ typically has a single maximum at a finite wavenumber $k^* > 0 $, where the instability is strongest. The zero mode $\Lambda(0) = A r_1 - B r_2$ corresponds to the growth of spatially uniform perturbations and must be negative to ensure mass conservation. The fastest-growing mode $k^*$ sets a preferred spatial scale, and thus a characteristic inter-center distance.

In a finite domain of size $L$, the number of emergent centers~$H$ (peaks in $\rho_b(x,t)$) scales proportionally to system size as
\begin{equation}
    H \sim L\, k^*,
\end{equation}
analogous to pattern formation in Turing-type reaction–diffusion systems. The wavelength $\lambda^* = 2\pi/k^*$ acts as a spontaneously selected length scale arising from the competition between agglomeration and dispersion forces.

This model provides a mechanistic explanation for the endogenous emergence of multiple business centers (or `edge cities') in large metropolitan areas and gives quantitative predictions concerning the polycentricity of the city (discussed in~\ref{subsec:shape}). However, while the model predicts a linear scaling of the number of centers with city size, empirical studies \cite{louf2013modeling} have shown that the number of subcenters scales sublinearly with population. This apparent discrepancy suggests that additional factors—such as land-use constraints, transport infrastructure bottlenecks, planning policies, or multi-scale agglomeration effects—must be included to accurately capture real urban dynamics. Nonetheless, the edge-city model remains a valuable analytical framework for understanding symmetry-breaking and spatial structure formation in economic geography.

\section{Statistical physics models} 
\label{chap:statphys}

In this chapter, we turn to models originating in statistical physics that were not initially conceived for urban applications but have since been adapted to the study of city growth. A paradigmatic example is diffusion-limited aggregation (DLA), introduced to generate clusters with highly irregular boundaries characterized by non-trivial fractal dimensions. The relevance of DLA for urban studies lies precisely in this property: the fractal geometry of its clusters resonates with empirical observations of urban perimeters, which has motivated numerous analyses of the fractal dimension of cities. Beyond DLA, we will review extensions such as the dielectric breakdown model (DBM), as well as the Eden model, Markov random fields, and correlated percolation, all of which capture different aspects of spatial growth and morphology. We will also discuss growth patterns shaped by human mobility behavior, and finally turn to empirical models that describe the evolution of the number of buildings. Together, these approaches illustrate how concepts from statistical physics provide a versatile toolkit for exploring the stochastic, collective, and often fractal-like dynamics of urban expansion.

\subsection{Diffusion limited aggregation} \label{subsec:dla}
\label{sec3A}

\subsubsection{The original DLA model}

Diffusion-limited aggregation (DLA) was introduced by Witten and Sander~\cite{witten1983diffusion} as a kinetic model for the irreversible formation of clusters in systems where diffusion is the rate-limiting step. The model has broad applications, including soot formation, dendritic crystal growth, electrodeposition, and colloidal aggregation. In DLA, particles undergo random walks (modeling diffusion) and irreversibly stick upon contact with a growing aggregate. This process naturally produces highly branched, scale-invariant structures with fractal geometry.

\begin{figure}[h!]
\centering
\includegraphics[width=0.5\textwidth]{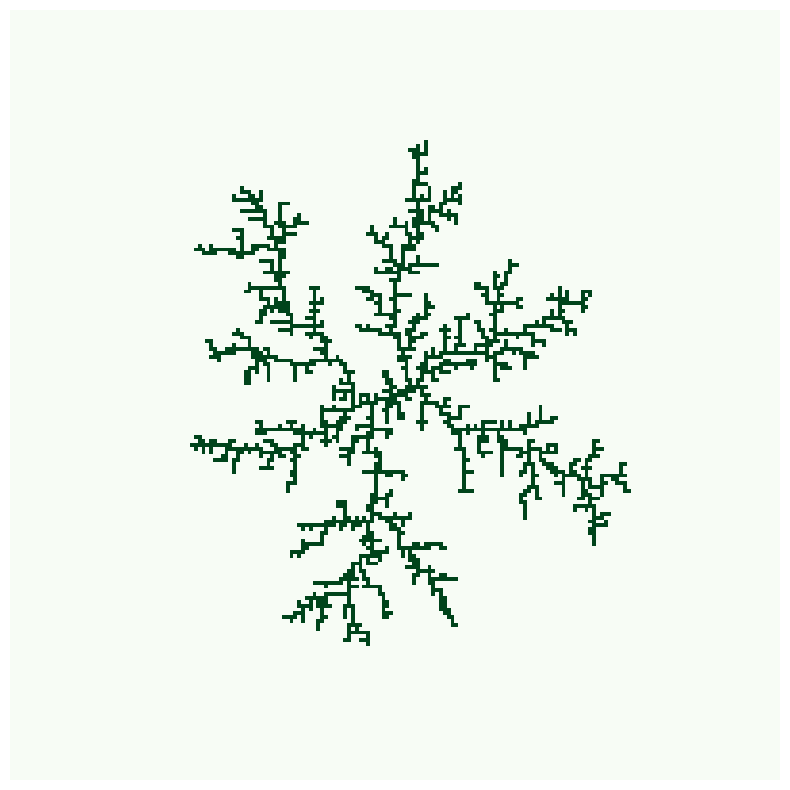}
\caption{DLA cluster formed by 500 particles on a \(200 \times 200\) square lattice.}
\label{fig:dla1}
\end{figure}

More precisely, DLA describes a process in which particles diffuse through a medium until they encounter and adhere irreversibly to a growing cluster. This mechanism captures key features of various natural and industrial processes, from dust coagulation to dendrite and aerosol formation. The resulting patterns exhibit strong ramification and self-similarity, due to instabilities at the growth interface (see an example in Fig.~\ref{fig:dla1}).

The process begins with a seed particle at the origin of a lattice. Subsequent particles perform random walks until they reach the cluster and stick irreversibly. As more particles are added, a highly branched fractal structure emerges. The dynamics are governed by the diffusion equation
\begin{equation}
    \frac{\partial u}{\partial t} = D \nabla^2 u,
\end{equation}
where $u$ is the concentration (or probability field) of diffusing particles, and $D$ is the diffusion coefficient. Aggregation occurs at the absorbing boundary $u = 0$, analogous to a conducting boundary in electrostatics, thus coupling diffusion to the moving growth interface.

DLA structures exhibit scale invariance, as captured by the two-point density-density correlation function
\begin{equation}
    \langle \rho(\vec{r}_1) \rho(\vec{r}_2) \rangle \propto |\vec{r}_1 - \vec{r}_2|^{-A},
\end{equation}
where $A$ is a correlation exponent related to the Hausdorff (fractal) dimension $D$ via \cite{witten1981diffusion}
\begin{equation}
    D = d - A,
\end{equation}
with $d$ the embedding spatial dimension. Numerical studies yield $D \approx 1.71$ in two dimensions, confirming the fractal nature of DLA aggregates. This value is robust, largely independent of microscopic details, and defines a universality class of growth processes. Analytical tools (e.g., scaling arguments, Laplacian growth theory) and extensive simulations have explored the morphology, growth laws, and instabilities of DLA clusters~\cite{meakin1983formation, meakin1998fractals}.



Simulations of the DLA model reveal several fundamental properties that underscore its role as a prototype of non-equilibrium pattern formation. A notable feature is the screening effect, wherein the inner regions of the aggregate become effectively shielded from incoming diffusing particles. This enhances growth at the outermost protrusions, resulting in the characteristic ramified structures. Another key property is universality: the fractal dimension $D$ of the resulting clusters is remarkably insensitive to the underlying lattice geometry or boundary conditions, indicating that DLA belongs to a broad universality class of growth processes. Scale invariance in these aggregates is further evidenced by correlation functions, with measurements of the density correlations and the radius of gyration $R_g$ typically obeying the scaling law
\begin{align}
R_g \propto N^{1/D},
\end{align}
where $N$ is the number of particles in the cluster.

This scale invariance manifests in various observable quantities. For instance, the structure factor derived from scattering experiments exhibits a power-law dependence directly related to the fractal dimension $D$. The growth kinetics of the aggregate also follow nontrivial scaling laws, with the cluster mass increasing over time as
\begin{align}
N \propto t^{d/(D+1)},
\end{align}
in spatial dimension $d$. At sufficiently high densities, independent aggregates eventually intersect and merge, leading to gelation phenomena that follow distinct scaling behaviors.

Overall, DLA provides a robust framework for understanding a wide array of aggregation processes, producing fractal structures with universal statistical properties. In contrast to classical equilibrium critical phenomena, DLA notably lacks an upper critical dimension, emphasizing its fundamentally non-equilibrium character. 


\subsubsection{DLA and urban growth}

The study proposed in \cite{fotheringham1989diffusion} investigates diffusion-limited aggregation as a novel framework for modeling urban growth. By emphasizing the fractal geometry of urban structures, it connects spatial density gradients and branching development patterns to a simple, diffusion-driven process. Through simulations and statistical analyses, the study suggests that urban forms generated by DLA display features such as negative density gradients and a form of ordered complexity, challenging classical interpretations of urban density distributions.

Traditionally, urban growth has been examined through frameworks like the AMM model, central place theory or von Thünen's model of concentric land use (see chapter~\ref{chap:3}), which primarily attribute density gradients to economic factors such as land values and transportation costs. In contrast, this DLA-based approach introduces a concept from statistical physics—originally used to model phenomena such as coral growth and frost patterns—as a compelling analogy for urban morphogenesis. The growth begins from a single developed site. New units are randomly introduced at the periphery and perform stochastic movements until they adhere to the existing structure. This generates clusters with characteristic tentacle-like extensions and voids, resembling urban expansion along infrastructure corridors.

A central outcome of this model is the emergence of fractal structures, described by a dimension $D$ that quantifies how the extent of developed land scales with distance from the center. Typically, for urban-like structures generated in this manner, $D$ lies below two, reflecting the increasingly sparse occupation of space with growing radius. This relationship is formalized through
\begin{equation}
    N(r) \propto r^D,
\end{equation}
where $N(r)$ denotes the number of developed parcels within radius $r$, leading to a corresponding density function
\begin{equation}
    \rho(r) = \frac{N(r)}{A(r)} \propto r^{D-2},
\end{equation}
which inherently produces a negative gradient for $D<2$. This framework naturally captures allometric growth, where the developed area expands more slowly than the total spatial area.

\begin{figure}
\centering
\includegraphics[width=0.5\textwidth]{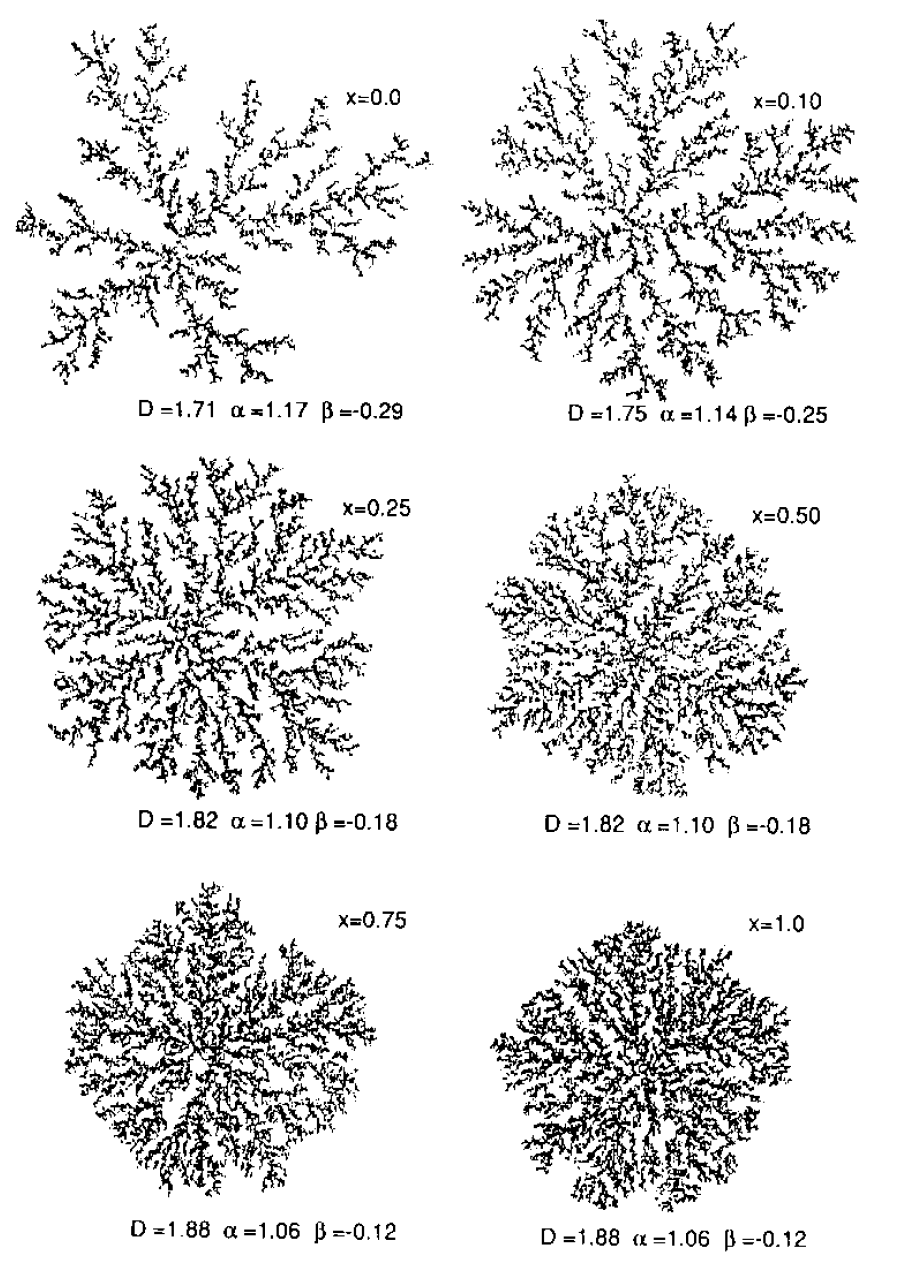}
\caption{Variations in DLA structures with different degrees of forward motion. Source: From \cite{fotheringham1989diffusion}.}
\label{fig:dla2}
\end{figure}

Explorations under varying assumptions show that purely random movement yields sparse, ramified structures with a fractal dimension close to $1.71$, while introducing directional biases—such as drift toward an urban center—produces more compact morphologies and higher $D$, as can be seen in the Fig.~\ref{fig:dla2}. More precisely, in \cite{fotheringham1989diffusion}, the diffusion process includes a directional drift toward the urban center. This is implemented by biasing the random walk so that steps directed toward the center occur with higher probability than outward steps, producing an average inward motion of particles.
Additionally, simulations identify a critical radius beyond which boundary effects become significant, flattening observed density gradients and suggesting that some temporal changes in urban gradients may stem from geometric constraints rather than fundamental shifts in growth dynamics.

Overall, the study of Fotheringham~\cite{fotheringham1989diffusion} demonstrates that complex urban patterns—such as the well-documented negative density gradients—can arise from simple stochastic aggregation processes, without invoking explicit factors like land prices or transportation costs. This approach offers an alternative to conventional economic models by highlighting how spatial organization may emerge from generic growth dynamics. Although it may appear as a straightforward adaptation of the DLA model to urban systems, the work marked a significant conceptual shift at the time: it advocated for minimal statistical models that foreground the emergence of collective behavior and the identification of relevant macroscopic variables.

\subsubsection{Dielectric breakdown model} 
\label{sec:dbm}

\begin{figure}
\centering
\includegraphics[width=0.8\linewidth]{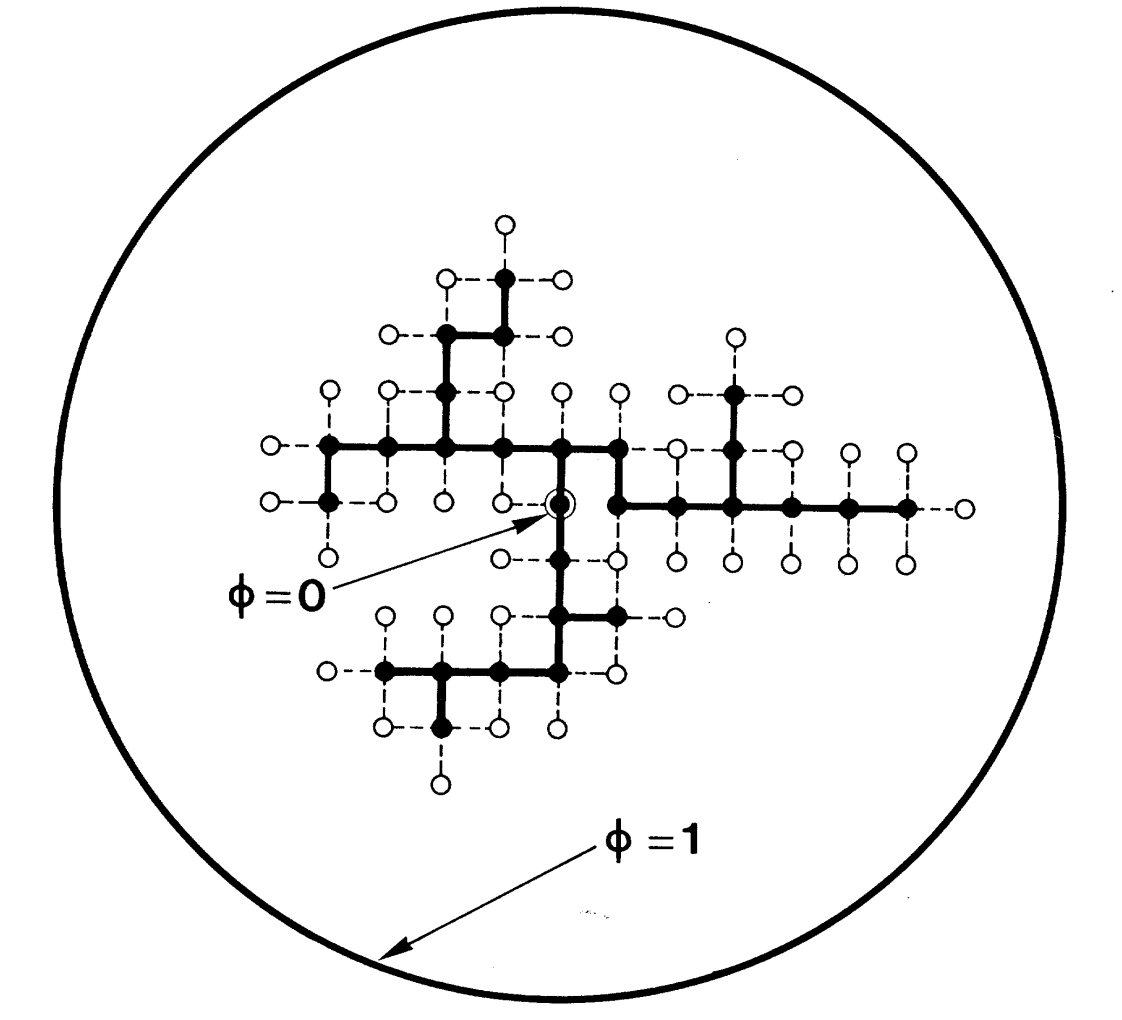}
\caption{Stochastic model of electric breakdown on a lattice, proposed in \cite{niemeyer1984}. The central point represents one electrode, while the other is modeled as a distant circular boundary. The discharge pattern is shown by black dots connected with thick lines and is treated as equipotential. Dashed bonds indicate all possible growth directions, with transition probabilities proportional to the local electric field.~$\phi$ is the electric potential: the discharge pattern is equipotential ($\phi=0$) -- the other boundary condition imposes~$\phi=1$ beyond the external circle. Source: From \cite{niemeyer1984}.}
\label{fig:niemeyer}
\end{figure}

An extension of the DLA model, focusing on the patterns of electric discharge, was proposed by Niemeyer et al \cite{niemeyer1984}. The diffusion field $u$ becomes an electric potential field $\phi$, for which the center of the plane is the point of discharge of the field and the dielectric breakdowns occur in the direction of the highest potential. More precisely, everywhere on the discharge pattern, $\phi=0$ while $\phi=1$ outside an external circle of radius $R_t$ (see Fig.~\ref{fig:niemeyer}).

There are two steps in the growth of the discharging pattern. First, at each step a bond is added to the pattern (possible bonds are dashed bonds in Fig.~\ref{fig:niemeyer}). Then, a site neighboring the discharging pattern (and the associated edge) is added to the pattern with probability
\begin{equation}
    p((i,k) \rightarrow (i', k')) = \frac{\left( \phi_{i',k'} \right)^\eta}{\sum_{\mathcal{C}} \left( \phi_{i',k'} \right)^\eta}
\end{equation}
where~$i,k$ are the coordinates of a site with no potential difference ($\phi_{ik}=0$, black dot in Fig.~\ref{fig:niemeyer}),~$(i',k')$ is a site connected to the black sites (white sites in Fig.~\ref{fig:niemeyer}),~$\mathcal{C}$ represents all the possible neighboring sites and $\eta$ modulates the dependence between local potential difference and probability. As in the case of the DLA, the electric field satisfies the Laplace equation
\begin{equation}
    \nabla^2\phi = 0
\end{equation}
with the boundary equations described previously. In the case $\eta=1$, the probability is proportional of aggregation is proportional to the potential and simulations reveal a fractal dimension $d_f$ about ~$1.75\pm0.02$, similar to the DLA. For $\eta=0$, the growth is compact and $d_f=2$. When $\eta \gg 1$, the produced shapes tend to be linear and $d_f \rightarrow 1$. 

\begin{figure*}
\centering
\includegraphics[width=0.9\linewidth]{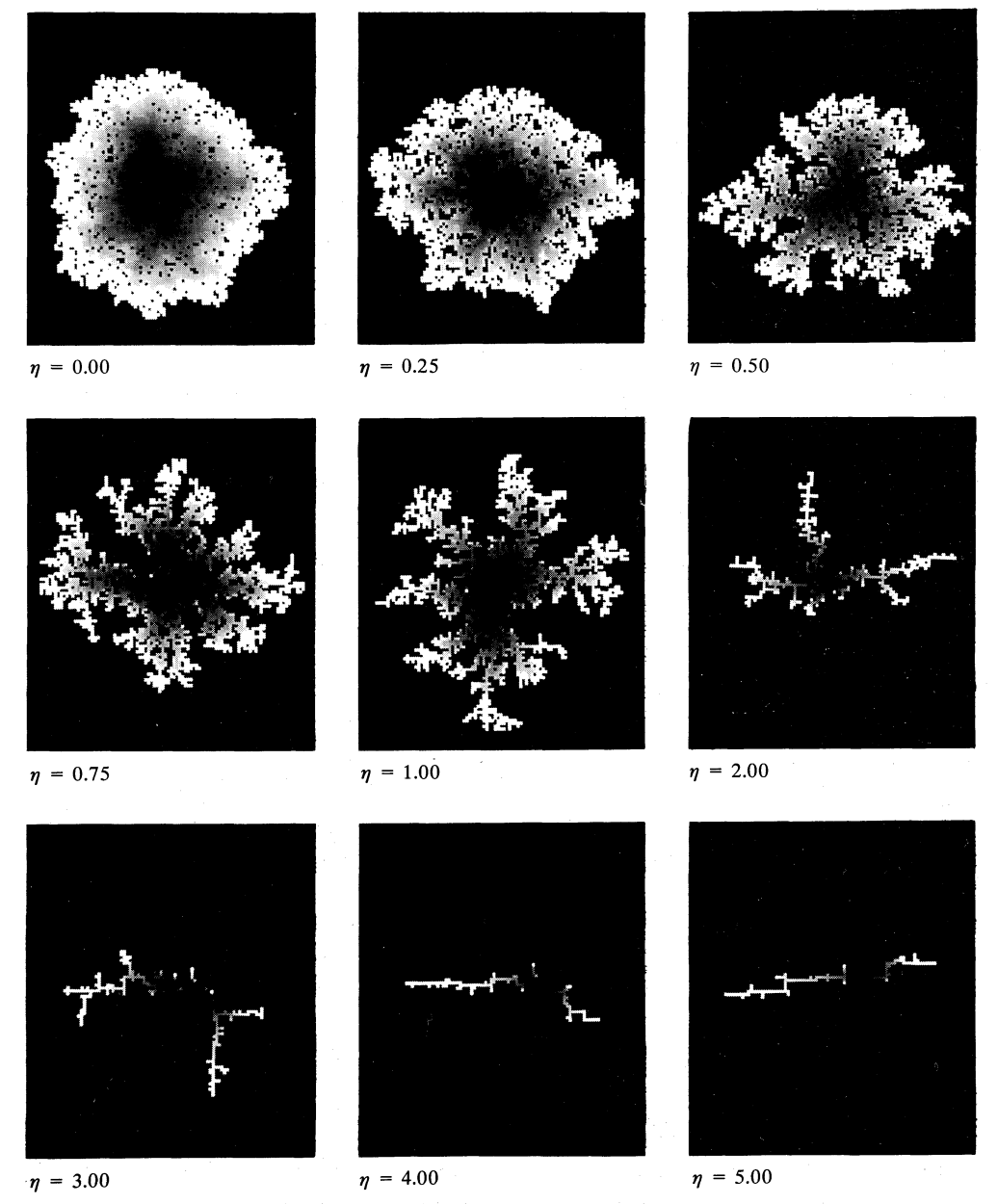}
\caption{Simulations of the city shape for increasing values of the `planning constraint' $\eta$. Source: From \cite{batty1991}.}
\label{fig:battyDB}
\end{figure*}

Batty \cite{batty1991} applied this framework to simulate urban form by drawing an analogy between city growth and the discharge of electrical potential. In this context, $\eta$ acts as a planning control parameter, with different values generating a range of urban morphologies (see Fig.~\ref{fig:battyDB}).
The resulting forms vary from linear structures with fractal dimension close to 1 to concentric patterns approaching dimension 2 (typical fractal dimension for cities are discussed in~\ref{subsec:fractal}). The model was further applied to urban development in Cardiff.

\subsection{Extension of the DLA}

In~\cite{rybski2013distance}, the authors introduce the \emph{Stochastic Gravitation Model} (SGM), which simulates urban growth through a probabilistic mechanism. The core idea is that urban expansion is more likely to occur near already urbanized areas. This growth dynamic is governed by a single parameter: an exponent that controls how strongly the attraction to urbanized sites decays with distance. The model operates iteratively, allowing existing clusters to expand while also permitting the emergence of new ones. The primary objectives of the SGM are to reproduce the power-law size distribution of urban clusters and to capture the fractal nature of the boundary of the largest cluster.

The central quantity in the model is the probability $q_i$ that a site $i$ transitions from a non-urban state $w_i = 0$ to an urbanized state $w_i = 1$, defined as
\begin{equation}
    q_i = \frac{\sum_j w_j \, d(i,j)^{-\gamma}}{\sum_j d(i,j)^{-\gamma}} ,
\label{eq:ryb}
\end{equation}
where  $d(i,j)$ denotes the Euclidean distance between sites $i$ and $j$, and $\gamma$ is a decay parameter that controls how rapidly the probability of urbanization decreases with distance. Thus, $q_i$ reflects the likelihood that site $i$ becomes urbanized, weighted by its proximity to existing urbanized sites.

\begin{figure*}
     \centering
     \includegraphics[width=1\linewidth]{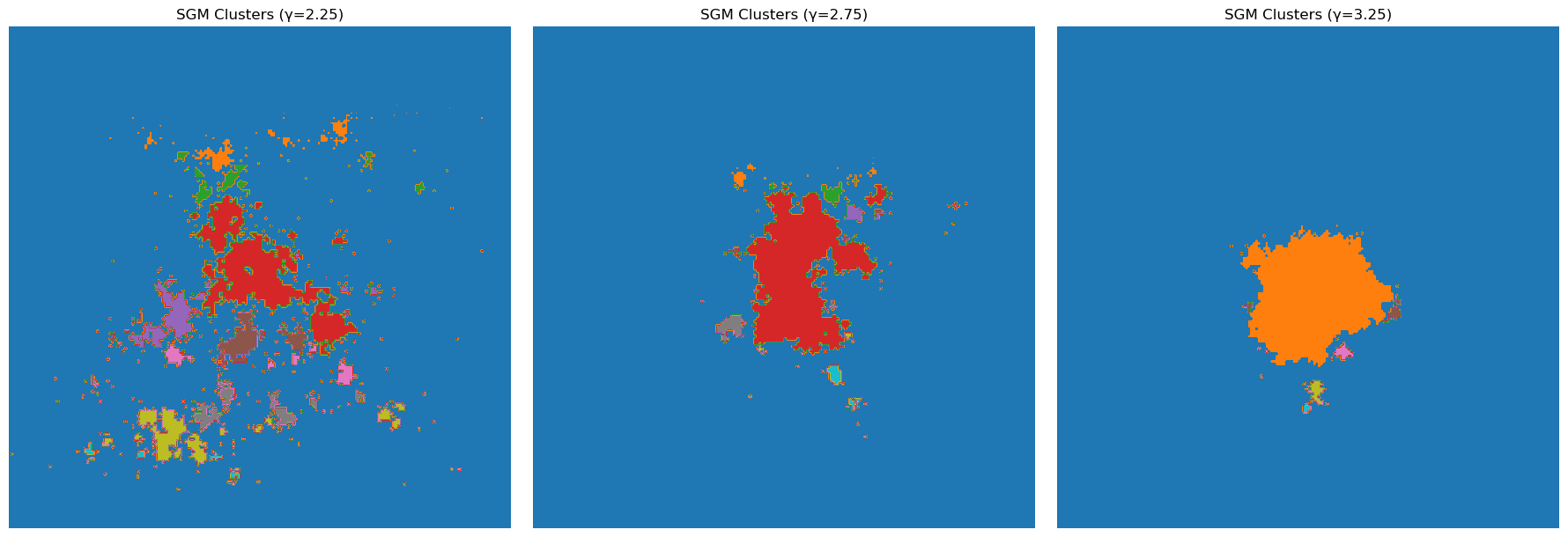}
     \caption{Simulation of growth dynamics using the SGM model proposed in~\cite{rybski2013distance}, for different values of $\gamma = 2.25, 2.75, 3.25$ from left to right (on a $200 \times 200$ lattice). Colors indicate distinct connected clusters of urbanized sites.}
     \label{fig:sgm}
\end{figure*}

The model is implemented on a two-dimensional square lattice. Nodes are scanned sequentially, and each site may become urbanized with probability $q_i$. After each full scan, all probabilities are updated. To ensure that the values remain within the interval 
$[0,1]$, they are normalized as
\begin{align}
\tilde{q}_i = \frac{q_i}{\max(q_i)},
\end{align}
so that the site with the highest raw probability has $\tilde{q}_i = 1$, while the relative differences between probabilities are preserved. Simulation results for different values of $\gamma$ are shown in Fig.~\ref{fig:sgm}.

The model captures key features of urban morphology, including the power-law distribution of cluster sizes, $P(S) \sim S^{-\zeta}$ with $\zeta \approx 2$ (see~\ref{subsec:component}), and the fractal geometry of the boundary of the largest cluster. While the cluster size distribution depends on both the decay exponent $\gamma$ and the number of iterations, the fractality of the boundary appears to depend primarily on the latter. Empirical analysis of land-cover data for Paris and its surroundings yields an estimated decay parameter of 
$\gamma \approx 2.5$.\\


In~\cite{rybski2021modeling}, Rybski et al. propose a hybrid framework for urban morphology that unifies the DLA model with the Stochastic Gravitation Model (SGM). In their earlier formulation, the probability that a site $i$ transitions from a non-urban to an urban state is given by $q_i$ (Eq.~\ref{eq:ryb}). In this subsequent model—termed Diffusion-Limited Gravitation (DLG)—they combine the attraction-based growth dynamics of the SGM with the random walker mechanism characteristic of DLA. The resulting structures appear visually less scattered than those produced by the SGM alone.

\begin{figure*}
    \centering
    \includegraphics[width=0.9\linewidth]{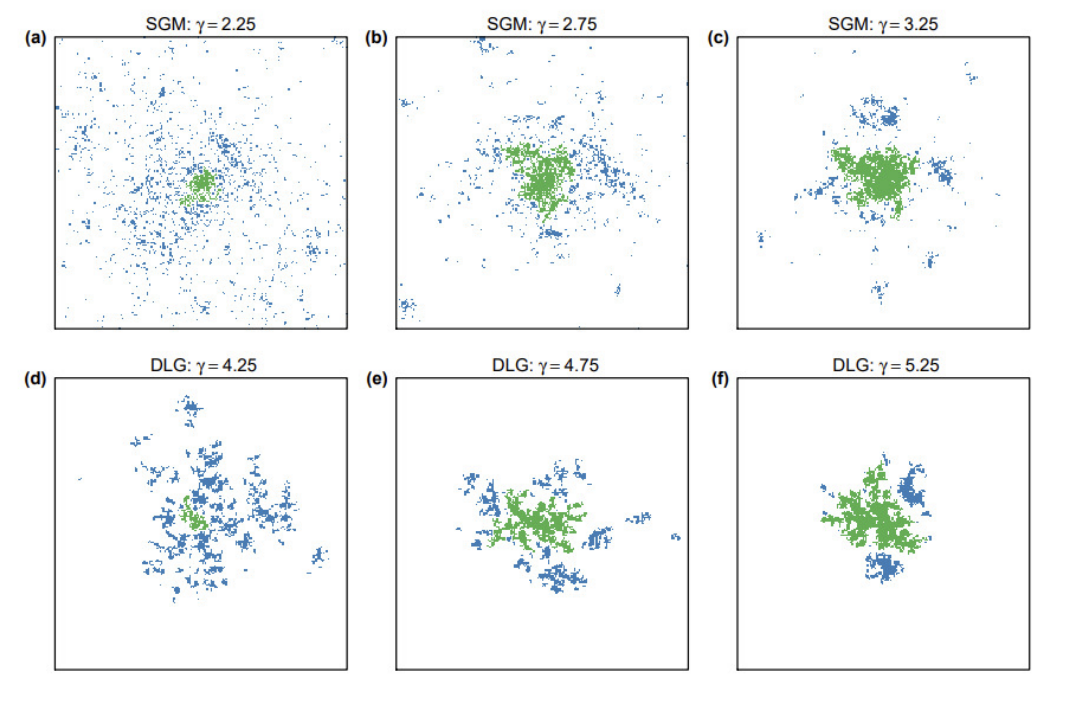}
    \caption{Examples of urban morphologies generated by the Stochastic Gravitation Model (SGM) (a–c) and the Diffusion-Limited Gravitation model (DLG) (d–f), for varying values of 
    $\gamma$. Green indicates the central cluster; blue indicates peripheral urbanized sites. Source: From \cite{rybski2021modeling}.}
    \label{fig:DLG_patterns}
\end{figure*}

The model operates on a square lattice, starting with a single occupied site at the center. Urban growth proceeds via the action of random walkers: each walker begins at a randomly chosen location on the grid, performs a random walk, and halts with probability $q_i$, defined as before. If the stopping condition is met, the site is marked as urbanized. This process repeats iteratively until the target number of urban sites is reached. The DLG model thus combines the dendritic, diffusion-driven expansion of DLA with the gravitation-based clustering of SGM. Figure~\ref{fig:DLG_patterns} illustrates the resulting morphologies for different values of the decay parameter $\gamma$. For small $\gamma$, growth remains dispersed and produces multiple clusters. For larger $\gamma$, a dominant cluster forms, reflecting the spatial primacy often observed in real urban systems. By balancing spatial diffusion and attraction, the DLG offers a more realistic and flexible framework for modeling urban morphologies, addressing limitations of both SGM and DLA.

\subsection{The Eden model}

\subsubsection{Definition}

The Eden model, introduced by Murray Eden in 1961~\cite{Eden:1961}, is a prototypical stochastic growth process on a discrete lattice. Originally motivated by biological phenomena such as the proliferation of bacterial colonies, it has since become a central object of study in statistical physics, particularly for modeling irreversible aggregation and the emergence of kinetically roughened interfaces.

The model is typically defined on a $d$-dimensional regular lattice (most commonly $\mathbb{Z}^d$), where a growing cluster 
$C(t) \subset \mathbb{Z}^d$ evolves in discrete time $t$. Starting from a single occupied seed site at $t = 0$, the cluster grows by randomly occupying one of the unoccupied lattice sites adjacent to the existing cluster. Let $\partial C(t)$ denote the boundary of the cluster—i.e., the set of unoccupied sites with at least one occupied neighbor. At each time step, a site 
$x \in \partial C(t)$ is selected uniformly at random and added to the cluster
\begin{equation}
C(t+1) = C(t) \cup \{x\}, \quad x \in \partial C(t).
\end{equation}

This simple local rule generates compact clusters with approximately circular (in 2D) or spherical (in 3D) geometries in the continuum limit, characterized by fluctuating interfaces. A natural observable to describe the geometry of the growing interface is the interface height function $h(\vec{x}, t)$, defined as the distance from the origin along a direction 
$\vec{x}$. The roughness (or width) of the interface is defined as
\begin{equation} \label{eq:widtheden}
w(L, t) = \sqrt{\langle h^2 \rangle - \langle h \rangle^2},
\end{equation}
where the average is taken over spatial positions $\vec{x}$ (and possibly over ensemble realizations), and $L$ denotes the linear system size.

The Eden model has been shown \cite{roux1991direct} to belong to the celebrated Kardar–Parisi–Zhang (KPZ) universality class~\cite{kardar1986}. Based on symmetry arguments and key physical ingredients (such as local growth and stochastic fluctuations), Kardar et al. proposed that the continuum limit of the Eden model is governed by the nonlinear stochastic partial differential equation
\begin{equation}
\frac{\partial h}{\partial t} = \nu \nabla^2 h + \frac{\lambda}{2} (\nabla h)^2 + \eta(\vec{x}, t),
\end{equation}
where $\nu$ and $\lambda$ are constants, and $\eta(\vec{x}, t)$ is a Gaussian white noise with zero mean and correlation
\begin{equation}
\langle \eta(\vec{x}, t) \eta(\vec{x}', t') \rangle = 2D\, \delta(\vec{x} - \vec{x}') \delta(t - t').
\end{equation}
This equation captures the essential features of stochastic growth with lateral correlations and nonlinearity in the growth rate. The connection between the Eden model and the KPZ equation was further reinforced by mapping it onto the problem of directed polymers in random media at zero temperature. In this mapping, the interface corresponds to the ground-state configuration of the polymer, and the stochastic growth rules of the Eden model translate into a noise term with a specific distribution. This mapping provides not only a conceptual link but also quantitative predictions, including universal scaling exponents.

Interestingly, while the Eden model always exhibits KPZ scaling, it contrasts with the behavior of directed polymers in very disordered media, where universality can break down. Thus, the Eden model serves as a robust example of a system that falls within a universal growth class, despite the simplicity of its local update rules.

The interface width exhibits Family-Vicsek dynamic scaling \cite{family1984} of the form
\begin{equation}
    w(L, t) \sim
\begin{cases}
t^\beta & \text{for } t \ll t_\times, \\
L^\alpha & \text{for } t \gg t_\times,
\end{cases}
\quad \text{with } t_\times \sim L^z, 
\end{equation}
and $z=\alpha/\beta$. The exponent $\alpha$ is the roughness exponent, $\beta$ is the growth exponent, and $z$ is the dynamic exponent. In $1+1$ dimensions, the Eden model exhibits exponents $\alpha = 1/2$, $\beta = 1/3$, and $z = 3/2$, consistent with exact KPZ  predictions~\cite{krug1997}. 

Parisi and Zhang~\cite{parisi1984} analytically studied the Eden model in high spatial dimensions $d$, and demonstrated that its mean square radius~$R_n$ depends on the number of particles~$n$ and on the dimension~$d$ like
\begin{equation}
    \langle R_n^2 \rangle \sim 2 \log(n) \left( 1+ \frac{3}{2d} \right) 
\end{equation}
for large $d$, which is markedly different from the naive expectation $\langle R_n^2 \rangle \sim n^{2/d}$. Indeed, if one assumes that the increment in particle number is $dn$, then a simplistic geometric argument based on volume growth $R^{d-1} \, dR \sim dn$ would suggest $R \sim n^{1/d}$. The logarithmic correction found by Parisi and Zhang thus reveals subtle collective effects in the growth process that are not captured by such mean-field arguments.

Due to its simplicity, the Eden model allows for efficient numerical simulation (see e.g. Chapter 10 of \cite{landau2014}) and serves as a baseline for exploring more complex growth scenarios. Numerous generalizations have been proposed, including anisotropic growth rules, nonuniform deposition probabilities, and growth on disordered or dynamic substrates. These extensions provide insight into crossover behavior and the robustness of KPZ-class universality under varying physical constraints. Thereafter, we discuss a variant of the Eden model for city growth.

\subsubsection{Application to city growth}


The standard DLA model fails to capture two essential features of real urban morphologies. First, it lacks the formation of a compact urban core. Second, it does not reproduce the typical spatial structure of a city consisting of a central cluster surrounded by smaller, detached clusters, rather than a single, continuous component. To address these limitations, Benguigui~\cite{benguigui1995} proposed a variant of the Eden model that introduces a parameterized mechanism for generating disconnected sub-clusters while preserving stochastic growth. As in the Eden model, the process begins with a seed site, and growth proceeds by selecting unoccupied neighboring sites. However, a key modification is introduced: each candidate site is only occupied once it has been selected $p$ times. This integer parameter $p$ controls the threshold for occupation and allows for the emergence of multiple disconnected clusters, depending on its value.

This $p$-model can be studied through several observables. These include the radius of gyration $R_G$, which characterizes the spatial extent of the aggregate; the time $t(N)$ required to reach a given number of sites $N$; and the density of occupied sites as a function of distance from the center, denoted by $\rho(r)$.

\begin{figure}
    \centering
    \includegraphics[width=0.8\linewidth]{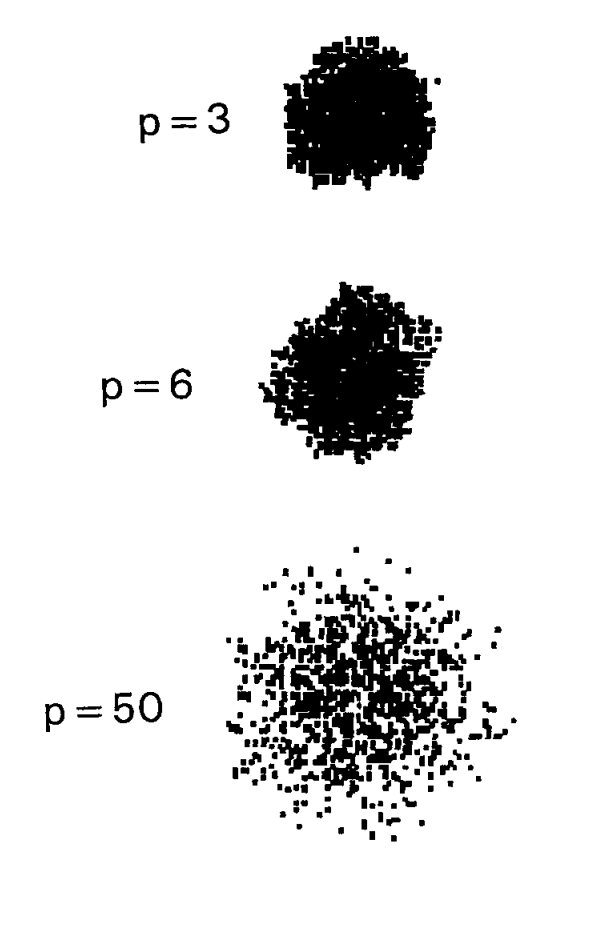}
    \caption{Urban clusters generated by the $p$-model~\cite{benguigui1995} for values $p = 3$, $6$, and
    $50$, with $N = 900$. The critical value is $p_c = 5$. Source: From~\cite{benguigui1995}.}
    \label{fig:benguigui1}
\end{figure}

A critical threshold $p_c$ separates two regimes of growth. For fixed aggregate size $N$, the radius of gyration scales as
\begin{align}
R_G \sim
\begin{cases}
    p^2 & \text{if } p < p_c, \\
    p^\beta & \text{if } p > p_c
\end{cases}
\end{align}
with $\beta \approx 0.18$. Numerical simulations indicate that $p_c$ itself scales with system size as 
\begin{align}
p_c(N) \sim N^\alpha,     
\end{align}
where $\alpha \approx 0.45$. When keeping $p$ fixed and varying $N$, no anomalous behavior is observed. In that case, the radius of gyration follows 
$R_G(N) \sim N^\gamma$, with $\gamma \approx 0.43$, close to the Eden model exponent of $1/2$ (see Fig.~\ref{fig:benguigui1} for examples of cluster generated with this model).

The parameter $p_c$ thus marks a transition between two types of aggregation processes. Below this threshold, the cluster is compact and dense, with central density $\rho_0 = 1$. Above 
$p_c$, the growth becomes more dispersed and the central density decreases. In the subcritical regime, the density profile exhibits a central plateau: $\rho(r) = 1$ for $r \leq r_1$, followed by a decay for $r > r_1$. The transition radius $r_1$ scales as 
$r_1 \sim (p_c - p)^\delta$, with $\delta \approx 1.7$.

In the supercritical regime, the density profile collapses when plotted in scaled coordinates, $\rho(r)/\rho_0$ versus 
$r/r_{\max}$, revealing a universal scaling function of the form 
$\rho(r) = \rho_0 f(r/r_{\max})$. Assuming that 
$\rho_0 \sim N^a p^b$ and $r_{\max} \sim N^c p^d$, and using the conservation relation
\begin{equation}
    N \sim \int_0^{r_{\max}} \rho(r) \, r\, dr \sim N^a p^b \int_0^{r_{\max}} f\left( \frac{r}{N^c p^d} \right) r\, dr,
\end{equation}
a change of variables $u = r / (N^c p^d)$ yields
\begin{equation}
    N \sim N^{a + 2c} p^{2d + b} \int_0^1 f(u) u \, du,
\end{equation}
which leads to the scaling constraints:
\begin{equation}
    a + 2c = 1, \qquad 2d + b = 0.
\end{equation}

A similar argument applies to the radius of gyration:
\begin{equation}
    R_G^2 \sim \int_0^{r_{\max}} \rho(r) \, r^3 \, dr,
\end{equation}
implying
\begin{equation}
    R_G \sim N^{(4c + a - 1)/2} p^{(4d + b)/2} \sim N^c p^d \sim r_{\max}.
\end{equation}
From numerical results, the exponents are estimated as 
$a = 0.15 \pm 0.02$, $b = 0.32 \pm 0.01$, $d = 0.12 \pm 0.02$, and $c = \gamma \approx 0.43$. The exponent~$c$ can also be understood as a fractal measure (discussed for cities in~\ref{subsec:fractal}).

Benguigui also explored the potential application of this model to real urban growth. To reproduce the radial population profiles 
$\rho(r)$ of cities (discussed in~\ref{subsec:clark}) such as Baltimore, Paris, and London, he introduced a dependence $p = p(n)$, where $n$ is the number of particles already in the aggregate. The function $p(n)$ must satisfy two constraints: it should be close to zero for small $n$, to ensure a dense urban core, and become much larger than $p_c$ at large $n$, to account for the formation of peripheral clusters. With appropriate functional forms, the model was able to reproduce empirical $\rho(r)$ curves in good agreement with the data. However, the resulting
morphologies differ significantly from those observed in real cities. This discrepancy likely arises from the assumption of a single growing cluster, whereas actual urban systems grow through the simultaneous expansion and
aggregation of multiple sub-clusters. In this respect, the model captures one important stylized fact of urban morphology—namely the radial decay of density $\rho(r)$ and the associated
mass scaling $N(r)$—but it fails to reproduce another key feature of real
cities: the coexistence of multiple centers and secondary clusters that
emerge during the growth process.

\begin{figure*}
    \centering
    \includegraphics[width=0.45\linewidth]{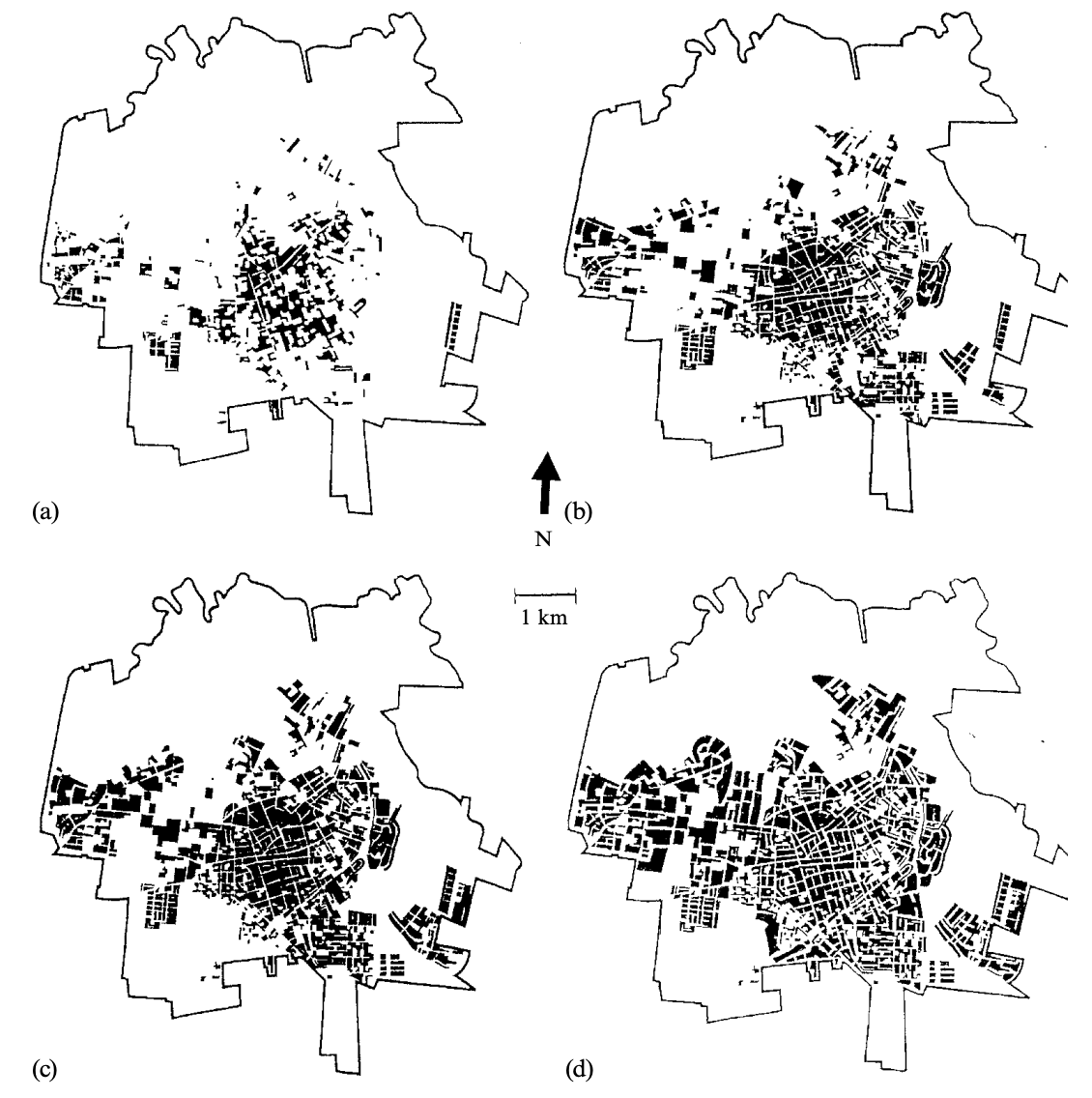}
    \includegraphics[width=0.45\linewidth]{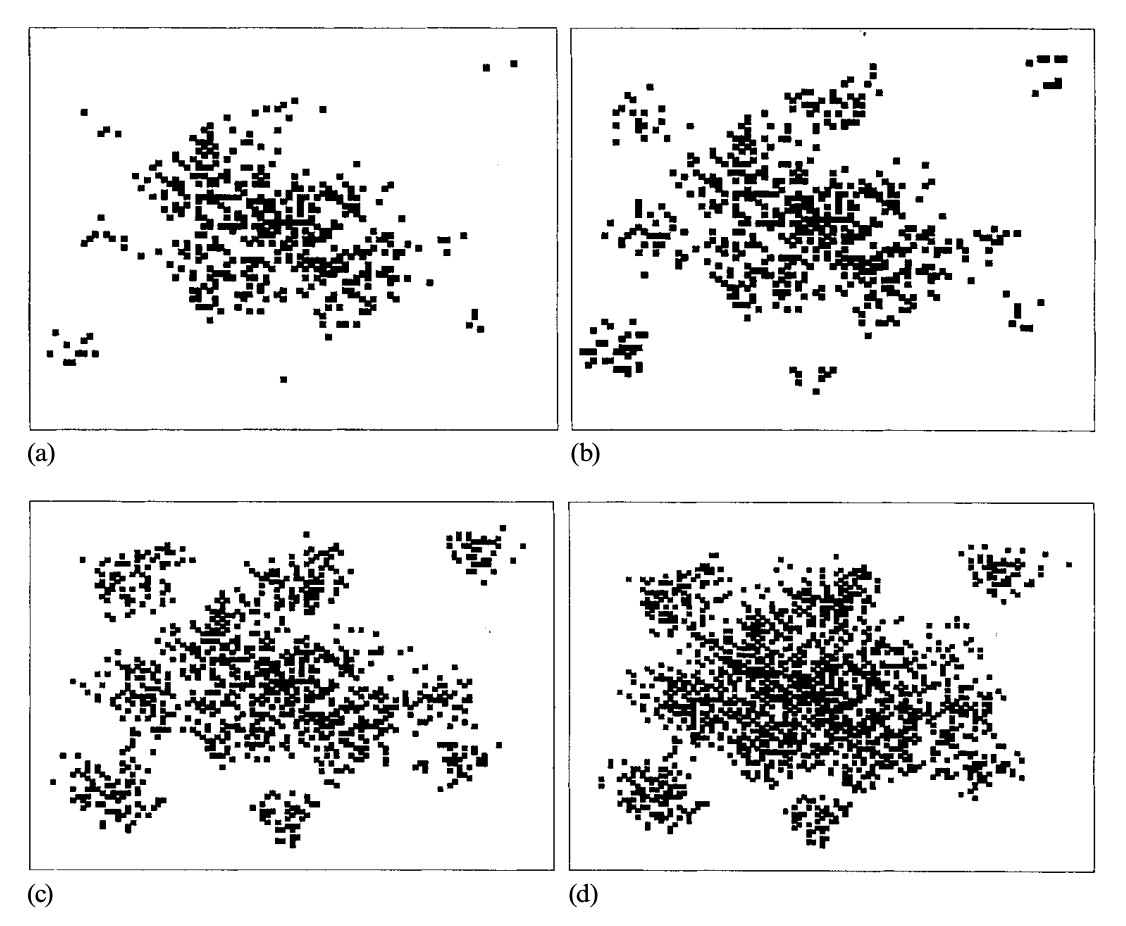}
    \caption{(Left) Built-up area of Petah Tikvah (Israel) for the years (a) 1949, (b) 1961, (c) 1971, and (d) 1996. (Right) Simulated aggregates generated by the \textit{p}-model corresponding to the same years. Source: adapted from~\cite{benguigui2001dynamics}.}
    \label{fig:benguigui2}
\end{figure*}

In a subsequent study~\cite{benguigui2001dynamics}, the model was applied to simulate the historical
growth of Petah Tikvah, a small town in the Tel Aviv metropolitan area. The simulated clusters shown in Fig.~\ref{fig:benguigui2} using the $p$-model display good agreement with historical built-up area maps from 1949 to 1996, both in
their radial density profile and their overall morphological evolution.

\subsubsection{Ciamarra-Coniglio model}

 In \cite{ciamarra2006randomwalkclustergrowth}, Ciamarra and Coniglio introduce the random walk growth (RWG) model, inspired from DLA and the Eden model. These authors proposed this model to describe the growth of compact clusters, characterized by fluctuating
and growing interface, and argued that this could be applied to real cities, which grow as new buildings are constructed,
usually in the suburbs. 

For both the Eden and DLA models, numerical simulations suggest that the radially averaged probability $P(r, N)\,dr$ — that the 
$(N + 1)$th unit is deposited within a shell of width 
$dr$ at a distance $r$ from the center of mass of the cluster — is well approximated, for large $r$ and $N$, by a Gaussian distribution
\begin{equation}
    P(r,N) = \frac{1}{\sqrt{2 \pi} \sigma_N} \exp \left( - \frac{(r - r_N)^2}{2 \sigma_N^2} \right),
\end{equation}
where $r_N$ is the mean radial position of the deposition shell and $\sigma_N$ is its standard deviation behaving as
\begin{equation}
     r_N \propto N^\nu, \sigma_N \propto N^{\nu'}
\end{equation}
($\nu$ and $\nu'$ are model-dependent). The density 
is then given by
\begin{equation}
    \rho(r,N)  = \frac{1}{S_d r^{d-1}} \int_0^N P(r,N') dN'
\end{equation}
where $S_d = 2 \pi^{d/2}/\Gamma(d/2)$ is the surface of the $d$-dimensional unit sphere. In the RWG, the goal is to grow compact clusters, of fractal dimension $D_f = d$, hence the growth probability becomes
\begin{equation}
    P_{\text{RWG}} = \frac{r^{d-1}}{\mu_N^{(d-1)}} P(r,N)
\end{equation}
where $\mu_N^{(d-1)}$ is the normalizing constant. This is based on the simple idea that the cluster radius grows as a
random walker subject to a drift, which gives a growing probability distribution $P_{RWG}(r, N) \propto r^{d-1}P(r, N)$. This leads to exponents
\begin{equation}
    \nu = \frac{1}{d},\, \nu' = \frac{1}{2d}.
\end{equation}
These findings relate to homothetic scaling of the density (see~\ref{subsec:homo}).
In the two-dimensional case, the model is exactly solvable. For large clusters with $\frac{r_N}{\sigma_N} \gg 1$, the radial density profile can be well-approximated by
\begin{equation} \label{eq:densityciamarra}
    \rho(r, N) \approx \rho_\infty + \frac{\rho_{\max} - \rho_\infty}{2} \, \text{Erfc} \left( \frac{r - r_N}{\sqrt{2} \, \sigma_N} \right),
\end{equation}
where $r_N$ is the characteristic radius of the cluster, 
$\sigma_N$ its standard deviation, $\rho_{\max}$ the density at the city center, and $\rho_\infty$ the asymptotic density in the outskirts.

\begin{figure}
    \centering
    \includegraphics[width=0.6\linewidth]{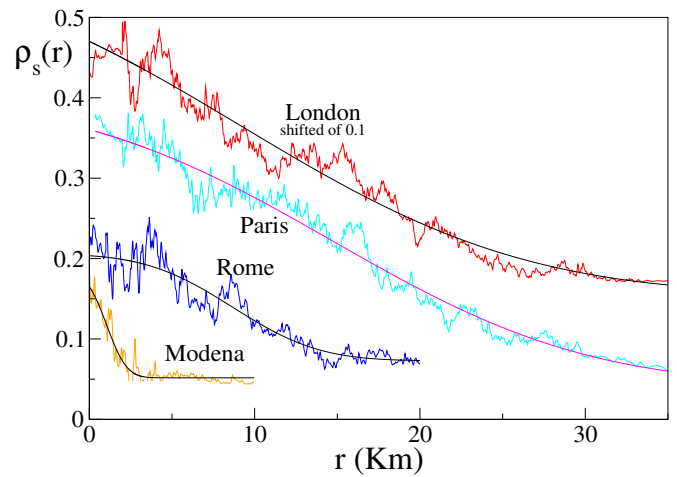}
    \caption{Density of streets as a function of distance from the city center, with best fit using Eq.~\eqref{eq:densityciamarra}. The fitted parameters are $ r_N $, $\lambda = \sigma_N^2 / r_N$, $\rho_{\max}$, and $ \rho_\infty$. Source: From \cite{ciamarra2006randomwalkclustergrowth}.}
    \label{fig:ciamarraconiglio_density}
\end{figure}

Although this type of model essentially focuses on the population density, Ciamarra and Coniglio considered the street density profiles in different cities. They tested their model on empirical data for the cities of Modena, Rome, Paris, and London (see Fig.~\ref{fig:ciamarraconiglio_density}), by fitting the observed street density profiles to Eq.~\eqref{eq:densityciamarra}. From the fitted parameters, one can extract key urban morphological indicators: the city radius $r_N$, the variance $\sigma_N^2$ of the deposition profile, the characteristic growth length 
$\lambda = \sigma_N^2 / r_N$, the central street density $\rho_{\max}$, and the peripheral density 
$\rho_\infty$.
The parameter $r_N$ acts as a proxy for city size and varies from one city to another. Similarly, $\rho_{\max}$ reflects the intensity of the street network in the urban core. Paris and London, characterized by dense central infrastructures and wide avenues, exhibit higher values of $\rho_{\max}$ than Rome and Modena. In contrast, the parameter $\rho_\infty$, which measures the typical density in the urban periphery, appears to be approximately constant across all cities studied, suggesting a unifying feature of suburban street networks.

Finally, the growth length $\lambda$ takes comparable values for Paris, Rome, and Modena, but a significantly larger value for London. Since $\lambda$ controls the variation of the distribution's variance via the relation 
$\sigma_N^2 = \lambda r_N$, a larger $\lambda$ implies a more diffuse urban expansion. This result leads to the conclusion that London is less compact than the other European cities considered.

The model manages to reproduce well street-network density radial patterns, while staying simple -- it only uses  four parameters, which is advantageous for tractability and interpretability. However, several limitations remain. First, the radial framework assumes isotropic growth, potentially overlooking anisotropic patterns observed in real cities~\cite{marquis2025universalroughnessdynamicsurban}. Second, the model provides limited insight into the dynamics of urban expansion or the detailed geometry of street networks and building layouts. Third, despite its focus on radial profiles, it does not provide information about spatial correlations $\langle \rho(r) \rho(r') \rangle - \langle \rho(r) \rangle \langle \rho(r') \rangle$. Finally, while the model has been tested on street networks, its connection to key urban quantities—such as population density or total built-up area—remains unclear, leaving open questions about its applicability to predictive urban modeling.

\subsection{Transient dynamics of urban growth}

The paper \cite{manrubia1998intermittency} studies a reaction-diffusion model that leads to a strongly inhomogeneous, spatiotemporally intermittent density field. 

In this model, cities result from the interplay of two competing transport processes that shape population distribution in opposing ways. First, a reaction process where people tend to cluster in cities to benefit from concentrated resources, economic opportunities, and social interactions. This attraction effect grows with city size. In the model, this is represented by multiplicative growth events, where, with a certain probability, the population of specific regions increases by a factor. This growth is sustained by population transport from other areas, leading to spatial inhomogeneities, while maintaining the total population on average. The localized nature of these events allows them to be interpreted as reaction-like processes in a reaction-diffusion framework.
Second, a diffusion process: in order to counterbalance excessive concentration, a spreading mechanism redistributes population across local neighborhoods, preventing extreme density accumulation. This is modeled as a diffusion process, ensuring a degree of local demographic homogenization. In numerical simulations, diffusion is implemented locally, but in analytical treatments, global redistribution is also considered. This allows for a theoretical analysis of how diffusion affects the heterogeneous patterns formed by the reaction process.

Together, these processes generate a self-organized, intermittent population distribution, balancing local clustering and global dispersal.

Mathematically, the model evolves on a two-dimensional lattice with discrete time steps. The population density is denoted by $\rho(x,t)$ at site $x$ and time $t$. The authors assume that $\rho(x,0)=1$ for all $x$. 
Each time step is divided in two substeps, at which the reaction and diffusion mechanisms are
successively applied. In the following, we specify the detailed form of these processes.

For the reaction process, at each time `substep' $t<t'<t+1$, this density evolves as
\begin{align}
\rho(x,t')=\xi(x,t)\rho(x,t)
\end{align}
where $\xi$ is a stochastic process defined as
\begin{align}
    \xi(x,t)=\begin{cases}
        (1-q)/p,\;\text{with probability}; p\\
        q/(1-p),\;\text{with probability}; 1-p
    \end{cases}
\end{align}
The population thus evolves via a stochastic multiplicative rule with parameters $p,q\in [0,1]$. This multiplicative process is a generalization of the Zeldovich process for intermittency which corresponds to $p=1/2$ and $q=0$. Under the action of this process, the higher moments of the density defined as 
\begin{align}
    \langle \rho(x,t)^k\rangle=\sum_x\rho(x.t)^k
\end{align}
diverge for large time \cite{zanette1994intermittency}.
This divergence is the signature of intermittency and the emergence of strong heterogeneities in $\rho(x,t)$ with sharp spikes appearing at locations where the probability $p$ accumulates and the population is multiplied at each step by $(1-q)/p>1$ (in other sites whose number becomes the majority the population decreases). This multiplicative process gives naturally rise to a log-normal distribution for the population frequency (ie. the probability distribution $f(N)$ to observe a population $N$), and is unable to reproduce observed empirical power laws (although we now know that this distribution is not in general a power law \cite{verbavatz2020growth}). This process is thus able to produce strong heterogeneities but fails to account for statistical properties of populations, which is the reason why they introduce a second process based on diffusion. 

For the diffusion process, the population then redistributes locally (from time $t'$ to $t+1$) according to
\begin{equation}
        \rho(x,t+1) = (1 - \alpha) \rho(x,t') + \frac{\alpha}{k} \sum_{x' \in \Gamma(x)} \rho(x',t'),
\end{equation}
where $\Gamma(x)$ represents the neighborhood of site $x$ and whose cardinal is $k$.

Depending on the definition of the neighborhood $\Gamma(x)$ the results are different. When it is extended to the whole system, the diffusion equation reads
\begin{align}
    \rho(x,t+1)=(1-\alpha)\rho(x,t')+\alpha n_0
\end{align}
where $n_0$ is the constant average population per site initially fixed at $n_0=1$. In this case, including diffusion stabilizes the population frequency $f(N)$ which becomes for large $N$
\begin{equation}
    f(N) \propto N^{-z}, \quad z = 1 + \frac{\ln p}{\ln(p / (1 - \alpha))}.
\end{equation}

The diffusion is thus able to modify the exponent $z$ of the power law distribution $f(N)$ and ensures its stationarity. This exponent $z$ is not universal and depends on both $p$ and $\alpha$. 

From these results, it follows that in low-dimensional systems, local diffusion (when $\Gamma(x)$ is small) is insufficient to prevent the formation of strong heterogeneities in $\rho(x,t)$. Instead, diffusion primarily acts to redistribute the population, transferring individuals from highly populated sites to those with lower population densities.

For local diffusion, it can be shown that the solution is $f(N)\sim A/N^2$ that corresponds better to empirical measures. 

The numerical simulations of the reaction–diffusion model show that it reproduces a wide range of empirical statistical properties of urban systems. The distribution of city areas follows a power law $f(N) \propto N^{-1.93 \pm 0.03}$, while the relationship between population and area scales nearly linearly as $m \propto a^{\beta}$ with $\beta \approx 1$. The model also yields an exponential decay of population density with distance from the city core, $\rho(d) \propto e^{-\lambda d}$, where $\lambda \sim 10^{-2}$, and generates city boundaries with fractal dimensions in the range $D \approx 1.15 - 1.35$, consistent with empirical estimates between $1.2$ and $1.4$.  

When compared with real-world urban distributions, the model successfully reproduces the universal statistical regularities observed across cities. In particular, the population frequency follows Zipf’s law with $f(n) \propto n^{-r_o}$ and exponent $r_o \approx 2$, while the distribution of city areas behaves as $f(a) \propto a^{-s_o}$ with $s_o \approx 2$. Furthermore, the exponential decay of urban density from the compact city core, $\rho(d) \propto e^{-\lambda_o d}$, is consistent with field data, and the geometry of urban boundaries exhibits fractal properties with dimensions in the interval $D_o \in [1.2,1.4]$.  

Taken together, these results demonstrate that the stochastic interplay of reaction-like growth and diffusive redistribution provides a robust explanation for the emergence of universal patterns in urban systems. The model shows how urban development self-organizes into scale-invariant distributions, combining local clustering with global dispersal, and thereby accounts for Zipfian size distributions, exponential density profiles, and fractal urban boundaries within a single theoretical framework.
\begin{figure}
\centering
\includegraphics[angle=0, width=0.45\textwidth]{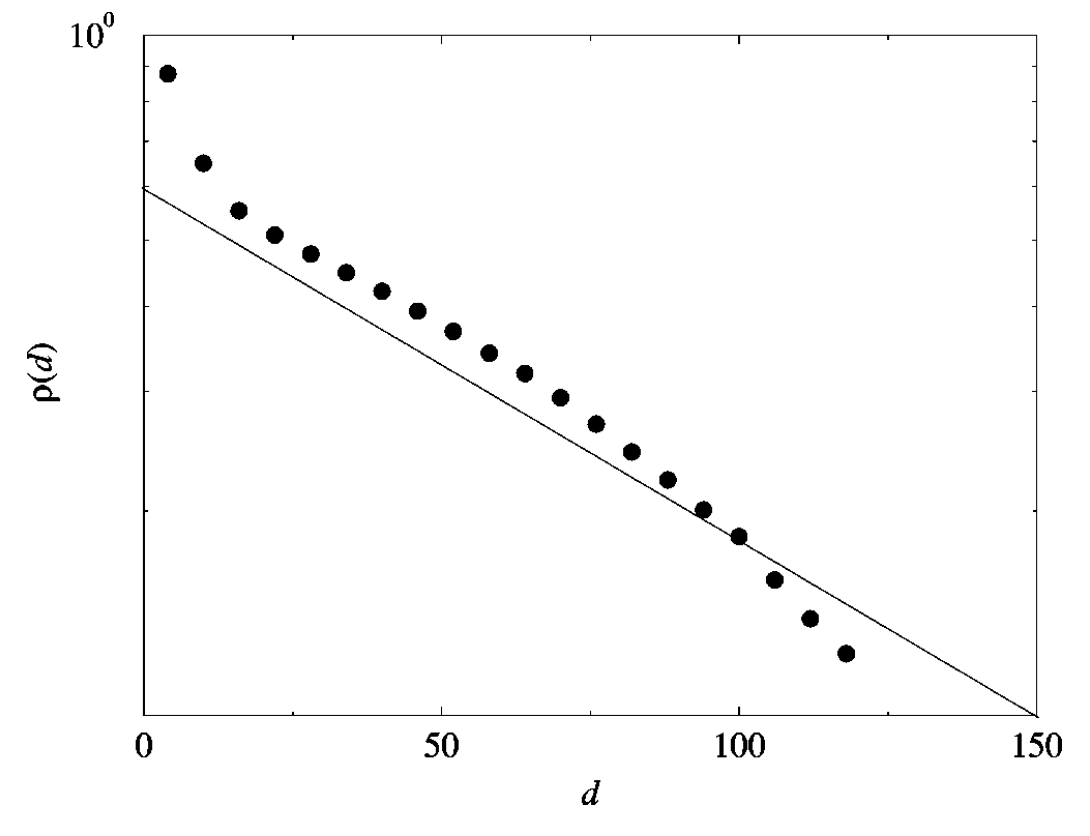}  
\includegraphics[angle=0, width=0.45\textwidth]{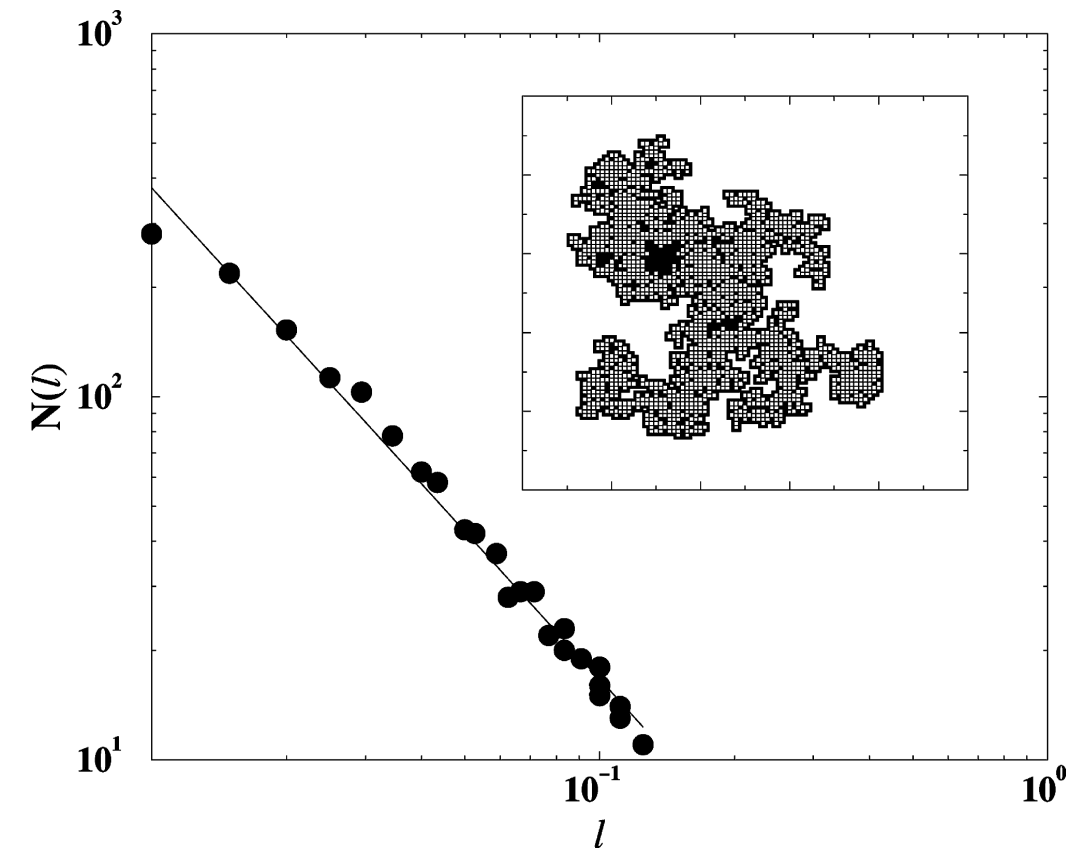}  
\caption{Top: Exponential decay of the probability of urbanization with distance from the compact city core, averaged over many realizations. Bottom: fractal dimension of the boundary of a typical large cluster, with slope $D \approx 1.3$. Across many realizations, the fractal dimension lies between $1.15$ and $1.35$, matching empirical observations. Source: From \cite{manrubia1998intermittency}.} 
\label{fig:zanette}
\end{figure}


\subsection{Correlated percolation}
\label{sec3B}


In 1998, Makse et al.~\cite{Makse2, Makse:1998} proposed a percolation-based model of city morphology, where built-up areas correspond to the occupied (also called wet or open) sites in a site-percolation process (see~\cite{Stauffer:1992} for an introduction to percolation theory), designed to overcome the limitations of the earlier DLA approach. The latter fails to capture essential features of real urban systems: it predicts the emergence of a single dominant cluster rather than a hierarchy of cities of different sizes, and it yields a population density that decays as a power law, $\rho(r) \sim r^{D-2}$, whereas empirical studies show an exponential decay \cite{Clark:1951} of the form
\begin{equation} \label{eq:density_cgr}
    \rho(r) = \rho_0 \exp(-\lambda r),
\end{equation}
where $r$ is the distance to the central business district (CBD), $\rho_0$ is the central population density, and $\lambda$ is the density gradient.
 
In the classical Bernoulli percolation model, sites are present with a given fixed probability $p$, independently from one another. This cannot reflect urban systems correctly as in real-world cities, new built sites are not placed independently at random. Rather, it is more likely for a new built site to be located close to already built sites, with a probability decreasing as the distance to these sites increases. This type of interactions can be incorporated in a correlated percolation model \cite{Prakash:1992}. Following these principles, it is possible to design a model considering these factors using correlated percolation \cite{Prakash:1992, Coniglio:1979} and gradient percolation \cite{Sapoval:1985} models. The details of sampling from the correlated percolation ensemble are not provided here; it suffices to note that the procedure involves manipulating a random Gaussian sequence \(u(r)\), transforming it into Fourier space to impose correlations of the form 
\[
C(\ell) = (1+\ell^2)^{-\alpha/2},
\] 
and then transforming it back to real space to obtain a sequence \(\eta(r)\) of long-range correlated numbers. These numbers can then be mapped to binary variables according to the local density 
\[
p(r) = \frac{\rho(r)}{\rho_0}.
\]
where~$\rho$ is defined in Eq.~\ref{eq:density_cgr}.
\begin{figure}[htb]
    \includegraphics[width=1.0\linewidth]{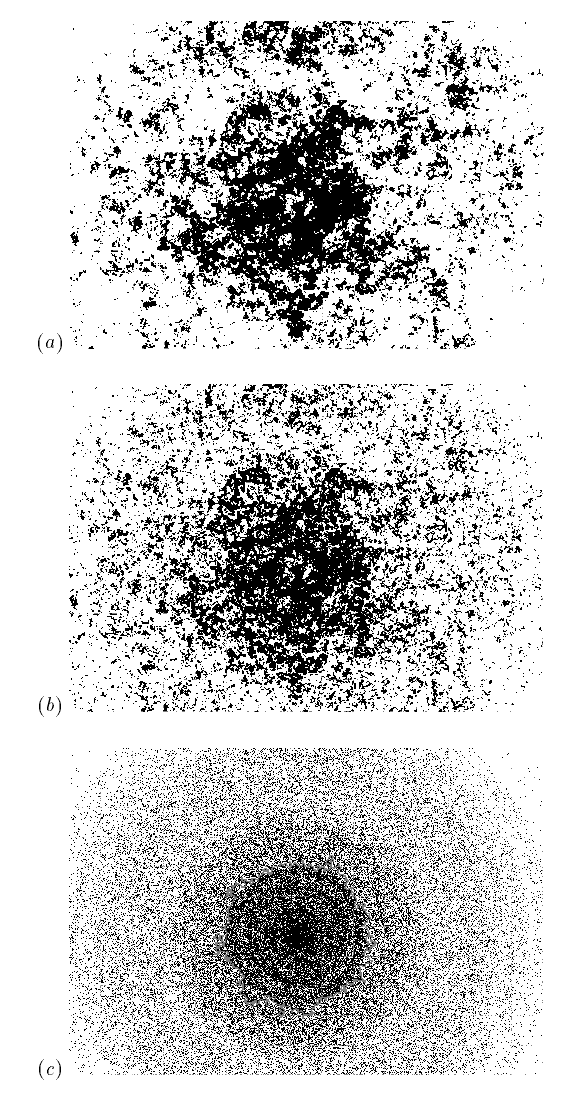} 
    \caption{Simulations of urban growth with varying correlation exponent $\alpha$. 
    (a)–(b): Correlated development for $\alpha = 0.6$ and $\alpha = 1.4$. Development units are placed with probabilities decaying exponentially with distance and influenced by neighboring occupancy; stronger correlations (small $\alpha$) yield more compact clusters. (c): Uncorrelated case (random placement), illustrating a fragmented morphology with scattered towns, unlike real urban patterns. All panels use a fixed density gradient $\lambda = 0.009$. Source: From \cite{Makse:1998}.}
    \label{disrup_examp_1}
\end{figure}

The correlated gradient percolation (CGP) ensemble only possesses two parameters, beside the system size ($L$) : $\alpha$, quantifying the correlations decay, and~$\lambda$, the density gradient. Realizations with different parametrizations, varying from strongly correlated ($\alpha\approx0$) to uncorrelated ($\alpha \geq2$), can be examined in Fig.~\ref{disrup_examp_1}. 

Firstly, consider the radius~$r_f$ at which the occupation probability is equal to the critical threshold for standard site-percolation on the two-dimensional lattice,~$p_c$,
\begin{equation}
    p(r_f) = p_c
\end{equation},
implying $r_f = -\frac{\log p_c}{\lambda}$. A giant connected component should span the area $r<r_f$, whereas beyond~$r_f$, disconnected clusters, forming a system of smaller and smaller cities as~$r$ increases, should appear.

These clusters are characterized by a correlation length $\xi(r)$ and close to the largest cluster fringe, the correlation length behaves as 
\begin{equation}
    \xi(r) \sim |p(r) -p_c|^{-\nu}
\end{equation}
Note that $\nu$ depends on the correlation exponent $\alpha$. At $\alpha \to 2$, $\nu$ tends to the usual value $4/3$ in the uncorrelated case.

Secondly, the urban fringe, perimeter of the giant cluster, is characterized by its average distance~$r_f$, but also by its width 
\begin{equation}
    \sigma_f = \langle (r-r_f)^2 \rangle^{1/2} \,,
\end{equation}
and also the number of sites along the perimeter $N_f$. Note that~$\sigma_f$ corresponds exactlty to the system-size width (Eq.~\ref{eq:widtheden}) also measured for e.g. the Eden model and in roughening studies. The study of diffusion front \cite{Sapoval:1985} yields
\begin{equation}
    \sigma_f \sim \lambda^{-\nu/{(1+\nu)}} 
\end{equation}
and
\begin{equation}
    N_f \sim  \lambda^{-\nu(d_f-1)/(1+\nu)}
\end{equation}
where $d_f$ is the fractal dimension of the largest cluster defined through a mass-cluster scaling relation
\begin{equation}
    M(r) \sim r^{d_f}
\end{equation}
at scales smaller than the width, at the border of the giant cluster. The fractal dimension $d_f \simeq 1.89$ is independent of the correlation exponent $\alpha$.
On the contrary, looking at the length of the perimeter in a box of length $l$ gives a fractal dimension $D_c$ such that
\begin{equation}
    L(l) \sim l^{D_c}
\end{equation}
which depends on the correlations. At $\alpha \approx 0$ (strong correlations), $D_c \simeq 1.4$ while $D_c \simeq 1.33$ in the uncorrelated case ($\alpha \simeq 2$). The presented model is fractal only at the border given that beyond this distance, clusters are defined by a correlation length while before $p(r) > p_c$ and the cluster is compact.

It can be shown that this model captures effectively the distribution of cluster sizes of real cities. Denote $N(A)$ the number of clusters covered by $A$ sites. Since all small clusters are at distance $r>r_f$, see that
\begin{equation}
    N(A) = \int_0^{p_c} n(A,p) dp 
\end{equation}
where $n(A,p)$ is the average number of clusters containing $A$ sites for a given $p$ (corresponding to a distance $r$). In particular,
\begin{equation}
    n(A,p) \sim A^{-\tau} g(A/A_0)
\end{equation}
where $\tau$ is the usual Fisher exponent, characterizing the cluster size distribution and $g$ is a scaling function decreasing sharply beyond $1$. $A_0$ is the maximum typical area for a cluster at distance $r$. In particular,
\begin{equation}
    A_0 \sim \xi(r)^{d_f} \sim |p(r) -p_c|^{- \nu d_f}
\end{equation}
Combining these equations gives
\begin{equation}
    N(A) \sim A^{-\tau + 1/d_f\nu}.
\end{equation}

The value of the exponent is $2.45$ for uncorrelated systems while, while strongly correlated systems approach $2.06$. The authors argue that these exponents resemble exponents measure on real-world data of Berlin, London and the whole Great Britain. This model can be used to simulate dynamics of real-world city by fitting the density gradient $\lambda(t) = \ln(p_c^{-1}) / r_f(t)$. 

While the correlated percolation framework provides an useful baseline for sampling urban form surrogates for urban form -- they display the density decay (\ref{subsec:clark}), power-law cluster distribution (\ref{subsec:component}) and rough boundary (\ref{subsec:surfacegrowth}) and fractal dimension (\ref{subsec:fractal}) -- it also does not yield a natural description of the urban growth process. In particular, its interpretability is extremely limited. Correlations and density gradient are imposed, rather than emergent properties. Moreover, one of its two parameters,~$\alpha$, is extremely arduous to estimate from real data, and makes informed predictions for real development difficult.






\subsection{Markov random fields}

Although diffusion-limited aggregation reproduces edge-driven accretion, it suffers from unrealistic assumptions and predicts a single cluster without clear correspondence to real urban processes. Correlated percolation provides a better match with empirical data, as it reflects the tendency of growth to attract further growth. Nevertheless, these models still rely on ad hoc prescriptions, such as imposing a predefined urban core with a density gradient, and thus lack a realistic microscopic dynamics grounded in actual urban decision-making.  Andersson et al.~\cite{andersson2002urban} developed a general model of urban growth to reproduce some characteristic urban morphologies from simple microscopic rules. Unlike earlier physical analogies such as diffusion-limited aggregation, dielectric breakdown, or correlated percolation, their approach is rooted in `first principles' designed to reflect human decision-making. The focus is particularly on urban sprawl, where low-density developments spread outward in ways that generate significant environmental and social problems.

The model is based on a two-dimensional Markov Random Field (MRF) with recursive mean-field interactions. Land is represented as a square lattice of $N$ cells, where each cell corresponds to a land-use type (e.g., undeveloped, residential, commercial). The total number of possible land-use classes is denoted by $C$. The state of each site is determined by an energy function
\begin{equation}
E_a(x) = \sum_{d \leq R} \sum_{b \in C} w_{ab}(d),
\end{equation}
where $w_{ab}(d)$ encodes the influence of state $b$ at distance $d$ on state $a$. Probabilities are assigned via a Gibbs weight
\begin{equation}
q_a(x) = \frac{e^{-\beta E_a(x)}}{\sum_{b \in C} e^{-\beta E_b(x)}},
\end{equation}
where $\beta = 1/T$ represents an inverse temperature that controls the disorder strength. In contrast to standard MRFs, the authors include long-range interactions by using mean-field approximations and define state transitions globally rather than locally. Growth then proceeds by allocating new urbanization to the locations with the highest `fitness' as defined by these probabilities.

To adapt the framework to urban growth, Andersson et al. introduced the `Unwilling Neighbor (UN) rule' as the central microscopic mechanism. This rule captures the dual forces of stimulation and inhibition that govern development: proximity to existing built areas is attractive because it provides access to infrastructure and reduces costs, while excessive local density creates competition, leading to higher land prices and congestion, which discourages further development nearby. Unlike standard MRF formulations that only consider immediate neighbors, their model accounts for both short-range and long-range interactions. Each undeveloped site is assigned a `fitness' or energy that balances these effects, such that growth is drawn to the edges of existing development to exploit infrastructural connections, yet repelled from overly dense areas where competition dominates. This tension between edge attraction and local repulsion constitutes the UN rule, which drives the emergence of realistic, sprawl-like urban morphologies.

The model generates clusters that, at low densities, resemble diffusion-limited aggregation but evolve into more compact and realistic urban clusters as the lattice fills. Measurements confirm the presence of scaling laws similar to those observed in real urban areas. The relation between built area $A$ and radius $r$ follows
\begin{align}
A(r) \sim r^D,
\end{align}
with fractal dimension $D \approx 1.8$ (see~\ref{subsec:fractal}for discussions of the fractality of real cities), consistent with empirical urban data. 

\begin{figure}[h]
    \centering
    \includegraphics[width=0.9\linewidth]{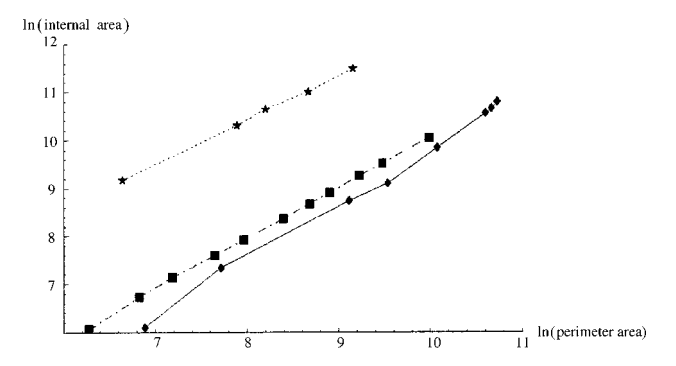}
    \caption{Comparison of the scaling relationship between developed area and perimeter for Sioux Falls (USA), a simulation using the UN rule, and Washington/Baltimore (USA). The curves, from top to bottom, correspond to empirical measurements for Sioux Falls, the simulation, and Washington/Baltimore, respectively. A nontrivial scaling exponent (i.e., distinct from that of a uniformly expanding disk) reflects the presence of distributed, sparse growth. Both axes are shown on a logarithmic scale. Source: From \cite{andersson2002urban}.}
\label{fig:andersson}
\end{figure}

A second test considers the scaling relation between the developed area and its perimeter (the area is the number of developed cells, while the perimeter is the set of undeveloped cells adjacent to development, defining the primary growth zone). Simulations reproduce power-law behavior close to that observed in the historical growth of cities such as Sioux Falls and Washington/Baltimore (see Fig.~\ref{fig:andersson}).

The framework demonstrates that the balance between stimulation and inhibition at different scales can reproduce urban sprawl. The parameter $\beta$ reflects uncertainty in decision-making: low values lead to edge-driven growth, while high values yield denser, more centralized development. The model is flexible, allowing for multiple land-use types, infrastructure, and topography, making it potentially useful for planning scenarios. However, it has clear limitations: the energy function is ad hoc, parameter choices are loosely linked to real data, and socio-economic factors are not explicitly modeled. Its complexity and input requirements contrast with the limited empirical validation, and the analogy with statistical physics remains suggestive rather than rigorous. Further analysis is needed to assess its robustness and relevance beyond stylized cases.

\subsection{Growth patterns from human mobility behavior}

Xu et al.~\cite{Xu2021} investigate how different human mobility models shape urban population density and compare their simulations with empirical built-area distributions. Their objective is to reproduce three classical yet debated empirical regularities of urban systems: the distribution of cluster sizes, which follows approximately a power-law of exponent~$2$ (see~\ref{subsec:component}); the superlinear scaling of population with urban area, which is not universal but has been observed in the United States with $P \sim A^{1/0.85}$ \cite{Barthelemy:2016}; and the exponential radial density profile $\rho(r) \sim \exp(-\lambda r)$ (see~\ref{subsec:clark}), which breaks down for large polycentric cities.

To explain these patterns, the authors propose the Collective Mobility Model (CMM), which is both socially interactive and memory-aware. Unlike previously studied mechanisms such as random L\'evy flights for human movements that are  characterized by a fat-tailed jump-size distribution of the form 
\begin{equation}
    P(\vec{r} | \vec{r}_0) \sim \frac{1}{|\vec{r}-\vec{r}_0|^{d+\alpha}},
\end{equation}
where $P(\vec{r} | \vec{r}_0)$ is the transition probability from location $r_0$ to $\vec{r}$ which leads, when social interactions and memory effects are neglected, to the following equation that governs the evolution of the population density $\rho(\vec{r},t)$
\begin{equation}
    \frac{\partial \rho(\vec{r},t)}{\partial t} = -D (-\Delta)^{\alpha/2} \rho(\vec{r},t),
\end{equation}
(see \cite{Zaburdaev_2015} for a full review of L\'evy walks and \cite{Barbosa_2018}, Chapter IV, for applications to human mobility). 

When we take into account the fact that the traffic flow between two locations depends on their populations, we have for instance the Gravity model that suggests the following form for the transition probability
\begin{equation}
    P(\vec{r} | \vec{r}_0) \approx \frac{\rho(\vec{r})+\rho_0}{|\vec{r}-\vec{r}_0|^{d+\alpha}},
\end{equation}
or for the Individual Mobility Model (IMM),
\begin{equation}
    P(\vec{r}_i) \propto f(\vec{r}_i),
\end{equation}
where the return probability is proportional to the historic visitation frequency $f$. The CMM explicitly incorporates both long-term memory and social interactions, and individuals return to previously visited locations with probability
\begin{equation}
    P_\text{ret} = 1 - \delta S^{-\gamma},
\end{equation}
where $S$ is the number of visited sites, $\gamma$ tunes the decay of return probability, and $\delta$ sets its initial value. When exploring new locations, individuals follow a gravity-like rule, with interactions modulated by the coupling constant $\rho_0^{-1}$.  

\begin{figure*}
    \centering
    \includegraphics[width=1\linewidth]{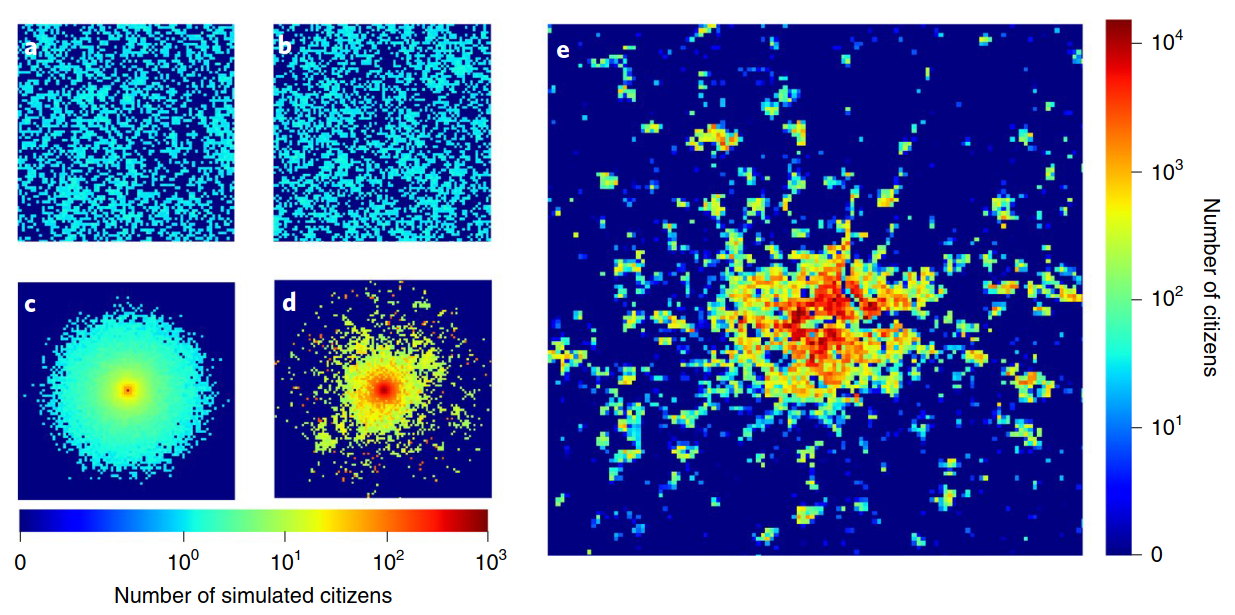}
    \caption{Morphologies retrieved from different models compared with London: a) Lévy flights, b) gravity model, c) Individual Mobility Model, d) Collective Mobility Model, e) London. Simulations were run until population distributions converged. Source: From \cite{Xu2021}.}
    \label{fig:morphologies_mobility}
\end{figure*}

Figure~\ref{fig:morphologies_mobility} compares morphologies generated by different models with that of London. While L\'evy flights and the gravity model produce nearly uniform density distributions and the IMM yields circular structures, the CMM uniquely reproduces fractal urban fringes and detached peripheral clusters, reminiscent of correlated percolation growth. Importantly, only the CMM captures all three empirical urban laws simultaneously.

A sensitivity analysis shows that the CMM is strongly influenced by the mobility scaling parameter $\alpha$ and the coupling constant $\rho_0^{-1}$, while remaining robust to variations in population density. These findings suggest that the model bridges mobility science and urban theory, providing a bottom-up mechanism for city formation and evolution. By moving beyond analogies with physical processes and relying instead on agent-based dynamics with social and memory effects, the CMM offers an interesting framework for explaining emergent urban patterns.
\begin{figure*}
    \centering
    \includegraphics[width=1\linewidth]{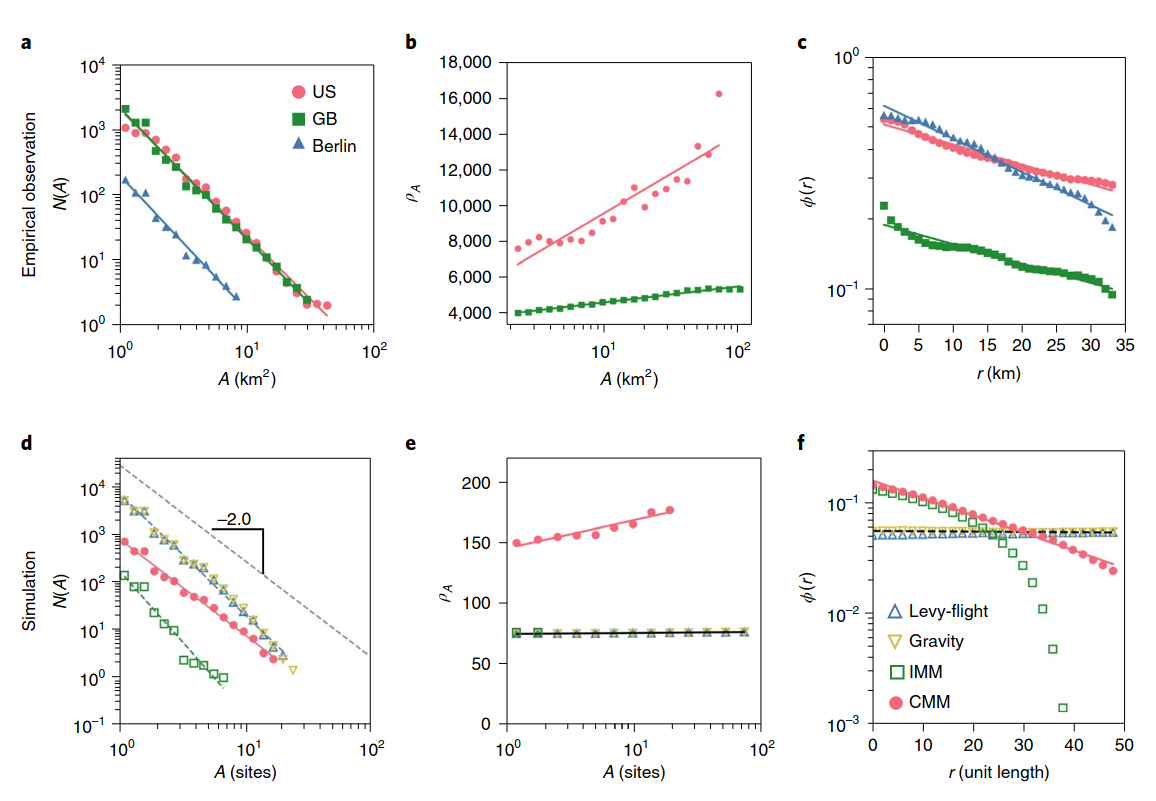}
    \caption{Comparison between empirical morphology statistics (upper row) and model simulations (lower row). Panels a) and d): cluster size statistics; b) and e): city density versus size; c) and f): radial density profiles. Source: From \cite{Xu2021}.}
    \label{fig:morphology_statistics}
\end{figure*}

Although interesting, the study could benefit from a clearer specification of the type of mobility being modeled. Since different forms of mobility play different roles, it may be helpful to distinguish between short-term visitation patterns, relevant for activities such as shopping, and relocation mobility, that is, the choice of a new place of residence, which has a more direct impact on long-term urban structure.

\subsection{Evolution of the number of buildings}

The growth of a city fundamentally depends on the development of its infrastructure, beginning with the expansion of roads and streets that structure accessibility, and followed by the construction of buildings that accommodate residential, commercial, and industrial activities. In this sense, transportation networks provide the backbone for urban expansion, while the built environment fills in around them, together shaping the spatial form of the city. 

\begin{figure}
    \centering
    \includegraphics[width=0.9\linewidth]{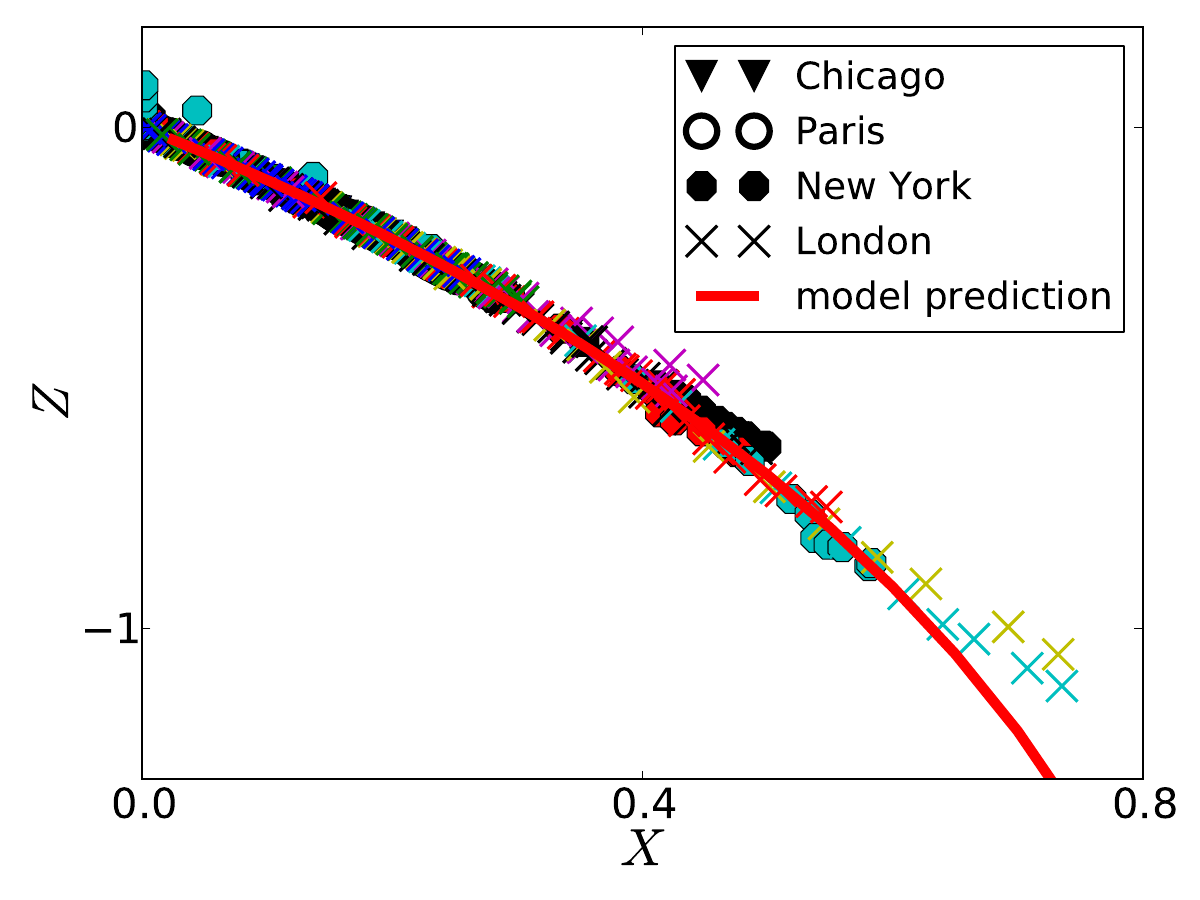}
    \caption{Collapse for the rescaled variable $Z$ and $X$ for all the 47 saturated districts of all cities considered in \cite{carra2019}. Each city is characterized
by a different symbol and each district by a different color. The continuous red line is the theoretical prediction given by Eq.~\ref{eq:collapse}. Source: From \cite{carra2019}.}
    \label{fig:carracollapse}
\end{figure}

In \cite{carra2019}, the authors analyzed the relation between buildings and population across neighborhoods in Chicago, London, New York, and Paris, and identified four phases: pre-urbanization, urbanization, conversion, and re-densification. Buildings first fill vacant lots until saturation, after which their functions shift (e.g., residential to commercial), leading to population decline. Eventually, neighborhoods re-densify, as seen in New York, Paris, and London, a process likely driven by external factors. The empirical findings are discussed in~\ref{subsec:buildings}. The authors also propose a simple model for the transition between urbanization and conversion. In this model, a neighborhood is modeled by a two-dimensional space of total area $A$ divided in lots $N_{\max}$ of size $a_\ell$, which can be either occupied or empty. On a lot $i$, there are $h_r(i)$ floors dedicated to residential use and $h_c(i)$ to commercial use such that $h(i) = h_r(i) + h_c(i)$. At each time step, a random cell~$i$ is chosen, time advances of $\Delta t$ and population can either increase of decay of $\Delta P$. If the chosen cell is empty, a residential floor is built : $P$ increases of $\Delta P$, $N_b$ increases of $1$, $h_r(i) = 1$, $h_c(i)=0$. If a building already sits on it, there are two possibilities, with probability $p_h$ a residential floor is built, while with probability $p_c$ a residential floor is converted to commercial use (and nothing happens with the complementary probability). The mean-field equations for this model can be written
\begin{align}
\frac{dH_r}{dt} &= \frac{N_b}{N_{\max}} (p_h - p_c) + \left(1 - \frac{N_b}{N_{\max}}\right) ~,\label{eq:M_H}\\ 
\frac{dN_b}{dt} &= 1 - \frac{N_b}{N_{\max}} ~, \\ 
\frac{dP}{dt} &= \Delta P \frac{dH_r}{dt} ~. 
\label{eq:M_P} 
\end{align}
These equations can be easily integrated, leading to  
\begin{align}
& N_b(t) = N_{\max} \left( 1 - e^{-t/N_{\max}} \right)~, \label{eq:Nb_t}
  \\
\nonumber
& P(t) = \Delta P [(p_h - p_c) t \\
& + N_{\max}(1 + p_c - p_h)(1 - e^{-t/N_{\max}})]~. 
\label{eq:P_t1}
\end{align}
The saturation point $(N^* _b, P^*)$ corresponding to the urbanization-conversion phase can be written 
\begin{align}
&N_b^* = \frac{N_{max}}{1 + p_c - p_h}~, \label{eq:N_star} \\
&\frac{P^*}{\Delta P N_{max}}= (p_h - p_c) \log \left(\frac{1 +
      p_c - p_h}{p_c - p_h} \right) + 1 ~. \label{eq:P_star}
\end{align}
Notice that this point exists only if the saturation value of the number of lots is smaller than the total amount of lots $N_{\max}$, which implies $p_c>p_h$. By writing, $n_b^* =N_b^* /N_ {\max}$ and $p^* = P^* /N_{\max}$, we have
\begin{equation}
    p^* = \Delta P \left[ 1 + (1/n_b^* -1 ) \log(1 - n_b^* ) \right]
\end{equation}
and then different neighborhood can be compared by looking at the values of
\begin{align}
&X(t) = \frac{N_b(t)}{ N_{\max} } \\
&Z(t) = \frac{ \frac{P(t)}{\Delta P N_{\max}} - \frac{N_b(t)}{N_b^*} }{
  \frac{1}{n_b^{*}} - 1}
\label{eq:rescaled}
\end{align}
which should collapse on the curve
\begin{align}
Z = \log(1-X)
\label{eq:collapse}
\end{align}
which is indeed the case (see Fig.~\ref{fig:carracollapse}) when tested on neighborhoods (restricted on the fact that they have reached saturation, in order to be able to estimate all the parameters of the model). 

Interestingly, this model is able to capture essential features of the urbanization process with zero parameter fit, as shown by the data collapse. In particular, it assumes uncorrelated vertical growth between buildings inside a neighborhood, as well as for conversions from  residential to commercial use, which is counter-intuitive.

\section{Dispersal models}
\label{chap:5}

\subsection{Dispersal Kernels in Ecology}

Dispersal is a key process in ecology, shaping species spread, gene flow, and population connectivity \cite{clobert2012dispersal}. Through dispersal, organisms colonize new habitats, maintain genetic diversity across fragmented landscapes, and buffer populations against demographic and environmental stochasticity. While this process has traditionally been examined in natural or semi-natural ecosystems, dispersal dynamics also provide valuable insights into human-dominated landscapes. In particular, the mechanisms and consequences of animal dispersal may hold important analogies—and sometimes even direct influences—on urban growth and spatial expansion.

Conceptually, dispersal as a driver of range expansion and network connectivity parallels the development of urban infrastructure. Just as individual-level movements and settlement decisions aggregate into collective spatial patterns, the local dispersal of organisms leads to emergent population-level range dynamics. This analogy highlights the shared logic of small-scale movements producing large-scale structures.


Mathematical models typically describe dispersal either as a diffusive process or as a probabilistic process characterized by dispersal kernels. A dispersal kernel represents the probability distribution of movement distances $x$. Formally, a dispersal kernel $k(x)$ satisfies the normalization condition
\begin{equation}
    \int_{-\infty}^{\infty} k(x) \, dx = 1.
\end{equation}
Two widely used forms are the Gaussian kernel, which arises naturally from diffusion processes, and the L\'evy flight kernel, which exhibits fat-tailed behavior allowing for occasional long-distance movements. The Gaussian kernel models Brownian motion, assuming individuals undergo random movement with short-distance bias. It is given by
\begin{equation}
    k(x) = \frac{1}{\sqrt{4 \pi D t}} e^{- \frac{x^2}{4Dt}},
\end{equation}
where $D$ is the diffusion coefficient, and $t$ is the time interval. This kernel decays exponentially, meaning that most individuals move over a short distances, with very few long-distance dispersers.

Many organisms (e.g., insects, marine predators, and human foragers) exhibit long-distance dispersal better described by power-law distributions \cite{viswanathan2011physics}. The L\'evy flight kernel is given by
\begin{equation}
    k(x) \propto |x|^{-\mu}, \quad 1 < \mu < 3.
\end{equation}
For $ 1 < \mu < 3 $, the variance is infinite, leading to frequent long-distance movements. This property is observed in animal movement data and optimal for searching in sparse environments.

The Gaussian kernel rapidly decays, while the L\'evy flight kernel maintains a fat tail, allowing for long-distance dispersal. The choice of the kernel has thus significant consequences: Gaussian kernels lead to localized spread and predict continuous range expansion, while L\'evy flight kernels enable long-distance colonization, increasing invasion speed and ecosystem connectivity. Empirical studies of ecological systems confirm that species exhibiting fat-tailed dispersal are more likely to adapt to fragmented habitats and colonize new territories \cite{hastings2005spatial}.

\subsection{Modeling stratified diffusion in theoretical ecology}


The paper by Shigesada et al. \cite{shigesada1995modeling} investigates the spread of biological invasions via stratified diffusion, a process that combines neighborhood diffusion (short-range dispersal) with long-distance jump dispersal. Such mechanisms might also be relevant for understanding urban growth, where local developments and distant new settlements jointly shape the spatial footprint of cities. 

Traditional diffusion models, such as Skellam's model \cite{skellam1951random}, predict a linear expansion of the invasion front over time. However, empirical observations often reveal accelerating range expansions, which cannot be captured by purely diffusive frameworks and instead suggest the influence of long-distance dispersers.

Range-versus-time curves (where the range is defined, following Andow et al. \cite{andow1993spread}, as either the square root of the invaded area or the average total expanding length) can be qualitatively classified into three types. All three exhibit an initial establishment phase, during which expansion is not yet apparent, followed by an expansion phase, and potentially ending in a saturation phase if geographic limits are reached. These distinctions are important because they highlight how deviations from simple diffusive spread can arise from ecological processes such as long-distance dispersal events and the establishment of new colonies beyond the advancing front.

Focusing on the expansion phase, type 1 curves show consistently linear growth, as seen for example in muskrat populations. Type 2 patterns, like those of the European starling, display an initial slow spread followed by faster linear expansion. Type 3, exemplified by the rice water weevil and cheat grass, involves an expansion rate that continually increases over time, resulting in a convex curve.

There are multiple reasons why an establishment phase may occur. One possibility is that newly introduced organisms are initially ill-adapted to the environment, persisting only at low densities until evolutionary changes (such as the appearance of more fecund offspring) trigger expansion. Alternatively, a small number of organisms may disperse widely from the introduction point and remain undetected until their descendants reach sufficient densities. The latter scenario motivates the mathematical treatment presented by the authors.

To formally analyze these dynamics, Shigesada et al.~\cite{shigesada1995modeling} develop a stratified diffusion model that integrates processes at multiple spatial scales. They incorporate the Skellam equation to model local or neighborhood diffusion, capturing how populations spread via random movements in continuous space. Simultaneously, they use a von Foerster equation to describe the temporal growth of isolated colonies arising from long-distance dispersal. To close the model, explicit colonization rates for long-distance dispersers are included, linking local growth and rare, long-range establishment events that can drive accelerating expansions. In the following we describe in more details these approaches.

\subsubsection{Skellam’s Model for Neighborhood Diffusion}

We first describe the classical Skellam model \cite{skellam1951random}, which captures population spread through diffusion coupled with local reproduction. The model is governed by the partial differential equation
\begin{equation}
    \frac{\partial \rho}{\partial t} = D \nabla^2 \rho + \epsilon \rho,
    \label{eq:skellam}
\end{equation}
where $\rho(x,t)$ denotes the population density at location $x$ and time $t$, $D$ is the diffusion coefficient, and $\epsilon$ is the intrinsic per capita growth rate of the population. This formulation predicts that an initially localized population expands with an asymptotic velocity given by
\begin{equation}
    v = 2 \sqrt{D\epsilon}.
\end{equation}
Such models typically produce range expansions that advance at a constant speed, resulting in a linear relationship between distance invaded and time. However, many empirical studies of biological invasions reveal deviations from this prediction, often showing accelerating spread rates due to the influence of long-distance dispersers.




\subsubsection{Stratified Diffusion and the Scattered Colony Model}

To explain accelerating invasions, Shigesada et al.~\cite{shigesada1995modeling} propose a stratified diffusion framework that explicitly incorporates both local diffusion and long-distance jump dispersal. In the scattered colony model, long-distance migrants establish new colonies at some distance $L$ from the parent population. Each of these new colonies then grows and spreads according to Skellam’s local diffusion dynamics.

The distribution of colony sizes, denoted by $p(r,t)$, where $r$ represents the colony radius, evolves according to the Von Foerster equation
\begin{equation}
    \frac{\partial p}{\partial t} + c \frac{\partial p}{\partial r} = -\lambda(r) p(r,t),
\end{equation}
where $c$ is the speed of local diffusion within a colony, and $\lambda(r)$ specifies the rate at which long-distance migrants are produced as a function of colony size. The Von Foerster equation can be simply derived by noting that for a change in time $dt$, and change in radius $dr$, the distribution of sizes is
\begin{align}
    p(r+dr,t+dt) = [1-\lambda(r)dt]p(r,t), 
\end{align}
that is, during a time period $dt$, the size distribution decreases of a factor~$\lambda(r)dt$. 

The total area invaded at time $t$ is given by
\begin{align}
    A(t)=\int_0^{2\pi}d\theta \int_0^\infty rdn(r,t)
\end{align}
where $dn(r,t)=p(r,t)rdr$ is the number of colonies of size in $[r,r+dr]$. This expression simplifies to
\begin{equation}
    A(t) = \int_0^\infty 2 \pi r^2 p(r,t) \, dr.
\end{equation}
The initial condition assumes a single nucleus of radius zero exists at $t=0$,
\begin{equation}
p(r,0) = \delta(r),
\end{equation}
and a boundary condition reflecting the birth of new colonies by long-distance dispersers
\begin{equation}
c\, p(0,t) = \int_0^\infty \lambda(r) p(r,t)\, dr.
\label{eq:boundary}
\end{equation}
This boundary condition indicates that the number of nuclei newly created at time $t$ per unit
time is the total rate at which the long-distance migrants succeed in colonization.

Shigesada et al. consider three prototypical forms of the colonization rate  
\begin{align}
\begin{cases}
\lambda (r) &=\lambda_0 \\
\lambda (r) &= \lambda_1 r \\
\lambda (r) &= \lambda_2 r^2 \,.
\end{cases}
\end{align}

In the constant colonization rate, each colony produces long-distance migrants at a constant rate, independent of size. The total area occupied evolves as
\begin{align}
A(t) = \frac{2\pi c^2}{\lambda_0} \left( \frac{1}{\lambda_0}(\mathrm{e}^{\lambda_0 t} - 1)-t \right),
\end{align}
growing exponentially as $\exp{\lambda_0 t/2}$. 

In the case where the colonization rate proportional to circumference, which corresponds to a situation where long-distance dispersers are produced mainly at the periphery of each colony. The solution is
\begin{align}
A(t) = \frac{\pi c}{\lambda_1} \left( \mathrm{e}^{\sqrt{c\lambda_1} t/2} - \mathrm{e}^{-\sqrt{c\lambda_1} t/2} \right)^2 \,,
\end{align}
which exhibits an accelerating expansion of order $\exp \left[ \sqrt{c\lambda_1}t \right]$.

Finally, in case where the colonization rate proportional to area which assumes that long-distance dispersers arise uniformly across the area of a colony, the total area grows according to
\begin{equation}
    A(t) = \frac{2\pi c^2}{3\omega^2}
    \left(
    \mathrm{e}^{\omega t}+2\mathrm{e}^{\omega t/2}\sin(\frac{\sqrt{3}}{2}\omega t-\frac{5\pi}{6})\right),
\label{eq:area_quad}
\end{equation}
(where $\omega=(2c^2\lambda_2)^{1/3}$) showing strong exponential acceleration, scaling as $\exp \left[(2c^2\lambda_2)^{1/3}t \right]$.

Different assumptions about how $\lambda(r)$ depends on $r$ lead then to different large-scale expansion behaviors. When $\lambda(r)$ is constant, the system exhibits a type 1 linear expansion. If $\lambda(r)$ is proportional to $r$, the model produces biphasic (type 2) expansion. Finally, when $\lambda(r)$ scales with $r^2$, corresponding to the area of the colony, the invasion undergoes type 3 accelerating expansion (see \cite{Shigesada:1997}). This modeling framework successfully explains observed spread patterns of various invasive species. 

In cases where new colonies are established close enough to the parent population, they eventually coalesce, altering the global expansion dynamics. This is captured by the coalescing colony model that we will discuss in detail in the next section.


\subsection{A growth-coalescence model for urban expansion}

In \cite{carra2017coalescing}, the authors analyze a model in which a primary colony grows while intermittently emitting secondary colonies that spread outward and eventually coalesce with the original cluster. Originally developed to describe processes such as population proliferation in theoretical ecology or tumor growth, this framework is also highly relevant for modeling urban expansion. Dispersal models have long been employed to study the spread of animal colonies in ecology~\cite{Shigesada:1997, Clark:2001} and have served as simplified representations of cancerous tumor growth~\cite{Iwata:2000,Haustein:2012}. By analogy, they provide a natural candidate for capturing the dynamics of urban built-up areas \cite{Barthelemy:2016}, especially since extensive empirical data on city growth are now available over long timescales \cite{Angel:2005} (see also Chapter~\ref{chap:2}).

The main feature of dispersal models is the
concomitant existence of two growth mechanisms. The first process is the growth of the main -- so-called primary -- colony, which occurs for example via a reaction-diffusion process (as described by a FKPP-like equation
\cite{Fisher:1937,Shigesada:2002}) and leads to a constant growth with velocity $c$, depending on the details of the system. The second ingredient is random dispersal from the primary colony, which represents the emergence of secondary settlements in the framework of animal ecology, the development of metastatic tumors, or, in the urban expansion case, the creation of small towns in the periphery of large cities. In the real world, dispersion follows privileged directions under the effect of external forces such as blood vessels, winds and rivers, or transportation networks for cities but in a first approach, these anisotropic effects will be neglected. It is also assumed that secondary colonies grow at the velocity $c$ and will eventually coalesce with the primary colony, leading to a larger primary colony whose time-dependent size will depend on the emission rate.

A classical way to study dispersal is through the dispersal kernel representing the probability distribution of dispersal distances and various forms  for these kernels have been  discussed~\cite{Lewis:2016}. A different approach has been introduced by Kawasaki and Shigesada in~\cite{Shigesada:1997, Shigesada:2002} who
proposed the use of simple models to tackle this challenging
problem and which is the point of view adopted here. The resulting
coalescing colony model considers a primary colony
grows at radial velocity $c$ and  emits  a
secondary colony at a rate $\lambda$ 
 and  at a distance $\ell$ from its border (long-range
dispersal). The variable $\ell$ can be drawn  from a
probability distribution $P(\ell)$ but it is assumed here that the secondary colonies are emitted at a constant distance $\ell_0$ from the
boundary of the primary colony
\begin{align}
    P(\ell) = \delta(\ell - \ell_0) \,.
\end{align}
 Besides, it is assumed that each secondary colony  also grows with the same radial speed $c$ and does not emit tertiary 
 colonies. The dependence of  the emission rate  on the colony size is taken into account by the functional form  
\begin{equation}
\lambda(r) = \lambda_0 r^{\theta}~,
\end{equation}
 $r$ being  the radius of the primary colony and 
 $\theta \geq 0$. When $\theta=0$ the growth rate is independent
from the primary colony size,  for $\theta=1$ it is proportional to its
perimeter and for $\theta=2$ to  its area. 

\begin{figure}
\begin{center}
\includegraphics[angle=0, width=0.5\textwidth]{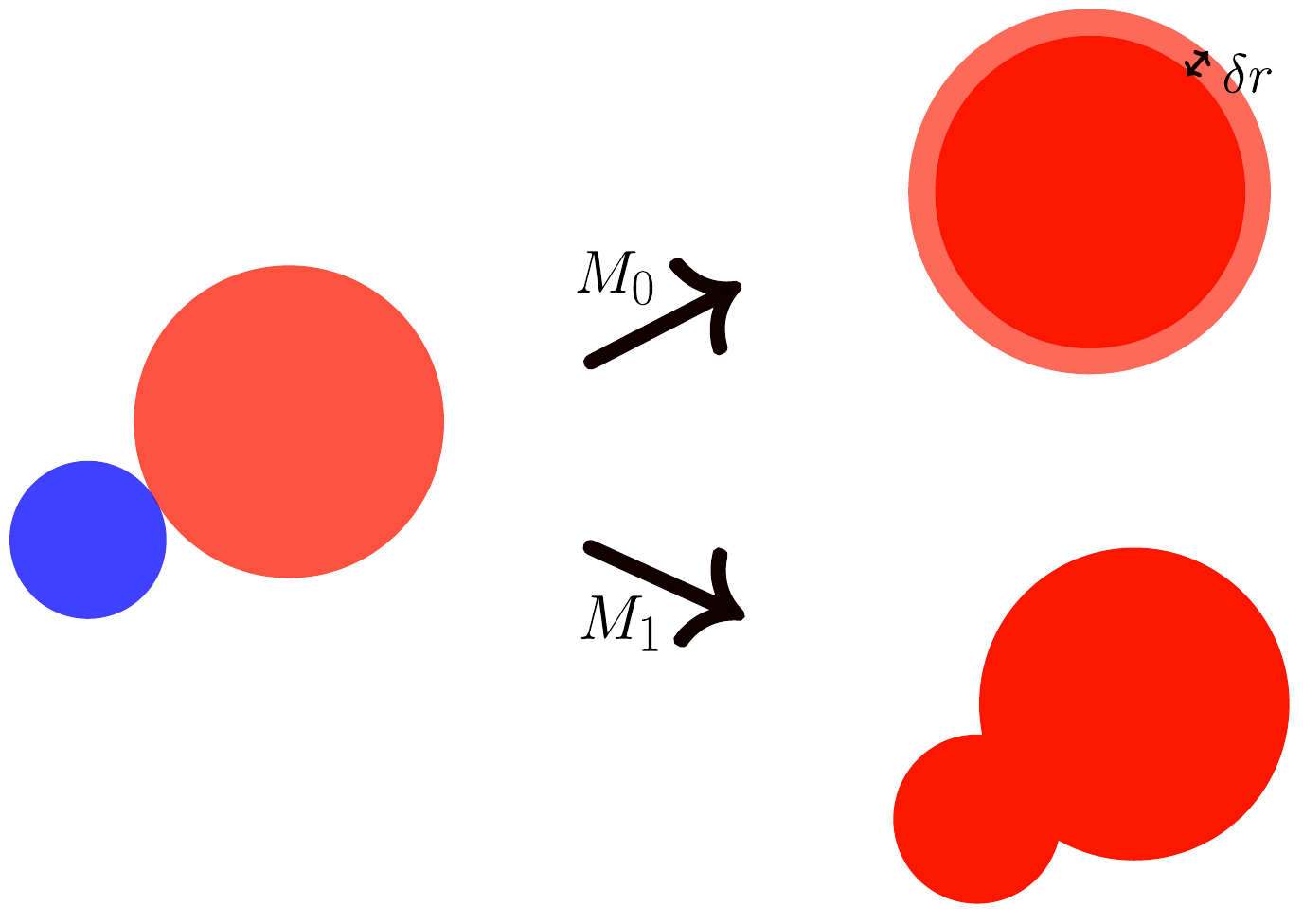} 
\end{center}
\caption{\textbf{Illustrating the~$M_0$ and~$M_1$ model} When the primary colonmy(in red in the left figure) encounters a secondary colony (in blue), its mass is redistributed such that the main colony remains a disk in the model~$M_0$, while in the model~$M_1$, the colonies are simply joined together. Source: From \cite{carra2017coalescing}.}
\label{fig:i}
\end{figure}

Coalescence happens when a secondary colony of radius $r_2$ intersects with the primary one, of radius $r$, and becomes part of the latter. We will discuss here two variants of the process.  In the first version of the model, denoted by the $M_0$ model, the primary
colony remains circular after coalescence (see Fig.~\ref{fig:i}), and has a new radius $r'$ given by
\begin{equation}
{r'}^2=r^2+r_2^2~.
\end{equation}
This interesting model was discussed in \cite{Shigesada:2002} but a full quantitative understanding of the radius $r(t)$ is still
lacking. In \cite{carra2017coalescing}, the authors present a microscopic derivation of the dynamics of the $M_0$ model, in the mean-field approximation, and study its solutions as a function of the parameter $\theta$. 


\subsubsection{Circular colony}

In the model~$M_0$, after absorption of a secondary colony, its mass is spatially redistributed such that the main colony remains circular. Denote $\lambda(t_i)dt_i$ the probability to emit a colony in the temporal interval $[t_i, t_i + dt_i]$ and ${t_i}'$ the time of coalescence of a colony emitted at time $t_i$. Writing 
\begin{equation}
\label{eq:coal}
    r({t_i}') + c {t_i}' = \ell_0 + r(t_i) + c t_i~,
\end{equation}
the coalescence condition, also enables to define a formal function~$f$ such that
\begin{equation}
{t_i}' =  f(t_i)~.
\end{equation}

The mean-field approach used in \cite{carra2017coalescing} consists in 
ignoring the fluctuations of the function $f(t)$ and to assume that it is identical for all the secondary colonies. Then, one can simply write the evolution of the area of the main colony
\begin{equation}
\label{eq:dA}
\frac{dA}{dt} = 2 \pi r c + \int d{t_i} \lambda(t_i) {\delta}(t - f(t_i)) \pi c^2 (t - t_i)^2~,
\end{equation}
where~$2\pi \, r \,  c$ represents the local growth of the main colony and the integral term accounts for coalescence with emitted secondary colonies. It is possible to rewrite Eq.\ref{eq:dA} 
\begin{equation}
\label{eq:dA_2}
\frac{dA}{dt} = 2 \pi r c + \lambda(f^{-1}(t))\left| \left[f^{-1}(t)\right]' \pi c^2 (t - f^{-1}(t))^2 \right|~.
\end{equation}

Let $x(t) = c \,(t - f^{-1}(t))$ denote the radius of the colony absorbed at time $t$ . Substituting this expression into Eqs.~\eqref{eq:coal} and \eqref{eq:dA_2} leads to the Kawasaki-Shigesada system of equations \cite{Shigesada:2002}
\begin{empheq}[left = \empheqlbrace]{align}
\frac{dr}{dt} &= c + \frac{\lambda_0 \,\big[r\!\left(t - \tfrac{x(t)}{c}\right)\big]^{\theta}}{2\pi r(t)} 
                  \left(1 - \tfrac{\dot{x}(t)}{c}\right)\pi x(t)^2 
\label{eq:Shi_1}\\[6pt]
\ell_0 &= r(t) - r\!\left(t - \tfrac{x(t)}{c}\right) + x(t) \,.
\label{eq:Shi_2}
\end{empheq}

In the limit  $t \gg x(t) / c$,  the equations Eqs.~\eqref{eq:Shi_1}, \eqref{eq:Shi_2} can be simplified
\begin{empheq}[left = \empheqlbrace]{align}
               \frac{dr}{dt} &= c + \frac{\lambda_0 r^{\theta-1} }{2} x(t)^2  \label{eq:Shi_1_s}\\
               x(t) &= \frac{\ell_0}{1 + \frac{\dot{r}}{c}}\label{eq:Shi_2_s} \,.
\end{empheq}


For the different values of $\theta$ we then obtain the following results. When~$\theta = 0$, (ie. $\lambda=\lambda_0$), the dominant contribution for $t \to \infty$ is 
\begin{equation}
r(t) \sim a + ct + \frac{\mathscr{C}}{c}\log{\left(\frac{c^2
      t}{\mathscr{C}} + 1 \right)}~,
\label{eq:rM0_teta_0}
\end{equation}
with $\mathscr{C} = \frac{\lambda l_0^2}{8}$.  This result was confirmed by numerical simulations \cite{carra2017coalescing}.

When $\theta>1$, assuming $\dot{r} \gg c$, the equations simplify Eqs.~\eqref{eq:Shi_2_s} and the system is ruled by the equation
\begin{equation}
    A r(t)^{\theta - 1} \simeq \dot{r(t)}^2 \left( \dot{r(t)} - c \right)
\label{eq:simple}
\end{equation}
with $A = \frac{\lambda_0}{2}c^2 \ell_0^2$. This nonlinear
differential equation captures the physics of the coalescence and
allows us to extract the large-time behavior of the main quantities of
interest in this problem. In particular, assuming scaling laws at
large times of the form $r(t) \sim at^{\beta}$ and $x(t) \sim dt^{-\alpha}$,
Eq.~\ref{eq:simple} yields
\begin{equation}
\beta = \frac{3}{4 -\theta}~, \qquad \qquad  \alpha = \beta - 1~.
\label{eq:beta}
\end{equation}
 Note  that for $\theta \rightarrow 4 $, we have $\beta \rightarrow
 \infty$, the radius grows faster than a power law  and 
explodes exponentially. For $\theta = 1$, we obtain
$\alpha =0, \beta =1$ which means that  we have  $x(t) = x^*$, the average radius of a colony at absorption, independent of $t$ and a linear behavior of $r(t)$. From Eq.~\eqref{eq:Shi_2} the radial velocity
$c'$ is given by 
\begin{equation}
c' = c + \frac{\lambda_0}{2} {x^*}^2
\end{equation}
and the value of $x^*$ can be obtained by solving 
 Eq.~\eqref{eq:Shi_2} that can be written as
\begin{equation}
\label{eq:c_primo}
\frac{\lambda_0 }{2c} {x^*}^3 + 2{x^*} - \ell_0 = 0~.
\end{equation}
For the specific case $\theta = 1$, this result was first derived by Shigesada and Kawasaki~\cite{Shigesada:1997} and tested numerically, showing excellent agreement.

For $\theta > 1$, the leading-order dynamics, characterized by the scaling law of Eqs.~\eqref{eq:beta}, should be observable within the intermediate time window $t_{\min} \ll t \ll t_{\max}$, where the bounds depend on $\theta$ and on the parameter $\eta^2 \lambda / 2$. These results were verified numerically, with good agreements at~$\theta\approx1$, and deviations at large~$\theta$, likely caused by the small extent of the scaling windows~$[t_{\min}, t_{\max}]$.

\subsubsection{Concatenating secondary colonies}

A modified version of the coalescence process, denoted the $M_1$ model \cite{carra2017coalescing}, is considered next. In this model, after coalescence, the secondary colony merges into the primary colony, and the primary colony no longer retains a circular shape. This key distinction between models $M_0$ and $M_1$ is illustrated in Fig.~\ref{fig:i}. This variant provides insight into the role of the circular approximation and its influence on the observed scaling behaviors.

\begin{figure}[h!]
\includegraphics[angle=0, width=0.45\textwidth]{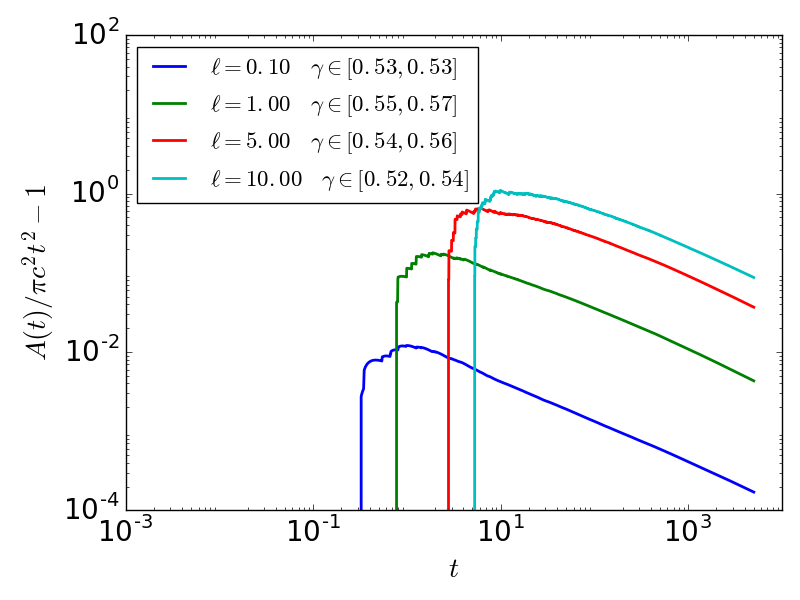}
\includegraphics[angle=0, width=0.45\textwidth]{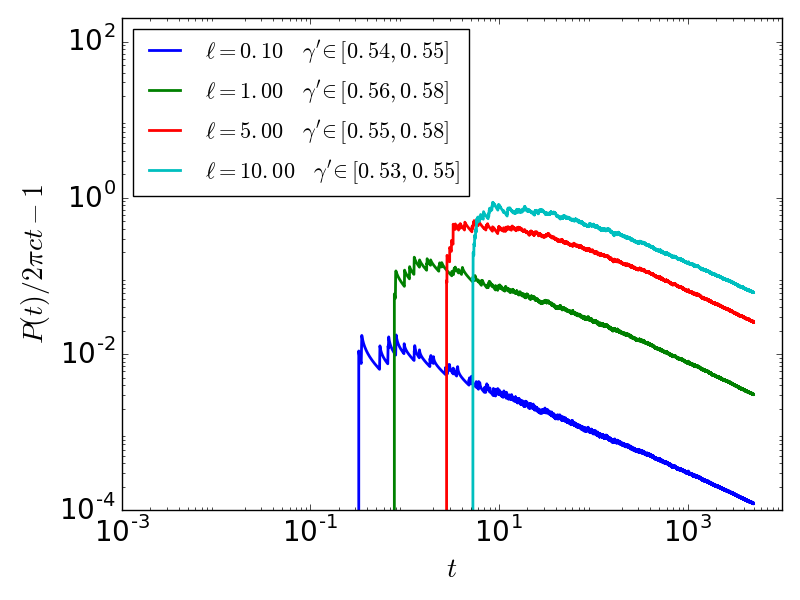} 
\caption{Deviation versus $t$ from disk-like growth $A(t)/(\pi c^2t^2) - 1$ (top), and $P(t)/(2\pi ct)- 1$ (bottom). Deviations decrease like power-laws of exponents~$\approx 0.5$, both for the area and the perimeter, independently of~$\ell$ (see insets for confidence intervals). Source : From \cite{carra2017coalescing}.}
\label{fig:carra2b}
\end{figure}

The case of a constant emission rate, $\lambda(r) = \lambda_0$, is first examined, assuming that the area $A$ and the perimeter $P$ follow power-law scaling:  
\begin{equation}
A(t) \sim t^{\mu}, \qquad P(t) \sim t^{\nu}~.
\label{eq:beh}
\end{equation}
A power-law fit to the empirical data yields $\mu \approx 2$ and $\nu \approx 1$, which can be compared with the corresponding results from model $M_0$. Figure~\ref{fig:carra2b} shows $A(t)/(\pi c^2t^2) - 1$ and $P(t)/(2\pi ct) - 1$ as functions of $t$. Both quantities vanish as $t \to \infty$, showing that at long times the dominant behavior in the $M_0$ and $M_1$ models converges to $A(t) \sim \pi c^2 t^2$ and $P(t) \sim 2 \pi c t$. Consequently, for $\theta = 0$ and sufficiently large $t$, the circular approximation--captured by the physics of the $M_0$ model--remains asymptotically valid.

Assuming that the sub-dominant corrections follow the scaling forms  
\begin{align}
\frac{A(t)}{\pi c^2 t^2} - 1 \sim t^{-\gamma}, \qquad \frac{P(t)}{2\pi c t} - 1 \sim t^{-\gamma'},
\end{align}
the numerical results indicate $\gamma \approx 0.5$ and $\gamma' \approx 0.5$, suggesting that in model $M_1$ the corrections to the leading behavior decay as a power law, at variation with the logarithmic corrections observed in model $M_0$.\\

\begin{figure}
\begin{minipage}{.5\linewidth}
\centering
\subfloat[]{\label{main:a}\includegraphics[angle=0, width=1.0\textwidth]{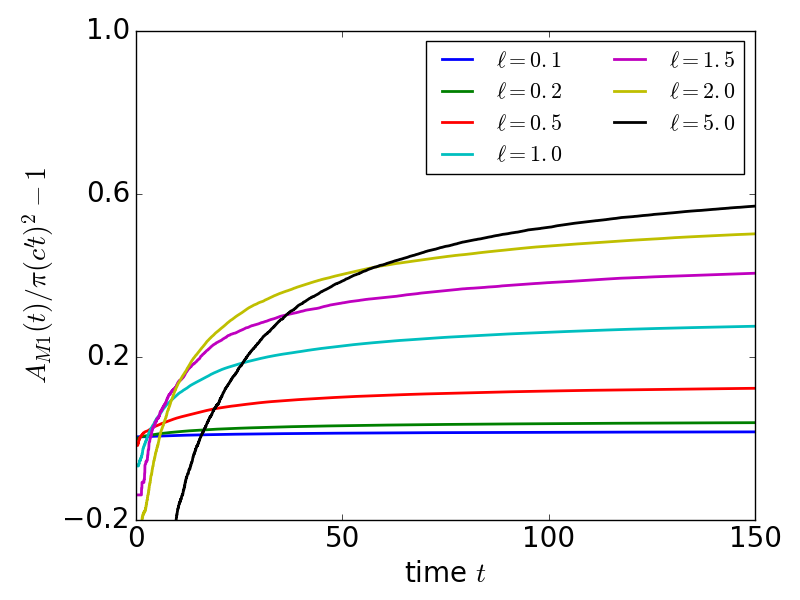}}
\end{minipage}%
\begin{minipage}{.5\linewidth}
\centering
\subfloat[]{\label{main:b}\includegraphics[angle=0, width=1.0\textwidth]{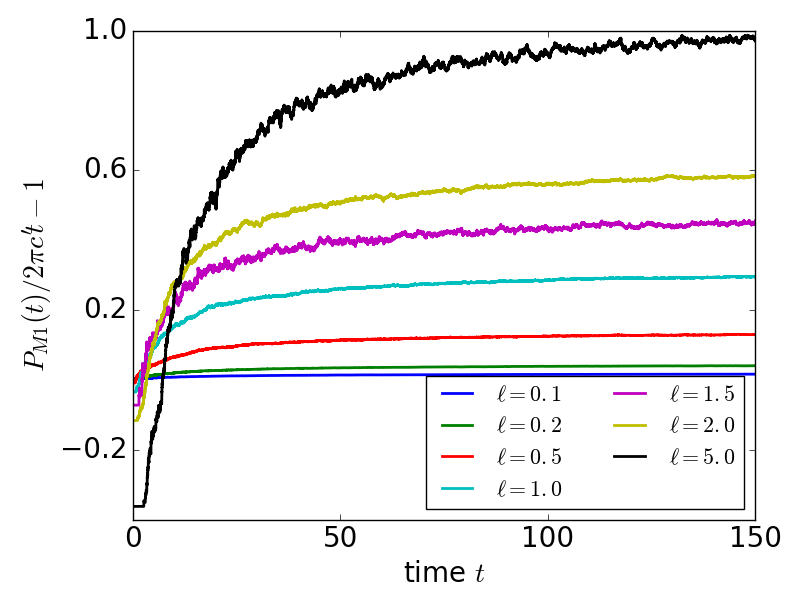}}
\end{minipage}\par\medskip
\centering
\subfloat[]{\label{main:c}\includegraphics[angle=0, width=0.25\textwidth]{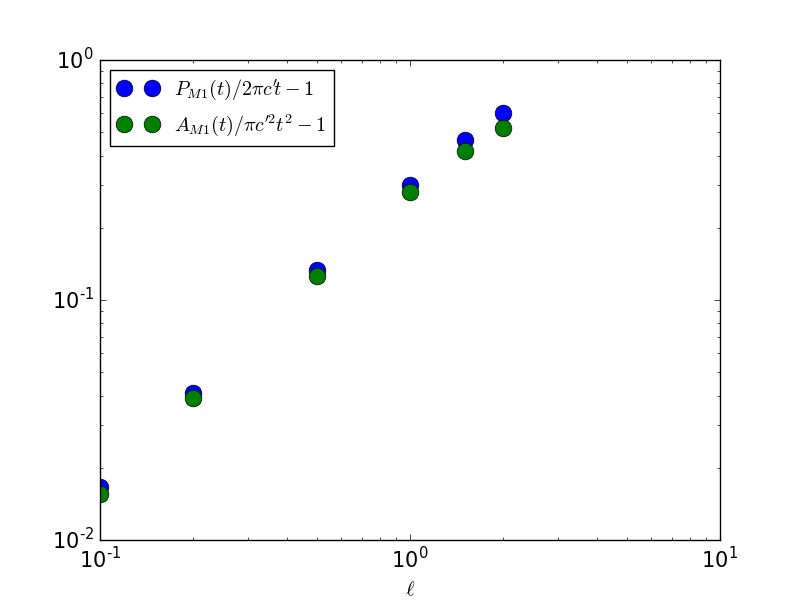}}
\caption{Model $M_1$ for $\theta=1$. (a) $A(t)/(\pi c^2t^2) - 1$ versus  $t$ for different values of
  $\ell$. (b) $P(t)/(2\pi ct)- 1$ versus  $t$ for different values of
  $\ell$. (c) Plot of $f_1(\ell)$ and $f_2(\ell)$ versus $\ell$ for $t = 200$. These results are
  obtained by averaging over $100$ simulations. Source : From~\cite{carra2017coalescing}.}
\label{fig:M1_1_prefactor}
\end{figure}
Second, the $M_1$ model is considered with an emission rate defined as  
\begin{equation}
\lambda(t) = \lambda_0 P(t),
\end{equation}
where $P(t)$ represents the total perimeter of the primary colony at time $t$, corresponding to the case $\theta=1$ in model $M_0$. Simulations show~$A(t)\sim t^\mu$ and~$P(t)\sim \nu$ with~$\mu\approx2$ and~$\nu\approx 1$, as expected from the circular model. Beyond the exponent, the authors examine the value of the prefactor. In Fig.~\ref{fig:M1_1_prefactor}(c), the quantities $A(t)/(\pi c'^2 t^2) - 1$ and $P(t)/(2 \pi c' t) - 1$, where~$c'$ represents the effective radial velocity of the main colony, are shown. As can be observed, these quantities do not go to~$0$, with time, indicating variations with the physics of the~$M_0$ model. Instead, these quantities converge to constants of time, dependent of~$\ell$. Carra et al. write
\begin{align}
A(t) &= \pi c'^2(1 + f_1(\ell)) t^2\\
P(t) &= 2\pi c'(1 + f_2(\ell)) t~.
\end{align}
and find numerically that $f_1\equiv f_2$ (see Fig.~\ref{fig:M1_1_prefactor}(c)).

\begin{figure*}
\begin{center}
\begin{tabular}{cc}
\includegraphics[angle=0, width=0.45\textwidth]{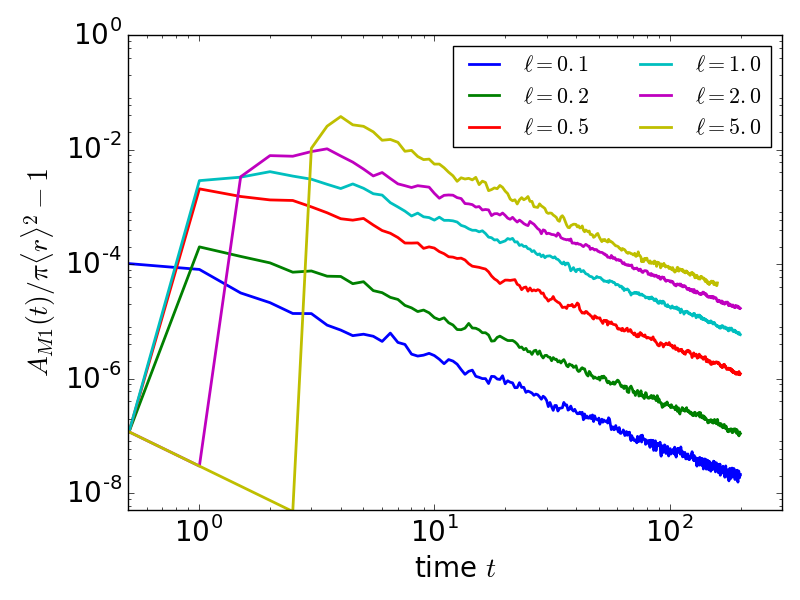} &
\includegraphics[angle=0, width=0.45\textwidth]{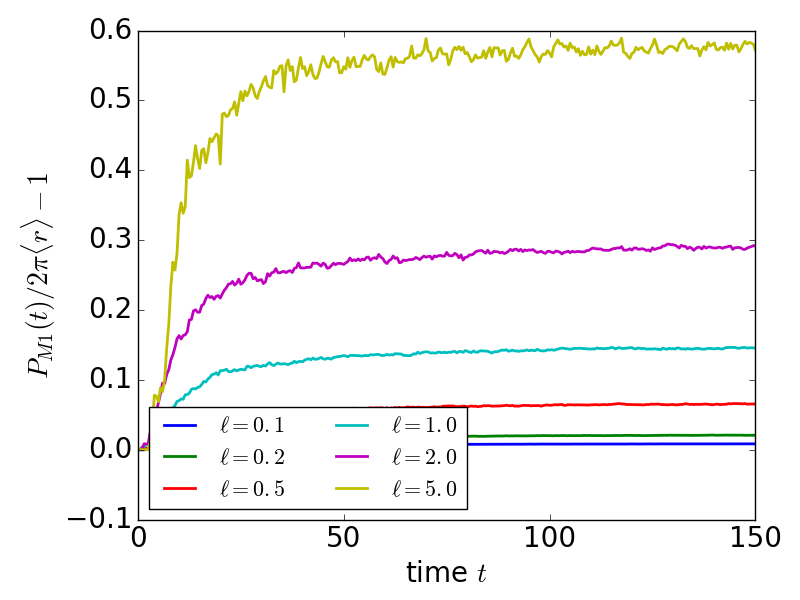}
\end{tabular}
\end{center}
\caption{Model $M_1$ for $\theta=1$. (Left) $\frac{\pi {\langle r \rangle}^2}{A(t)} -1$ versus $t$ for
  different values of $\ell$. (Right)
  $\frac{P(t)}{2\pi {\langle r \rangle}} -1 $ for different values of
  $\ell$. The results are obtained averaging over $100$ simulations. Source : From~\cite{carra2017coalescing}.}
\label{fig:A_cor_2_1}
\end{figure*}
In Fig.~\ref{fig:A_cor_2_1}, the quantities $\frac{A(t)}{\pi {\langle r \rangle}^2} -1$ and $\frac{P(t)}{2\pi {\langle r \rangle}} -1$ are displayed ($\langle r \rangle$
denotes the average radius of the primary colony).

Even though the area resembles asymptotically the area of a circle, as shown in Fig.~\ref{fig:A_cor_2_1} (left), its perimeter show large deviations, as it roughens with time, with deviations increasing with~$\ell$ (see Fig.~\ref{fig:A_cor_2_1} (right)).

Consider a circle of radius~$\langle r \rangle$ (playing the role of the primary colony), and attach it~$n$ semi circles of radius~$\delta < \langle r \rangle$. By spatial constraint,~$n<N = \frac{\pi \langle r\rangle}{\delta}$, and
\begin{equation}
\frac{P}{2\pi \langle r \rangle} - 1 = \frac{n}{N}\left(\frac{\pi}{2} - 1 \right),
\end{equation}
consistently with the results observed in Fig.~\ref{fig:A_cor_2_1} (right).

The observable 
\begin{equation}
S(t) = P(t) / (2\sqrt{\pi A(t)}) - 1~
\end{equation}
quantifies the rugosity of the main colony.
By construction, $S(t)=0$ for a circle, while $S(t) > 0$ quantifies the deviations. Simulations show that for $\theta = 0$, $S(t)$ is positive but decays towards zero as $t$ increases, in agreement with the previous analysis. In this regime, the $M_0$ model thus offers a reliable approximation. In contrast, for $\theta = 1$ , $S(t)$ grows and eventually exceeds unity at long times, signaling persistent rugosity and demonstrating that the $M_0$ model no longer provides an adequate description.

%

This framework is very general and allows for a theoretical analysis of growth and coalescence processes, which are highly relevant to city expansion. This discussion focused on quantitative predictions for a simplified model where the emission rate depends on the exponent $\theta$, while the distance $\ell$ remains constant and the process is isotropic. To further refine this framework, additional factors such as anisotropy and random emission distances should be incorporated to assess their impact on the results. The circular approximation, which facilitates the analytical approach, appears justified when the emission rate does not increase too rapidly with the size of the primary colony. However, when this condition is not met, accounting for the colony’s geometry becomes essential, significantly complicating the theoretical treatment. Given its generality and adaptability, this model holds strong potential for studying the dynamics of complex systems, including urban expansion.
        
\section{Diffusion-based approaches}
\label{chap:6}

In this chapter, we explore various mathematical approaches to modeling urban expansion, incorporating a diffusion term alongside other key factors. While reaction-diffusion models technically fall within this category, we discuss them separately in the following section.  

Despite the importance of these factors in urban expansion, the number of mathematical approaches that explicitly incorporate them remains surprisingly limited. Several key ingredients play a crucial role in shaping urban expansion dynamics, including congestion effects, which influence the movement of individuals and businesses; migration dynamics, which govern population redistribution in response to urban density; the role of services, which affect attractiveness and settlement patterns; and the coevolution of the transport network, where infrastructure development and urban expansion mutually influence each other. Integrating these elements into mathematical models is essential for a more comprehensive understanding of urban growth, yet such approaches are still relatively scarce.  

As discussed in Chapter~\ref{chap:1}, there are various ways to characterize a city and to monitor its growth. However, in the approaches considered here, the city is described through the local population density, $\rho(x,t)$, and its temporal evolution. The  objective of these approaches is to formulate an evolution equation of the form  
\begin{align}
\frac{\partial \rho(x,t)}{\partial t} = F(\rho, \nabla^k \rho, x, t, \dots)
\end{align}
where $\nabla^k \rho$ denotes the $k$-th spatial derivative of $\rho$, and $F$ is a function encapsulating the mechanisms that drive urban evolution. Typically, the first term in such equations corresponds to a diffusive process. In this chapter, we primarily focus on equations of the form  
\begin{align}
\frac{\partial \rho(x,t)}{\partial t} = \nabla^2 (D\rho) + F(x, t, \rho, \dots)
\end{align}
where $D$ is the diffusion coefficient, which may itself depend on the local density.

The earliest models \cite{ishikawa1980new,bracken1992simple} were primarily developed to explain and replicate one of the most significant empirical observations of the time: the decline in population density from the city center, typically described by an exponential function $\exp(-br)$ \cite{Clark:1951} (see Chapter~\ref{subsec:clark}). These models were inherently constrained, as they relied on the assumptions that cities are monocentric and isotropic—assumptions that contemporary urban studies now recognize as unrepresentative of most real cities. Subsequent work has incorporated additional ingredients, such as congestion effects, the role of services, migration dynamics, and the coevolution of the transport network, as in \cite{simini2015discovering,capel2024angiogenic}.

\subsection{Isotropic and monocentric cities}
\label{subsec:ishikawa}

\subsubsection{Isolated city}

Ishikawa~\cite{ishikawa1980new} was among the first to propose a partial differential equation (PDE) approach to modeling urban population density. The goal was to refine Clark’s exponential model~\cite{Clark:1951}, which captures the general decrease of density with distance from the city center but fails to explain several empirical observations. In particular, as city populations grow, density gradients gets less steep; gradients decrease with historical time; daytime densities are steeper than resident densities; and, perhaps most strikingly, Clark’s model cannot reproduce the so-called `density crater',  where the maximum residential density occurs in a ring surrounding the center.

Ishikawa introduced a model in which births and deaths are ignored and changes in residence arise from two processes. The first is a centripetal movement toward the center, driven by the demand for accessibility and described by a potential $U(r)$ with a minimum at the city center. In this view, without such a potential no city could ever form. The second is an isotropic outward movement generated by the avoidance of crowding, represented by a density-dependent diffusion. This `crowding pressure effect’ parallels population-pressure effects in ecology.

Formally, the population density flux $J(r,t)$ is defined as
\begin{equation}
    J(r,t) = -\nabla(D\rho) - \rho \nabla U \,,
\end{equation}
where the first term describes isotropic diffusion and the second a centripetal drift toward the center. The diffusion coefficient is assumed linear in the density
\begin{equation}
    D(\rho) = \alpha + \beta \rho \,,
\end{equation}
with $\alpha>0$ the intrinsic diffusion coefficient and $\beta$ the strength of crowding avoidance. Conservation of population implies the continuity equation, 
\begin{equation}
    \frac{\partial \rho}{\partial t} + \nabla \cdot J = 0 \,
\end{equation},
with boundary condition 
\begin{equation}
    \qquad J(r\to\infty)=0 \,,
\end{equation} 
which yields the governing PDE
\begin{equation}
    \frac{\partial \rho}{\partial t} = \nabla^2(D\rho)+\nabla(\rho\nabla U)\,.
\end{equation}
The first term represents density-dependent diffusion, while the second encodes the force generated by the potential gradient.

In one dimension, stationary solutions $\rho_s$ satisfy
\begin{equation}
    \log \rho_s(x) + \frac{2\beta}{\alpha}\rho_s(x) + \frac{U(x)}{\alpha} = C \,.
\end{equation}
Without crowding ($\beta=0$), this reduces to
\begin{equation}
    \rho_s(x) \propto e^{-U(x)/\alpha} \,,
\end{equation}
which coincides with Clark’s exponential form for a linear potential $U(x)\propto |x|$ -- see~\ref{subsec:clark}. With crowding, however, central densities are reduced and the profile deviates from a pure exponential. Similar results hold in two dimensions for isotropic potentials $U(r)$.

Ishikawa’s contribution was pioneering in linking behavioral mechanisms—convenience-driven centripetal movement and crowding-induced dispersal—to macroscopic urban population patterns within a PDE framework. The model demonstrates how aggregated individual choices can generate measurable spatial structures. Yet, it relies on strong assumptions, such as an unspecified form of the potential and a simple linear crowding effect, and remains limited in explanatory power. 

\subsubsection{Including congestion}

In the 1990s, Bracken and Tuckwell \cite{bracken1992simple} continued the effort towards description of population density dynamics through a continuous-space partial differential equation (PDE) framework. Their formulation brings together four fundamental processes that are understood to shape the evolution of a growing city: (i) local growth, which accounts for demographic expansion at a given site; (ii) diffusion, representing the spatial spread of population into adjacent areas; (iii) congestion effects, which act as a regulatory mechanism slowing growth in densely populated regions; and (iv) migration flows, which are introduced via suitable boundary conditions. In their model, the urban area is idealized as being monocentric and isotropic, developing symmetrically around a central business district (CBD).

The starting point is the description of local growth. At each spatial location, population density $\rho$ is assumed to evolve according to a logistic law
\begin{align}
\frac{\partial\rho}{\partial t} = k\rho(\sigma - \rho),
\end{align}
where $k$ represents the intrinsic growth rate, and $\sigma$ denotes a local carrying capacity.

In order to capture spatial expansion, Bracken and Tuckwell introduced a diffusion term. This modification yields a Fisher–KPP type equation \cite{Fisher:1937,kolmogoroff1988study}, a well-known PDE in population dynamics
\begin{align}
\frac{\partial\rho}{\partial t} = D \Delta \rho + k\rho(\sigma - \rho),
\end{align}
where $D$ is the diffusion coefficient and $\Delta$ is the Laplacian operator.

The third ingredient of the model is congestion. As the central areas of the city densify, congestion is assumed to hinder further growth. To formalize this effect, the cumulative population within a given radius $r$ is introduced
\begin{align}
N(r,t) = \int_0^r 2\pi r' \rho(r',t),dr',
\end{align}
which measures the number of inhabitants contained within the circle of radius $r$. The negative influence of congestion is then incorporated through a nonlinear inhibition term proportional to $\rho N(r)$, producing the governing equation
\begin{equation}
\frac{\partial \rho}{\partial t} = D \nabla^2 \rho + k \rho(\sigma - \rho) -\beta \rho \int_0^r \rho(r',t),dr',
\label{eq:bracken}
\end{equation}
where $\beta$ controls the strength of congestion effects.

Finally, migration flows are implemented through boundary conditions that allow for net inflows at the city center. Specifically,
\begin{equation}
\lim_{r \to 0} r \frac{\partial \rho}{\partial r} = \alpha,
\end{equation}
with $\alpha$ prescribing the boundary flux. The system is initialized with some prescribed density profile $\rho(r,0) = \phi(r)$.

To make the analysis more tractable, Bracken and Tuckwell considered a one-dimensional reduction of the model, for instance to represent development along a coastline or linear corridor (for example due to topographic constraints)
\begin{align}
\begin{cases}
\frac{\partial \rho}{\partial t} = D \frac{\partial^2 \rho}{\partial x^2} + k\rho(\sigma - \rho) - \beta \rho \int_0^x \rho(x'),dx'\\
\rho(x,0) = \phi(x)\\
\frac{\partial \rho}{\partial x}(0,t) = \alpha.
\end{cases}
\label{eq:brackendyn_1d}
\end{align}

Insight can already be gained by examining the model in the absence of diffusion ($D=0$). In this case, the stationary state satisfies
\begin{equation}
k(\sigma - \rho) = \beta \int_0^x \rho(x'),dx'
\quad \Rightarrow \quad \rho(x) = \sigma e^{-\beta x / k},
\end{equation}
which corresponds to an exponentially decaying density profile (see sec.~\ref{subsec:clark}). The total population in this solution is $P = \sigma k / \beta$, so an increase in the total population effectively “flattens” the profile. Extending the calculation to two dimensions (still with $D=0$) produces a Gaussian profile,
\begin{align}
\rho(r) = \sigma e^{-\pi(\beta/k)r^2},
\end{align}
which has the appealing feature of showing depressed densities at the very center, in closer agreement with empirical urban density distributions \cite{Clark:1951}.

\begin{figure}
\centering
\includegraphics[angle=0, width=0.45\textwidth]{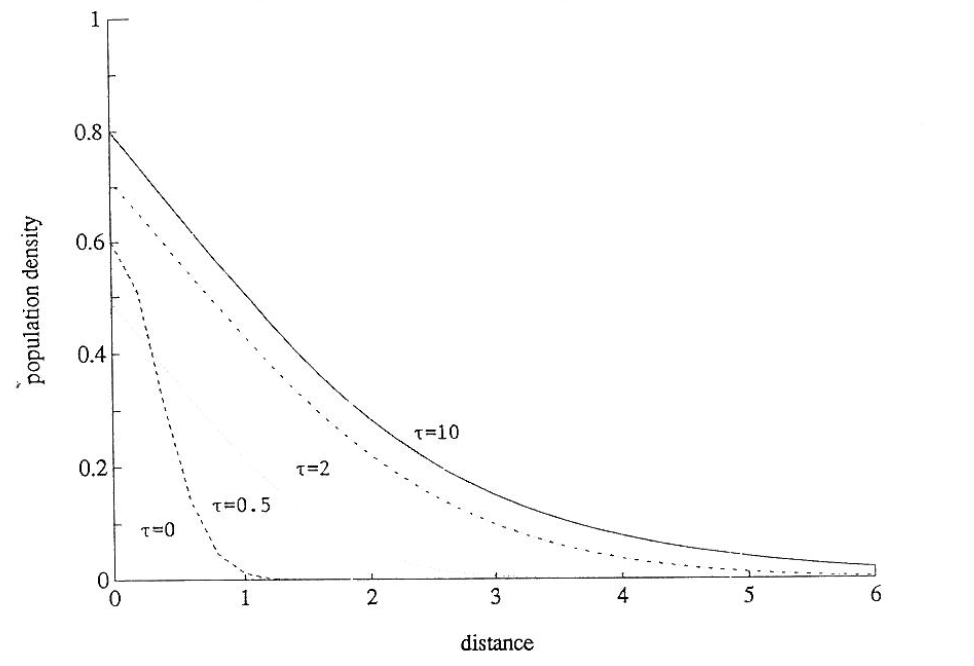}
\caption{Time evolution of population density in the Bracken–Tuckwell model. The simulations display convergence to a stationary profile with a gradually declining density gradient. Source: From \cite{bracken1992simple}.}
\label{fig:bracken_time}
\end{figure}

When diffusion is reintroduced ($D > 0$), the steady-state profile in one dimension must satisfy
\begin{align}
D \frac{d^2\rho}{dx^2} + k\rho(\sigma - \rho) - \beta\rho\int_0^x \rho(x'),dx' = 0.
\end{align}
One particular solution to this nonlinear equation is
\begin{align}
\rho(x) = \left( \sigma + \frac{D\beta^2}{k^3} \right) e^{-(\beta/k)x},
\end{align}
although in general the system requires numerical integration. Numerical experiments, illustrated in Fig.~\ref{fig:bracken_time}, show that the density profile converges to a stationary form over time, with the gradient becoming progressively flatter. This behavior is consistent with empirical findings on urban population distributions \cite{Clark:1951,ishikawa1980new}, further supporting the plausibility of the model.

\begin{figure}
\includegraphics[angle=0, width=0.4\textwidth]{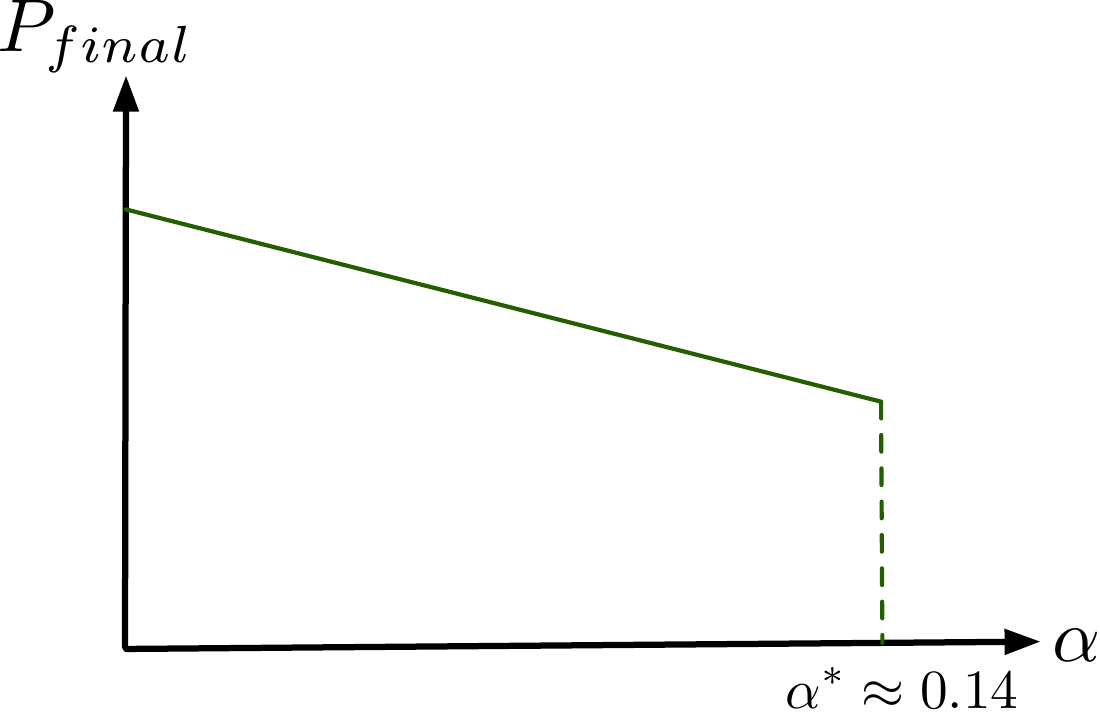}  
\caption{Final population $P_{\text{final}}$ vs emigration rate $\alpha$. Below $\alpha^* \approx 0.14$, the city persists. Above, it collapses. From \cite{bracken1992simple}.}
\label{fig:bracken}
\end{figure}

The model also explores other phenomena such as city persistence versus extinction by varying emigration rate $\alpha$. For initial condition
\begin{align}
\rho(x,0) = \sigma \, c \, \exp\left( -(\frac{\beta}{k})^2 x^2 / \gamma \right),
\end{align}
numerical results show a threshold $\alpha^* \approx 0.14$: for $\alpha > \alpha^*$, the city collapses (see Fig.~\ref{fig:bracken}), while $\alpha < \alpha^*$ yields survival.
Initial conditions also matter: low initial density can lead to extinction, while slightly higher values ensure survival. These results suggest a critical line in the $(c,\alpha)$ phase space.

While the model primarily aims to reproduce exponential density decay, its PDE framework with congestion and migration demonstrates how minimal ingredients can yield rich spatial dynamics. Despite limited calibration, it extends earlier approaches~\cite{ishikawa1980new} and provides a flexible foundation for modeling urban evolution.

\subsection{Including services}

The study by Whiteley et al.~\cite{whiteley2022modelling} investigates the emergence of urban structures using integro-differential equations that couple the spatial dynamics of population and service densities. The central assumption is that spatial proximity benefits both residents and services, leading to self-organized urban patterns. The focus is not on urban growth per se, but on the organization of multiple cities and the emergence of a inter-city characteristic length scale. Analyses based on Fourier transforms and spatial autocorrelation reveal that the typical spacing between cities in the UK is about 45 km, and one of the aims of the model is to explain this order of magnitude.

More precisely, the model consists of equations describing the evolution of the population density $\rho(x,t)$ and a service density $s(x,t)$ at spatial location $x$ and time $t$. The key variable is the attractiveness $A(x,t)$ of a location, which increases when services are available nearby but decreases when the location itself is saturated with services. The attractiveness is defined as
\begin{equation}
A(x,t) = \bigl(1 - s(x,t)\bigr)\int K(x-y)\,s(y,t)\,dy \,,
\end{equation}
where the factor $1-s(x,t)$ accounts for the reduced appeal of service-congested areas, and the integral represents the contribution of surrounding services weighted by a kernel $K(x)$ that decreases with distance,  which is assumed to follow a Gaussian form
\begin{equation}
K(x) = \frac{1}{\beta_1\sqrt{2\pi}} e^{-\frac{x^2}{2\beta_1^2}}.
\end{equation}

Individuals can relocate from position $y$ to $x$, with probability
\begin{align}
P(y\to x)&=\rho(y)A(x)F(d(x,y)),
\end{align}
where $F(d)$ is a decreasing function of distance $d$. The evolution of population density is then governed by
\begin{equation} 
\frac{d\rho}{dt} = D \int \left[A(x) \rho(y) - A(y) \rho(x)\right] F(x - y)dy. 
\end{equation}

The new ingredient in this study is the service density, which follows logistic growth with a carrying capacity that depends on the population density
\begin{equation}
\frac{ds}{dt} = (f + gs) (\sigma(P) - s),
\end{equation}
with
\begin{equation}
\sigma(P) = 1 - e^{-\left( \frac{P}{\lambda} \right)^{\mu}}.
\end{equation}
Here $\lambda$ sets the population scale for saturation and $\mu$ controls the rate of convergence toward $\sigma(P)\to 1$ for large $P$.

\begin{figure*}
\centering
\includegraphics[angle=0, width=0.9\textwidth]{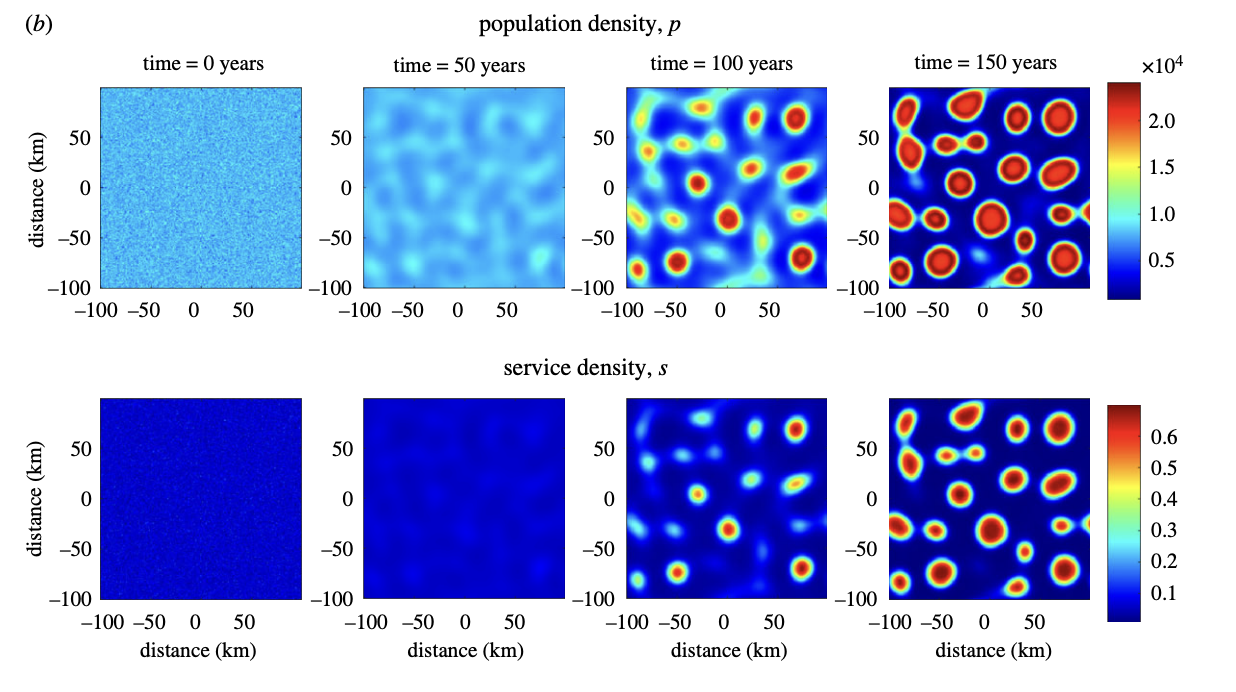}  
\caption{Simulation of population density (top) and service density (bottom). The initial population is random. After 100 years, city structures emerge, which are further strengthened after 150 years. Source: From \cite{whiteley2022modelling}.}
\label{fig:whiteley_time}
\end{figure*}

Simulations of the model (see Fig.~\ref{fig:whiteley_time}) illustrate the parallel development of multiple urban centers together with the emergence of service hotspots. Starting from a random initial population, city structures appear after about 100 years and become further consolidated after 150 years.

\begin{figure}
\centering
\includegraphics[angle=0, width=0.45\textwidth]{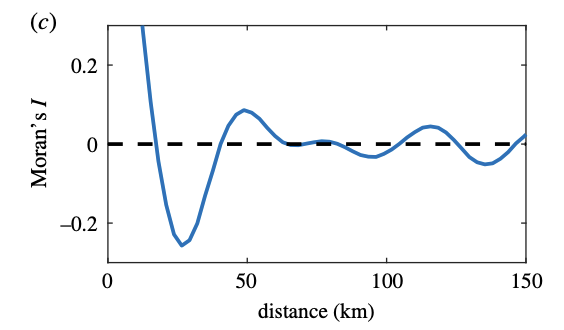}  
\caption{Autocorrelation plot from simulations. The characteristic length scale between cities is about 50 km. Source : From\cite{whiteley2022modelling}.}
\label{fig:whiteley_autocorr}
\end{figure}

In order to gain analytical insight and to understand the organization of multiple cities, closed-form solutions are not available and, as in previous models, one resorts to linear stability analysis. Perturbations of the homogeneous steady state are governed by the Jacobian $J(k)$ in Fourier space
\begin{align}
\nonumber
J(k) &=\\
&\begin{bmatrix}
-D s_0(1 - s_0)(1 - \hat{F}(k)) & D p_0(1 - \hat{F}(k))\\
(1 - s_0)\hat{K}(k) - s_0 & (f + g s_0)(\sigma'(p_0)\hat{F}(k) - 1)
\end{bmatrix}.
\end{align}
Pattern formation occurs when at least one eigenvalue of $J(k)$ has a positive real part. The analytical treatment is involved, but the characteristic length scales can be estimated from numerical simulations (see Fig.~\ref{fig:whiteley_autocorr}). These simulations reveal a typical inter-city spacing of about 50 km, in close agreement with empirical observations for UK cities ($45$--$50$ km).

To model intra-urban structure, two additional mechanisms are introduced. First, residents prefer to be near but not too close to services, which modifies the attraction kernel $K$. Second, competition for space between services and population modifies the service growth equation to
\begin{equation}
\frac{ds}{dt} = \bigl(H(\sigma(P) - (s + \alpha_1 p)) (f + gs)\bigr) (\sigma(P) - (s + \alpha_1 p)).
\end{equation}
Competition strength is parametrized by~$\alpha_1$. Note that there is now a term~$\bigl(H(\sigma(P) - (s + \alpha_1 p))$ where~$H$ is the Heaviside function ensuring that~$s\geq0$.
When competition is sufficiently strong, out-of-phase patterning emerges, with services and population occupying distinct areas.
This model explains the emergence of urban patterns through simple spatial interactions. The analysis predicts the formation of distinct cities at characteristic length scales, the appearance of secondary patterning within cities due to competitive interactions, and highlights the fundamental role of spatial interactions in shaping urban structure. The framework can be extended to incorporate dynamic parameters and heterogeneous populations, broadening its relevance for urban modeling.

However, this approach relies on a relatively large set of parameters—one for each of the three kernels, the population scale parameter~$\sigma$, the steepness parameter, among others. The presence of so many parameters, with only one testable feature, the inter-city length scale, raises concerns about the robustness and explicative power of the model.

\subsection{Migration effects}

As in the study by Whiteley et al.~\cite{whiteley2022modelling}, Simini and James~\cite{simini2015discovering} (published only on arXiv) focus on the structure of systems of cities rather than on city growth itself. Nevertheless, their framework can serve as a useful starting point for modeling urban growth, and for the sake of completeness we briefly discuss it here. The model explores numerically the consequences of an integro-differential equation composed essentially of two terms: migration and natural growth. Following earlier  approaches~\cite{ishikawa1980new,bracken1992simple}, the natural growth component is described by a logistic equation with carrying capacity $\rho_0$ and growth rate $g$. Migrations are described by continuous version of the gravity model or the intervening opportunity model, which both describe how the migration flows vary with distance. The out-migration is denoted by $T^{out}$ and the increase of density due to migration is denoted by $T^{in}$. The density evolution equation can then be written as
\begin{align}
    \frac{\partial\rho(x.t)}{\partial t}=g\rho(x,t)[1-\rho(x,t)/\rho_0] - T^{out}+T^{in}
\end{align}

The functions $T^{\text{out(in)}}$, representing the flows of individuals leaving or relocating to location $x$, describe migration processes and can be specified using different models. In this study, the authors considered two alternatives: the gravity model \cite{Erlander:1990} and the intervening opportunities model \cite{stouffer1940intervening}.

\subsubsection{Gravity Model for Migration}

Migration between locations $i$ and $j$ is modeled using the Gravity model \cite{Erlander:1990}, which assumes that migration likelihood depends on population size, opportunities, and distance between locations. The probability that an individual relocates from location $i$ to $j$ is given by
\begin{align}
    P_{i\to j} = C\, \rho(i)^\alpha \rho(j)^\beta f(r_{ij}) \,,
    \label{eq:gravity}
\end{align}
where $\rho(i)$ is the local density at $i$, $\alpha$ and $\beta$ are positive exponents, and $C$ is a normalization constant ensuring that probabilities sum to unity over the domain $D$ (i.e., $\sum_{j\in D} P_{i\to j} = 1$). The deterrence function $f(r)$ captures the effect of distance, typically chosen as a continuous decreasing function such as $f(r)\sim e^{-r/r_0}$ or $f(r)\sim r^{-\gamma}$. 

This formulation implies $P_{i\to j} = 0$ if the density at $j$ vanishes, but in fact, the probability for an individual to relocate depends also on the availability of resources or opportunities at the new location. If we denote these resources by $w$, we then obtain
\begin{equation}
P_{i\to j} \propto [\rho(j) + w(j)]f(r_{ij})
\end{equation}
where $w(j)$ represent the opportunities or resources independent of population (e.g., natural resources). Note that the term $\rho(i)$ simplifies with the normalization. 

In this framework, if we assume a constant average migration rate (ie. the fraction of individuals that will relocate per unit time from $i$), then we can write
\begin{align}
T^{out}=T\rho(i).
\end{align}
The average number of individuals relocating in $i$ is given by 
\begin{align}
\nonumber
    T^{in}&=\int_D\mathrm{d}j T\rho(j)P_{j\to i}\\
    &=[\rho(i)+w(i)]\int_D\mathrm{d}j T\frac{\rho(j)f(r_{ij})}{\int_D\mathrm{d}i[\rho(i)+w(i)]f(r_{ij})}
\end{align}

For a 1d line, the dynamic equation then becomes
\begin{align}
\nonumber
&\frac{\partial \rho(x, t)}{\partial t} = g\rho(x, t)\left(1 - \frac{\rho(x, t)}{\rho_0}\right) - T\rho(x, t)\\
&+ T[\rho(x,t)+w]\int_{-\infty}^\infty \frac{\rho(x-x',t)f(x')}
{\int_{-\infty}^\infty f(y)[\rho(x-y,t)+w]dy}dx'.
\end{align}
The homogeneous solution $\rho(x,t)=\rho_0$ is a stationary state for this equation, and it is natural to look at the stability of this solution by assuming small perturbations around this solution: $\rho(x,t)=\rho_0+\delta\rho(x,t)$. Introducing the Fourier transform
\begin{align}
h_k(t)=\int \mathrm{d}x\mathrm{e}^{-ikx}\delta\rho(x,t)    
\end{align}
we get an equation of the form \cite{simini2015discovering}
\begin{align}
    \frac
    {\partial h_k(t)}
    {\partial t}
    =h_k(t)\Lambda_k(\rho_0,g,T,w,f)
\end{align}
where $\Lambda_k$ is the growth rate of mode $k$ and which depends on the all the constants of the problem and on the deterrence function $f$. In this approach, for a city to develop we need to have $\Lambda_k>0$ for some $k$. Indeed, if $\rho_0$ is an unstable equilibrium then an initially small perturbation can grow leading to
the formation of zones of high population density (cities). For a deterrence function of the form $f(r)=\exp{-(r/r_0)}$, we get
\begin{align}
    \Lambda_k=-g-T\frac{\rho_0}{\rho_0+w}+T
    \frac{w+(kr_0)^2(\rho_0+w)}{(1+(kr_0)^2)^2(\rho_0+w)}
    \label{eq:lambdak}
\end{align}
We  find the maximum by writing $d\Lambda_k/dk=0$ and obtain the most unstable mode as
\begin{align}
    k_m=\frac{1}{r_0}\sqrt{\frac{\rho_0-w}{\rho_0+w}}
    \label{eq:km}
\end{align}
This value can be interpreted as the number of cities per unit length (provided that $\Lambda_{k_m}>0$), or equivalently, as the inverse of the typical distance between cities. For $\rho_0\gg w$, we obtain $k_m\sim 1/r_0$ implying that the density of cities is determined by the characteristic travel length $r_0$, and independent from the other variables such as growth and migration rates.

Inserting the value of $k_m$ (Eq.~\ref{eq:km}) in Eq.~\ref{eq:lambdak}, the condition $\Lambda_{k_m}>0$ reads
\begin{align}
    \Lambda_{k_m}>0 \Rightarrow T>4g\rho_0\frac{\rho_0+w}{(\rho_0+w)^2}
\end{align}
and if $\rho_0\gg w$, the condition simply reads as $T>4g$. This means that cities will form if the population is sufficiently mobile (with a migration rate $T$  sufficiently larger than the growth rate $g$). We note that similar results were obtained in the case of a power law deterrence function and also in the 2-dimensional case \cite{simini2015discovering}.

\subsubsection{Intervening Opportunities Model}

Another migration model was considered in \cite{simini2015discovering}: the intervening opportunities model which is based on the idea that the number of individuals going a given distance is proportional to the number of opportunities at that distance and decreasing with the number of opportunities within that distance \cite{stouffer1940intervening}. We thus have 
\begin{align}
P_{i\to j} = [\rho(j)+w(j)]f\left(\int_{B_i(r_{ij})} [\rho(z) + w(z)] \, dz\right),
\end{align}
where $B_i(r_{ij})$ is the region containing opportunities closer to $i$ than $j$ (and where the normalization condition $\int djP(i\to j)=1$ is imposed as a condition on the function $f$).

For this model in 1d, the dynamical equation is the following \cite{simini2015discovering}
\begin{align}
\nonumber
&\frac{\partial \rho(x, t)}{\partial t} = g\rho(x, t)\left(1 - \frac{\rho(x, t)}{\rho_0}\right) - T\rho(x, t)\\
\nonumber
&+ T[\rho(x,t)+w]\int_{-\infty}^\infty \mathrm{d}x' \rho(x-x',t)\\
&\times f\left(\int_{B(x,x')} [\rho(z) + w(z)] \, dz\right),
\end{align}
The same stability analysis as above can be performed on this equation and for an exponential deterrence function $f(r)=1/r_0\mathrm{e}^{-r/r_0}$, the result reads for $\rho_0\gg w$
\begin{equation}
T > 3g.
\end{equation}
The most unstable mode is of the form
\begin{align}
    k_m=\frac{w}{r_0}F(\rho_0/w)
\end{align}
where $F(x)$ is a known function \cite{simini2015discovering} and behaves as $F(x)\to x$ for $x\gg 1$. This is a distinctive difference between the gravity model and the intervening opportunities model: in the gravity model, for $\rho_0\gg w$, the number of cities per unit length tends to a finite value controlled by the typical length of migrations ($\sim 1/r_0$). In the intervening opportunities model, $k_m$ increases with $\rho_0/w$, and the number of cities will grow (results that are confirmed by numerical simulations in \cite{simini2015discovering}).
\begin{figure}[htp]
\includegraphics[angle=0, width=0.45\textwidth]{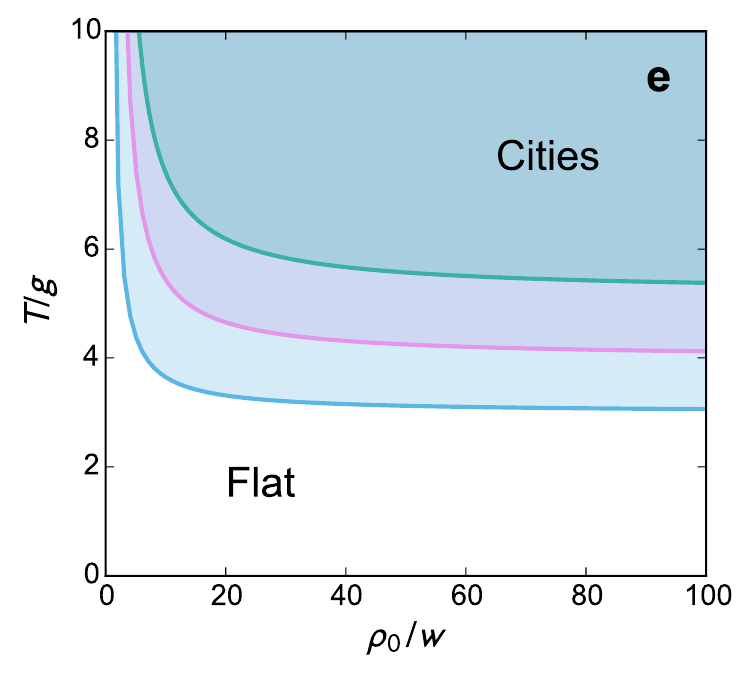}  
\caption{Simulation result obtained in \cite{simini2015discovering}. Cities emerge if the population
is sufficiently mobile, i.e. if the migration rate T is sufficiently higher than the population growth
rate g. The three curves show the conditions for the equilibrium state of uniform population to be
unstable for the exponential IO model, Gravity models, and a power-law IO model (from bottom to
top).}
\label{fig:simini}
\end{figure}

This study formalizes some empirical laws observed for cities, and examines the impact of migration dynamics through both the Gravity and the Intervening Opportunities models. The analysis shows that the Gravity model produces a fixed number of cities, independent of the carrying capacity $\rho_0$, whereas the Intervening Opportunities model predicts that city density increases with $\rho_0$, thereby capturing the effect of higher population densities. In both cases, thresholds for city formation are identified: cities emerge only when the migration rate $T$ exceeds a critical value proportional to the growth rate $g$. Although these results provide a valuable starting point for analytical approaches to urbanization, the study lacks direct empirical validation. Bridging this gap would be essential to assess the robustness of the theoretical predictions and to connect them more closely with observed urban dynamics.


\subsection{Coevolution of the population density and the transportation network}

The coevolution of population density and the transportation network is a dynamic process where urban growth and infrastructure development mutually influence each other over time \cite{Barthelemy:2009,raimbault2016hybridnetworkgridmodelurban}. As population density increases in certain regions, the demand for efficient transportation systems grows, prompting expansions in road networks, public transit, and other mobility solutions. Conversely, improvements in transportation infrastructure can drive changes in population distribution by making previously remote areas more accessible, leading to urban sprawl or densification. This feedback loop is shaped by economic, social, and environmental factors, including land-use policies, travel behavior, and viability considerations. Computational models and network theory are often used to study this interplay, capturing how cities evolve through self-organizing mechanisms and optimizing transport efficiency while balancing congestion and emissions. Understanding this coevolution is crucial for designing resilient and adaptive urban environments that meet future mobility and durability challenges.

In particular, the paper \cite{capel2024angiogenic} explores the `angiogenic' growth of cities, drawing an analogy between urban development and biological processes such as angiogenesis. It introduces a reaction-diffusion (RD) model to describe the coevolution of urban populations and transport networks. The study examines the long-term development of London (1831--2011) and Sydney (1851--2011), demonstrating the interplay between population dynamics, transport infrastructure, and economic constraints. 

\subsubsection{Coupling network and density growth: a first step}

As cities expand, congestion and accessibility constraints drive the emergence of new subcenters. To explain how these locations arise, the study \cite{Barthelemy:2009} proposes a generative model in which the placement of new centers—such as homes, businesses, or other activity hubs—is governed by the trade-off between two competing factors: rent costs, which increase with local population density, and accessibility, which depends on network centrality. In their framework, urban centers are either distributed randomly or follow underlying population gradients, and a growing street network evolves to accommodate them. At each step, unconnected centers stimulate the addition of new road segments, which are iteratively extended to connect efficiently to the existing network. This minimal mechanism captures both the spatial distribution of activity centers and the self-organized growth of the transportation infrastructure that links them.

For the rent cost, the city is divided into sectors of area $S$. If the sector $i$ comprises $N(i)$ centers, the local density is
\begin{align}
\rho(i) = \frac{N(i)}{S} \,.
\end{align}
The rent cost is assumed to be proportional to the local density
\begin{align}
C_R(i) = A \rho(i)
\end{align}

In order to get an expression for the accessibility, they use the betweenness centrality $g(v)$ of a node $v$ and which is given by \cite{Freeman:1977}
\begin{align}
g(v) = \frac{1}{N(N-1)} \sum_{s \ne t} \frac{\sigma_{st}(v)}{\sigma_{st}}
\end{align}
where $\sigma_{st}$ is the number of shortest paths from $s$ to $t$ and $\sigma_{st}(v)$ is this number for shortest paths going through $v$. The average centrality in sector $i$ is then obtained by
\begin{align}
g(i) = \frac{1}{N(i)} \sum_{v \in S_i} g(v) \,.
\end{align}
The transportation cost is supposed to decrease with increasing centrality and a simple form for the cost function is
\begin{align}
C_T(i) = B(g_m - g(i)) \,,
\end{align}
where $B$ and~$g_m$ are positive constants.

\begin{figure}
  \centering
  \includegraphics[width=0.99\linewidth]{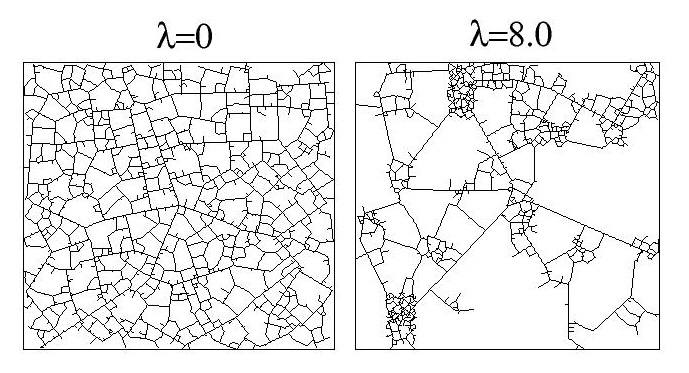}
  \caption{Networks obtained for different values of $\lambda$ (and for $N = 500$ and $\beta = 1$). On the left, $\lambda = 0$ and only the density plays a role and we obtain a uniform distribution of centers. On the right, we show the network obtained for a large value $\lambda = 8$. In
    this case, the centrality is the most important factor leading
    to a few dominant areas with high density. Source : From \cite{Barthelemy:2009}.}
  \label{fig:flammin1}
\end{figure}

Assuming that all individuals have income $Y$, the net income in sector $i$ is
\begin{align}
K(i) = Y - C_R(i) - C_T(i)
\label{eq:Y}
\end{align}
and the probability for a new arriving center to choose sector $i$ is  given by
\begin{align}
P(i) = \frac{e^{\beta K(i)}}{\sum_j e^{\beta K(j)}} = \frac{e^{\beta A(\lambda g(i) - \rho(i))}}{\sum_j e^{\beta A(\lambda g(j) - \rho(j))}}
\label{eq:Pi}
\end{align}
where $\lambda = B/A$ controls the trade-off between accessibility and density. For $\lambda \approx 0$, the probability reduces to $\propto \exp[-\beta A \rho(i)]$, so that sectors with lower density are preferred, and the city evolves toward a statistically uniform distribution. In contrast, for $\lambda \gg 1$, the probability becomes $\propto \exp[\beta A g(i)]$, with sectors of higher centrality dominating, leading to a highly heterogeneous structure. Figure~\ref{fig:flammin1} illustrates these two limiting cases. When $\lambda$ is small, density governs the location of new centers: they preferentially appear in low-density areas, smoothing out random fluctuations and producing a uniform distribution. Conversely, when $\lambda$ is very large, centrality drives the process, causing all centers to cluster in a limited area. In this sense, density promotes dispersion, while centrality favors concentration.

The main point of this model is to incorporate, in a simple way, both the network structure and the population density through the probability of selecting a new center (Eq.\ref{eq:Pi}), where density and accessibility—estimated via betweenness centrality—combine to define a composite commodity (Eq.\ref{eq:Y}). This type of ingredient will reappear in the more recent and realistic model of Capel-Timms et al. \cite{capel2024angiogenic}, which we describe in the next section.

\subsubsection{Coupling network and density growth: going further}

In \cite{Barthelemy:2009}, the authors proposed a minimal model for the co-evolution of road networks and population density, later extended in \cite{capel2024angiogenic}. The model reproduces key empirical features of street networks and shows how accessibility shapes urban growth. High-centrality nodes attract settlement, which further increases their centrality, creating a feedback loop that yields an exponential decay of density from the core, with its size determined by the balance between transport and rent costs. Empirical support for such feedback was provided by Levinson~\cite{levinson2008density}, who showed that rail stations promote higher densities, while denser areas encourage more rail construction. Motivated by this, the model studies how road networks and densities co-evolve under simple assumptions, with the city discretized into sectors where the probability $P(i)$ of new centers depends on rent and accessibility, in line with urban economics~\cite{brueckner1986structure}.

The variable chosen is the population density $\rho(x,t)$ which is modeled as a continuous function over time $t$ and space $x$ (which is here a two dimensional vector) within a domain $V \subset \mathbb{R}^2$. The governing equation incorporates diffusion, population redistribution, and external factors under the form
\begin{equation}
\frac{\partial \rho(x,t)}{\partial t} = D\nabla^2 \rho + \eta(x, t) \xi(t) + R(x, t) - S(x, t),
\end{equation}
where $D$ is a diffusion coefficient which 
is assumed to be uniform and constant, 
$\eta(\overrightarrow{X}, t)$ is the local attractiveness, $\xi(t)$ (in $cap/km^{2}/yr$) the total population growth rate (distributed in space via the local attractiveness $\eta$). This quantity is assumed to follow a logistic model with carrying capacity $K$ and growth rate $r$
\begin{equation}
\xi(t) = r P(t) \left( 1 - \frac{P(t)}{K} \right),
\label{eq:xi}
\end{equation}
where $P(t)$ is the total population.

The quantity  $R(x, t)$ denotes the internal population redistribution, and $S(x, t)$ represents an external population sink. The quantity $R(x,t)=R_-(x,t+R_+(x,t)$ represents `internal' migrations of residents from sources to sinks (distributed according to $R_+$ and $R_-$), such that 
\begin{align}
    \int_\Omega R(x,t)\mathrm{d}^2x = 0
\end{align}

In contrast, $S(x, t)$ is an `external' sink that removes population from the domain $\Omega$ under consideration. This term represents exogenous perturbations that accounts for population decrease. 

The quantity $\eta$ governs the attractiveness of a location for new inhabitants. Following \cite{Barthelemy:2009}, this factor is written under the form
\begin{align}
    \eta(x,t)=C\mathrm{e}^{\beta Y(x,t)}
\end{align}
where $C$ is a normalization factor, and $Y(x,t)$ is the net income at location $x$ at time $t$. This quantity is obviously complex and depends on various socio-economic processes, but the authors propose a simplified model capturing the important effects of rental and transportation costs, as in \cite{Barthelemy:2009}. 

Furthermore, in the case of London discussed in \cite{capel2024angiogenic}, there is a first phase of densification when the transportation (railway) network is less developed (`early densification phase') for $t<t_I$ where $t_I$ is the year defining the end of the densification phase, and for which $Y$ depends on the base per capita $Y_0$ and a densification exponent $b$: $Y=Y_0(\rho(x,t))^b$. After this phase, transportation networks become relevant, and suburbanization starts as distances shorten. The net income becomes then a balance between gross income $Y_{\text{gross}}$, and living cost $C_L$ and rent cost $C_R$ as discussed above 
\begin{align}
Y(x, t) = Y_\text{gross} - C_L(x, t) - C_T(x, t)
\end{align}
In summary, we  have
\begin{align}
\begin{cases}
    Y=Y_0(\rho(x,t))^b&\mathrm{for}\; t<t_I\\
    Y=  Y_\text{gross} - C_L(x, t) - C_T(x, t)& \mathrm{for}\; t>t_I \,.
\end{cases}
\end{align}

Living costs are calculated according to 
\begin{align}
    C_L(x,t)=\kappa\left(\int_0^t\rho(x,t)\right)^{\tau}
\end{align}
where $\kappa$ and $\tau$ are model parameters. The authors considered here the cumulative population density rather than $\rho$ (as done in \cite{Barthelemy:2009}), since it serves as an indicator of the availability of buildings and the centrality associated with prior population growth. Consequently, living costs do not necessarily decrease with a declining population, thereby explaining the increase in rent and property values observed in most historical centers despite their decreasing residential population.

The transportation costs $C_T$ depend on the transportation network structure. The network is denoted by $G=(V,E)$ where $V$ is the set of stations and $E$ the set of edges between consecutive stations (in the study \cite{capel2024angiogenic}, the railway network is considered). In order to quantify the accessibility of a given location, one can characterize the centrality of the closest node on the transportation network. Betweenness \cite{Barthelemy:2009} or closeness centrality can be used, but in \cite{capel2024angiogenic}, the authors used the definition proposed in \cite{wang2009spatiotemporal}. This measure describes the interaction between each node 
and the rest of a network as the ratio of total travel distance between all other nodes and the network
average travel distance. More precisely, the accessibility $A_i$ of node $i$ is calculated as follows. For each node, one computes the total distance $D_i$ which is the sum of all distances $d(i,j)$ (on the graph) between $i$ and all other nodes $j$ 
\begin{align}
    D_i=\sum_{j\in V}d(i,j)
\end{align}
The total distance in the graph is $D=\sum_{i,j}D(i,j)=\sum_i D_i$. In order to get a greater value of $A_i$ for higher accessibility (and vice versa), Capel-Timms et al., used the the inverse of the metric of \cite{wang2009spatiotemporal}
\begin{align}
    A_i=\left(\frac{D_i}{DN}\right)^{-1}
\end{align}
where $N=|V|$ is the number of nodes (stations). More accessible stations have a
higher value and therefore a lower transport cost. Stations with average accessibility have $A_i\approx 1$. Also, in order to account for multiple nodes in
an area, the transport cost $C_T(x)$ is calculated using the accessibility value of the nearest node of $x$.

The cost $C_T$ is then estimated from the accessbility but also considering the minimum distance $d_{\min}$ to the nearest station to account for commuting to/from the railway
\begin{align}
    C_T(x)=\mu d_{\min}(x)+\nu(h-A_v(x))
\end{align}
where $d_\text{min}$ is the distance to the nearest transport node, $A_v$ is node accessibility of the nearest node (or the average over all nodes in the same grid cell), and $h$ is a scaling factor. The quantities $\mu$ and $\nu$ are model parameters whose ratio governs the balance between the costs to reach the transport network and to use it. 

In order to estimate the redistribution functions $R$ and $S$, the authors of \cite{capel2024angiogenic} argued that 
areas with higher $Y$ are more likely to retain population due to their attractiveness, and areas with lower~$Y$ might lose their population to either other areas of the domain or to outside of the domain. Specifically, internal redistribution is estimated based on the economic attractiveness of an area.

Negative values of $Y$ imply that costs are higher than the gross income, prompting residents to move to more convenient locations. Hence, the flux of residents leaving an area X is calculated as
\begin{align}
    R_-(x,t)=
    \begin{cases}
r_R\rho(x,t) & \mathrm{if}\;\;Y(x,t)<0\\
0 & \mathrm{otherwise}
    \end{cases}
\end{align}
where $r_R$ is the sink rate specific to $R_-$. To account for migration within the domain, individuals leaving with rate $R_-$ are redistributed according to
\begin{align}
R_+(x,t)=\eta(x,t)\frac{1}{A}\int_{\Omega}R_-(x,t)\mathrm{d}^2x
\end{align}
where $A$ is the area of the domain $\Omega$. 

However, a migrating population can also leave the domain entirely due to exogenous factors affecting the city's population, such as post-war population decline or the removal of industries. This process is captured by the sink term $S(x, t)$, defined as:
\begin{align}
    S(x,t) = r_S \rho(x,t) \eta_S(x,t),
\end{align}
where $r_S$ represents the sink rate for $S$. If we assume that areas with lower $Y$ experience a higher sink $S$—as they are perceived as less attractive not only for new settlers but also for potentially migrating populations—then $\eta_S(x,t)$ is given by:
\begin{align}
    \eta_S(x,t) =
    \frac{\mathrm{e}^{-\beta_S Y(x,t)}}
    { \frac{1}{A} \int_\Omega \mathrm{e}^{-\beta_S Y(x,t)} \,\mathrm{d}^2x }.
\end{align}

We almost have defined all the ingredients of this model except for the transport network. This network expands in response to population changes: new stations are added at location $x$ with a probability $Q(x,t)$ that depends on the local population density (see \cite{capel2024angiogenic} for details). This probability increases with the number of individuals in the vicinity of $x$, weighted by a power-law kernel that decreases with distance. Newly added nodes are then connected to the existing network based on their proximity to the core. The accessibility feedbacks influence future population dynamics by altering $C_T$, which recursively modifies $K$ and $\rho$.

\begin{figure*}
  \centering
  \includegraphics[width=0.99\linewidth]{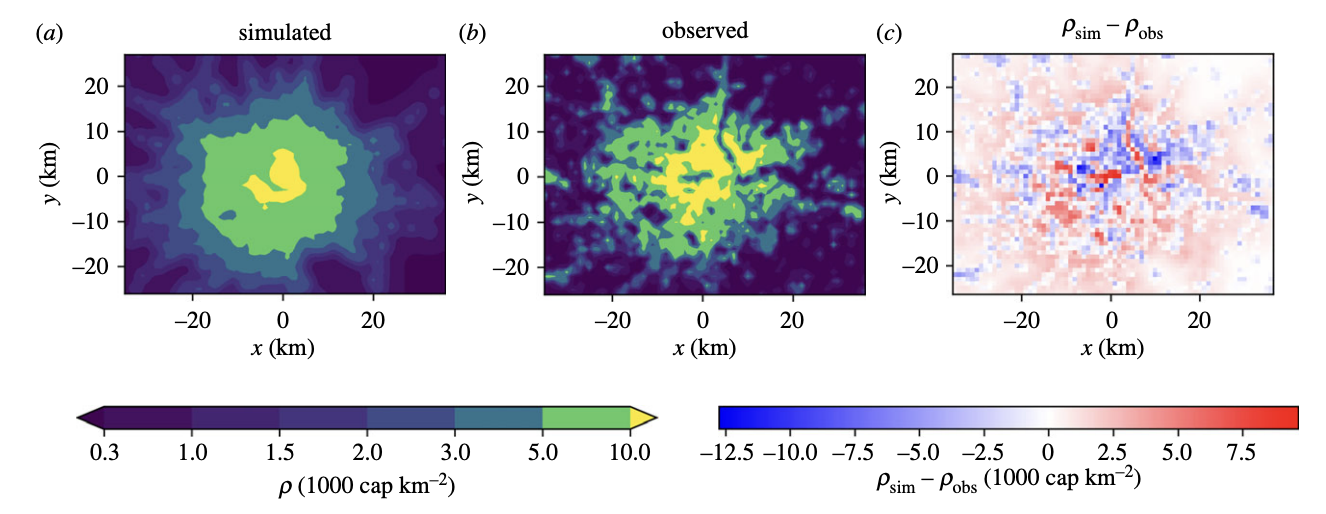}
  \caption{Simulated and observed population density in London for 2011. In (a) the simulated density, in (b) the observed density and in (c) the difference. Source: From \cite{capel2024angiogenic}.}
  \label{fig:capel}
\end{figure*}

The authors validate the model using historical data on population and rail infrastructure. Simulations for London and Sydney reproduce stylized facts such as radial expansion, polycentricity, and the emergence of commuter belts (see the results for the population density in London 2011 in Fig.~\ref{fig:capel}). The rail networks exhibit hierarchical branching, consistent with empirical centrality patterns and core-periphery structures observed in transport geography \cite{strano2012elementary, barthelemy2011}.

This work represents an important step toward a theory that integrates the explicit dynamics of infrastructure growth with land-use feedbacks. It offers a flexible framework for scenario exploration, policy testing, and long-term forecasting of urban form and mobility infrastructure. These insights are crucial, as discussed in~\ref{subsec:street} and~\ref{subsec:impact}. It would also be interesting to explore simplified versions of this model that allow for analytical insights, in order to quantify the impact of infrastructure changes.

\section{Two species reaction-diffusion models}
\label{chap:7}

Turing’s seminal work on morphogenesis \cite{turing1952chemical} introduced a mathematical framework to explain how biological patterns, such as stripes on animal coats or leaf arrangements in plants, emerge spontaneously from initially uniform conditions. This theory rests on the idea that stationary solutions of reaction-diffusion equations can become unstable under small perturbations, leading to the formation of spatial patterns—a phenomenon now known as the Turing instability (or diffusion-driven instability). 

In its original formulation, Turing considered two interacting chemical species: an activator, which promotes its own production, and an inhibitor, which suppresses the activator. The concentrations of these species, denoted by $u(x, t)$ and $v(x, t)$ respectively, evolve according to a system of partial differential equations
\begin{align}
\begin{cases}
\frac{\partial u}{\partial t} &= D_u \nabla^2 u + f(u,v) \,, \\
\frac{\partial v}{\partial t} &= D_v \nabla^2 v + g(u,v)
\end{cases}
\end{align}
where $f$ and $g$ describe the local reaction kinetics, and $D_u$, $D_v$ are the diffusion coefficients of the activator and inhibitor. Crucially, pattern formation requires that the inhibitor diffuses significantly faster than the activator ($D_v \gg D_u$), which allows local concentrations of the activator to be quickly suppressed over larger regions, thereby generating spatial heterogeneity. Turing analyzed the linear stability of spatially homogeneous steady states $(u^*, v^*)$ under small perturbations, showing that while these steady states may be stable in the absence of diffusion, the addition of differential diffusion can induce an instability. This counterintuitive mechanism—where diffusion, typically a homogenizing process, acts to destabilize uniform solutions—explains how stable, non-uniform spatial patterns of localized peaks of activator (spots/stripes) surrounded by inhibitors can spontaneously emerge from homogeneous initial conditions.

Given that urban systems also exhibit rich spatial patterns and structures--such as dense centers, peripheral sprawls, and corridor-like developments--it is tempting to adapt this framework to study urban growth. In such models, the reaction terms can represent local processes like population growth, economic attraction, or congestion effects, while diffusion terms capture spatial spillovers due to infrastructure expansion, or the diffusion of amenities and innovations. This analogy has led to various reaction-diffusion-inspired models in urban science, aiming to explain phenomena ranging from the emergence of polycentric urban structures to the segregation of different land uses.

However, applying reaction-diffusion equations to urban growth poses significant challenges. Unlike chemical morphogens, human mobility is not purely diffusive and often involves long-distance moves driven by expectations and economic incentives. Moreover, the parameters governing urban `diffusion' and `reaction' are harder to quantify and may vary with local geography, planning policies, and socioeconomic heterogeneities. Despite these complexities, reaction-diffusion models offer a valuable lens to understand how local interactions coupled with spatial processes can produce large-scale urban patterns. They provide a mathematically tractable way to explore how instabilities in uniformly distributed populations or activities can give rise to structured urban forms, complementing approaches based on economic optimization or agent-based simulations.

\subsection{Built versus non-built areas}

Traditional urban planning methods struggle to regulate urban growth, particularly in developing countries, motivating the study of urban self-organization. The article \cite{schweitzer1997urban} explores urban growth using self-organization principles and reaction-diffusion models. The core assumption of the model is a non-linear feedback mechanism between the existing urban aggregation and its further growth. This feedback is mediated by an attraction field generated by the urban settlement itself.

To describe the aggregation process, they introduce two distinct species:
\begin{itemize}
    \item $C_1$, representing already aggregated (immobile) particles, corresponding to the built-up area at time $t$, with a concentration field $c_1(r,t)$ that describes their spatial distribution;
    \item $C_0$, representing growth units (mobile particles) that have not yet aggregated, with their concentration given by $c_0(r,t)$. The transformation of a moving growth unit $C_0$ into a non-moving built-up unit $C_1$ follows the symbolic reaction $C_0 \to C_1$ which occurs at a specific reaction rate.
\end{itemize}

The authors also assume that the existing urban aggregate generates a spatio-temporal attraction field, denoted as $h(r,t)$, which influences the movement and aggregation of growth units. The exact nature of this attraction may arise from various factors such as economic opportunities, political dynamics, cultural appeal, or general urban desirability. Within this model, the authors make the following assumptions about the dynamics of this attraction field: it is continuously produced by the built-up area at a rate $q$; it fades over time if not actively maintained, at a rate $\mu$; and it spreads into surrounding areas through a diffusion process, characterized by the diffusion constant $D_h$. The evolution of the field is then governed by the following equation
\begin{align}
\frac{\partial h(r,t)}{\partial t} 
&= -\mu h + q c_1 + D_h \nabla^2 h \,.
\end{align}

The growth units can precipitate and
transform into a build-up unit -- either by attachment to an existing cluster or by formation of a new
one. The probability $\gamma$ of transformation of the growth units, should depend on the normalized
local attraction and we have
\begin{align}
\frac{\partial c_1(r,t)}{\partial t} = \gamma(r,t) c_0
\label{eq:c1}
\end{align}
with $\gamma(r,t)=h(r,t)/h_{\max}(t)$. 
The build-up units create an attraction field, which affect the movement of moving units, which are later converted into build-up units, further increasing urban aggregation. 

The demand for new build-up areas can however not always be satisfied at a given location due to the local depletion of empty space. Empty space is an important variable of urban growth and we represent its density by $a(r,t)$ (at initial time, it is assumed that free space is uniformly distributed with density $a(r,0)=a_0=A_0/A$ where $A_0$ is given and is the initial free space and $A$ the surface of the area under consideration). Another important quantity is the demand for build-up areas with density denoted by $b(r,t)$. Initially the demand is uniformly distributed with density $\beta$. Not all demand can be satisfied at some location and it has to match available free space. The authors of \cite{schweitzer1997urban} assume that the demand diffuses with constant $D_b$ as long as it meets free space, and leading to the creation of a growth unit $C_0$ symbolized by the reaction 
\begin{align}
A+B\xrightarrow{\alpha}C_0 \,,
\end{align}
where $\alpha$ denotes the free space disappearing rate. This can be written as a action mass equation of the form
\begin{align}
\frac{\partial a(r,t)}{\partial t} &= -\alpha a(r,t) b(r,t)
\end{align}
and for the demand
\begin{align}
\frac{\partial b(r,t)}{\partial t} &= \beta - \alpha a b + D_b \nabla^2 b \,.
\end{align}
The authors then assume a current 
\begin{align}
J=\lambda \frac{\partial h}{\partial r}+D_c\frac{\partial c_0}{\partial r}
\end{align}
which indicates a movement towards regions with a large concentration of grown units and/or a large value of the attractivity field. The dynamics for the concentration $c_0$ is then described by the following equation
\begin{align}
\frac{\partial c_0(r,t)}{\partial t} &= -\nabla J + \alpha a b - \gamma c_0 \,.
\end{align}
In summary, the dynamics of the urban growth in this framework is described by 4 equations for the quantities $h$, $a$, $b$, and $c_0$ \cite{schweitzer1997urban}

\begin{align}
\begin{cases}
\frac{\partial h(r,t)}{\partial t} 
  &= -\mu h + q c_1 + D_h \nabla^2 h \,, \\
\frac{\partial a(r,t)}{\partial t} 
  &= -\alpha a b \,, \\
\frac{\partial b(r,t)}{\partial t} 
  &= \beta - \alpha a b + D_b \nabla^2 b \,, \\
\frac{\partial c_0(r,t)}{\partial t} 
  &= -D_c \frac{\partial^2 c_0}{\partial r^2}
     - \lambda \frac{\partial^2 h}{\partial r^2}
     + \alpha a b - \gamma c_0 \,,
\end{cases}
\end{align}
where $h$ represents an attraction field, $a$ denotes available free space, $b$ represents demand for new settlements, $c_0$ models mobile growth units (and $c_1$ the density of built-up units given by Eq.~\ref{eq:c1}

Obtaining analytical predictions for this system is challenging, and its simulation is equally complex due to numerous uncertainties and difficulties in estimating initial conditions. The authors simulated Berlin's evolution between 1910 and 1920, 
however, they did not provide any quantitative comparison between their simulation and empirical data.  Nevertheless, the reaction-diffusion approach successfully captures key features such as the depletion of free space, shifting growth zones, and attraction-driven clustering.

\subsection{The Gray-Scott reaction-diffusion model}

\subsubsection{The original model}

The Gray-Scott model \cite{gray1984autocatalytic} is a reaction-diffusion system that describes the evolution of two interacting chemical substances, $U$ and $V$, over time due to both reaction and diffusion processes. It has been applied in various fields such as in biology for explaining  pattern formation in animal skins, morphogenesis, and cell differentiation, in chemistry (autocatalytic chemical reactions, such as the Belousov-Zhabotinsky reaction), in physics for the study of nonlinear waves and turbulence, in computer science for generative models of textures and synthetic patterns, and in urban science in \cite{friesen2019reaction}.

This model is governed by the following partial differential equations
\begin{align}
\begin{cases}
 \frac{\partial u}{\partial t} &= D_u \Delta u + R \left[ - u v^2 + F(1 - u) \right] \,,\\
    \frac{\partial v}{\partial t} &= D_v \Delta v + R \left[ u v^2 - (F + k) v \right] \,,
\end{cases}
    \label{eq:GS}
\end{align}
where $u(x,t)$ and $v(x,t)$ are the concentrations of two chemical substances $U$ and $V$ at position $x$ and time $t$. The quantities $D_u$ and $D_v$ are the diffusion constants for substances $U$ and $V$, $R$ is the reaction rate constant, $F$ is the feed rate, representing the external supply of $U$, $k$ is the decay rate of $V$.

Each equation contains different terms:
    \begin{itemize}
        \item{} Both substances diffuse across the spatial domain which is described by the terms of the form $D_u \Delta u$ and $D_v \Delta v$. These terms tend to smooth out concentration differences.

        \item{} A nonlinear term $uv^2$ that governs the local chemical reaction: it represents an  autocatalytic reaction, where substance $V$ catalyzes its own production at the expense of $U$. More precisely, substance $V$ consumes $U$ and catalyzes its own production via the reaction:
        \begin{equation}
            U + 2V \rightarrow 3V.
        \end{equation}
        
        \item The term $(F + k)v$ accounts for the decay of $V$.
        
        \item The term $F(1 - u)$ describes the external feed of $U$.
    \end{itemize}

The Gray-Scott model exhibits rich pattern formation depending on the different parameters of the model \cite{gray1984autocatalytic}. In particular, it can display `spots' that are isolated regions of high $V$ concentration, `stripes' and `labyrinths' (interconnected filaments of high $V$ concentration), `waves' and `spirals' (complex oscillatory patterns), or `localized Structures' (stable, self-replicating patterns). These patterns emerge essentially due to the interplay between the chemical reaction which creates inhomogeneities and diffusion which spreads them out.

Introducing the dimensionless parameters $d=D_u/D_v$, $\gamma=RL^2/D_v$, $t^*=D_vt/L^2$, $x^*=x/L$ where $L$ is the typical length of the system ($t^*$ and $x^*$ are then normalized using characteristic length and diffusion timescales), these equations Eqs.~\ref{eq:GS} can be rewritten as
\begin{align}
\begin{cases}
    \frac{\partial v}{\partial t^*} &= \Delta v + \gamma \left[ uv^2 - (F + k)v \right] \,, \\
    \frac{\partial u}{\partial t^*} &= d\Delta u + \gamma \left[-uv^2 + F(1 - u)\right] \,,
\end{cases}
\end{align}
where it is assumed that the solutions satisfy the usual zero-flux Neumann boundary conditions $\frac{\partial u}{\partial\mathbf{n}} =\frac{\partial v}{\partial\mathbf{n}}=0$.

Three steady states exist for this system,
\begin{align}
\begin{cases}
(u_1,v_1)&=(1,0)\\
(u_2,v_2)&=\frac{1}{2}\left((1-\sqrt{p}),\frac{F}{F+k}(1+\sqrt{p})\right)\\
(u_3,v_3)&=\frac{1}{2}\left((1+\sqrt{p}),\frac{F}{F+k}(1-\sqrt{p})\right)
\end{cases}
\end{align}
where $p=1-4(F_k)^2/F>0$. The state 1 is shown to be always stable, the state 3 is always unstable and the state 2 can be stable for a range of parameters. Thus, this state 3 is the only steady state that can lead to diffusion driven instability (the Turing instability). The necessary and sufficient condition for this Turing instability occurs for specific conditions on the parameters $k$ and $F$ essentially.

\subsubsection{Urban application}

Friesen et al. proposed a reaction-diffusion model for describing urban morphogenesis and applied it to the development of US cities \cite{friesen2019reaction}. Inspired by Turing's pioneering work on pattern formation, they adapt the Gray-Scott model \cite{gray1984autocatalytic}, a well-known reaction-diffusion system originally formulated for autocatalytic chemical reactions, to simulate urban phenomena such as segregation and gentrification. Their approach modifies the classical biological framework to capture processes driving the spatial evolution of urban structures. Using census data from 2000 and 2010 for several US cities experiencing significant economic growth, they calibrate the model parameters and perform stability analyses to assess whether reaction-diffusion dynamics can reproduce observed patterns of urban change. By systematically exploring the parameter space and comparing simulated morphologies to empirical urban patterns, the study illustrates how reaction-diffusion mechanisms—originally developed to explain biological morphogenesis—can be fruitfully extended to investigate urban spatial processes.

The interpretation of the Gray-Scott model in the context of cities is the following. The authors focused essentially on migration, segregation and  gentrification as the main drivers of urban growth. The densities $u$ and $v$ represent the density of poor ($U$) and wealthy ($V$), respectively. The diffusion terms naturally represent migration and the nonlinear term $u v^2$ represents segregation where the wealthier population is slowly `squeezing' the poorer population out of their neighborhood. The wealthy population decreases due to the term $-(F+k)v$ showing that it increases less in areas where there is already a high concentration of rich people ($v\approx 1$).
The poorer population moves out of its neighborhood to  close-by districts which is described by a higher value of diffusion ($d>1$), or can relocate further away which is described by the refill term $F(1-u)$. Unfortunately, no wealth data was available at the time of this study and ethnicity data was chosen (as it
shows the highest correlation to wealth of the accessible data \cite{shapiro2013roots}).

\begin{figure*}[htp]
\centering
\includegraphics[angle=0, width=0.9\textwidth]{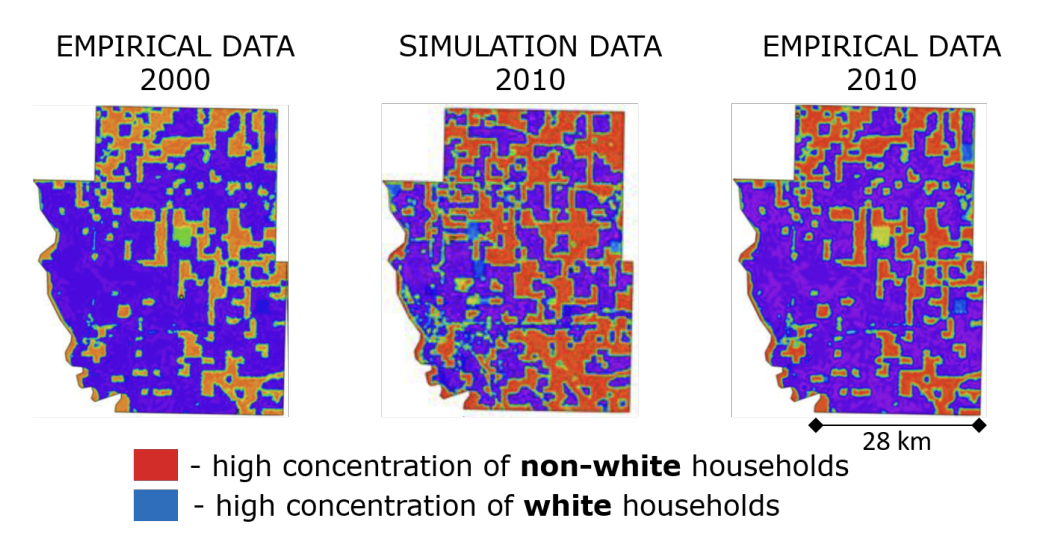}  
\caption{Comparison between empirical census data and simulation results for the city of Bismarck (US). Source: adapted from \cite{friesen2019reaction}.} 
\label{fig:bismark}
\end{figure*}

Simulations were conducted using census block data for Midland/Odessa, TX, Bismarck, ND, and Victoria, TX. The model parameters were varied to optimize the quality function  
\begin{equation}
    Q_F = \sum_{i=1}^{B_{\text{max}}} \big| AC_{2010,i} - AC_{\text{Sim},i} \big|,
\end{equation}
where $AC_{2010,i}$ and $AC_{\text{Sim},i}$ denote activator concentrations from census data and simulations, respectively. Figure~\ref{fig:bismark} illustrates an example comparison between empirical and simulated data for Bismarck.  

The simulations for Midland/Odessa and Bismarck indicated that the best agreement between data and model was obtained for small values of $F$ and $k$, combined with relatively large values of $d$. The diffusion parameters can be interpreted in sociological terms: higher values of $d$ (the ratio $D_u/D_v$) correspond to greater mobility among non-white or lower-income populations, and thus capture processes such as gentrification and segregation. Victoria, TX was used to test the robustness of the calibrated parameters, and the results confirmed that the optimized values provided consistent and reasonable predictions.  

By transferring a framework originally developed for chemical and biological systems to the context of urban morphogenesis, Friesen et al. demonstrated that reaction-diffusion equations can be interpreted sociologically, capturing fundamental structural changes. While a more detailed mapping of urban structures might be achievable using reaction-diffusion equations with locally optimized parameters, it remains uncertain whether the patterns observed in real cities can be reliably reproduced by the Gray-Scott system explored here. Moreover, it is yet to be determined whether effective parameter combinations identified in certain cities can meaningfully generalize to others. This question was preliminarily examined using two example cities, suggesting the need for further studies to assess the extent to which future urban developments might be anticipated by historically calibrated models. This approach does not claim that simple mathematical equations can fully capture the complexity of urbanization. Rather, it seeks to explore whether (i) there exist dominant fundamental processes in urban growth that allow secondary effects to be neglected, and (ii) whether these processes can be meaningfully represented through comparatively simple mathematical formulations.

\subsection{Diffusion of residents and rents}

Zhang \cite{zhang1988pattern} develops a dynamic framework to study how residential density and land rent interact and diffuse across urban space, aiming to explain the formation and evolution of urban patterns. The study emphasizes that urban systems display tendencies toward aggregation or regionalization, shaped both by local interactions between socioeconomic variables and by diffusion processes that redistribute population and economic activity. Unlike static approaches, this model explicitly integrates feedbacks and spatial dynamics.

The urban system is represented as consisting of three components: the central business district (CBD), where the main socioeconomic activities are concentrated; the suburban area, which accommodates residential development and additional activities; and the urban boundary, separating the city from agricultural land. Two state variables describe the system: the residential density $D(x,t)$ and the land rent $R(x,t)$ at distance $x$ from the CBD and time $t$. 

The dynamics combine the interaction structure of Dendrinos and Mullally’s predator–prey model with diffusion processes introduced by Vinod and Ishikawa~\cite{vinod1979,ishikawa1980new}. The coupled equations are
\begin{align}
\begin{cases}
    \frac{\partial D(x, t)}{\partial t} &= \alpha \,(R_0 - R(x, t))\,D(x, t) + A_1 \frac{\partial^2 D(x, t)}{\partial x^2}, \\
    \frac{\partial R(x, t)}{\partial t} &= \beta \,(D(x, t) - D_0)\,R(x, t) + A_2 \frac{\partial^2 R(x, t)}{\partial x^2},
\end{cases}
\end{align}
where $\alpha$ and $\beta$ are interaction parameters, $A_1$ and $A_2$ are diffusion coefficients, and $R_0$, $D_0$ denote equilibrium values. Neumann boundary conditions enforce zero flux at the city limits, so that neither residents nor rents diffuse beyond the urban boundary:
\begin{align}
    \frac{\partial D}{\partial x}\Big|_{x=0,x_g} = 0, \quad \frac{\partial R}{\partial x}\Big|_{x=0,x_g} = 0,
\end{align}
with $x_g$ the distance to the urban edge.

The results of the analysis highlight the crucial role of diffusion. When diffusion dominates, density and rent converge toward homogeneity, erasing spatial heterogeneities. In the absence of diffusion, the system reduces to Dendrinos’ predator–prey interactions, producing neutral cycles in the $(D,R)$ plane. Once spatial diffusion is introduced, however, these periodic trajectories disappear: Zhang proves that the combined system admits no periodic solutions, showing that spatial effects stabilize the dynamics and disrupt the cyclic behavior predicted by purely interaction-based models. Linear stability analysis further reveals that perturbations around the steady state decay over time, ensuring that the homogeneous equilibrium is asymptotically stable.

Overall, the model demonstrates that the integration of spatial diffusion with urban interaction dynamics yields qualitatively new behavior. Whereas earlier approaches captured either predator–prey type oscillations (Dendrinos and Mullally~\cite{dendrinos1985urban}) or pure diffusion smoothing (Vinod~\cite{vinod1979}, Ishikawa~\cite{ishikawa1980new}), Zhang’s framework shows that their combination drives urban systems toward stable homogeneous distributions of residents and rents. This theoretical contribution not only resolves limitations of prior models but also establishes a foundation for analyzing urban growth processes where both socioeconomic interactions and spatial dispersion mechanisms are essential. 

\subsection{The scale of slum sizes}

\begin{figure*}[htp]
\centering
\includegraphics[angle=0, width=0.9\textwidth]{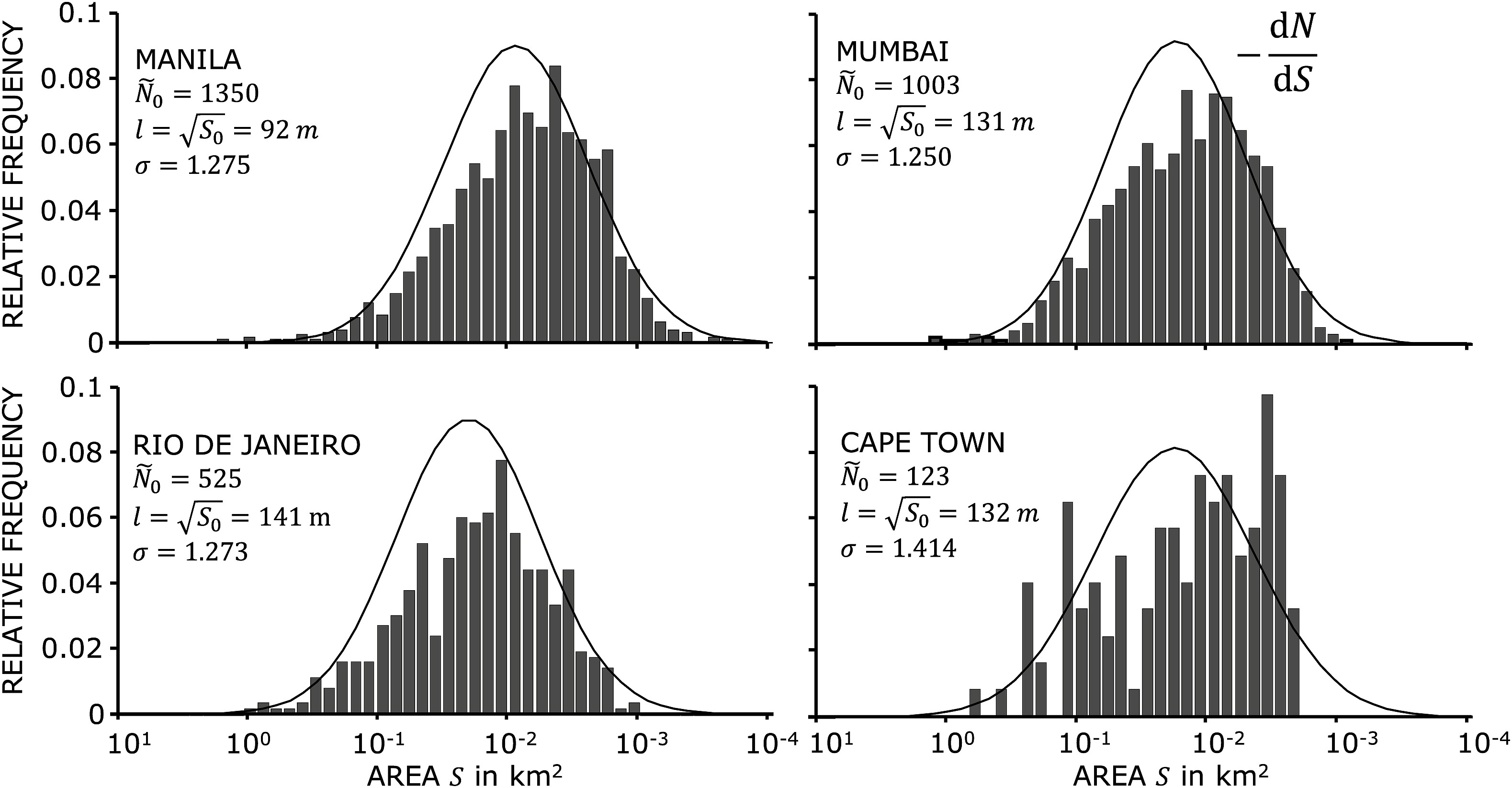}  
\caption{Histograms and log-normal distribution of morphological slums in Manila, Mumbai, Rio de Janeiro, and Cape Town (without townships). Insets : distribution parameters. Source: From \cite{friesen2018similar}.}
\label{fig:slums}
\end{figure*}

In another study based on reaction-diffusion equations, Friesen et al.~\cite{friesen2018similar} investigate the number and sizes of slums in different countries. They observe that both cities with relatively few morphological slums, such as Cape Town ($N_0 = 123$), and those with substantially larger numbers, like Manila (with over 1000 morphological slums), exhibit similar distributions of slum sizes. These findings relate to the cluster size distribution discussed in~\ref{subsec:component}. Specifically, the geometric mean of slum areas is approximately $S_0 \approx 10^{-2}$~km$^2$, with most slums falling within a range of $10^{-3}$ to $10^{-1}$~km$^2$, as illustrated by the histograms in Figure~\ref{fig:slums}. Assuming a square footprint, this corresponds to edge lengths between roughly $30$~m and $300$~m. Remarkably, despite stark differences in historical, cultural, and economic contexts, the average area of morphological slums across these diverse urban systems is found to be $1.6 \times 10^{-2}$~km$^2$, corresponding to an edge length of about $120$~m.

A key question arising from these findings concerns their implications for planning essential infrastructure such as water supply, sanitation, and electricity. The study underscores that beyond the widely recognized large slums, there exists a multitude of smaller slum units that often share similar dimensions. The typical slum size (with $l \approx 126.5$~m) is roughly equivalent to a football field. For these numerous yet relatively small units, decentralized infrastructure solutions may be particularly effective. For example, water provision could rely on smaller filling stations supplied by trucks, while decentralized energy systems such as solar kiosks might support mobile phone charging and lighting. Consequently, urban planning should not focus exclusively on large, well-documented slum areas but must also account for the far more numerous small-scale slum settlements that significantly shape the urban fabric.

Interestingly, Cape Town exhibits a distinct pattern in its distribution of morphological slums. Unlike other cities where a log-normal distribution is observed, Cape Town's slum distribution appears more random (see Fig.~\ref{fig:slums}). This deviation may be attributed to the presence of planned townships for the urban poor, which reduce pressure on informal land occupation. However, when these planned townships are included in the analysis, the expected log-normal distribution emerges, highlighting the substantial influence of slum classification methodologies on analytical outcomes. Additionally, the analysis of morphological slum distributions in Rio de Janeiro, derived from remote sensing, revealed similar distributional characteristics to data from the Brazilian Institute of Geography and Statistics (IBGE), though the potential time lag between these datasets remains an open question for future investigation.

Building on these empirical observations, Pelz, Friesen, and Hartig \cite{pelz2019similar} propose that the characteristic slum size of approximately $16,000$~m$^2$, consistently observed in cities such as Mumbai, Manila, Cape Town, and Rio de Janeiro, arises from fundamental migration dynamics rather than from external socio-economic or political factors. They argue that slum formation can be understood as a consequence of a Turing instability in migration behavior. By modeling migration as a reaction-diffusion system involving two social groups—the rich and the poor (as in \cite{friesen2019reaction})—they suggest that slums emerge as self-organized patterns rather than being solely the result of external constraints. More precisely, they hypothesize that the interaction between these populations follows a Turing mechanism, leading to slum formations of similar sizes across different urban contexts. 

The key elements of their approach include: (i) Treating long-distance migration as a reaction term and short-distance migration as diffusion, (ii) developing a system of reaction-diffusion equations for the density of rich and poor populations, and (iii) performing a stability analysis to determine conditions under which a homogeneous population distribution becomes unstable, leading to pattern formation.

Their model describes the densities of rich ($u_2$) and poor ($u_1$) populations with reaction-diffusion equations:
\begin{align}
\begin{cases}
    \frac{\partial u_1}{\partial t} &= f_1(u_1, u_2) + D_1 \nabla^2 u_1 \,, \\
    \frac{\partial u_2}{\partial t} &= f_2(u_1, u_2) + D_2 \nabla^2 u_2 \,,
\end{cases}
\end{align}
where $D_1$ and $D_2$ are diffusion coefficients and $f_1, f_2$ describe the interaction dynamics between the two populations. The system exhibits a Turing instability when diffusion drives pattern formation, resulting in slum emergence. Linearizing the system around the homogeneous equilibrium $U_i$ 
\begin{align}
    u_{i}=U_i+\delta u_i
\end{align}
and introducing an ansatz for the perturbation of the form
 $\delta u_i = R[\delta\hat{u}_i \exp(\sigma t + i \mathbf{k} \cdot \mathbf{x})]$, results in an eigenvalue problem. Solving the dispersion relation leads to the two following eigenvalues
\begin{equation}
    2\sigma_{1,2} = b_{ii} \pm \sqrt{b_{ii}^2 - 4 \det (b_{ij})},
\end{equation}
where $b_{ij} = a_{ij} - d_{ij} k^2$ and $a_{ij}$ is the Jacobian of $f_i$ ($a_{ij}=\partial f_i/\partial u_j$).
A Turing instability occurs if:
\begin{equation}
    a_{11} d + a_{22} > 2\sqrt{d \det(a_{ij})}.
\end{equation}
This instability leads to pattern formation, with a dominant wavenumber (when $a_{11}+a_{22}/d>0$) given by 
\begin{equation}
    k_{\text{dom}}^2 = \frac{1}{2} \left( a_{11} + \frac{a_{22}}{d} \right).
\end{equation}
This analysis demonstrates that when the mobility of the wealthy (diffusion coefficient $D_2$) is sufficiently larger than that of the poor ($D_1$), the homogeneous distribution of populations becomes unstable. In the author's terms, a stable social system may become Turing unstable when `the generalized attraction of poor' dominates
the `generalized repulsion of rich'. When $d$ exceeds this
threshold, a Turing instability resulting in a Turing pattern
with the dominant wave number $k_{\text{dom}}$ will emerge and leads to the self-organization of slums with a characteristic spatial scale given by
\begin{equation}
    \lambda = \frac{2\pi}{k_{\text{dom}}}.
\end{equation}
However, no concrete estimate have been given for this $\lambda$, but this study suggests that the slum formation is a universal pattern emerging from basic migration dynamics rather than specific city policies or external constraints. 

The interpretation of this work is that the combined effects of short- and long-distance migration are sufficient to produce the self-emergence of slums from an initially homogeneous distribution of rich and poor populations. Crucially, self-organized pattern formation in this model only arises when short-distance migration is explicitly accounted for. According to this framework, slums form only when individuals exhibit a concentration-dependent tendency to relocate away from neighbors of the same social group in their immediate surroundings. An interesting outcome of the model is that the interaction behavior within the `poor' group varies with distance: while poor individuals attract other poor individuals over long distances, they tend to repel each other locally when local concentration becomes high. These findings provide quantitative insights and perspective on city growth mechanisms (see \ref{subsec:shape} and~\ref{subsec:surfacegrowth}).

\subsection{Street network growth}

Tirico et al.~\cite{tirico2021morphogenesis} propose a reaction–diffusion framework to simulate the morphogenesis of street networks, extending Turing’s seminal theory of pattern formation \cite{turing1952chemical}. Street networks, understood as the structural backbone of urban systems \cite{marshall2004streets}, are not only shaped by centralized planning but also by decentralized, self-organizing processes driven by local interactions and feedback mechanisms. The model formalizes these dynamics using three coupled layers embedded in an external environment: (i) a cellular automaton that seeds the spontaneous emergence of morphogens, (ii) a reaction–diffusion (RD) layer generating spatial patterns, and (iii) a dynamic vector field that translates these patterns into network growth. Crucially, the resulting network modifies the RD layer, creating feedback between network formation and morphogen organization.

At the microscopic scale, the dynamics are governed by two interacting morphogens: an activator $A$ and an inhibitor $B$. The activator promotes its own production as well as that of the inhibitor, while the inhibitor suppresses the activator and diffuses faster. Their interactions are captured by a Gray–Scott-type RD system \cite{gray1984autocatalytic}
\begin{align}
\begin{cases}
    \frac{\partial A}{\partial t} &= D_A \nabla^2 A + \rho A^2 B - \alpha A \,, \\
    \frac{\partial B}{\partial t} &= D_B \nabla^2 B + \beta - \rho A^2 B - \gamma B \,,
\end{cases}
\end{align}
where $D_B > D_A$ (meaning that $B$ diffuses faster), $\rho$ is the reaction rate, $\alpha$ and $\gamma$ are decay parameters, and $\beta$ is the feed rate. These equations generate a variety of heterogeneous spatial structures—spots, stripes, labyrinths, or soliton-like patterns \cite{pearson1993complex}—which act as morphogenetic fields guiding street growth.

Streets are iteratively generated when the concentration of the activator $A$ exceeds a threshold $A_c$. New edges grow along the direction of a vector field
\begin{equation}
    \mathbf{V} = \mathbf{V}_{RD} + \mathbf{V}_{env},
\end{equation}
where $\mathbf{V}_{RD}$ derives from RD gradients and $\mathbf{V}_{env}$ incorporates exogenous constraints such as topography, existing land use, or political restrictions. The emergent street network feeds back into the RD process by locally altering morphogen parameters, thus reshaping subsequent patterns and capturing co-evolution between networks and their underlying drivers. 

To summarize, the model proposed by Tirico et al.~\cite{tirico2021morphogenesis} consists of three coupled layers embedded in an external environment:  
\begin{enumerate}
    \item A \textbf{cellular automaton layer}, which governs the spontaneous emergence of morphogen concentrations.  
    \item A \textbf{reaction–diffusion layer}, based on the Gray–Scott model \cite{gray1984autocatalytic}, where two morphogens (an activator and an inhibitor) interact and diffuse to generate spatial patterns.  
    \item A \textbf{dynamic vector field layer}, which translates the morphogenetic patterns into directional influences that guide the growth of the street network.  
\end{enumerate}

This layered setup captures how morphogens—abstract representations of drivers such as population, economic activity, or policy decisions—interact, diffuse, and feed back into the evolving street network, thus formalizing urban morphogenesis as a self-organizing process shaped by both endogenous dynamics and environmental constraints.

\begin{figure*}
    \centering
    \includegraphics[width=0.99\textwidth]{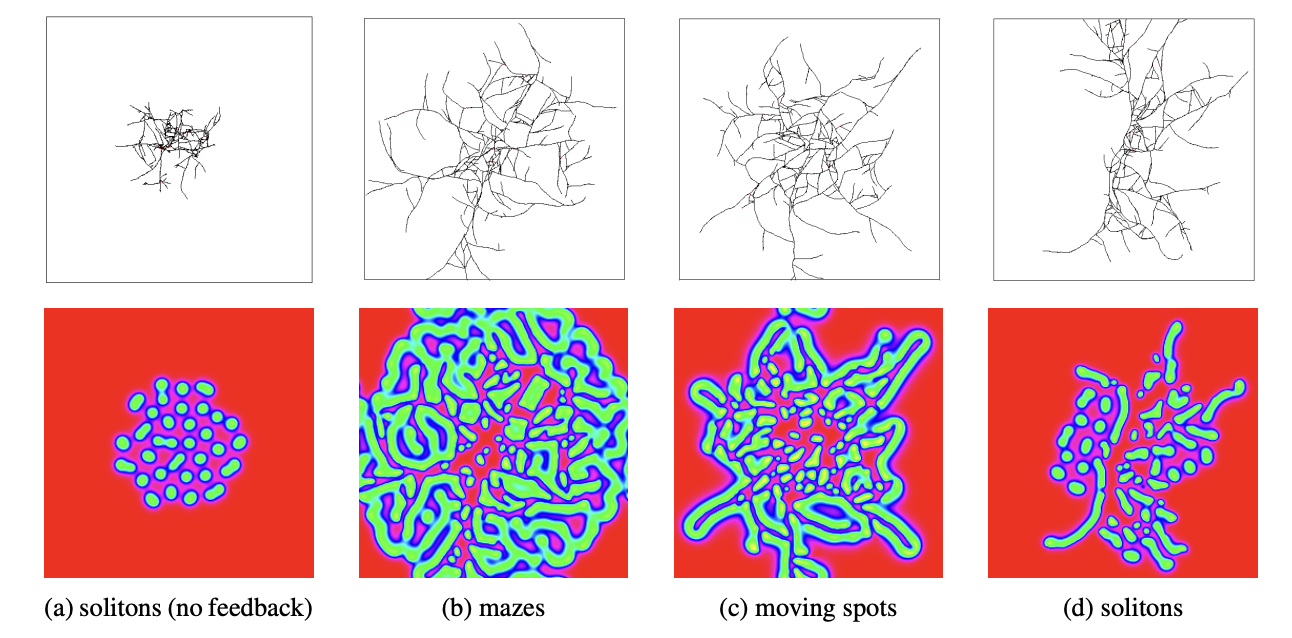}
    \caption{Examples of street network morphogenesis driven by reaction–diffusion dynamics. Different RD parameterizations generate varied morphogen fields (spots, mazes, solitons) which, when translated into vector fields, guide the formation of distinct network structures. Source: From  \cite{tirico2021morphogenesis}.}
    \label{fig:tirico}
\end{figure*}

As illustrated in Fig.~\ref{fig:tirico}, different RD parameterizations generate distinct morphogen fields (spots, mazes, solitons) which, when coupled to the vector-field layer, lead to the formation of networks with varied structures.

The model was applied to the case of Fécamp, a town in Normandy, France. Here, the environment was represented by orographic maps and land-use constraints: network growth was penalized in green areas or densely built-up zones to mimic planning restrictions. Simulations across several RD parameter sets produced networks that preserved the backbone of Fécamp’s original structure while generating organic extensions rich in bifurcations. Statistical analysis revealed an increase in degree-3 nodes, typical of urban intersections, and a more hierarchical distribution of betweenness centrality, both in agreement with real-world street networks.  

This work demonstrates that street networks can emerge as self-organized morphogenetic structures from the interplay of activator–inhibitor dynamics, environmental constraints, and feedback effects. By grounding street-network growth in a reaction–diffusion framework, Tirico et al. challenge purely top-down planning views and highlight morphogenesis as a driver of urban form. Future research may extend this approach to other cities, incorporate explicit demographic and economic variables as morphogens, and explore its potential for guiding viable planning.

\section{Discussion and perspectives}

In this article, we reviewed a wide range of quantitative approaches to describe the spatial evolution of urban areas, often referred to as urban expansion (and sometimes urban sprawl). The diversity of these models—spanning geography, economics, statistical physics, and ecology—reflects the inherent complexity of urban growth. While some models are motivated by the need to replicate observed expansion patterns, others aim at explaining empirical data or uncovering mechanisms behind stylized facts.

Frameworks range from cellular automata and agent-based simulations, widely used in geography, to equilibrium-based economic models rooted in the Alonso–Muth–Mills tradition, and further to statistical physics analogies such as diffusion-limited aggregation, Eden growth, or percolation. Partial differential equation (PDE)-based approaches integrate diffusion and additional mechanisms such as congestion, migration, and the co-evolution of density with transportation networks. Reaction–diffusion formulations in particular have been fruitful in linking urban morphogenesis to broader classes of pattern-forming systems.

A recurring theme is the importance of empirical regularities and stylized facts. Observed patterns include the exponential decay of population density from the urban core, the approximate Zipf-like scaling of city sizes, the fractal geometry of built-up areas, the scaling of radial density profiles, and the self-affine roughness of urban boundaries. These benchmarks—together with those that will emerge from the increasing availability of high-resolution data—are essential: models that fail to reproduce them cannot be regarded as realistic descriptions of urban evolution. The analogy with surface growth in physics is particularly promising, since urban expansion displays radial anisotropy, roughness exponents, and coalescence dynamics reminiscent of universality classes in interface growth. Whether urban expansion falls into an established universality class, or instead defines a new one, remains an open question.

Despite significant progress, limitations persist. Many approaches are tailored to specific cases and lack generality; others capture heterogeneity but fail to connect with data at large scales. Moreover, models often isolate mechanisms—diffusion, clustering, transportation, or socio-economic drivers—whereas real cities evolve from the interplay of all these factors. Bridging empirical observations and theoretical models thus requires integrated frameworks capable of combining multiple processes while remaining analytically tractable.

Future research should advance along several complementary directions. A first priority is systematic empirical validation against high-resolution datasets such as GHSL, GUF, or WSF Evolution, which are essential for testing models consistently across different cities and time periods. Another promising avenue lies in the development of hybrid frameworks that combine the strengths of statistical physics, urban economics, and transportation science, with the aim of producing more robust and predictive models. Finally, closer attention must be paid to vertical growth and volumetric scaling, since cities are increasingly expanding upward, challenging the traditional two-dimensional paradigms that have long dominated urban modeling.

In summary, mathematical modeling of urban expansion is still fragmented, but converging toward a synthesis where empirical stylized facts and theoretical principles mutually reinforce one another. A promising avenue is to explicitly connect urban growth with the theory of surface growth and universality classes, leveraging the analogy with physical systems to uncover invariant mechanisms of city expansion. Such advances hold the potential not only to improve our fundamental understanding of cities as complex systems but also to inform urban planning strategies in an era of rapid global urbanization.

\vskip0.5cm

{\centerline{\bf Acknowledgments}}
MB gratefully acknowledges many stimulating discussions over the past years with many colleagues in many different disciplines, and in particular with M. Batty, and H. Berestycki, whose discussions on this topic have been especially valuable. UM acknowledges Elsa Arcaute and CASA for their hospitality.

\clearpage
\newpage


\bibliographystyle{abbrv}	                 

\bibliography{bibfile_physrep_revised}		         

\end{document}